\documentclass[a4paper,book,10pt,twoside]{combine}

\usepackage{a4wide, amssymb, epsfig, graphicx, fancyhdr}

\usepackage{longtable} 
\usepackage{amsmath,amsfonts} 
\usepackage{psfrag,booktabs} 
\usepackage{wrapfig} 
\usepackage{multirow} 
\usepackage{url}
\usepackage{tocloft}

\newcommand{\Section}[1]{\section{#1}\vspace{-8pt}}
\newcommand{\Subsection}[1]{\vspace{-4pt}\subsection{#1}\vspace{-8pt}}

\begin{document}
\pagestyle{fancy}
\fancyhead[LO,RE]{D.~Grzonka, S.~Schadmand, B.~Hoistadt and C.~F.~Redmer}
\fancyhead[LE,RO]{PrimeNet Workshop 2011,  J\"{u}lich}
\fancyfoot{}
\fancyfoot[LE,RO]{\thepage}
\newcommand{\authortochead}[1]{%
\coltocauthor{#1}
\fancyhead[LO,RE]{#1}
}
\setlength{\toctitleindent}{1.5em}
\setlength{\tocauthorindent}{3em}

\pagestyle{empty}
\begin{center}
\vspace*{5cm}
\vspace {0.5cm} {\bf \Huge International PrimeNet Workshop}\\

\vspace{2.6cm} {\Large September, 26 -- 28, 2011, J\"{u}lich, Germany\\}

\vspace{5cm} {\huge Summary of Contributions}
\end{center}

\newpage
\mbox{  }\\
\newpage
\tableofcontents 
\clearpage

\pagestyle{fancy}
\setcounter{page}{1}
\section*{Introduction}\label{intro}
\addcontentsline{toc}{part}{\sffamily Introduction}

\vspace{0.75cm}
\noindent This workshop is part of the activities in the project “Study of Strongly Interacting Matter”
(acronym HadronPhysics2), which is an integrating activity of the Seventh Framework Program of EU. This HP2
project contains several activities, one of them being the network PrimeNet having the focus on Meson Physics
in Low-Energy QCD.  This network is created to exchange information on experimental and theoretical ongoing
activities on mainly $\eta$ and $\eta^\prime$ physics at different European accelerator facilities and
institutes. 

\vspace{0.75cm}

\noindent The present workshop included the three general topics:
\begin{enumerate}
 \item $\eta, \eta^\prime$ and $\phi$ decays from experimental and theoretical perspectives.
 \item Meson production in photon-induced reactions and from NN collisions as well as e$^+$e$^-$ production
from pp collisions.
 \item Interaction of $\eta$ and $\eta^\prime$ with nucleons and nuclei.
\end{enumerate}

\noindent The talks covered the very recent achievements in the respective fields from the experimental
facilities KLOE at DAPHNE, Crystal Ball at MAMI, Crystal Ball and TAPS at ELSA, and WASA-at-COSY, as well as
from different theory institutes. Electronic versions of the talks can be found on the PrimeNet homepage  
\url{http://www.fz-juelich.de/ikp/primenet}\\

\noindent The detailed program was arranged by a program committee led by Dieter Grzonka and Susan Schadmand. 

\vspace{0.75cm}

\noindent The workshop was held  Sept 26-28, 2011, at the Forschungszentrum J\"ulich, enjoying kind
hospitality and support from the Forschungszentrum.

\vspace{0.75cm}

\noindent Financial support is gratefully acknowledged from the European Commission under the 7th Framework
Programme through the 'Research Infrastructures' action of the 'Capacities' Programme;  Call:
FP7-INFRASTRUCTURES-2008-1, Grant Agreement N. 227431.

\vspace{1.75cm}

{\large Dieter~Grzonka, Susan~Schadmand, Bo~H\"oistad and Christoph~Redmer}
\cleardoublepage

\pagestyle{plain}
\begin{center}
 \vspace*{0.3\textheight} {\Huge\textbf{\boldmath $\eta, \eta^\prime$ and $\phi$ Decays}}\\
 \vspace{2cm} {\Large\textbf{from Experimental and Theoretical Perspectives}}
\addcontentsline{toc}{part}{\sffamily \boldmath $\eta, \eta^\prime$ and $\phi$
Decays from Experimental and Theoretical Perspectives}
\end{center}
\cleardoublepage
\pagestyle{fancy}

\begin{papers}
\coltoctitle{\boldmath Introduction: Rare $\eta$ Decays and Tests of Fundamental Symmetries}
\authortochead{A.~Wirzba}
\label{WA}

\setcounter{chapter}{1}
\setcounter{section}{0}
\maketitle

\vspace{2ex}\Section{Introduction}
The organizers asked me to give an introductory talk for this meeting. Because of time  limitations and also
my ignorance about other subjects, I interpreted this task in a somewhat restricted sense by only
concentrating on the physics of rare $\eta$ decays. This is of course  only a small subset of the various
topics covered in this meeting. Many other aspects would have deserved dedicated introductions  as well. Let
me also remark that I heavily relied on Ref.~\cite{Kupsc:2011ah} in preparing this talk, a joint paper with
Andrzej Kupsc about tests of fundamental symmetries in rare $\eta$ decays. 

After these disclaimers, I would like to start with two questions: Why is the $\eta$ meson interesting in
general and why are the  rare $\eta$ decays especially interesting? Well, the $\eta$ is special, because of
essentially three  reasons: it is a Pseudo-Goldstone boson of the strong interaction part of the Standard
Model (SM), {\it i.e.} its mass and all its interactions vanish in the chiral limit, the limit of vanishing
quark masses and external momenta. It is a neutral meson with respect to all its internal quantum numbers.
Finally, it is the heaviest of the octet of pseudoscalar mesons and therefore many decay channels are
energetically accessible. The last two arguments do hold of course for its heavier ``nonet-brother", the
$\eta'$ meson  as well which, however, is not a Pseudo-Goldstone boson, because  the  $U_A(1)$ symmetry is
{\em explicitly} broken by the non-perturbative instanton physics.

Although both mesons possess many open decay channels, all the decays --- with the exception of  the allowed
decay $\eta' \to \eta \pi\pi$  --- are suppressed in one way or the other:  by the conservation of Noether
currents (in the chiral limit) which  in turn may be broken by (quantum) anomalies, by the conservation of 
parity, $C$-parity, $CP$, $G$-parity etc. Since the $\eta$ and $\eta'$ are eigenstates of $G$, $P$, $C$ and
$CP$, namely  $I^G (J^{PC}) = 0^+(0^{-+})$, both mesons can serve as ``test laboratories'' for the
conservation or rather the breaking of these discrete symmetries. Moreover, since there is mixing, namely the
singlet-octet mixing of the $\eta$ and $\eta'$ mesons because of the $U_A(1)$ breaking, and --- to a lesser
degree -- the mixing with the $\pi^0$ meson because of isospin breaking, there is also some access to these 
mixing and breaking phenomena as well.

In responding to the second question one should keep in mind that the pertinent SM-mechanisms of rare decays
have to be small to begin with. Moreover, symmetry-violating decays are rare by nature, as otherwise we would
not speak about a symmetry. Thus, by studying rare $\eta$ decays, we  have the chance to encounter a window of
opportunity for experimental searches for physics beyond the SM. Since many different final states are
energetically accessible, a variety of different symmetry tests is possible. These tests include searches for
rare of forbidden decays, for asymmetries among the decay products in not-so-rare decays, and for light
dark-matter particles if there is a discrepancy between SM expectations and experimental results. Since the
$\eta$ and $\eta'$ mesons belong to the class of flavor-neutral (light) mesons,  especially flavor-conserving
symmetry violating mechanisms are tested -- complementary to the flavor-changing $CP$ tests with kaons or
$B$-mesons.

\vspace{2ex}\Section{\boldmath $CP$ tests}
Most of the $CP$ tests in $\eta$ decays  are modeled in analogy to $CP$-violating decays for the $K_L$ meson.
Whereas the latter are flavor-changing, the former decays test  -- as mentioned above -- a {\em
flavor-conserving} $CP$ violating mechanism.   

\Subsection{\boldmath$\eta \to \pi\pi$} \label{sec: eta_pi_pi}
As reviewed in Ref.\,\cite{Jarlskog}, theory predicts very tiny branching ratios for the  $\eta\to \pi\pi$
decays. Since the flavor is conserved, these $CP$-violating decays  have to be second order weak decays in the
SM (governed by the CKM matrix), {\it i.e.} they are proportional to  $G_F^2 \sin^2\theta_C$. In addition,
there are dynamical suppressions such that the upper bound on the  branching ratio is given by \mbox{${\cal
B}({\rm CKM}_{\rm SM}) \leq 2 \times 10^{-27}$}. The contribution of the the $CP$-violating  QCD $\theta$
term to this decay is indirectly constrained by the empirical upper bound on the electric dipole moment $d_n$
of the neutron which in turn limits the QCD vacuum angle, such that \mbox{${\cal B}(\theta_{QCD}) \leq 3
\times 10^{-17}$}, whereas the `extra Higgs' models allow a slightly larger upper bound, \mbox{${\cal B}({\rm
extra\, Higgs}) \leq 1.2 \times 10^{-15}$}. According to Gorchtein\,\cite{Gorchtein}, the empirical bound on
$d_n$ can be used directly to determine an upper bound for the  branching ratio of $\eta\to\pi\pi$, no matter
what the underlying $CP$-violating mechanism will turn out to  be:  for a larger value of the branching ratio
than $3.5 \times 10^{-14}$  the $CP$-violating part of the $\eta$-loop contributions alone would already
violate the bound on $d_n$. 

The empirical numbers for the upper bounds are much larger: ${\cal B}(\eta \to \pi^+\pi^-) \leq 1.3 \times
10^{-5}$ from the KLOE collaboration \cite{KLOE_2005} and ${\cal B}(\eta \to \pi^0\pi^0) \leq 3.5 \times
10^{-4}$ from GAMS-4$\pi$ \cite{GAMS}.

\Subsection{\boldmath $\eta$ decay asymmetries}
The decay $\eta\to \pi^+\pi^-\gamma$ is governed by the $CP$-allowed $M_1$ amplitude which can be traced back
to the chiral triangle and box anomalies. Still, this decay can be used to test the $CP$ violation in $\eta$
decays, if an interference with any $CP$-violating (direct) $E_1$ term can be made visible. In principle,
there might exist the interference with the Bremsradiation-modified decay $\eta\to \pi^+\pi^-$ as well, but as
shown in Sect.\,\ref{sec: eta_pi_pi} this is very unlikely because of the tiny branching ratio of the
$\eta\to\pi\pi$ decay. Geng, Ng and Wu\ \cite{Geng} suggested to measure the (linear) polarization state of
the outgoing photon, since differential cross section measurements are insensitive to the interference
patterns. In the same year, Gao \cite{Gao} proposed to concentrate -- in analogy to the $CP$ experiment in the
$K_L$ system -- on the  $\eta \to \pi^+ \pi^- e^+ e^-$ decay and to measure rather the asymmetry between the
pion and lepton decay planes which is sensitive to the inference term between the $M_1$ and $E_1$ decay
amplitudes. Gao estimated the theoretical bound on the asymmetry as  ${\sf A}_{\rm theo} \leq 0.02$ by
invoking an unconventional flavor-conserving $CP$-violating electric dipole mechanism (originally suggested in
 Ref.\,\cite{Geng}), which is especially sensitive to the $s \bar s$ quark content of the $\eta$ and therefore
not constrained but other $CP$ tests. The KLOE collaboration\,\cite{KLOE_2009} measured ${\sf A}_{\rm emp} =
(-0.6 \pm 2.5 \pm 1.8) \times 10^{-2}$ as empirical value which is both compatible with the theoretical bound
and with zero of course.

\Subsection{\boldmath New types of $CP$ tests without kaon-analog case}
Whereas the upper cases were modeled according to $K_L$ decays, Nefkens\,\cite{Nefkens} suggested a $CP$ test
involving the $\eta$ meson which has no analog in the kaon or $B$-meson sectors, namely the decay $\eta\to 4
\pi^0$  which is characterized by a very low excess energy of 7.9\,MeV. It is forbidden if all pion pairs are
in relative $S$ or $P$-wave states. In the SM, it  can only proceed by $D$-wave or higher interactions which
are -- because of phase space limitations -- extremely suppressed. For more information on this and other
four-pion decays of the $\eta$ and $\eta'$, please consult the talk of  Bastian Kubis\,\cite{Kubis} in this
conference. Since the decay is characterized by a unique 8 photon signal and as there is a very low background
from direct pion production, this decay  in fact is the most sensitive of any of the measured $\eta$ decays:
${\cal B}(\eta\to 4\pi^0) < 6.9 \cdot 10^{-7}$ \cite{CBball}. The experimental limit on the branching ratio of
the $\eta'\to 4 \pi^0$ decay, however, is larger: $5 \cdot 10^{-4}$\,\cite{Alde}.

\Section{\boldmath Tests of the $C$ symmetry in rare $\eta$ decays}
Any decay of the $\eta$ (or $\eta'$) meson into neutrals including an {\em odd} number of photons in the final
state would be a signal for $C$ violation. The simplest example, $\eta \to 3 \gamma$ is heavily suppressed as
each 2-photon-pair has to be at least in a relative $J=2$ state. The $J=0$ case would correspond to a
radiative $0\to 0$ transition which is forbidden  because of angular momentum conservation\,\cite{Sakurai},
and the $J=1$ case would violate the Bose symmetry. The experimental upper limit for the branching ratio is
$1.6\cdot 10^{-5}$\,\cite{Aloiso}. The next simplest case, $\eta\to \pi^0\gamma$ is again an example for the
forbidden radiative $0\to 0$ transition. The empirical upper limit on its branching ratio is  $9\cdot 10^{-5}$
from the Crystal Ball experiment\,\cite{Nefkens:2005a}. The Crystal Ball collaboration\,\cite{Nefkens:2005b}
gives also the best limits on the branching ratios ${\cal B}(\eta\to \pi^0\pi^0\gamma) <5\cdot 10^{-4}$ and
${\cal B}(\eta \to 3\pi^0\gamma)<6\cdot 10^{-5}$, respectively\,\cite{Nefkens:2005b}.  

\Subsection{\boldmath The decay $\eta \to \pi^0 e^+ e^-$ as $C$ symmetry test}
The branching ratio for the  $C$-allowed process  $\eta \to \pi^0 \gamma^* \gamma^* \to \pi^0 e^+ e^-$, which
involves  two virtual photons, was estimated to be of the order  $10^{-8}$\,\cite{Cheng}. Therefore any
measurement with a branching ratio ${\cal B}(\eta\to \pi^0 e^+ e^-)$ larger than this value would be a clear
signal for  $C$ violation. Note, however, that the decay mechanism with one off-shell photon is not only
suppressed by $C$, but also by a vanishing transition form factor in the on-shell limit, since the ``parent''
decay $\eta\to \pi^0 \gamma$ is -- as mentioned above -- a radiative $0\to 0$ transition. The  current upper
limit of the branching ratio ${\cal B}(\eta \to \pi^0 e^+ e^-)$ is equal to $4.5\cdot 10^{-5}$ 90\,\% CL, from
an experiment performed in the 70s of the last century\,\cite{Jane}. For further details, especially about the
measurement of the WASA-at-COSY collaboration, please consult the presentation of Florian
Bergmann\,\cite{Bergmann}. 

The $C$ invariance can also be tested in, {\it e.g.},  $\eta \to \pi^+ \pi^- \gamma$ and in $\eta\to \pi^+
\pi^- \pi^0$ decays by $\pi^\pm$  asymmetry measurements in the $\eta$ rest frame. Most of the measurements
date again back to the 70s of the last century. For the asymmetries in the $\eta\to \pi^+ \pi^-\pi^0$ decay
there exist recent upper bounds from the KLOE collaboration with $10^{-3}$ sensitivity\,\cite{Antonelli}.  

\Section{Dark particle searches}
Note that rare dilepton decays of  $\eta$  mesons and pseudoscalar mesons $P=\pi^0,\eta,\eta'$ in general can
be used as candidates for  light dark-matter particle searches. Since the tree-level decay via a virtual $Z$
boson is minuscule, the leading order decay process is $P \to \gamma^* \gamma^* \to l^+ l^-$. The latter is
suppressed not only by two powers of the fine-structure constant $\alpha$, but also by the helicity mismatch
of the outgoing leptons, {\it i.e.} ${\cal B}(P \!\to\! l^+ l^-)/ {\cal B} (P\!\to\! \gamma\gamma) \sim
(\frac{\alpha}{\pi}\frac{m_l}{m_P})^2$ where $m_P$ and $m_l$ are the masses of the decaying meson and the
leptons $l^\pm=e^\pm$ (or also $\mu^\pm$ in the $\eta,\eta'$ cases), respectively. For the $\pi^0\to e^+ e^-$
decay, the branching ratio is  of order ${\cal O}(10^{-7})$. In fact, the best  experimental\,\cite{KTeV}  and
theoretical\,\cite{Dorokhov} values of ${\cal B}(\pi^0 \to e^+ e^-)$, namely $(7.48\pm 0.29_{\rm stat} \pm
0.25_{\rm syst})\cdot 10^{-8}$ and $(6.23\pm 0.09) \cdot 10^{-8}$, respectively, differ by three standard
deviations.

Kahn and collaborators\,\cite{Kahn} suggested as a possible explanation of this excess the tree-level 
exchange of an off-shell neutral vector boson $U$ (pioneered by Fayet\,\cite{Fayet}) of  mass $m_U$\
$\sim$\,(10 -- 100)\,MeV. Acccording to the light  dark matter  models of Refs.\,\cite{BoehmFayet,Boehm}, this
$U$ boson mediates the annihilation reaction $\chi \chi \to e^+ e^-$ of a neutral scalar dark matter  particle
$\chi$ of (1 --  10)\,MeV mass, such that the produced positrons could account for the bright 511\,keV line
emanating from the center of the Galaxy. Kahn et al.\ could resolve the mismatch between experiment and theory
of the $\pi^0\to e^+e^-$ decay by assuming that the neutral vector meson $U$ would couple  to the $u$ and $d$ 
quark fields of the  $\pi^0$ and the $e^+e^-$ dilepton pair with a common axial-vector coupling $g_A\equiv 
g_A^u-g_A^d\equiv g_A^e$ of strength $g_A  = (2.0  \pm 0.5)  \cdot 10^{-4}  \cdot m_U/(10\,{\rm  MeV})$. If
the octet axial-vector quark-coupling is assumed to be of the same order as above, the $U$ boson contribution 
to the branching ratio BR$(\eta\to e^+e^-)$ is estimated as $\sim 10^{-9}$, which is of the same order as the
predictions of Dorokhov and collaborators\,\cite{Dorokhov} and much smaller  than the experimental  bound $2.7
\cdot 10^{-5}$ of the CELSIUS/WASA  collaboration\,\cite{Celsius}. However, the branching ratio $\cal
B(\eta\to \mu^+\mu^-)$ is then predicted to be nearly an order of magnitude larger than the measured  value
${\cal B}(\eta\to\mu^+\mu^-) \sim (5.7 \pm 0.9)\cdot 10^{-6}$\,\cite{Abegg}, unless the axial-vector coupling
of
the $U$ meson to the muon is smaller than $g_A^e$ or the octet axial-vector quark coupling is smaller than 
$g_A^u-g_A^d$ or both. This limits of course the predictive power of the above presented $U$ boson exchange
mechanism.

Another variant of dark matter $U$ gauge bosons, which acquires a mass scale of 1 MeV to a few GeV via a `dark
Higgs mechanism' and which couples vectorially to all SM charged fields with a strength $\epsilon \sim
10^{-3}$ or less, was suggested by Reece and Wang\,\cite{Reece} (see also \cite{Fayetalt}). This type of
$U$-boson would also couple to SM weak neutral currents; however, the corresponding coupling constant would be
suppressed by a factor of order $m_U^2/m_Z^2$. Therefore, this variant cannot explain the missing excess of
the $\pi^0\to e^+e^-$ decay. First searches for $U$ bosons of the Reece and Wang type are on the way, carried
out by the KLOE collaboration. The relevant channels are, {\it e.g.}, the decays $\phi \to \eta U,\  U \to 
e^+ e^-$  and $\eta  \to  \gamma U,\  U \to  e^+ e^-$. A signature for  a $U$  boson would  be a peak  in the 
pertinent Dalitz decay $\phi \to  \eta e^+ e^-$ or $\eta \to  \gamma e^+e^-$, while the background  for  such
searches  is  the  standard  Dalitz decay.  

Finally note that there may be attractive alternatives for $U$ boson searches which are not affected by
resolving a narrow peak structure on a large conversion background: namely the $C$-forbidden $\eta$ decays,
{\it e.g.} the above mentioned  decay $\eta\to \pi^0 e^+ e^-$ would be an ideal candidate from the background
point of view.

\coltoctitle{\boldmath Search for Exotic Effects in the $\eta\rightarrow e^+e^-$ Decay}
\authortochead{M.~Ber{\l}owski}
\label{BeM}

\setcounter{chapter}{1}
\setcounter{section}{0}
\maketitle

\Section{Introduction}
The goal the preliminary analysis is presented here the search for the very rare $\eta$ meson decay
$\eta\rightarrow e^+e^-$. From the experimental point of view this decay is very hard to observe due to the
highly suppressed branching ratio (BR). In the Standard Model (SM) this fourth-order electromagnetic
transition, predicted (see Ref.~\cite{AD09}) at the level of $\sim10^{-9}$, is described by the leading order
Feynman amplitude in Fig.~\ref{fig:ll}. The branching ratio is thus suppressed with respect to the
$\eta\rightarrow\gamma\gamma$ decay by:
\begin{center}
$BR_{theo}[\eta\rightarrow e^+e^-] \sim BR[\eta\rightarrow\gamma\gamma]\cdot\alpha^2\cdot(m_e/m_\eta)^2$,
\end{center}
where $\alpha$ is the fine structure constant and the $(m_e/m_\eta)^2$ dependance stems from helicity
conservation~\cite{QC08}. Currently the best limit $BR_{exp}<2.7\cdot10^{-5}~at~CL=90\%$ comes from the
CELSIUS/WASA experiment~\cite{MB08} and it is five orders of magnitude bigger then the value expected from SM
calculations.
\begin{figure}[htb]
 \centerline{\epsfig{file=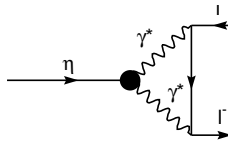,totalheight=2cm,clip=}}
 \caption{\label{fig:ll}The conventional mechanism for neutral meson decay into an lepton-antilepton pair.}
\end{figure}

\Section {\boldmath The $\eta\rightarrow e^+e^-$ decay as a probe of new physics}
The $\eta$ meson decay into an $e^+e^-$ pair became one of the most interesting topics in low-energy hadron
physics, since there is a possible admixture of non-SM processes that can enhance the BR of this decay. The
most recent and very precise experimental value of $BR(\pi^\circ\rightarrow
e^+e^-)=(7.49\pm0.29\pm0.25)\cdot10^{-8}$ determined by the KTeV collaboration exceeds the latest theoretical
calculations from Dorokhov and collaborators~\cite{AD07} by 3 standard deviations. The suggestion from
Kahn~{\it et al.}~\cite{YK07} is that the possible explanation can be the exchange of an off-shell neutral
boson $U$ of mass $m_U\sim(10-100)MeV$. The $U$ boson was previously proposed by Boehm~{\it et
al.}~\cite{CB03} in the light dark matter models to mediate the annihilation of a neutral scalar dark matter
particle $\chi$ of (1-10)~MeV mass in the reaction $\chi\chi\rightarrow e^+e^-$. This models were used as an
explanation for a 511~keV line emanating from the center of the Galaxy due to excess positrons produced in
this annihilation process~\cite{YK07}. An observation of a signal above the level predicted in the SM
calculations could be evidence for an unconventional process which enhances this decay rate.

\Section{Data analysis}
The experiment was performed using the WASA detector~\cite{wasa} at the COoler SYnchrotron COSY, located in
the Forschungszentrum J\"ulich, Germany. The $\eta$ mesons were produced in the reaction $pp\rightarrow
pp\eta$ at 1.4 GeV proton beam energy. We used a special trigger in which a high energy deposit in both sides
of the electromagnetic calorimeter was demanded. This type of trigger should equally favor electromagnetic
decays of $\eta$ mesons like $\eta\rightarrow\gamma\gamma$, $\eta\rightarrow e^+e^-\gamma$, $\eta\rightarrow
e^+e^-$. Using the $\eta\rightarrow\gamma\gamma$ decay channel we estimate the number of $\eta$ mesons
collected in the 2 week data sample to be $\sim4.4\cdot10^7$. This number is approximately forty times bigger
than the data sample used in the previous analysis~\cite{MB08}. It was found that the main sources of
background are $pp\rightarrow pp\pi^+\pi^-$ due to a 100 time larger cross section, $pp\rightarrow
p(\Delta(1232)\rightarrow p[\gamma^*\!\rightarrow e^+e^-])$ having the same particles in the final state as
$\eta\rightarrow e^+e^-$ and $pp\rightarrow pp (\eta\rightarrow e^+e^-\gamma)$ due to the same final state
particles if the mass of the virtual photon is large and the photon is not observed.

One of the analysis highlights, the charged particle identification by energy deposited in the
electromagnetic calorimeter (SEC) is shown on the Fig.~\ref{fig:edeps}.
\begin{figure}[htb]
  \centerline{\includegraphics[width=0.45\textwidth]{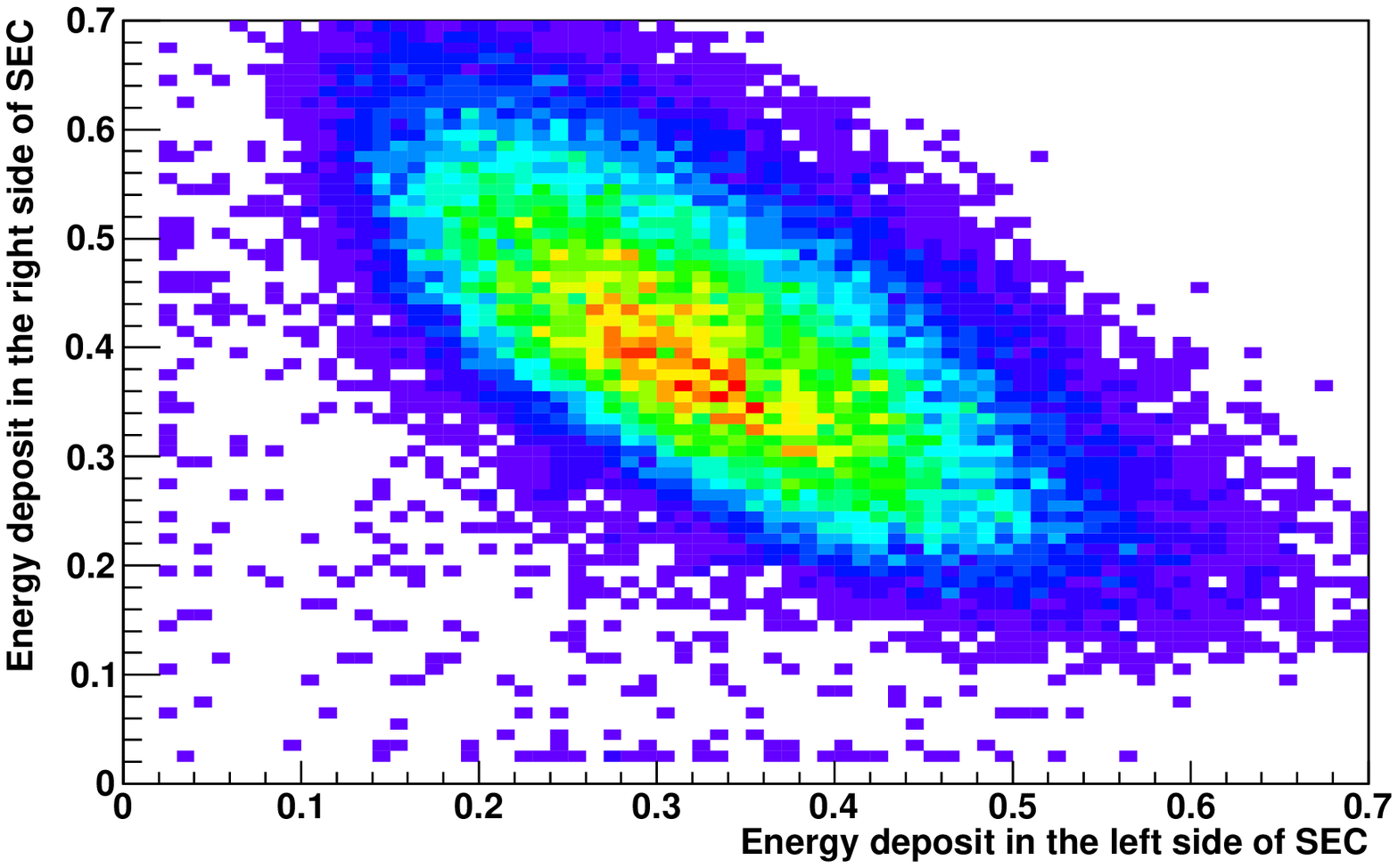}%
              \includegraphics[width=0.45\textwidth]{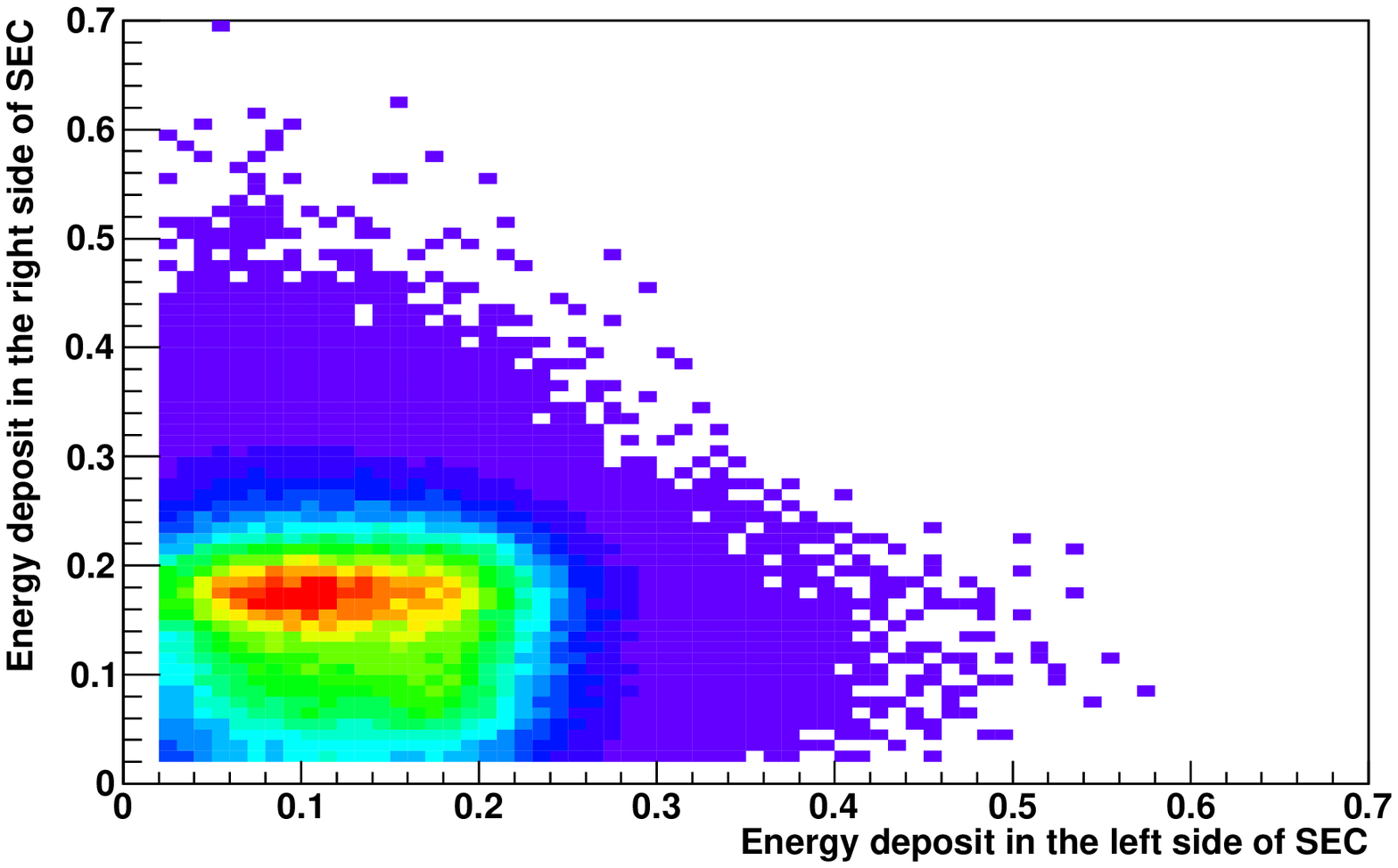}}
  \caption{\label{fig:edeps}Energy deposit in each side of the electromagnetic calorimeter for: MC simulated
$\eta\rightarrow e^+e^-$ signal (left panel), MC direct production of two charged pions (right panel).}
\end{figure}

\Section{Results and conclusions}
After the full data analysis we identified 182 event candidates in the entire data sample in the plot of
missing mass deducted from two protons before subtracting non-eta background. At this level the Monte Carlo
signal acceptance is estimated to be $4.8\%$. The background coming from $pp\rightarrow pp\pi^+\pi^-$ and
$pp\rightarrow p(\Delta(1232)\rightarrow p[\gamma^*\rightarrow e^+e^-])$ is subtracted by a polynomial fit on
the missing mass plot and the expected background is coming from the $\eta\rightarrow e^+e^-\gamma$ decay
channel with $5.6\pm1.1$ events. The remaining number of events after background subtraction is consistent
within errors with the expected number of events coming from $\eta\rightarrow e^+e^-\gamma$. No events of
$\eta\rightarrow e^+e^-$ corresponds to the BR upper limit better than the previously published
one~\cite{MB08} by an order of magnitude.

Based on $\eta\rightarrow e^+e^-\gamma$ we have proven the ability to detect and reconstruct $e^+e^-$ pairs
with different masses. At this stage of the analysis we are able to go an order of magnitude below the present
limit~\cite{MB08} for the BR of $\eta\rightarrow e^+e^-$ decay. Four times bigger statistics of the same
reaction with the same trigger is still available for analysis. We plan an additional eight weeks of data
taking using the same incident proton energy and improved experimental conditions.

\vspace{2ex}\noindent{\it This publication has been supported by the European Commission under the 7th
Framework Programme through the 'Research Infrastructures' action of the 'Capacities' Programme. Call:
FP7-INFRASTRUCTURES-2008-1, Grant Agreement N. 227431.}

\coltoctitle{\boldmath Analysis of the $C$-Violating Decay
$\eta\rightarrow\pi^0+\gamma^*\rightarrow\pi^0+\mathrm{e}^++\mathrm{e}^-$ \\ with \mbox{WASA-at-COSY}}
\authortochead{F.~Bergmann}
\label{BF}

\setcounter{chapter}{1}
\setcounter{section}{0}
\maketitle

\Section{Introduction}
The decay $\eta\rightarrow\pi^0+\gamma^*\rightarrow\pi^0+\mathrm{e}^++\mathrm{e}^-$ is forbidden since it
violates the $C$-parity conservation:
\begin{equation}
	C=C_{\pi^0}\cdot C_{\gamma^*}=(+1)\cdot(-1)=-1\neq+1=C_{\eta} .
\end{equation}
So far this decay has not yet been observed, but an upper limit for the branching ratio has been determined
in previous experiments under the assumption of a phase space distribution for the emission of the ejectiles:
\begin{equation}
  \mathrm{BR}(\eta\rightarrow\pi^0+\mathrm{e}^++\mathrm{e}^-)<4\cdot10^{-5} \cite{qu:1}
\end{equation}
Recent high statistics measurements with WASA-at-COSY~\cite{qu:2} aim at lowering the existing upper limit to
test $C$-parity conservation with increased sensitivity. For the $C$-violating decay
$\eta\rightarrow\pi^0+\gamma^*\rightarrow\pi^0+\mathrm{e}^++\mathrm{e}^-$ different decay model
assumptions exist which have to be taken into account for the analysis, e.g.: decay via a virtual photon,
vector meson dominance (see Fig.~\ref{fig:mod}).

\begin{figure}[htb]
 \begin{center}
   \includegraphics[width=0.7\textwidth]{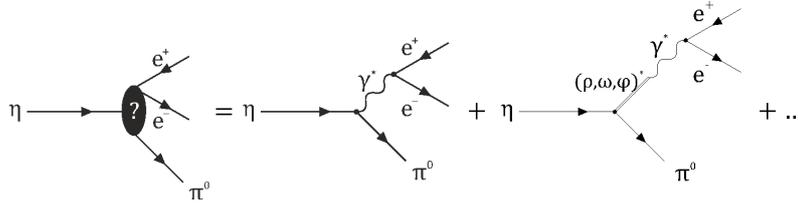}
 \end{center}
 \caption{\label{fig:mod}Feynman diagrams of the decay
$\eta\rightarrow\pi^0+\gamma^*\rightarrow\pi^0+\mathrm{e}^++\mathrm{e}^-$ assuming different models.}
\end{figure}

\Section{Analysis}
For the analysis two data sets are used. One with $10^7$ registered $\mathrm{p}+\mathrm{d}\rightarrow
{^3\mathrm{He}}+\eta$ events from a beam time in 2008 and the other one with $2\cdot10^7$ such $\eta$-events
from a beam time in 2009.

\noindent The first step in the analysis is a preselection on $\mathrm{p}+\mathrm{d}\rightarrow
{^3\mathrm{He}}+\mathrm{X}$ events via ${^3\mathrm{He}}$ identification using the $\Delta E$-$E$ method in the
forward detector system which consists of a tracking device and an array of thin plastic scintillators. The
next step is a preselection on the signature of the decay
$\eta\rightarrow\pi^0+\mathrm{e}^++\mathrm{e}^-\rightarrow\gamma+\gamma+\mathrm{e}^++\mathrm{e}^-$,
requiring two neutral particles in the central detector ($2 \gamma$), one positively and one negatively
charged particle in the central detector ($\mathrm{e}^+$, $\mathrm{e}^-$) and one charged particle in the
forward detector (${^3\mathrm{He}}$). In order to determine further selection conditions all possible
background reactions like (multi-)pion production and other $\eta$-decay channels are simulated. For
choosing further possible selection conditions the underlying mechanisms for the decay
$\eta\rightarrow\pi^0+\mathrm{e}^++\mathrm{e}^-$ have to be taken into account. This is illustrated in
Fig.~\ref{fig:IMee} for the invariant mass distribution of the lepton pair.

\begin{figure}[htb]
  \centerline{\includegraphics[width=0.35\textwidth]{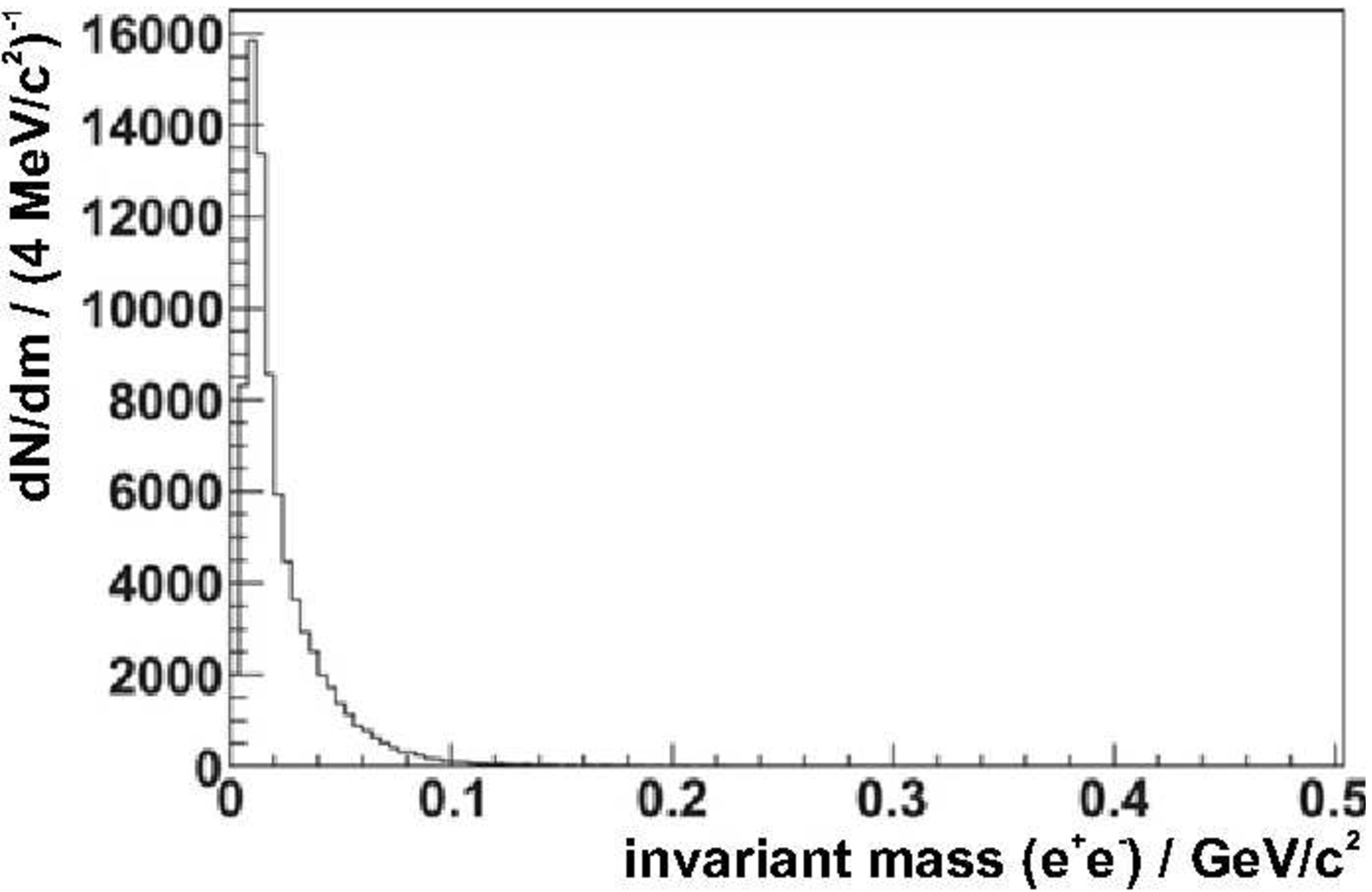}\hspace{0.15\textwidth}%
              \includegraphics[width=0.35\textwidth]{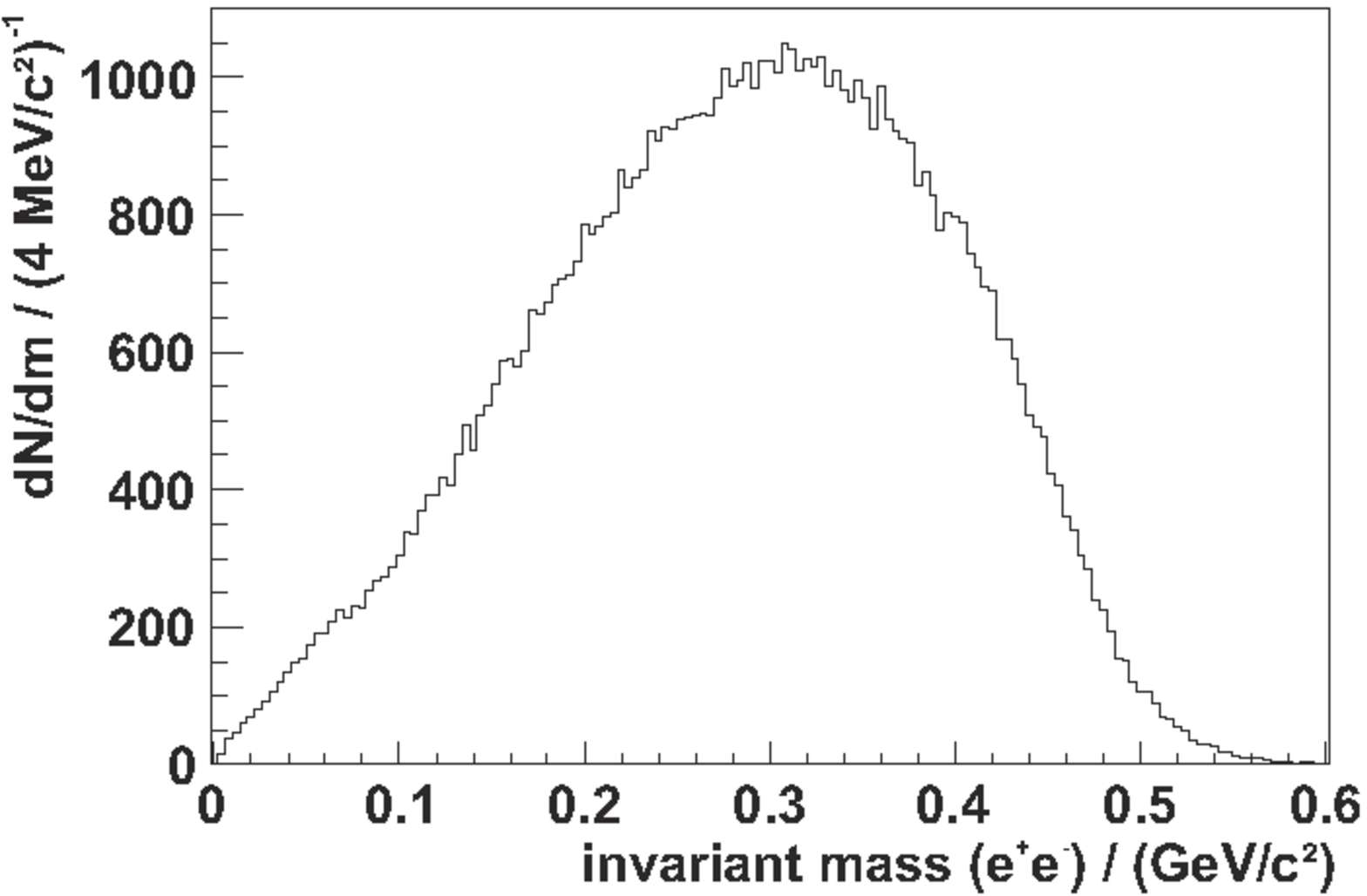}}
  \caption{\label{fig:IMee}Simulated invariant mass of the lepton pair for different decay mechanisms. Left:
$\eta\rightarrow\pi^0+\gamma^*\rightarrow\pi^0+\mathrm{e}^++\mathrm{e}^-$, right:
$\eta\rightarrow\pi^0+\mathrm{e}^++\mathrm{e}^-$; both assuming a phase space distribution. Further
simulations of decay mechanisms which, e.g., consider vector meson dominance are planned.}
\end{figure}

\noindent Before searching for the optimal set of cuts one has to determine the number of events from each
background channel which are still left after the preselection. For this purpose the measured data (2008) has
been fitted by a Monte Carlo cocktail using the same fit parameters for all differential spectra. This results
in an accurate description of the data (See Fig.~\ref{fig:MM}).

\begin{figure}[htb]
  \centerline{\includegraphics[width=0.4\textwidth]{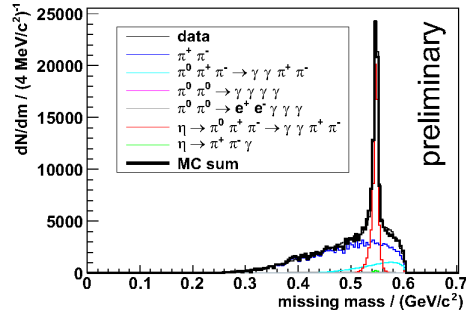}}
  \caption{\label{fig:MM}Missing mass of the measured data fitted with a Monte Carlo cocktail.}
\end{figure}

\noindent For the determination of the optimal selection conditions the evaluation function
$G=S_R\cdot\frac{S_R}{B_R}$ is used for all cuts, where $S_R$ is the relative signal and $B_R$ the relative
background after applying the cuts. Only Monte Carlo data are used to find the optimal set of cuts via varying
the cuts and maximizing the evaluation function $G$.

\Section{Preliminary Results and Outlook}
The 2008 data set has been analyzed by A. Winnem\"{o}ller~\cite{qu:3}: After applying the optimal set of cuts
9920 out of $10^6$ simulated $\eta\rightarrow\pi^0+\gamma^*\rightarrow\pi^0+\mathrm{e}^++\mathrm{e}^-$ events
remain and according to the simulations the background is reduced by a factor $1.7 \cdot 10^8$. In agreement
with the simulations one event candidate remains in the measured data after applying all cuts.

\noindent Currently further $2\cdot10^7$ $\mathrm{p}+\mathrm{d}\rightarrow {^3\mathrm{He}}+\eta$ events from
2009 are analyzed considering different model assumptions. Additionally, significantly more
$\mathrm{p}+\mathrm{p}\rightarrow \mathrm{p}+\mathrm{p}+\eta$ events are available for the analysis and
further $\mathrm{p}+\mathrm{p}\rightarrow \mathrm{p}+\mathrm{p}+\mathrm{X}$ beamtimes are planed. 

\vspace{4ex}\noindent\it{This publication has been supported by the European Commission under the 7th
Framework Programme through the 'Research Infrastructures' action of the 'Capacities' Programme. Call:
FP7-INFRASTRUCTURES-2008-1, Grant Agreement N. 227431.}

\coltoctitle{\boldmath Theoretical Perspectives on $\eta,\,\eta'\to 4\pi$ Decays}
\authortochead{B.~Kubis}
\label{KB}

\setcounter{chapter}{1}
\setcounter{section}{0}
\maketitle

\Section{Introduction: anomalous processes}
Processes in low-energy QCD that involve an odd number of (pseudo-)Goldstone bosons (and possibly photons),
which are, therefore, of odd intrinsic parity, are thought to be governed by the Wess--Zumino--Witten
term~\cite{WZW} via chiral anomalies.  While the so-called triangle- and box-anomaly terms are well-tested in
processes such as e.g.\ $\pi^0,\,\eta\to\gamma\gamma$ (triangle) or $\gamma\pi\to\pi\pi$,
$\eta\to\pi\pi\gamma$ (box), the pentagon-anomaly remains more elusive; the simplest possible process that is
usually cited is $K^+K^-\to\pi^+\pi^-\pi^0$, which however has not been experimentally tested yet. A different
set of processes involving five light pseudoscalars are the four-pion decays of $\eta$ and $\eta'$.
Experimental information about these is also scarce, only upper limits on branching ratios exist~\cite{PDG};
however, this may change in the near future for at least some of the possible final states.

In principle, the decays $\eta' \to 4\pi$ seem not terribly forbidden by approximate symmetries: they are
neither isospin-forbidden, nor electromagnetic. $\eta\to 4\pi$, in contrast, is essentially suppressed by tiny
phase space: only the decay into $4\pi^0$ is kinematically allowed ($M_\eta-4M_{\pi^0} = 7.9$\,MeV,
$M_\eta-2(M_{\pi^\pm}+M_{\pi^0}) = -1.2$\,MeV).  The fact that anomalous amplitudes always involve the totally
antisymmetric tensor $\epsilon_{\mu\nu\alpha\beta}$ furthermore shows that no two pseudoscalars are allowed to
be in a relative S-wave; the decays $\eta'\to  2(\pi^+\pi^-)$ and $\eta'\to \pi^+\pi^-2\pi^0$ are therefore
P-wave-dominated. As furthermore Bose symmetry forbids $2\pi^0$ to be in an odd partial wave, $\eta'\to
4\pi^0$ and $\eta\to  4\pi^0$ even require all $\pi^0$ to be at least in relative D-waves~\cite{KupscWirzba}.
This, combined with the tiny phase space available, leads to the notion of $\eta\to  4\pi^0$ being
CP-forbidden~\cite{PDG}, although strictly-speaking it is only S-wave CP-forbidden.

\Section{Charged final states: \boldmath{$\eta'\to  2(\pi^+\pi^-)$}, \boldmath{$\eta'\to \pi^+\pi^-2\pi^0$}}

\begin{figure}[t]
\centering
\includegraphics[width=0.73\linewidth]{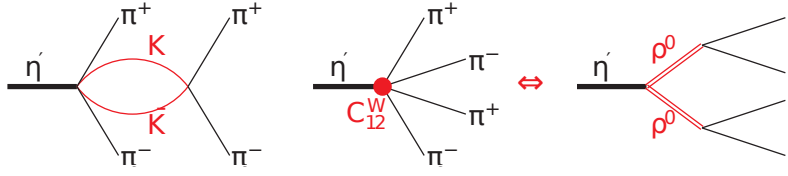}
\caption{Feynman graphs contributing to $\eta'\to  2(\pi^+\pi^-)$.}
\label{fig:charged}
\end{figure}
In the following, we assume the standard $\eta\eta'$ mixing in terms of singlet ($\eta_0$) and octet
($\eta_8$) states, i.e.\  $|\eta\rangle = {1}/{3}|\eta_0\rangle+{2\sqrt{2}}/{3}|\eta_8\rangle$, 
$|\eta'\rangle = {2\sqrt{2}}/{3}|\eta_0\rangle-{1}/{3}|\eta_8\rangle$. The flavor structure of the chiral
anomaly does not allow for direct contributions of the Wess--Zumino--Witten term to the decays
$\eta,\,\eta'\to 4\pi$. The leading amplitude to the two channels involving charged pions in the final state
occurs at one loop or $\mathcal{O}(p^6)$ in terms of chiral power counting, and is given as the sum of kaon
loops and local counterterms, see Fig.~\ref{fig:charged}. For instance, the decay amplitude for $\eta_8 \to
\pi^+(p_1)\pi^-(p_2)\pi^+(p_3)\pi^-(p_4)$ is given by~\cite{arXiv:Kub}
\begin{align}
\mathcal{A}(\eta_{8}&\to\pi^+\pi^-\pi^+\pi^-) 
= \frac{N_c\epsilon_{\mu\nu\alpha\beta}}{3\sqrt{3}F_\pi^5} p_1^\mu p_2^\nu p_3^\alpha p_4^\beta
\big[ F_8(s_{12})+F_8(s_{34})-F_8(s_{14})-F_8(s_{23}) \big] ~, \nonumber\\
F_8(s) &= \frac{1}{8\pi^2F_\pi^2}\bigg\{ \big(s-4M_K^2\big) \bar J_{KK}(s) - \frac{s}{16\pi^2}
\Big( 2\log\frac{M_K}{\mu} + \frac{1}{3} \Big) \bigg\} - 16 C_{12}^{Wr}(\mu)s ~, \nonumber\\
\bar{J}_{KK}(s) &=\frac{1}{8\pi^2}\big\{1-\sigma_K\operatorname{arccot}\sigma_K\big\}~, \quad
\sigma_K=\sqrt{\frac{4M_K^2}{s}-1}~, \label{eq:AmpOp6}
\end{align}
where $s_{ij} = (p_i+p_j)^2$.  The corresponding $\eta_0$ amplitude consists of local contributions only. The
scale dependence of the counterterm $C_{12}^{Wr}$ is known~\cite{BGT_Op6} and cancels the one of the loop
contribution. $\eta_8 \to \pi^+(p_1)\pi^0(p_2)\pi^-(p_3)\pi^0(p_4)$ is given by the same amplitude
Eq.~\eqref{eq:AmpOp6} up to a sign.

In order to estimate the finite part of the coupling constant $C_{12}^{Wr}$, we resort to resonance
saturation. While this has been studied in a more general setting recently~\cite{Kampf}, we employ the
hidden-local-symmetry (HLS) formalism to calculate vector-meson contributions, see Fig.~\ref{fig:charged}.  We
only need two (combinations of) coupling constants occurring in the anomalous sector~\cite{FKTUY}, $c_1-c_2$
and $c_3$. The two chiral constants contributing to five-pseudoscalar amplitudes at $\mathcal{O}(p^6)$ 
(only one of them contributes to $\eta'\to 4\pi$) are given in terms of these by
\begin{equation}
C_1^{Wr}(M_\rho) = -2 C_{12}^{Wr}(M_\rho) = \frac{3(c_1-c_2+c_3)}{128\pi^2M_\rho^2}~.
\end{equation}
The HLS couplings are often assumed to be given roughly by $c_1-c_2 \approx c_3 \approx 1$. A more extensive
phenomenological study finds $c_1-c_2 \approx 1.21$, $c_3 \approx 0.93$~\cite{Benayoun}. Comparing the
relative importance of counterterm (=vector-meson) contributions to kaon loops in Eq.~\eqref{eq:AmpOp6}, we
find that the decay amplitude is entirely dominated by the vector-meson terms.  Furthermore, as the maximum
allowed kinematical range in $\eta' \to 4\pi$ is $\sqrt{s_{ij}} \leq M_{\eta'}-2M_\pi \approx 680$\,MeV, it is
necessary to retain the full $\rho$ propagators for a phenomenologically reliable description.  This, at the
same time, takes care of the essential P-wave $\pi\pi$ final-state interactions.

We calculate the branching ratios as functions of the HLS couplings, before inserting the two sets of
numerical values quoted above. In the isospin limit, $2 \times \text{BR}(\eta'\to
2(\pi^+\pi^-))=\text{BR}(\eta'\to \pi^+\pi^-2\pi^0)$; we adjust for the correct phase space by using the pion
masses $M_{\pi^\pm}$ and $(M_{\pi^\pm}+M_{\pi^0})/2$, respectively.  We find
\begin{align}
\text{BR}\big(\eta'\to  2(\pi^+\pi^-)\big) &= \big[ 0.15  (c_1-c_2)^2 
+ 0.46 (c_1-c_2)c_3 
+ 0.37 c_3^2 \big] \times 10^{-4} 
= \big\{ 0.97  ,\, 1.04 \big\}\times 10^{-4} , \nonumber\\
\text{BR}\big(\eta'\to \pi^+\pi^-2\pi^0\big) &= \big[ 0.34  (c_1-c_2)^2 
+ 1.07 (c_1-c_2)c_3 
+ 0.85 c_3^2 \big] \times 10^{-4} 
= \big\{ 2.26  ,\, 2.44 \big\}\times 10^{-4}  .
\end{align}
These numbers are to be compared to the available experimental upper limits, $\text{BR}(\eta'\to 
2(\pi^+\pi^-))$ $< 2.4\times 10^{-4}$ and $\text{BR}(\eta'\to \pi^+\pi^-2\pi^0) < 2.6\times
10^{-3}$~\cite{PDG}.

\Section{Neutral final states: \boldmath{$\eta'\to  4\pi^0$}, \boldmath{$\eta\to  4\pi^0$}}
As we have mentioned in the introduction, the P-wave mechanism described in the previous section, proceeding
essentially via two $\rho$ intermediate resonances, cannot contribute to the $4\pi^0$ final states.  In fact,
we can show that the D-wave characteristic of $\eta,\,\eta' \to 4\pi^0$ suppresses these decays to
$\mathcal{O}(p^{10})$ in chiral power counting, that is to the level of three loops in the anomalous sector.
This is in particular due to the flavor and isospin structure of the anomaly, which does not contain
five-meson vertices including $2\pi^0$ at leading order ($\mathcal{O}(p^4)$), and to the chiral structure of
meson--meson scattering amplitudes, which only allows for S- and P-waves at tree level ($\mathcal{O}(p^2)$).
As a complete three-loop calculation would be a formidable task and is beyond the scope of our exploratory
study, we instead construct a decay mechanism that, we believe, ought to capture at least the correct order of
magnitude. This is built on the observation that we \emph{can} calculate the leading imaginary part of the 
decay amplitude at $\mathcal{O}(p^{10})$, which is given by a $\pi^+\pi^-$ intermediate state, see
Fig.~\ref{fig:4pi0}. As indicated, we thereby insert the full vector-meson amplitude for $\eta,\,\eta'\to
\pi^+\pi^-2\pi^0$ described in the previous section.  We then reconstruct the ``full'' D-wave $\pi\pi$
final-state amplitude by requiring that it is proportional to the Omn\`es function $\Omega_2^0(s)$ (assuming
dominance of isospin $I=0$), which near threshold behaves like
\begin{equation}
\text{Im}\,\Omega_2^0(s) \approx \sqrt{1-\frac{4M_\pi^2}{s}} \times t_2^0(s) ~,
\end{equation}
where $t_2^0$ is the appropriate $\pi\pi$ partial wave of isospin $I=0$ and angular momentum $\ell=2$. Well
above threshold, however, the Omn\`es function can well be approximated by the $f_2(1270)$ resonance,
$\Omega_2^0(s) \approx M_{f_2}^2 / (M_{f_2}^2-s)$.
\begin{figure}[t]
\centering
\includegraphics[width=0.73\linewidth]{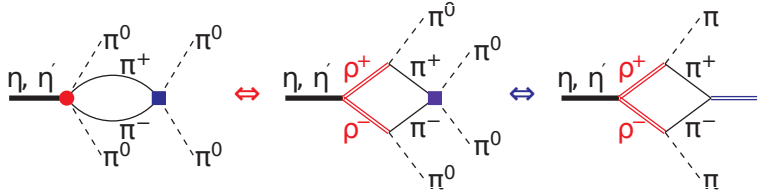}
\caption{D-wave mechanism for $\eta,\,\eta' \to 4\pi^0$.}
\label{fig:4pi0}
\end{figure}
The resulting decay mechanism is indicated in Fig.~\ref{fig:4pi0}.  Note that it is far from a real $f_2$
dominance picture, as the $\rho$ is not ``short-ranged'' on the scale of the $f_2$ mass. For the branching
ratios, we find
\begin{align}
\text{BR}\big(\eta'\to  4\pi^0\big) &= \big[ 0.4  (c_1-c_2)^2 
+ 1.6 (c_1-c_2)c_3 
+ 1.6  c_3^2 \big] \times 10^{-8} \hspace{4pt}
= \big\{ 3.6  , \, 3.8 \big\}\times 10^{-8} ~, \nonumber\\
\text{BR}\big(\eta\to  4\pi^0\big) &= \big[ 0.3  (c_1-c_2)^2 
+ 1.0 (c_1-c_2)c_3 
+ 1.0  c_3^2 \big] \times 10^{-30}  
= \big\{ 2.4  ,\, 2.5 \big\}\times 10^{-30} . \label{eq:4pi0}
\end{align}
While the D-wave mechanism therefore suppresses the decay $\eta'\to  4\pi^0$ by 3--4 orders of magnitude
compared to the charged final states, the suppression of $\eta\to  4\pi^0$ due to the combination of D-wave
and tiny phase space is enormous, and any experimental signal for this channel could indeed be interpreted
as a sign of CP violation.

\Section{CP violation through the QCD \boldmath{$\theta$}-term}
One possible source of strong CP violation is the so-called QCD $\theta$-term, linked to the $U(1)_A$ anomaly,
that can also be treated on an effective-Lagrangian level~\cite{CVVW}.  It induces effects such as an electric
dipole moment of the neutron, and also allows for the decay $\eta \to 2\pi$. In addition, it induces a
CP-violating S-wave mechanism for the decay $\eta\to  4\pi^0$~\cite{arXiv:Kub}, 
\begin{equation}
\mathcal{A} \big(\eta\stackrel{\not\!\text{CP}}{\longrightarrow} 4\pi^0 \big)
= - \sqrt{\frac{2}{3}}\,\frac{M_{\eta'}^2 }{3 F_\pi^3} \times \bar\theta_0 ~.
\end{equation}
The resulting branching ratio is given by
\begin{equation}
 \text{BR}\big( \eta\stackrel{\not\!\text{CP}}{\longrightarrow} 4\pi^0\big) = 5.2\times 10^{-5} \times
\bar\theta_0^2 ~,
\end{equation}
which, were $\bar\theta_0$ a quantity of natural size, would demonstrate the enhancement of the CP-violating
S-wave mechanism compared to the CP-conserving D-wave one, see Eq.~\eqref{eq:4pi0}. However, current limits
from neutron electric dipole moment measurements suggest $\bar\theta_0 \lesssim 10^{-11}$~\cite{Ottnad}, which
puts also this result beyond any experimental reach.

\coltoctitle{\boldmath Pluto: A Multi-Purpose Event Generator Framework for Rare $\eta$ Decays}
\authortochead{I.~Fr\"ohlich}
\label{FI}

\setcounter{chapter}{1}
\setcounter{section}{0}
\maketitle

\Section{Introduction}
In this contribution, the new design and recent developments of the simulation package Pluto~\cite{pluto_web}
have been presented. Pluto's main mission is to offer a modular framework with an object-oriented structure,
thereby making additions such as new particles, decays of resonances, new models up to modules for entire
changes easily applicable~\cite{pluto}. Overall consistency is ensured by a plugin- and distribution manager. 
Particular features are the support of a modular structure for physics process descriptions, and the
possibility to access the particle stream for in-line modifications. Additional configurations can be attached
by the user by a dedicated script language (``PlutoScript'') without re-compiling the package, which makes
Pluto extremely configurable.

\vspace{1ex}\Section{\boldmath Rare $\eta$ decays}
The rare $\eta$ decays have been implemented into Pluto via means of a dedicated ``plugin''. This ensures that
changes introduced by the new implementation do not affect simulations done by other users. 

The models follow basically calculations done by Wirzba and Petri~\cite{petri}, but leaving out tiny effects
like magnetic form factors.

\vspace{1ex}\Subsection{\boldmath $\eta \to e^+ e^- e^+ e^-$}
This decay implementation is done  via two virtual photons:
\begin{verbatim}
  PReaction my_reaction("_T1=2.2","p","d",
     "He3 eta[dilepton [e+ e-] dilepton [e+ e-]]");
\end{verbatim}

In the first step of the event generation Pluto samples the mass of the 2 dileptons. This is very similar to
the implementation of the $\eta$ Dalitz decay of the base package. The shape of the dilepton spectrum is based
on an implementation provided by A. Kupsc~\cite{kupsc}.

In a second step, the angular distributions are sampled via the rejection method. This can be done as the
correction is only about 20\%, which means that only a small fraction of events have to be resampled.
Moreover, the angular distributions of the ``simple'' generator are isotropic and therefore equally populated.

\vspace{1ex}\Subsection{\boldmath $\eta \to \pi^+ \pi^- e^+ e^-$}
This decay implementation uses the a ``genbod'' technique: Here, the first step of the 4-body decay is done
via a modified genbod algorithm, where the mass distribution $M(q^2) = F(q^2) / (8 \cdot q^2)$ is folded into
the genbod model (with $q$ as the four-momentum of the virtual photon). This is done, because a generation of
events in a 4-body space is numerically difficult and would result in large computing times. On the other
hand, a pure rejection method (using an unmodified genbod), as in the previous case, is not possible, because
the pole of $M(q^2)$ is located in a region where no events are generated by genbod.

For the form factor by default a "simple VMD model" is used:
\begin{equation}
  F(q^2) = m_v^2 / (m_v^2 - q^2) 
\end{equation}
with $m_v$ = 0.77GeV.

\noindent In total, the following matrix element is used:
\begin{equation}
  A^2 = M(q^2) * s_{\pi\pi} * \beta_\pi^2 \sin^2(\theta_\pi)
        * \lambda(m_\eta^2,s_{\pi\pi},q^2) (1+\beta_e^2*\sin^2(\theta_e)*\sin^2(\phi))
\end{equation}
with $s_{\pi\pi}$ as the invariant mass of the pion pair, and the factor $\beta_x =
\sqrt{1-\frac{4m^2_x}{s_{xx}}}$. In the case of the dilepton pair, $s_{ee}$ is the invariant mass of the
virtual photon.

\Section{Scripting with Pluto}
The idea behind scripting with Pluto is to combine objects like histograms and ntuples via a flexible batch
language with the main event loop, model objects, and the internal data base of Pluto. The commands of the
batch script are compiled by Pluto in advance and can be executed as often as required without additional
delay. This execution can be done inside the event loop which allows for modifications of particles, e.g.\ to
write acceptance filters or include smearing effects.

Furthermore, it is also possible to call many public available Pluto-methods of the utility class (for random
sampling) and all methods of the Pluto particle class, which is inherited from TLorentzVector. As the script
can combine ROOT-objects (histograms and ntuples) with the event loop, particle properties can be converted
very quickly to histograms and ntuples. And finally, the script can be packed together with acceptance and
resolution matrices to form a universal, common format for a detector description.

\Section{Examples}
Finally, the application of the script method is discused. Exchanging, e.g., the widely discussed $\omega$
form factor is a quite simple action\footnote{This is, just to keep it to be an example, the old form factor
of Landsberg.}:

\begin{verbatim}
  PSimpleVMDFF *ff = new PSimpleVMDFF("vmd_ff_dd@w_to_dilepton_pi0/formfactor",
                                      "VMD form factor",-1);
  ff->AddEquation("_ff2 = 0.17918225/( (0.4225- _q2)*(0.4225- _q2) + 0.000676)");
  makeDistributionManager()->Add(ff);
\end{verbatim}

The basic syntax is based on the ROOT TFormula class, but with a lot of extensions. For details, reference is
made to the full manual which is available at the Pluto web page~\cite{pluto_web}.

Considering, that the form factor model of rate  eta decays is a separated model, the form factor can be read
via:
\begin{verbatim}
  PSimpleVMDFF *ff = (PSimpleVMDFF*)makeDistributionManager()->
     GetDistribution("eta_ee_pipi_vmd_ff");
\end{verbatim}
and therefore exchanged via similar means

\coltoctitle{\boldmath Dalitz Plot Analysis for $\eta \rightarrow \pi^+ \pi^- \pi^0$  at
KLOE}
\authortochead{L.~Caldeira~Balkest\aa{}hl}
\label{CBL}

\setcounter{chapter}{1}
\setcounter{section}{0}
\maketitle

\Section{Theoretical Motivation}
The decay $ \eta \rightarrow \pi^+ \pi^- \pi^0 $ is an isospin-violating process sensitive to the light quark
mass double ratio:

\begin{equation}
Q^2= \frac{m_s^2-\hat{m}^2}{m_d^2-m_u^2} \mbox{, where } \quad  \hat{m}= \frac{1}{2}(m_d+m_u)\label{eq:Q}
\end{equation}

The  $ \eta \rightarrow \pi^+ \pi^- \pi^0 $ decay rate is proportional to $Q^{-4}$ and it has been calculated
at leading order (LO) and next-to-leading order (NLO) in $\chi$PT. The values obtained, $ \Gamma_{LO} \sim
70\mbox{ eV }$ and $\Gamma_{NLO}=160 \pm 50  \mbox{ eV }$, are significantly lower than the experimental value
$\Gamma_{exp}=296 \pm 16 \mbox{ eV }$, obtained from a fit to all the available data~\cite{pdg}. The
difference between LO and NLO calculations and the discrepancy with data at NLO are evidence of the slow
convergence of the $\chi$PT series, due to strong pion rescattering effects in the final state. These effects
can be  treated by means of dispersion relations as recently proposed in Ref.~\cite{Colangelo}. The relations
contain four parameters to be fixed either matching the LO $\chi$PT results or by a fit to experimental data.
Using the first approach G. Colangelo and coauthors~\cite{Colangelo} obtained $Q = 22.3 \pm 0.4$, in good
agreement with most of the results presented in Fig.~\ref{fig:Q2}, where the gray elliptic band corresponds to
the 2-$\sigma$ range of the $Q$-value above. Other plotted constraints come from lattice calculations.

\begin{figure}[htb]
\centering
\includegraphics[width=0.8\textwidth]{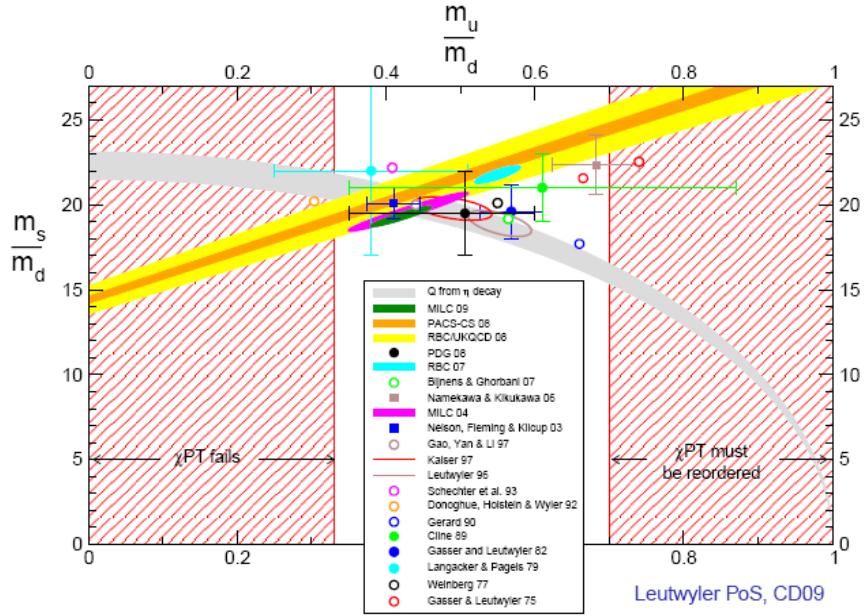}
\caption{\label{fig:Q2} Ratio of the light quark masses. The gray band corresponds to $Q = 22.3 \pm
0.8$~\cite{leutwyler}.}
\end{figure}

\Section{Data analysis}
The KLOE detector, at the DA$\Phi$NE collider, consists of a large volume drift chamber and a
lead-scintillating fibers calorimeter. The detector usually takes data at the $\phi$-meson peak
\mbox{($\sqrt{s}=1.019$~GeV)}. The experiment took data in years 2001-2002 and 2004-2006, integrating a total
luminosity $2.5 \mbox{ fb}^{-1}$ at the $\phi$-peak and $240 \mbox{ pb}^{-1}$ at the center-of-mass energy of
1~GeV. KLOE-2 has been approved to collect $\sim 20 \mbox{ fb}^{-1}$ in 3-4 years of data taking at the
upgraded DA$\Phi$NE facility, which is expected to operate with a factor $\sim 3$ improvement in luminosity. 

The KLOE collaboration published the Dalitz plot analysis of the $ \eta \rightarrow \pi^+ \pi^- \pi^0$ decay
in year 2008~\cite{kloepaper}. In the paper, based on $\sim 500 \mbox{ pb}^{-1}$ of integrated luminosity, the
Dalitz plot density shown in Fig.~\ref{fig:DP} was fitted by:
\begin{equation}
|A(X,Y)|^2 \simeq 1 + aY + bY^2 + cX + dX^2 + eXY +fY^3 + gX^3 + hX^2Y + lXY^2
\end{equation}
obtaining:
\[
a=-1.090 \pm 0.005(stat)^{+0.008}_{-0.019}(sys) \quad b=0.124 \pm 0.006(stat) \pm 0.010(sys)
\]
\[  
d=0.057 \pm 0.006(stat)^{+0.007}_{-0.016}(sys) \quad f=0.14 \pm 0.01(stat) \pm 0.02(sys)
\]
and the other parameters consitent with zero. The $c$ and $e$ parameters are expected to be zero from
C-invariance.

\begin{figure}[htb]
  \centerline{\includegraphics[width=0.65\textwidth]{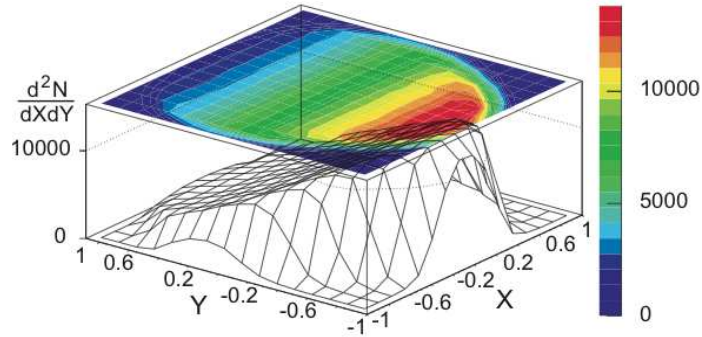}}
  \caption{\label{fig:DP}Dalitz plot density measured by KLOE\cite{kloepaper} on the basis of  $\sim 500
\mbox{ pb}^{-1}$ integrated luminosity.}
\end{figure}

The results have been recently used by S. Lanz and coworkers to fix the parameters of the dispersion
relations in the framework of the analysis presented in Ref.~\cite{Colangelo}. The resulting $Q$-value is $Q=
21.31 \pm 0.60$~\cite{LanzPrime} and the Dalitz plot density obtained is in better agreement with the
measurement performed on the $\eta \rightarrow  \pi^0 \pi^0 \pi^0$ sample~\cite{Ambrosinop0p0p0} than previous
calculations based on the matching of the dispersion-relation parameters with one-loop $\chi$PT results. The
KLOE results have also been used in analytic dispersive analysis~\cite{zdrahal} for the value of $R$
($R=\frac{m_s - \hat{m}}{m_d - m_u}$), both in a direct fit to the results, and in a correction to $\chi$PT
NNLO low energy constants.

More data on  $ \eta \rightarrow \pi^+ \pi^- \pi^0$ are needed to understand the origin of the residual
tension between data and theoretical calculations, concerning both, the comparison between neutral and charged
channel, and the $Q$-value.

\begin{figure}[htb]
  \centering
  \includegraphics[width=0.5\textwidth]{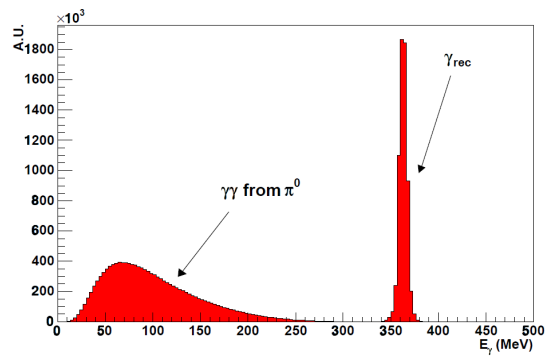}
  \caption{\label{fig:phoen}Energy distribution of the photons from $\phi \rightarrow \eta \gamma \rightarrow
\pi^+ \pi^- \gamma \gamma \gamma$.}
\end{figure}

On the experimental side, a new analysis of the  $ \eta \rightarrow \pi^+ \pi^- \pi^0$ sample
(Fig.~\ref{fig:phoen}) is in progress to improve on the statistical sample and to overcome some limitations of
the previous analysis. For the latter, the selection efficiency is measured directly from minimum bias events,
while in the old analysis it was evaluated from Monte Carlo simulation, giving a large contribution to the
systematics. Further improvement on the precision of the Dalitz plot distribution is expected: $i)$ replacing
the kinematical fit, requiring 4-momentum conservation and the speed of light for photons, with constraints
from the two-body $\phi$ decay kinematics; $ii)$ eliminating the request of vertex reconstruction and the cut
on the total number of tracks in drift chamber; $iii)$ introducing the particle identification via
time-of-flight measurement with the calorimeter; $iv)$ performing the analysis on an independent, larger data
sample.

\coltoctitle{\boldmath The $\eta\rightarrow\pi^+\pi^-\pi^0$ Decay With WASA-at-COSY}
\authortochead{P.~Adlarson}
\label{AP}

\setcounter{chapter}{1}
\setcounter{section}{0}
\maketitle

\Section{Introduction}
The decay $\eta\rightarrow\pi^+\pi^-\pi^0$ is important since it allows access to light quark mass ratios.
Since electromagnetic corrections are small~\cite{DS66, CD09} this decay is a probe for the strong isospin
violation. The calculated tree level decay width, $\Gamma_{tree}\approx 70$~eV~\cite{WB67}, deviates more than
a factor of four from the PDG value $\Gamma=296\pm16$ \cite{PDG}. Higher order diagrams are needed which take
into account $\pi\pi$ final state interactions~\cite{CR81,JG85, JB07}. Alternatively, $\pi\pi$ final state
re-scattering can be implemented in dispersive approaches~\cite{AA96,JK96,JB02, SL09, KK11} as well as other
approaches~\cite{SS11}. The decay width scales as
\begin{equation}
\Gamma=\left(\frac{Q_D}{Q}\right)^4 \bar{\Gamma},
\end{equation}
where $Q^2=(m_s^2-\hat{m}^2)/(m_d^2-m_u^2)$, $\hat{m}=\frac{1}{2}(m_u+m_d)$, and the decay width
$\bar{\Gamma}$ and $Q_D=24.2$ are calculated in the Dashen limit~\cite{RD69}. $Q$ gives access to light quark
mass ratios. The ratios serve also as an important input for lattice QCD~\cite{GC10}. To derive $Q$,
$\bar{\Gamma}$ has to be known reliably from theory which can be tested by comparing with the experimental
Dalitz plot distributions. As Dalitz plot variables 
\begin{equation}
x=\sqrt{3}\frac{T_+-T_-}{Q_{\eta}}, \ y=\frac{3T_0}{Q_{\eta}}-1.
\end{equation}
\noindent are used. Here  $T_+$, $T_-$ and $T_0$ denote the kinetic energies of $\pi^+$, $\pi^-$ and $\pi^0$
in the $\eta$ rest frame, and $Q_{\eta}$ is the sum of the three kinetic energies. The standard way to
parametrize the Dalitz plot density is a polynomial expansion around $x=y=0$
\begin{equation} 
\left|A(x,y)\right|^2 \propto 1+ay+by^2+dx^2+fy^3+gx^2y+...
\end{equation}
where $a, b, ...,g$ are called the Dalitz Plot parameters. The most precise experimental result is based on a
Dalitz plot containing 1.34$\cdot$10$^6$ events obtained by KLOE~\cite{FA08}. Parameters $b$ and $f$ in
\cite{FA08} deviate significantly from ChPT predictions. Independent measurements are therefore important and 
WASA-at-COSY~\cite{AA04} aims at providing two independent data sets with $\eta$ produced in $pp$ and $pd$
reactions. 

\Section{Experiment}
In 2008 and 2009  WASA-at-COSY collected $1\cdot10^7$ and $2\cdot10^7$ $\eta$ mesons respectively, in the
reaction $pd\rightarrow \mbox{}^{3}\mbox{He} X$ at a kinetic beam energy of 1~GeV. The missing mass with
respect to  $\mbox{}^{3}\mbox{He}$ is used to tag the $\eta$ in the reaction $pd\rightarrow
\mbox{}^{3}\mbox{He} X$. To obtain $\eta\rightarrow\pi^+\pi^-\pi^0$ candidates, at least two particles of
opposite charge and at least two photons with an invariant mass close to $\pi^0$ are required. To further
reduce time coincidental events, cuts on vertices and timing requirements are included. To reduce background
channels, mainly coming from $pd\rightarrow \mbox{}^{3}\mbox{He} \pi\pi$ reactions, conditions on the missing
mass calculated for $\mbox{}^{3}\mbox{He} \pi^+\pi^-$, $\mbox{}^{3}\mbox{He} \pi^0$ and $\pi^+\pi^-$ are
applied. The preliminary analysis gives 260 000 $\eta\rightarrow\pi^+\pi^-\pi^0$ candidates from the 2008 data
sample. The majority of the events populating the Dalitz plot is $\mbox{}^{3}\mbox{He}\pi^+\pi^-\pi^0$ and
$\eta\rightarrow\pi^+\pi^-\pi^0$. The $\eta$ content in each Dalitz plot bin is obtained  by performing a
four-degree polynomial fit over the direct 3$\pi$ background and the fitted polynomial is subtracted from the
signal region. The number of $\eta$ events in each bin is corrected for acceptance. 

\Section{Result}
The preliminary acceptance corrected experimental results with statistical errors for the $x$, $y$
projections of the Dalitz plot are compared in Fig.~\ref{ACCCORR} to the distributions of the
$\eta\rightarrow\pi^+\pi^-\pi^0$ with Monte Carlo weighted with the set of Dalitz plot parameters calculated
for the leading order (LO) in ChPT (solid blue line), next-to-next-to-leading order (NNLO) (dashed-dotted
green line) and the central values obtained by KLOE (dashed red line). The Monte Carlo results have been
normalized to the WASA experimental data. From the preliminary Dalitz plot projections the WASA data points
are in reasonable agreement with the KLOE experimental result. The work on the \textit{pd} data will be
continued in order to obtain the Dalitz plot parameters for $\eta\rightarrow\pi^+\pi^-\pi^0$. An effort will
be made to understand the systematical effects.

\begin{figure}[htb]
 \centerline{\includegraphics[width=1.0\textwidth]{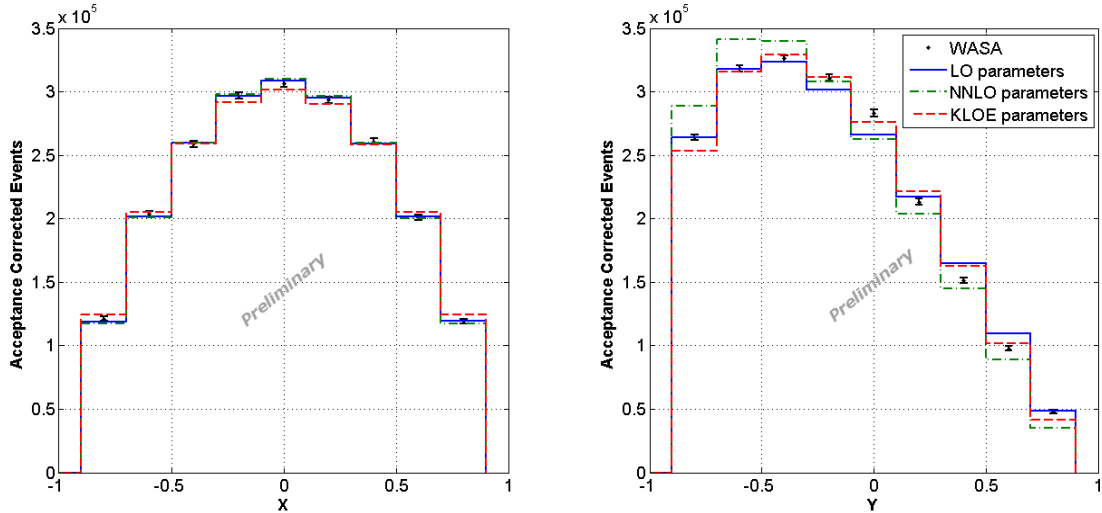}}
 \caption{\label{ACCCORR}Acceptance corrected data points with statistical errors, projected on x ({\bf left})
and y  ({\bf right}), compared to Monte Carlo weighted with the Dalitz plot parameters for the leading order
calculation (LO) in ChPT (solid blue line), next-to-next-to-leading order (NNLO) (dashed-dotted green line)
and the central values obtained by KLOE (dashed red line). The Monte Carlo results are normalized to the
experimental data.} 
\end{figure}

\vspace{3ex}\section*{Acknowledgements}
This publication has been supported by the European Commission under the 7th Framework Programme through the
'Research Infrastructures' action of the 'Capacities' Programme. Call: FP7-INFRASTRUCTURES-2008-1, Grant
Agreement N. 227431.

\coltoctitle{\boldmath Berne-Lund-Valencia Dispersive Treatment of $\eta\to 3\pi$}
\authortochead{S.~Lanz}
\label{LS}

\setcounter{chapter}{1}
\setcounter{section}{0}
\maketitle

\noindent The decay $\eta \to 3 \pi$ is of particular theoretical interest because it can only proceed through
isospin breaking. If the strongly suppressed electromagnetic contributions are neglected, the decay amplitude
is proportional to $(m_u - m_d)$ or, alternatively, to the quark mass double ratio
\begin{equation}
	Q^2 = \frac{m_s^2 - \hat{m}^2}{m_d^2 - m_u^2} \;, \quad \mbox{with} \quad \hat{m} = \frac{1}{2}(m_u+m_d) \;.
\end{equation}
Furthermore, this process involves two theoretical puzzles. The first one is the fact that the predictions for
the decay width from current algebra, but also from one-loop chiral perturbation theory ($\chi$PT), fail to
reproduce the experimental value. It is understood that this is due to large final state rescattering effects
that can be ideally treated using dispersion relations. Gasser and Leutwyler, having included one final state
rescattering process in the decay amplitude~\cite{Gasser+1985a}, found for the decay width of the charged
channel $\Gamma_c = 160 \pm 50~\mbox{eV}$. This was in perfect agreement with the then-current PDG average
$\Gamma_c = 197 \pm 29~\mbox{eV}$ and the case seemed to be settled. However, the authors already pointed out
a new experiment hinting at a larger decay width, which would ``resurrect the problem'' they ``claim[ed] to
solve''. Figure~\ref{fig:decayWidthPDG} clearly shows that over the following years, the PDG average grew out
of the error bars of the one-loop prediction. The second puzzle is related to the slope parameter $\alpha$ in
the neutral channel, where $\chi$PT, but also an older dispersive analysis~\cite{Kambor+1996}, fail to
reproduce the experimental finding.

\begin{figure}[h]
	\begin{center}\includegraphics[width=0.6\textwidth]{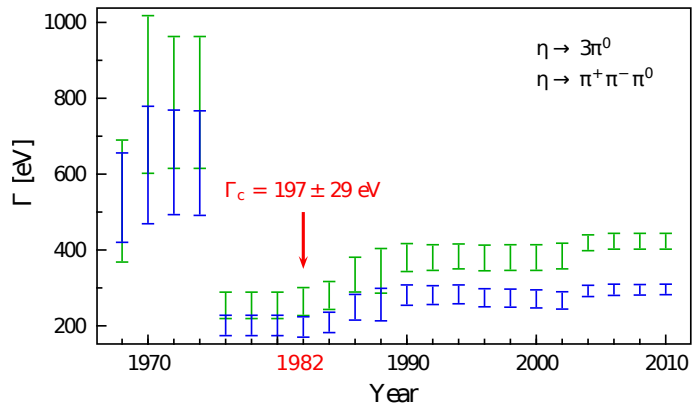}\end{center}
\vspace{-5mm}\caption{\label{fig:decayWidthPDG}PDG average for the decay width of the two $\eta \to 3 \pi$
channels since 1968. Gasser and Leutwyler used the value for the charged channel from 1982.}
\end{figure}

In 1996 two dispersive analyses of $\eta \to 3 \pi$ have been published~\cite{Kambor+1996,Anisovich+1996}. Our
analysis follows closely the method proposed in the second reference, where a detailed description and further
references can be found. Many developments in recent years, as well as current experimental and theoretical
activity by other groups, have made a new analysis worthwhile. The inputs that the dispersive analysis relies
on have been improved considerably: several groups have published accurate descriptions of the $\pi \pi$
scattering phase shifts~\cite{Ananthanarayan+2001,Descotes-Genon+2002,Kaminski+2009}, and a modern precision
measurement of $\eta \to 3 \pi$ is available~\cite{Ambrosino+2008}, which can be used in order to fix the
subtraction constants. Even more experimental data is to be expected in the near future~\cite{expEtaPoster}.
In addition, several theoretical works on this decay have appeared in the last few years (e.g.,
\cite{Bijnens+2007,Ditsche+2009,Schneider+2011,KolesarPN}). In particular, there is also an analytical
dispersive treatment~\cite{KampfPN} that was presented at this workshop.

The dispersive analysis relies on a decomposition of the amplitude into isospin amplitudes $M_I(s)$ that are
functions of one variable only,
\begin{equation}
	M(s,t,u) = M_0(s) + (s-u) M_1(t) + (s-t) M_1(u) + M_2(t) + M_2(u) - \frac{2}{3} M_2(s) \;.
	\label{eq:MIsospin}
\end{equation}
From unitarity and analyticity follow dispersion relations for the functions $M_I(s)$,
\begin{equation}
 M_0(s) = \Omega_0(s)\; \left\{ \alpha_0 + \beta_0 s + \gamma_0 s^2 + \frac{s^2}{\pi} \int \limits_{4 m_\pi^2}
^{\infty}
\frac{ds^\prime}{s^{\prime 2}} \frac{\sin \delta_0(s^\prime)\hat{M}_0(s^\prime)}{|\Omega_0(s^\prime)|
(s^\prime - s -i \epsilon)} \right\} \;,
\label{eq:dispRel}
\end{equation}
and similarly for $M_1(s)$ and $M_2(s)$. The functions $\hat{M}_I(s)$ are angular averages over all of the
$M_I$ such that the dispersion relations are coupled. The Omn\`es function is given by
\begin{equation}
 \Omega_I(s) = \exp \left\{ \frac{s}{\pi} \int \limits_{4 m_\pi^2}^{\infty} \frac{\delta_I(s^\prime)}{s^\prime
(s^\prime - s)}\; ds^\prime \right\} \;.
\end{equation}

The dispersion relations contain a total number of four subtraction constants: $\alpha_0$, $\beta_0$ and
$\gamma_0$ in the equation for $M_0$ and one more, $\beta_1$, in the equation for $M_1$. These are left free
by the dispersion relations and have to be determined otherwise. The dispersion relations are solved
numerically by an iterative procedure, starting at some initial configuration for the $M_I(s)$. If the
subtraction constants are determined by a matching to the one-loop result from $\chi$PT (as in
Ref.~\cite{Anisovich+1996}), it turns out that there is a clear deviation from the experimental result in
Ref.~\cite{Ambrosino+2008} for large $s$ (see Fig.~\ref{fig:dalitz}). Alternatively, we can determine the
subtraction constants by a combined fit to the measured Dalitz plot and to one-loop $\chi$PT around the Adler
zero of the amplitude, where the series is expected to converge best. Since the amplitude is proportional to
$Q^{-2}$, its normalisation cannot be fixed from the experimental data. For this, Chiral Perturbation Theory
remains the only source of information. The results from a preliminary fit are shown in Fig.~\ref{fig:dalitz}.
As expected, these curves agree well with the result from KLOE.

\begin{figure}[b]
 \centerline{\includegraphics[clip,width=7.25cm]{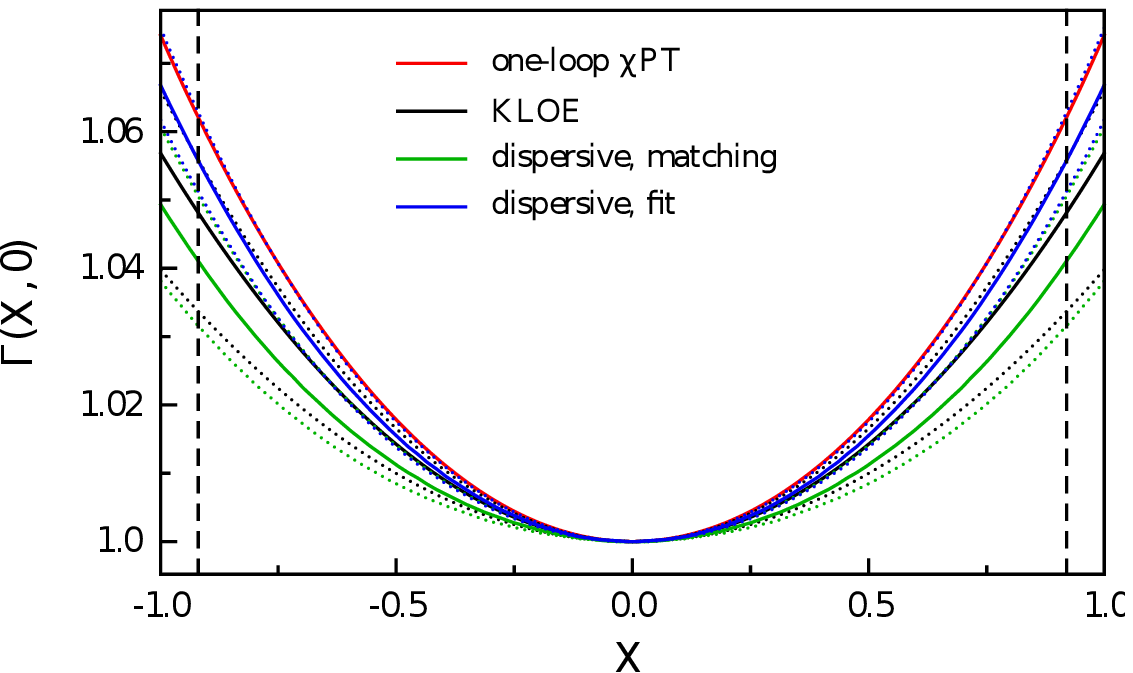}\hfill%
             \includegraphics[clip,width=7.25cm]{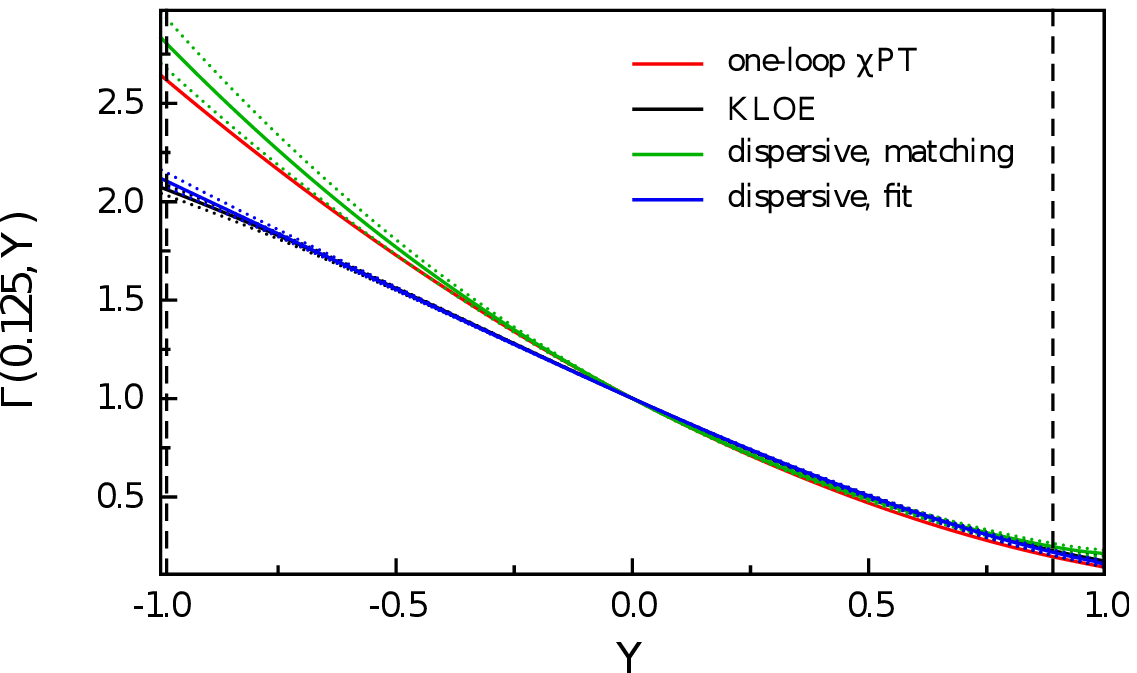}}
 \caption{\label{fig:dalitz}The squared amplitude for the decay $\eta \to \pi^0 \pi^+ \pi^-$ along the lines
$Y = 0$ (left) and $X = 0.125$ (right). $X$ and $Y$ are the usual Dalitz plot variables, the dashed lines
represent the limits of the physical region. The uncertainty band is marked by the dotted lines. One can
clearly see the disagreement between the dispersive result that relies only on one-loop $\chi$PT and
experiment for $Y < 0$ (which corresponds to large $s$).}
\end{figure}

Comparing the decay width that is calculated from the dispersive amplitude with the present PDG value of
$\Gamma_c = 296 \pm 16~\mbox{eV}$, we can extract a value for $Q$. Using the subtraction constants from the
pure matching to one-loop we get $Q = 22.7 \pm 0.7$, thus updating the analysis from
Ref.~\cite{Anisovich+1996}. The preliminary result using the subtraction constants from the fit to the KLOE
data is $Q = 21.3 \pm 0.6$. In Fig.~\ref{fig:Q}, these results are compared to a number of other values for
$Q$ from the literature. Note, in particular, the rather strong disagreement with the result from the analytic
dispersive calculation by Kampf et al. The origin of the discrepancy is not yet fully understood, but it lies
in the different procedure that is used in order to fix the normalisation, since $Q$ is very sensitive to the
latter. While we fit to the one-loop amplitude along the line $s=u$ in the vicinity of the Adler zero, the
authors of Ref.~\cite{KampfPN} fit along the line $t=u$. Doing the same with our dispersive representation
results in a very strong violation of the Adler zero and is thus not a viable course.

For the slope parameter in the neutral channel, we get $\alpha = -0.045 \pm 0.01$, in only mild disagreement
with the PDG average. This value is in excellent agreement with the finding of Kampf et al., again hinting
that the discrepancy in $Q$ is not due to the shape but to the normalisation of the amplitude, since $\alpha$
is independent thereof.

\begin{figure}[t]
  \centerline{\includegraphics[width=0.77\textwidth]{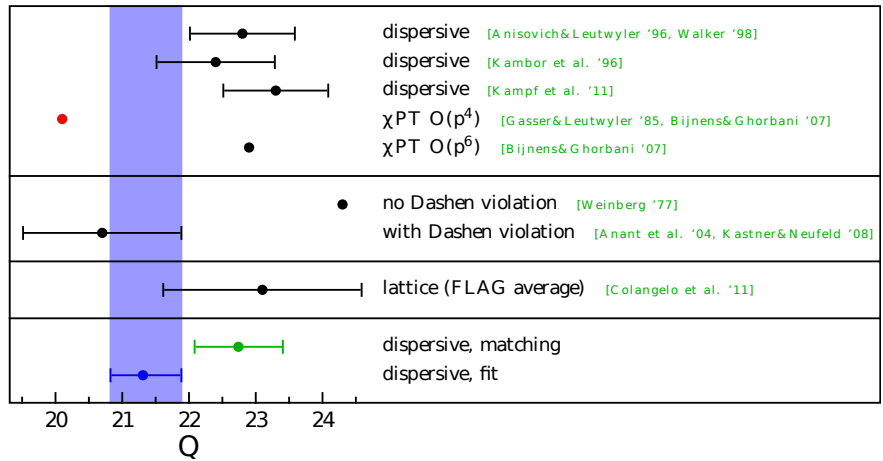}}
  \vspace{-3pt}\caption{\label{fig:Q}Comparison of several results for the quark mass ratio $Q$. The two last
dots represent our preliminary results.}
\end{figure}

Note that all results presented here are preliminary. Last refinements on the numerical analysis are in
progress and the final results will be made available soon.

\section*{Acknowledgement}\vspace{-8pt}
I would like to thank the organisers for an interesting workshop and Johan Bijnens, Karol Kampf, Bastian
Kubis, Sebastian Schneider, and Martin Zdr\'ahal for discussions. This work is supported by the Swiss Science
Foundation and in part by the European Community-Research Infrastructure Integrating Activity ``Study of
Strongly Interacting Matter'' (HadronPhysics2, Grant Agreement n. 227431) and the Swedish Research Council
grants 621-2008-4074 and 621-2010-3326.

\coltoctitle{\boldmath Prague-Lund-Marseille (Analytical) Dispersive Approach to $\eta\to3\pi$}
\authortochead{M.~Zdr\'ahal}
\label{ZM}

\setcounter{chapter}{1}
\setcounter{section}{0}
\maketitle

\vspace{2ex}%
\Section{Introduction}
As is obvious from the number of contributions dealing with $\eta\to3\pi$ decays presented in this workshop,
these processes in recent years attracted and still attract both theoreticians and experimentalists. The main
reasons are the following. These processes proceed via isospin breaking effects. Moreover, since the
electromagnetic contribution to them is strongly suppressed, to a good approximation they are proportional to
$(m_u-m_d)$ mass difference and thus by comparing the result of measurement of their decay rate with the
theoretically predicted ones we have a direct method for determination of this quantity, in our work appearing
as the quark mass ratio $R=\frac{m_s-\hat{m}}{m_d-m_u}$, where $\hat{m}=\frac12(m_u+m_d)$. Secondly, recent
measurements (cf.~e.g.~\cite{prakhov}) of the neutral decay $\eta\to3\pi^0$ start to be precise enough to
observe the effect of cusp occurring in this decay. Finally, there is still a theoretical challenge to explain
satisfactorily the observed discrepancy between the experimentally measured Dalitz plot parameters describing
the energy dependence of the amplitudes and their values predicted by the two-loop computation in the
framework of chiral perturbation theory (ChPT)~\cite{BG}.

This has inspired various studies exploring possible explanations of the discrepancy~\cite{vysvetleni}, namely
higher order final state rescatterings, influence of slow convergence of $\pi\pi$ scattering or of
$\eta\to3\pi$ amplitude itself, unexpectedly large electromagnetic contributions, effects of resonances or
with that connected possibility of incorrectly determined next-to-next-to-leading-order (NNLO) low-energy
constants of ChPT $C_i$s. On the experimental side there is substantial activity as well (for instance KLOE
and WASA-at-COSY at this workshop~\cite{experimenty} promised new data on charged decay, for which there
exists currently only one precise analysis on at least 4 Dalitz plot parameters,~\cite{KLOE}).

The complication of those analyses which are trying to avoid the above mentioned possible problems of ChPT by
using alternative approaches is that currently there exists no such alternative containing explicitly quark
masses. This means that even if one found using them a correct description of the energy dependence of the
$\eta$ decays, in order to extract any information on the quark masses (i.e.\ to fix the normalization), one
unavoidably needs to match it back to ChPT. In the case one used in his approach different assumptions than
those of ChPT, this matching brings another sources of errors into the game.

\vspace{3ex}%
\Section{Analytic dispersive parametrization}
For the task of finding the correct value of $R$ taking all this into account, we have employed in~\cite{KKNZ}
our analytic dispersive parametrization valid to two-loop order, which is constructed using just basic
assumptions of quantum field theories together with some observed hierarchy of various contributions to the
amplitude and therefore can reproduce very well also the NNLO chiral amplitude. It is possible to include full
isospin breaking effects connected with different masses of charged and neutral pions but at the current level
of precision we work in the limit $m_{\pi^\pm}= m_{\pi^0}$. In that limit the neutral decay is connected with
the charged one and the parametrization of the charged decay is in this case in the form
\begin{equation}\begin{split}
M_{x}(s,t,u)&=A_x M_\eta^2+B_x(s-\bar{s})+C_x(s-\bar{s})^2+D_x[(t-\bar{s})^2+(u-\bar{s})^2]+E_x(s-\bar{s})^3\\
&+F_x[(t-\bar{s})^3+(u-\bar{s})^3]+U(s,t,u),
\end{split}
\end{equation}
where the unitarity part $U(s,t,u)$ contains the same parameters $A_x,B_x,$ $C_x,D_x$ together with a few
additional parameters describing $\pi\pi$ scattering (which are held fixed) and is given analytically.

Using this parametrization and the data for charged decay from KLOE~\cite{KLOE}, we perform the following two
distinct analyses.

\vspace{3ex}%
\Section{\boldmath Analysis I: Correcting ChPT $\eta\to3\pi$ result}
By comparing the values of Dalitz plot parameters from the experimental determination with their predictions
from NNLO ChPT, we have found that despite the discrepancy for the individual parameters, they fulfill two
relations which are valid in the case the imaginary part of the chiral amplitude is determined correctly (the
third relation which is simply the value of the neutral parameter $\beta$ has not been experimentally measured
yet). This means that the data indicate the possibility that all the difference between the physically
measured amplitude and the one stemming from the ChPT computation can be described by a small real polynomial,
i.e.~it can also be included into the values of the parameters $C_i$. This inspired our first analysis, where
we have added such polynomial to our parametrization of the ChPT amplitude and fitted it from the KLOE data.
From that one could extract constraints on such phenomenological values of $C_i$ and more importantly for our
task we have obtained corrected value of $R=37.7\pm2.9$, where the error is estimated from the (slow)
convergence of chiral expansion in the first leading orders.

\vspace{3ex}%
\Section{\boldmath Analysis II: Direct fit to experimental $\eta\to3\pi$ data}
Our second analysis resigned on explaining the origin of the Dalitz plot discrepancy and instead assumed that
independently on the way one obtained the correct amplitude from QFT computations, it would fulfill the
general properties we have employed in the construction of our parametrization, i.e.~it will be reproduced
well by our parametrization. We have therefore fitted the experimental data on the parametrization.

As was pointed out in the introduction, in order to obtain any information on the quark masses, we need to
fix the normalization of the amplitude by matching to ChPT. Thanks to the fact we have analytic expressions,
it is sufficient to find just one point where we perform the matching, i.e.~ the point where the chiral
expansion of the amplitude converges fast. However, since there is no theoretical argument that would choose
one such point for the matching, we have used the assumption that if such point exists, it will not be an
isolated point but a region where the behavior of the NNLO chiral amplitude and the behavior of the amplitude
fitted from data would be similar and also various chiral orders behave in that region well. We have found a
prescription for such matching fulfilling these requirements and obtained as a result of this analysis
$R=37.8\pm3.3$, where the error is fully dominated by the error of the fit to KLOE data.

\vspace{3ex}%
\Section{Conclusions}
We have used two analyses based on different assumptions, which can be however fulfilled simultaneously by
the genuine physical amplitude. Since both of the analyses lead to the compatible values for $R$, we expect
that they provide us with some information about the amplitudes and also that the determined value of $R$ is
reasonable.

In~\cite{MZ} we have shown how one can combine this determination with the isospin symmetric studies on
lattice and those using sum-rules techniques in order to obtain individual quark masses. For example using
FLAG~\cite{FLAG} values, we obtain
\begin{equation}
m_u=2.23(13)\,\mathrm{MeV}, \qquad\qquad m_d=4.63(16)\,\mathrm{MeV},
\end{equation}
where these values are taken in the $\overline{MS}$ renormalization scheme at $\mu=2\,\mathrm{GeV}$.

Let us conclude with the comment that it will be very important in the forthcoming experimental measurements
of the charged $\eta\to3\pi$ decay to verify the validity of our assumptions since the current analysis is
based on only one experimental information on 5 Dalitz plot parameters which is at our disposal~\cite{KLOE}.

\coltoctitle{\boldmath $\eta\to$3$\pi$ in Resummed $\chi$PT: Dalitz Plot Parameters}
\authortochead{M.~Koles\'{a}r}
\label{KM}

\setcounter{chapter}{1}
\setcounter{section}{0}
\maketitle

We analyze the Dalitz plot parameters of $\eta\to3\pi$ decay in the framework of resummed chiral
perturbation theory. This approach allows us to keep the uncertainties in the NNLO and higher orders under
better control and estimate their influence. The compatibility of assumption of reasonably small higher order
remainders with experimental data and NNLO standard $\chi$PT results is investigated. We calculate the effect
of lowest order resonance exchange and $\pi\pi$ rescattering bubble corrections up to three loops.

The Dalitz plot parameters are usually defined as:

\vspace{-0.3cm} \begin{equation} \hspace{-1cm}
	\eta\to\pi^0\pi^+\pi^-:\,\ |A|^2 = A_0^2 (1+ay+by^2+dx^2+\dots)\quad
\end{equation}
\vspace{-0.5cm} \begin{equation} \hspace{-3.2cm}
	\eta\to 3\pi^0:\hspace{0.9cm} |\overline{A}|^2 = \overline{A}_0^2(1+\alpha z+\dots)
\end{equation}
\noindent%
where $x\,\sim\,u-t$, $y\,\sim\,s_0-s$, $z\,\sim\,x^2$+$y^2$ and $s_0$ is the Dalitz plot center
$s_0=1/3(M_\eta^2$+$3M_\pi^2)$. There is recent experimental information
available~\cite{Cr.Barrel+,Cr.Barrel,KLOE+,KLOE,Cr.Ball,CELSIUS,COSY,MAMI-B,MAMI-C}, results from
collaborations producing values for both channels can be found in table~\ref{tab1}, along with NNLO $\chi$PT
results~\cite{BijnEta}. 

As can be seen, a discrepancy can be suspected for the charged decay parameter $b$ and the neutral decay
parameter $\alpha$. Our aim is to try to understand whether the theory, by which we mean $\chi$PT as a low
energy representation of QCD, really does have difficulties explaining the data and if so, try to identify the
source of the problem.
\begin{table} \small
 \begin{center}
  \begin{tabular}{|c|c|c|c|c|}
    \hline \rule[-0.2cm]{0cm}{0.5cm} & $a$ & $b$ & $d$ & $\alpha$ \\
    \hline \rule[-0.2cm]{0cm}{0.5cm} Crystal Barrel \cite{Cr.Barrel+,Cr.Barrel} & $-1.22\pm0.07$ &
$0.22\pm0.11$ & $0.06\pm0.04$ & $-0.052\pm0.020$ \\
    \rule[-0.2cm]{0cm}{0.5cm} KLOE \cite{KLOE+,KLOE} & $-1.090\pm0.020$ & $0.124\pm0.012$ & $0.057\pm0.017$ &
$-0.0301\pm0.0050$ \\
    \hline \rule[-0.2cm]{0cm}{0.5cm} NNLO $\chi$PT \cite{BijnEta} & $-1.271\pm0.075$ & $0.394\pm0.102$ &
$0.055\pm0.057$ & $+0.013\pm0.032$ \\
    \hline
  \end{tabular}
  \caption{\label{tab1}Some of the available experimental and $\chi$PT results.}
 \end{center} 
\end{table}\normalsize

There is a long standing suspicion that chiral perturbation theory might posses a slow or irregular
convergence in the case of the three quark flavor series~\cite{Stern99}. An alternative approach, dubbed
resummed $\chi$PT~\cite{SternRes,Descotes}, was developed in order to express standard assumptions in terms of
parameters and uncertainty bands. The procedure can be very shortly summarized as:

\begin{itemize}	
  \item[-]  standard $\chi$PT Lagrangian and power counting \vspace{-0.2cm}
  \item[-]  possible irregularities in the expansion assumed \vspace{-0.2cm} 
  \item[-]  only expansions derived directly from the generating functional trusted \vspace{-0.2cm}
  \item[-]  explicitly to NLO, higher orders collected in remainders\vspace{-0.2cm}
  \item[-]  remainders not neglected, treated as sources of error \vspace{-0.2cm}
  \item[-]  manipulations done in non-perturbative algebraic way \vspace{-0.1cm}
\end{itemize}

Our calculation closely follows the procedure outlined in~\cite{Kolesar}, more details can be found
in~\cite{HS}. A comprehensive work is in preparation~\cite{prep}.

In accord with the method, leading order low energy constants (LECs) are expressed in terms of convenient
free parameters
\begin{equation} \hspace{-0.6cm}
	Z = \frac{F_0^2}{F_{\pi}^2}\, ,\ \
	X = \frac{2F_0^2B_0\hat{m}}{F_{\pi}^2M_{\pi}^2}\, ,\ \
	r = \frac{m_s}{\hat{m}}\, ,\ \
	R = \frac{(m_s-\hat{m})}{(m_d-m_u)}\, ,             
\end{equation}
\noindent where $\hat{m}$=$(m_u+m_d)/2$. To first order in isospin breaking the Dalitz plot parameters do not
depend on $R$. At next-to-leading order, the LECs $L_4$-$L_8$ are algebraically reparametrized using chiral
expansions of two point Green functions. The $O(p^6)$ and higher order LECs, notorious for their abundance,
are collected in a relatively smaller number of higher order remainders. 

We use several approaches to deal with the remainders. The first one is based on general arguments about the
convergence of the chiral series~\cite{SternRes}, which leads to
\begin{equation}
G\ =\ G^{(2)}+G^{(4)}+\Delta_G^{(6)},\quad \Delta_G^{(6)}\ \sim \pm 0.1 G,             
\end{equation}
\noindent where $G$ stands for any of our 2- or 4-point Green functions, which generate the remainders. This
is in principle an assumption. Hence we test the compatibility of this assumption of a reasonably good chiral
convergence of trusted quantities with experimental data in a statistical sense.

The framework of resummed $\chi$PT is well suited to include additional information about higher orders from
various sources~\cite{Kolesar}. An independent estimate of the remainders can on one side be an important
check of the validity of the statistical remainder estimate or could try to explain any deviations from this
assumption. Here we employ a specific higher order calculation, namely 3-loop bubble contributions to final
state $\pi\pi$ rescattering and an estimate of the influence of the lowest lying resonances using resonance
chiral theory~\cite{Ecker}.

Using the approach outlined above, we have found the experimental values consistent with the assumption of
small higher order remainders for the charged decay parameters. We cannot confirm the suspected discrepancy in
the case of parameter $b$, as small uncertainties in higher orders could accommodate the difference.
	
In the case of the neutral decay parameter $\alpha$, we confirm the positive-sign prediction of $\chi$PT  and
find the experimental value incompatible with the assumption of good convergence properties in the center of
the Dalitz plot for any scenario of spontaneous chiral symmetry breaking controlled by the free parameters
$X$, $Z$ and $r$.

Contributions from resonance exchange have turned out to be small. On the other hand, $\pi\pi$ rescattering
might explain the discrepancy in $\alpha$, especially for low values of the pseudoscalar decay constant in the
chiral limit. That could signal a failure of convergence of chiral series already at around 500MeV.

\begin{figure}[htb]
  \hspace*{-0.3cm}
  \epsfig{figure=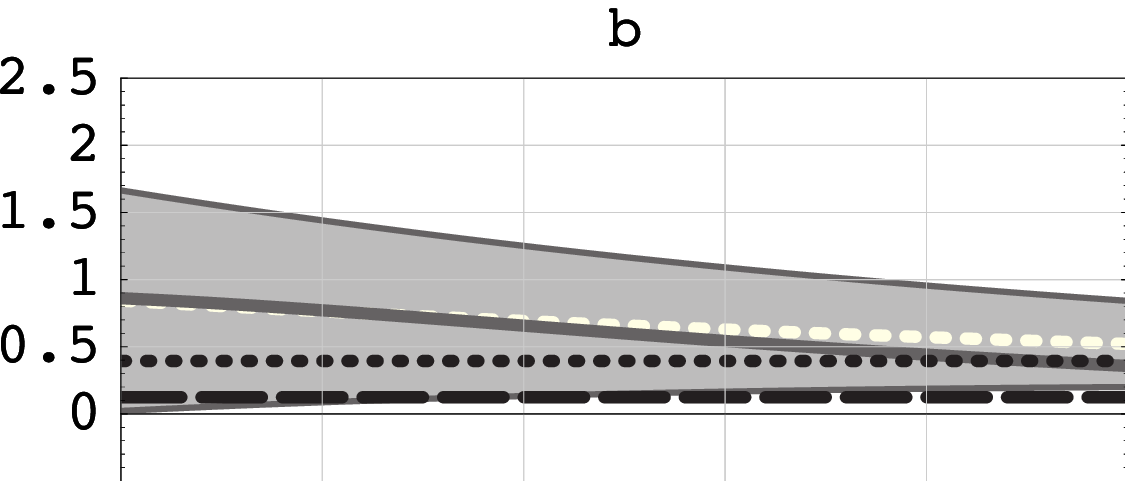,width=0.4865\textwidth}\hspace{0.02cm}
  \epsfig{figure=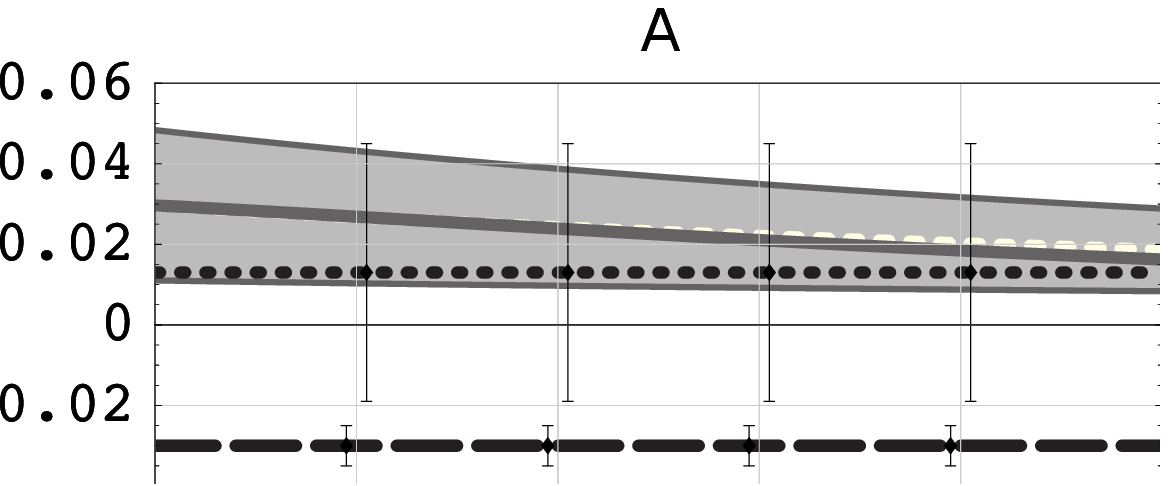,width=0.515\textwidth}\\
  \hspace*{-0.25cm}
  \epsfig{figure=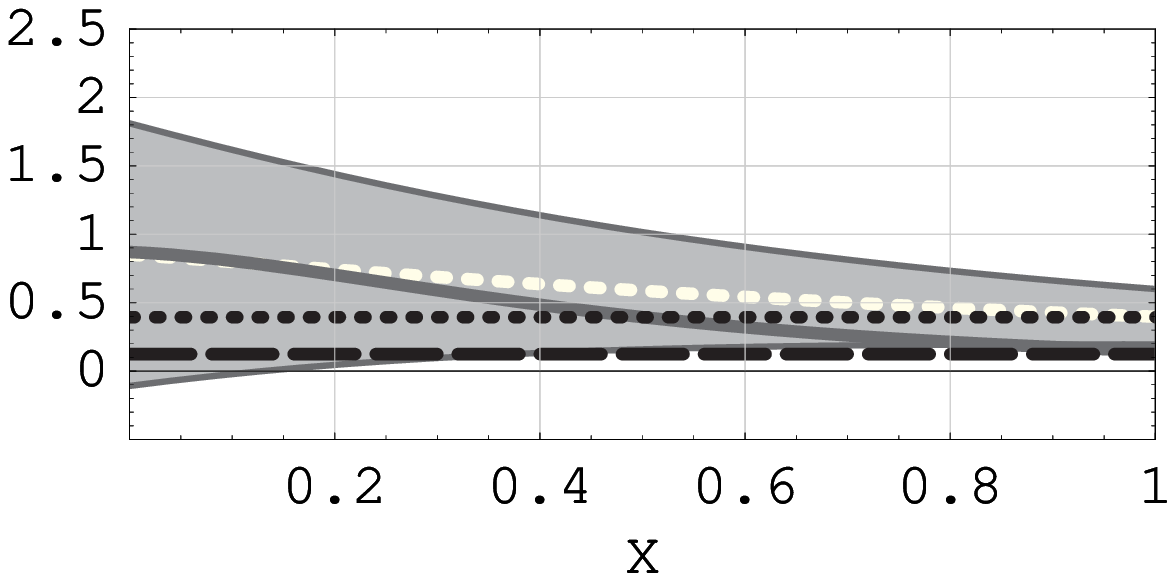,width=0.49\textwidth}
  \epsfig{figure=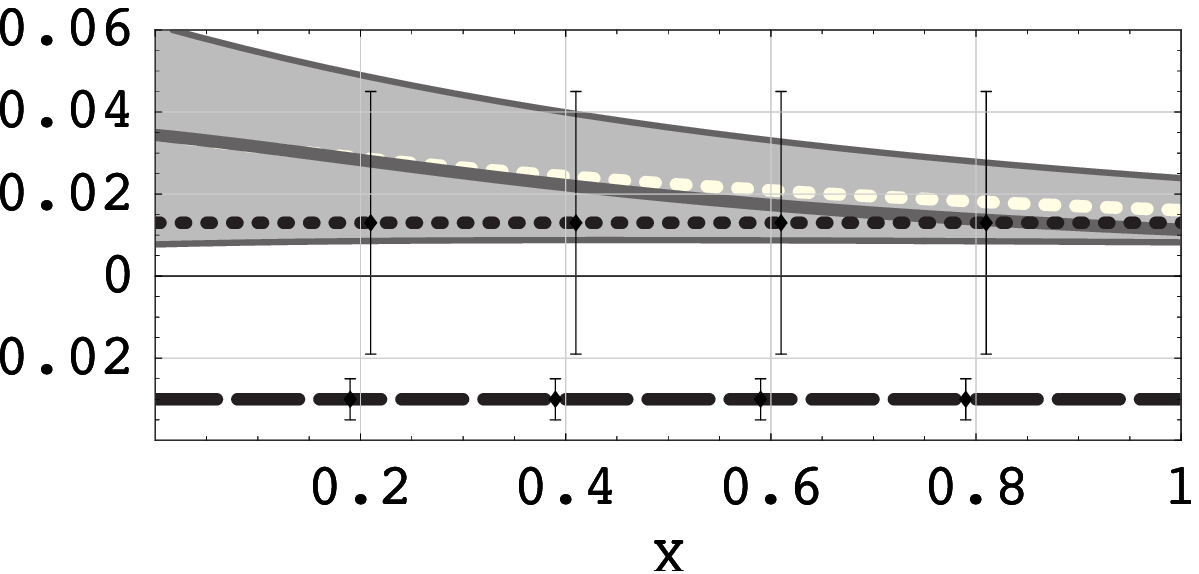,width=0.52\textwidth}
  \caption{\label{fig1}Dalitz plot parameters $b$ and $\alpha$ at $r$=25: top $Z$=1, bottom $Z$=0.5 . 
Horizontal lines: dashed - KLOE measurement \cite{KLOE+}; dotted - NNLO $\chi$PT \cite{BijnEta}.
Light band - statistical remainder estimate. Dark solid line - result including lowest lying
resonance multiplets.}
\end{figure}

\begin{figure}[htb]
  \hspace*{-0.3cm}
  \epsfig{figure=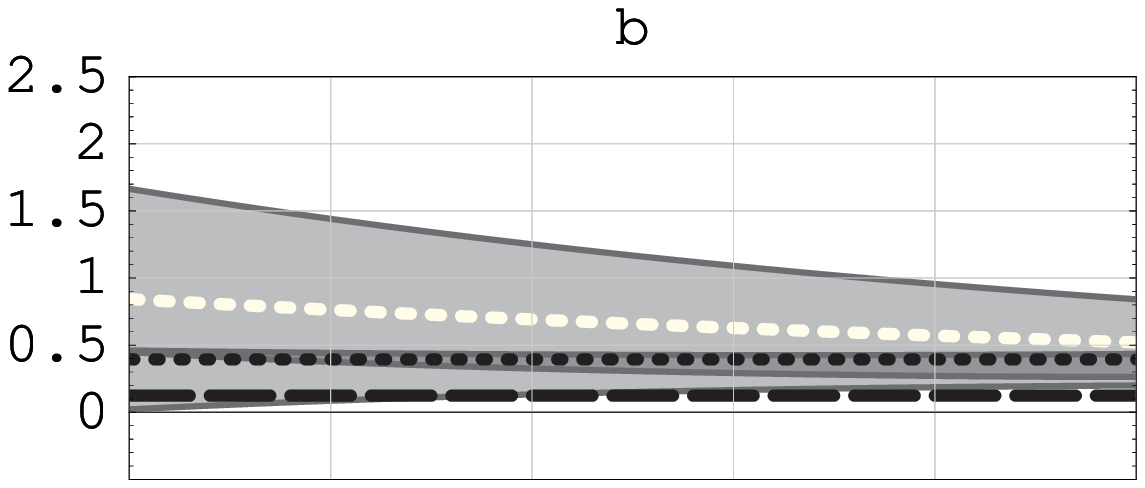,width=0.4865\textwidth}\hspace{0.02cm}
  \epsfig{figure=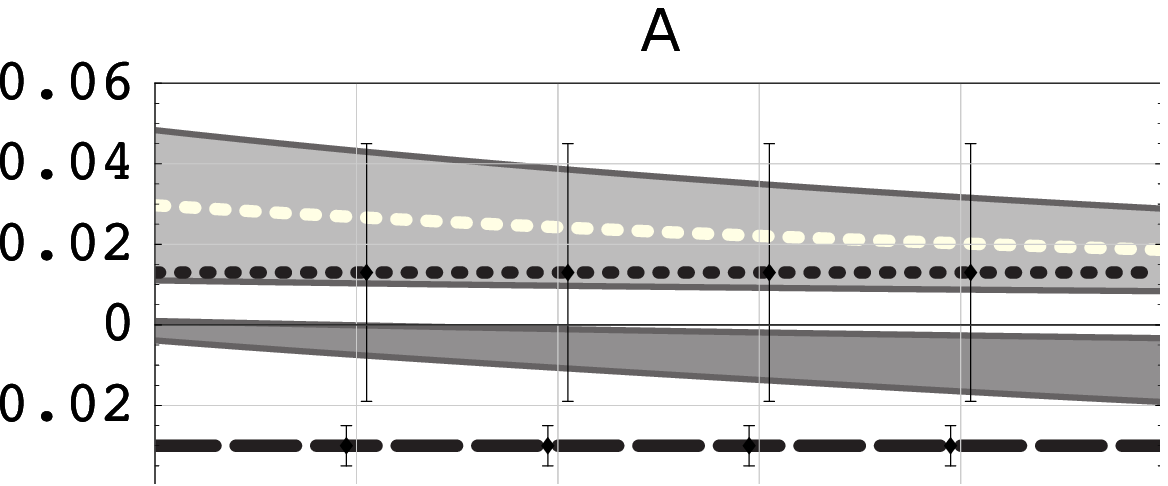,width=0.515\textwidth}\\
  \hspace*{-0.25cm}
  \epsfig{figure=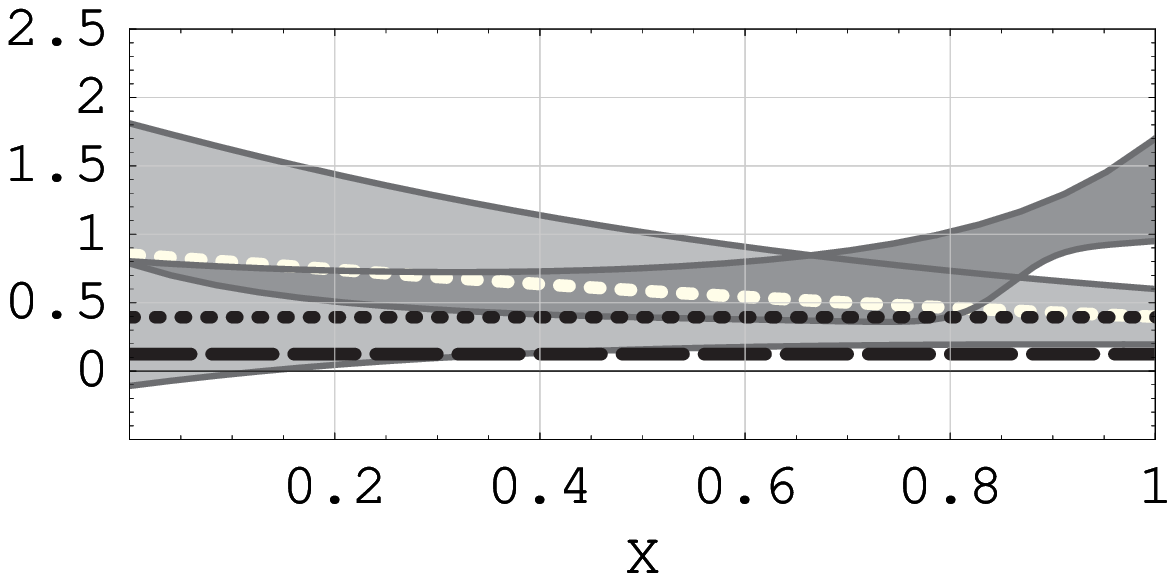,width=0.49\textwidth}
  \epsfig{figure=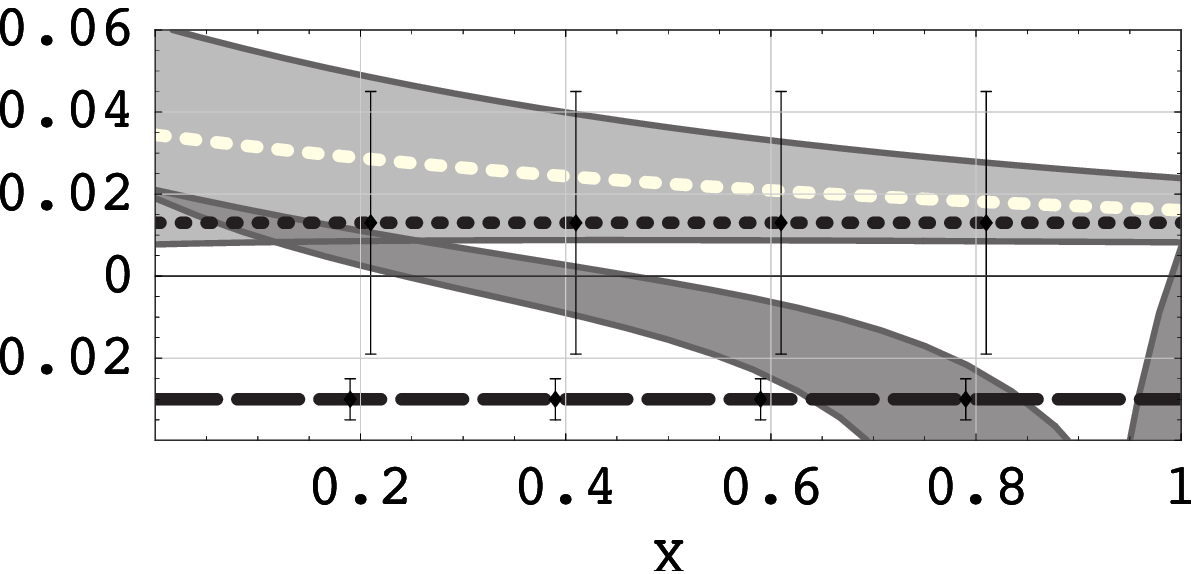,width=0.52\textwidth}
  \caption{\label{fig2}Dalitz plot parameters $b$ and $\alpha$ at $r$=25: top $Z$=1, bottom $Z$=0.5 .
Horizontal lines: dashed - KLOE measurement \cite{KLOE+}; dotted - NNLO $\chi$PT \cite{BijnEta}. Light
band - statistical remainder estimate. Dark band - result including $\pi\pi$ rescattering, depending on
ren.scale $\mu$=0.5$\div$1GeV}
\end{figure}

\coltoctitle{\boldmath Model-Independent Approach to $\eta\to \pi^+\pi^-\gamma$ and
$\eta^\prime\to\pi^+\pi^-\gamma$}
\authortochead{F.~Stollenwerk}
\label{SF}

\setcounter{chapter}{1}
\setcounter{section}{0}
\maketitle

\vspace{-4ex}\Section{Introduction}
We present a new, model-independent method \cite{ourpaper} to analyze radiative decays of mesons to a vector,
isovector pair of pions of invariant mass square below the first significant $\pi\pi$ threshold in the vector
channel. It is based on a combination of chiral perturbation theory and dispersion theory. This allows for a
controlled inclusion of resonance physics without the necessity to involve models, like vector meson
dominance, explicitly.

\vspace{-2ex}\Section{The approach}
For the radiative decays at hand, there is a significant deviation between the predictions~\cite{Bijnens:1989}
of chiral perturbation theory (ChPT) and data,the source of which is mostly the non-perturbative $\pi\pi$
final state interaction. Those are universally collected in the pion vector form factor $F_V(s_{\pi\pi})$,
whose analytic properties can be exploited in the form of a dispersion relation to give 
\begin{equation}
F_V(s_{\pi\pi}) = \exp \left( {\textstyle \frac{1}{6} } s_{\pi\pi} \langle r^2
\rangle + \frac{s_{\pi\pi}^2}{\pi}
\int_{4m_\pi^2}^\infty ds
\frac{\delta_{11}(s)}{s^{2}(s-s_{\pi\pi}-i\epsilon)} \right) \ ,
\label{eq:omnessub}
\end{equation}
where $\langle r^2 \rangle$ and $\delta_{11}(s)$ denote the mean square charge radius of the pion and the
elastic $\pi\pi$ phase shift in the vector--isovector channel, respectively, and $s_{\pi\pi}$ is the invariant
pion mass.

\begin{figure}[hctb]
 \centerline{\includegraphics[width=0.4\textwidth]{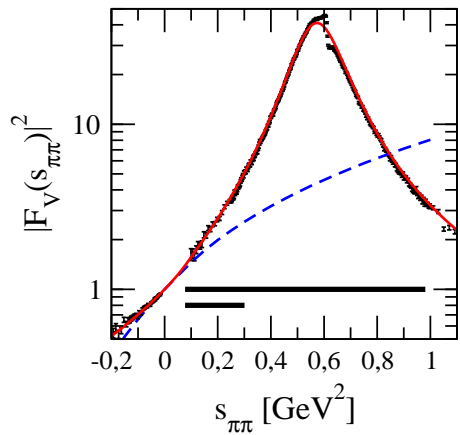}}
 \caption{The (red) solid band shows the form factor derived from Eq.\,(\ref{eq:omnessub}), the (blue) dashed
line the result from one--loop ChPT --- both with identical values for the pion radius. The time-like data are
from Refs.\,\cite{babarFF,KLOE}, whereas the space-like data are from Ref.\,\cite{Na7FF}. The short (long)
thick, horizontal bar denotes the kinematic range covered in the decay of the $\eta$ ($\eta'$). In the $\eta$
range, the form factor can be approximated by
$|F_V(s_{\pi\pi})|=1+(2.12\pm0.01)s_{\pi\pi}+(2.13\pm0.01)s_{\pi\pi}^2+(13.80\pm0.14)s_{\pi\pi}^3$.}
\end{figure}
Eq.~(\ref{eq:omnessub}) fixes the form factor only up to a multiplicative function without right-hand cuts. As
the transition amplitude features the same analytic structure as $F_V(s_{\pi\pi})$ apart from a kinematically
and chirally suppressed left-hand cut, we may factorize the amplitude according to
\begin{equation}
\frac{d\Gamma}{ds_{\pi\pi}} \sim \left | A \,
P\left(s_{\pi\pi}\right)F_V\left(s_{\pi\pi}\right)\right|^2
\ ,
\label{eq:main}
\end{equation}
and expand $P(s_{\pi\pi})$ in a Taylor series around $s_{\pi\pi}=0$:
\begin{equation}
P(s_{\pi\pi}) = 1 + \alpha^{(\prime)} s_{\pi\pi} + {\mathcal O}\left(s_{\pi\pi}^2\right)
\ .
\label{eq:poly}
\end{equation}
Assuming that the universal part $F_V(s_{\pi\pi})$ comprises {\em all} non-perturbative effects, the reaction
specific part $A \,P\left(s_{\pi\pi}\right)$ should be accessible to a perturbative treatment. Indeed, we find
that a linear polynomial (\ref{eq:poly}) with 
\begin{equation}
 \alpha        = (1.96 \pm 0.27 \pm 0.02)~{\rm GeV}^{-2} \ ; \quad
 \alpha^\prime = (1.80 \pm 0.49 \pm 0.04)~{\rm GeV}^{-2} \label{eq:alphavalues}
\end{equation}
is already sufficient to describe spectral decay data, see Fig.~\ref{fig:spectra}. In principle, $\rho-\omega$
mixing distorts the $\eta^\prime$ spectrum. If one assumes the mixing strength to be identical to that in the
vector form factor (based on the quark model the effect is much smaller), one finds the wiggly (green) line.
Clearly, the data is not yet sufficient to extract information on the magnitude of the effect.

\begin{figure}[htb]
 \centerline{\includegraphics[scale=0.5]{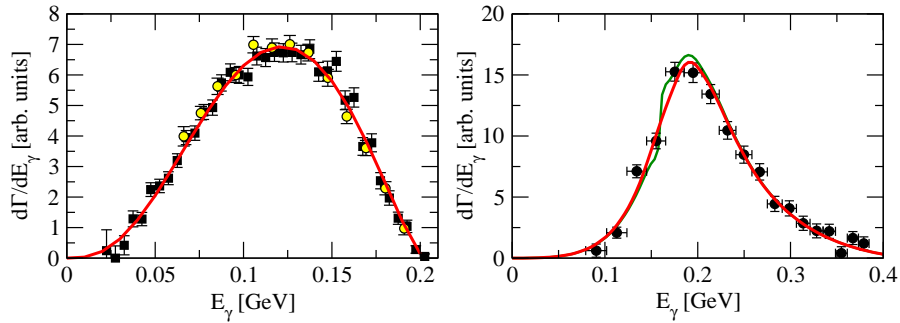}}
 \caption{\label{fig:spectra}Experimental data and error weighted fits for $\eta$ (left, data are from
Ref.\,\cite{wasa:2011} (filled squares) and Ref.\,\cite{Gormley:1970qz} (open circles)) and $\eta^\prime$
(right, data are from Ref.\,\cite{cry:1997}) to $\pi^+ \pi^- \gamma$ according to Eqs.\,(\ref{eq:main}) and
(\ref{eq:poly}) with $s_{\pi\pi} = m_{\eta^{(\prime)}} ( m_{\eta^{(\prime)}} -2 E_\gamma)$. The wiggly (green)
line in the right panel denotes the possible impact of $\rho-\omega$ mixing under the assumption that it
appears here with the same strength as in $F_V$.}
\end{figure}
The postulated decay amplitude (\ref{eq:main}) can be matched to one--loop $U(3)$ extended ChPT. Concerning
the $s_{\pi\pi}$-dependent part, this connects the slope parameters $\alpha^{(\prime)}$ to the low energy
constant $a_2$ giving
$$
a_2^\eta=9.70\pm 0.7 \ ; \qquad a_2^{\eta'} = 9.23 \pm 1.4 \ .
$$
Thus, the data sets are consistent with the assumption that the only non-perturbative part of the amplitude
originates from the pion vector form factor, and Eq. (\ref{eq:main}) provides a controlled way to disentangle
perturbative and non-perturbative effects in order to methodically extend perturbative calculations to the
resonance region. By invoking chiral Ward identity and large $N_c$ arguments, essentially half of the
parameter $\alpha^{(\prime)}$ can be physically interpreted. Further possible ways of interpretation are
subject of current research.

\coltoctitle{\boldmath Study of $\eta \to \pi^+ \pi^-\gamma$ Decay}
\authortochead{C.~Di~Donato}
\label{DDC}

\setcounter{chapter}{1}
\setcounter{section}{0}
\maketitle

\Section{Introduction}
We measured the ratio $R_{\eta}=\Gamma(\eta \to \pi^+\pi^-\gamma)/\Gamma(\eta \to \pi^+\pi^-\pi^0)$, analyzing
an integrated luminosity of 558 pb$^{-1}$, collected by KLOE at the DA$\Phi$NE $e^+ e^-$ collider. The $\eta
\to \pi^+\pi^-\gamma$ process is supposed to proceed  both via a resonant contribution, mediated by the $\rho$
meson, and a non resonant direct term, connected to the box anomaly. The presence of the direct term
affects the partial width value.\\
The Chiral Perturbation Theory (ChPT) provides accurate description of interactions and decays of light
mesons~\cite{GasserLut}. The decays $\eta \rightarrow \pi^+ \pi^- \gamma$ and $\eta^{\prime} \rightarrow \pi^+
\pi^- \gamma$ are expected to get contribution from the
anomaly accounted for by the Wess Zumino Witten (WZW) term into the ChPT Lagrangian~\cite{Ben2003}. According
to effective theory~\cite{Ben2003} the contribution of the direct term should be present together with VMD. In
case of $\eta \rightarrow \pi^+ \pi^- \gamma$ the $\rho$ contribution is not dominant, this makes the partial
width sensitive to the
presence of the direct term, while in case of $\eta^{\prime} \rightarrow \pi^+ \pi^- \gamma$ the partial width
is dominated by the resonance and
the direct term effect should be visible in the  dipion invariant mass distribution. The present world
average of the $\eta \rightarrow \pi^+ \pi^- \gamma$ partial width, $\Gamma(\eta \rightarrow \pi^+ \pi^-
\gamma) = (60 \pm 4)$~eV~\cite{PDG10}, provides strong
evidence in favour of the box anomaly, compared with value expected with and without the direct term,
respectively $(56.3 \pm 1.7)$~eV and $(100.9 \pm 2.8)$~eV~\cite{Ben2003}. Recently CLEO~\cite{Lopez07} has
measured the ratio $R_{\eta}= 0.175\pm 0.007_{stat} \pm 0.006_{syst}$, which differs by more  than $3\sigma$
from the average result of previous measurement~\cite{Gormley,Layter}, $R_{\eta}= 0.207\pm
0.004$~\cite{PDG06}. Here we present a new measurement with the highest statistics and the smallest systematic
error ever achieved. 

\Section{Event selection}
The analysis has been performed using 558 pb$^{-1}$ of the 2004-2005 data set acquired by KLOE at the
DA$\Phi$NE $e^+ e^-$ collider ($\sqrt{s} \simeq 1.02$ GeV). The final state under study is
$\pi^+\pi^-\gamma\gamma$, since the $\eta$ mesons are produced together with a monochromatic recoil photon
($E_{\gamma} = 363$ MeV) through the radiative decay $\phi \rightarrow \eta \gamma$. In the considered data
sample about $\simeq
25 \times 10^{6}\ \eta$'s  are produced. The main background comes from $\phi \to \pi^+\pi^-\pi^0,  \pi^0 \to
\gamma \gamma$ decaying to the same final state. Other backgrounds are $\phi \to \eta \gamma\to
\pi^+\pi^-\pi^0\to\pi^+\pi^- 3\gamma$ with one photon lost, and $\phi\to\eta\gamma,  \eta\to e^+ e^-\gamma$
when both electrons are mis-identified as pions.\\
As first step of the analysis, a preselection is performed, requiring at least two tracks with opposite charge
pointing to the interaction point (IP) and at least two neutral clusters in time (not associated to any
track), having energy $E_{cl} \ge 10$ MeV and polar angle in the range $(23^\circ -157^\circ)$. Tracks are
sorted according to the distance of the point of closest approach from the IP. The first two tracks are
selected. We require the most energetic cluster ($\gamma_{\phi}$) to have $E_{cl}>250$ MeV and we identify it
as the photon recoiling against the $\eta$ in the $\phi \to \eta \gamma$ decay. Moreover we ask for
$\gamma_{\phi}$ inside the calorimeter barrel ($55^{\circ}-125^{\circ}$), to avoid effects of cluster merging
between barrel and end-caps of the calorimeter. Other cuts are imposed to clean up the sample; cut on
cluster-track collinearity and identification by time of flight (TOF) are used to reject electrons. The cut
effectively rejects Bhabha background and other processes with electrons in the final state. To select $\eta$
decays we exploit the $\phi\to\eta\gamma$ two body decay kinematic  computing the $\gamma_{\phi}$
energy, using only the $\gamma_{\phi}$ polar angle:
\[
\vec{p}_{\phi}=\vec{p}_{\eta}+\vec{p}_{\gamma} \qquad
E_{\gamma_{\phi}}=
\frac{m_{\phi}^2-m_{\eta}^2}{2(E_{\phi}-|\vec{p}_{\phi}|cos\varphi)}
\] 
where $\varphi$ is the angle between the average $\phi$-meson momentum measured run by run with high accuracy
and $\gamma_{\phi}$. This allows us to improve the energy measurement of the recoil photon to $0.1\%$. We can
determine the direction of the photon from $\eta$ decay using $\phi$ and $\pi$-mesons information:
\[
\vec{p}_{\gamma_{\eta}} = \vec{p}_{\phi}- \vec{p}_{\pi^+} -
\vec{p}_{\pi^-}-\vec{p}_{\gamma_{\phi}}
\]
The photon direction is compared with the direction of each neutral cluster, $\Delta \varphi =
\varphi^{\mathrm{clu}} - \varphi_{\gamma_{\eta}}$. If no cluster within $\Delta \varphi < 8.5^{\circ}$ is
found the event is rejected. The cluster with the minimum value of $\Delta\varphi$ is selected for further
analysis. In order to reject the $\phi \to \pi^+ \pi^- \pi^0$ background, the angle between the two photons in
the $\pi^0$ reference frame,  evaluated using the $\phi$ and the $\pi$-mesons momenta,  is calculated and
rejected with an angular cut $\varphi_{\gamma\gamma}^{\pi^+\pi^-\gamma} <165^{\circ}$; in order to reduce the
systematics the angle is evaluated in the transverse plane. Finally we select events requiring $539.5$ MeV $<
M_{\pi^+ \pi^- \gamma} < 554.5$ MeV (Fig.~\ref{pppresults1}).

\begin{figure}[htb]
\centering \includegraphics*[width=0.75\textwidth]{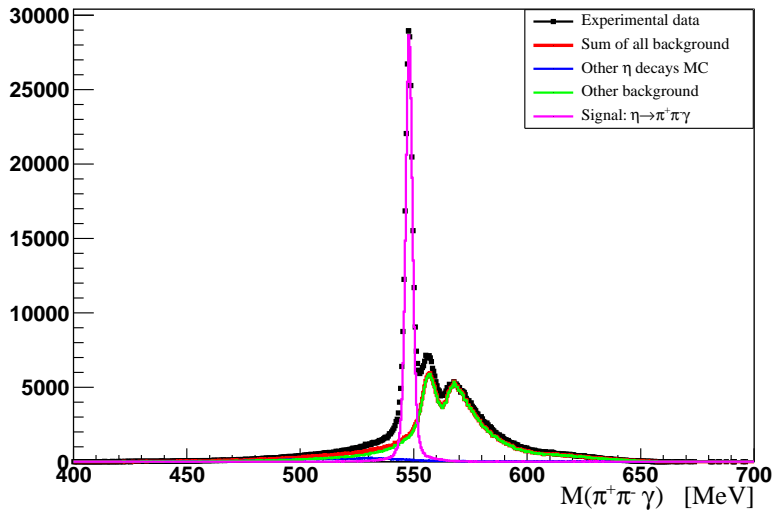}
\caption{The $\pi^+ \pi^- \gamma_{\eta}$ invariant mass distribution:
  Data-MC comparison. Dots are data, Magenta is MC signal $\eta \rightarrow \pi^+ \pi^- \gamma$,
  Red is all MC background contribution}
\label{pppresults1}
\end{figure}

\Subsection{\boldmath Normalization Sample: $\eta \to \pi^+ \pi^- \pi^0$}
The process $\phi \to \eta \gamma$ with $\eta \to \pi^+ \pi^- \pi^0$ represents a good control sample, due to
the similar topology. Moreover the ratio $\Gamma(\eta \to \pi^+ \pi^- \gamma)/\Gamma(\eta \to \pi^+
\pi^-\pi^0)$ is not affected by the uncertainties on the luminosity, the $\phi \to \eta \gamma$ partial width
and the $\phi$ production cross section cancel in the ratio. We use the same preselection as for the $\eta \to
\pi^+ \pi^- \gamma$ signal and calculate the missing four-momentum:
\[
\mathbb{P}_{miss}=\mathbb{P}_{\phi}-\mathbb{P}_{\pi^+}-\mathbb{P}_{\pi^-}-\mathbb{P}_{\gamma_{\phi}}
\]
where the variables in the formula represent the four-momenta of the $\phi$ meson and the products of the
decays. For the  $\eta \to \pi^+ \pi^- \pi^0$ signal, the missing mass peaks at the $\pi^0$ mass value
and we select events with $|M_{miss}-m_{\pi^0}|<15$ MeV. The remaining background is rejected very efficiently
by using an angular cut applied to the two photons from the $\pi^0$ decay,
$\varphi_{\gamma\gamma}^{3\pi}>165^{\circ}$; the angle is evaluated in the transverse plane. We select $N(\eta
\to \pi^+\pi^-\pi^0)= 1115805 \pm 1056$, with a selection efficiency of $\varepsilon = 0.2276\pm0.0002$ and a
background contamination of $0.65\%$.

\Section{Results}
The total selection efficiency of the $\eta \to \pi^+\pi^-\gamma$ signal is $\varepsilon = 0.2131 \pm 0.0004$.
Background contribution and the signal amount in the final sample are evaluated with a fit to the
$E_{miss}-P_{miss}$ distribution of the $\pi^{+} \pi^{-} \gamma_{\phi}$ system with the shapes from remaining
background and signal MC in the range $|E_{miss}-P_{miss}|<10$ MeV. We find $N(\eta \to \pi^+\pi^-\gamma)=
204950 \pm 450$ with a background contamination of $10\%$. Combining our results we obtain the ratio:
\begin{equation}
R_{\eta}=\frac{\Gamma(\eta \to \pi^+ \pi^- \gamma)}{\Gamma(\eta \to
  \pi^+ \pi^- \pi^0)}= 0.1838\pm0.0005_{stat} \pm 0.0030_{syst}
\end{equation}
to be compared with world average value $\Gamma(\eta \to \pi^+ \pi^-\gamma)/ \Gamma(\eta \to \pi^+ \pi^-
\pi^0)=0.202\pm0.007$~\cite{PDG10}. Our result is in agreement with a recent CLEO~\cite{Lopez07} measurement,
which differs by more  3 $\sigma$ from the average of previous results~\cite{Gormley,Layter,PDG06}.\\ 
The systematic uncertainties due to analysis cuts have been evaluated by varying the cuts on all variables and
re-evaluating the branching ratios. The relative variation for each source of systematic is in
Tab.~\ref{tab:systematics}. The total error is taken as the quadratic sum of all contributions.
\begin{table}[!t]
 \begin{center}
  {\scriptsize
    \begin{tabular}{cc}
    \hline Source of uncertainty    & Relative error \\ \hline
    $\varphi_{\gamma\gamma}^{\pi^+\pi^-\gamma} < 165^{\circ} \pm
    2^{\circ} $ & $\pm0.6\%$\\        
    $\Delta \varphi > 8.5^{\circ} \pm 2^{\circ}$ & $\pm 0.4\%$\\  
    $|M_{\pi^+ \pi^-\gamma}-M_{\eta}|< 7.5$ MeV $\pm$ 2 MeV & $\pm 0.6\%$\\  
    $|M_{miss} - M_{\pi^0}| < 15$ MeV $\pm$ 4 MeV & $\pm 0.4\%$\\  
    $\varphi_{\gamma\gamma}^{3\pi} >  165^{\circ} \pm 2^{\circ}$ & $\pm 0.1\%$\\  
    $E_{min}^{\gamma} > 10$ MeV $\pm$ 2 MeV  & $\pm 0.1\%$\\  
    $E_{clu}^{\gamma_{\phi}} > 250$ MeV $\pm$ 4 MeV  & $\pm 0.1\%$\\ 
    Fit $E_{miss}-P_{miss}$ & $\pm0.6\%$ \\ \hline 
    Total                 & $1.6\%$ \\ \hline    
    \end{tabular}}
   \caption{Summary table of systematic uncertainties.}
  \label{tab:systematics}
  \end{center}
\end{table}

The $M_{\pi^+ \pi^-}$ dependence of decay width has been parameterized in different approaches, in which VMD
has been implemented in effective Lagrangians~\cite{Ben2003,Picciotto92}. Preliminary comparison between
dipion invariant mass, with the most simple approach as from~\cite{Picciotto92}, shows a good agreement 
with data. Fit with model independent parameterizations, as from~\cite{Uppsala10}, is in progress.

\Section{Conclusions}
Using a data sample corresponding to an integrated luminosity of 558~pb$^{-1}$, we select $204950$ $\eta \to
\pi^+ \pi^- \gamma$ events and $1115805$ $\eta \rightarrow \pi^+ \pi^- \pi^0$, from the $\phi \to \eta \gamma$
decays. The corresponding
width ratio is: $ R_{\eta}= 0.1838\pm0.0005_{stat} \pm 0.0030_{syst}$ Our measurement is in agreement with the
most recent result from CLEO~\cite{Lopez07}, which is $R_{\eta}= 0.175\pm0.007_{stat} \pm 0.006_{syst}$.\\
Combining our measurement with the world average value $\Gamma(\eta \to \pi^+\pi^-\pi^0)= (295\pm
16)$~eV~\cite{PDG10}, we get  $\Gamma(\eta \to \pi^+\pi^-\gamma)=(54.2\pm 0.3)$~eV, which is in agreement with
the value expected taking into account the direct term~\cite{Ben2003}, providing a strong evidence in favour 
of the box anomaly.\\

\coltoctitle{\boldmath In Search of the Box Anomaly by Studying $\eta\rightarrow\pi^{+}\pi^{-}\gamma$}
\authortochead{D.~Lersch}
\label{LD}

\setcounter{chapter}{1}
\setcounter{section}{0}
\maketitle

\Section{Introduction \& motivation}
The decay channel $\eta\rightarrow \pi^{+} \pi^{-} \gamma$ provides the opportunity to study QCD anomalies at
the chiral limit.
\begin{figure}[htb]
  \centering
  \begin{minipage}[c]{0.5\textwidth}
    \includegraphics[width=\textwidth]{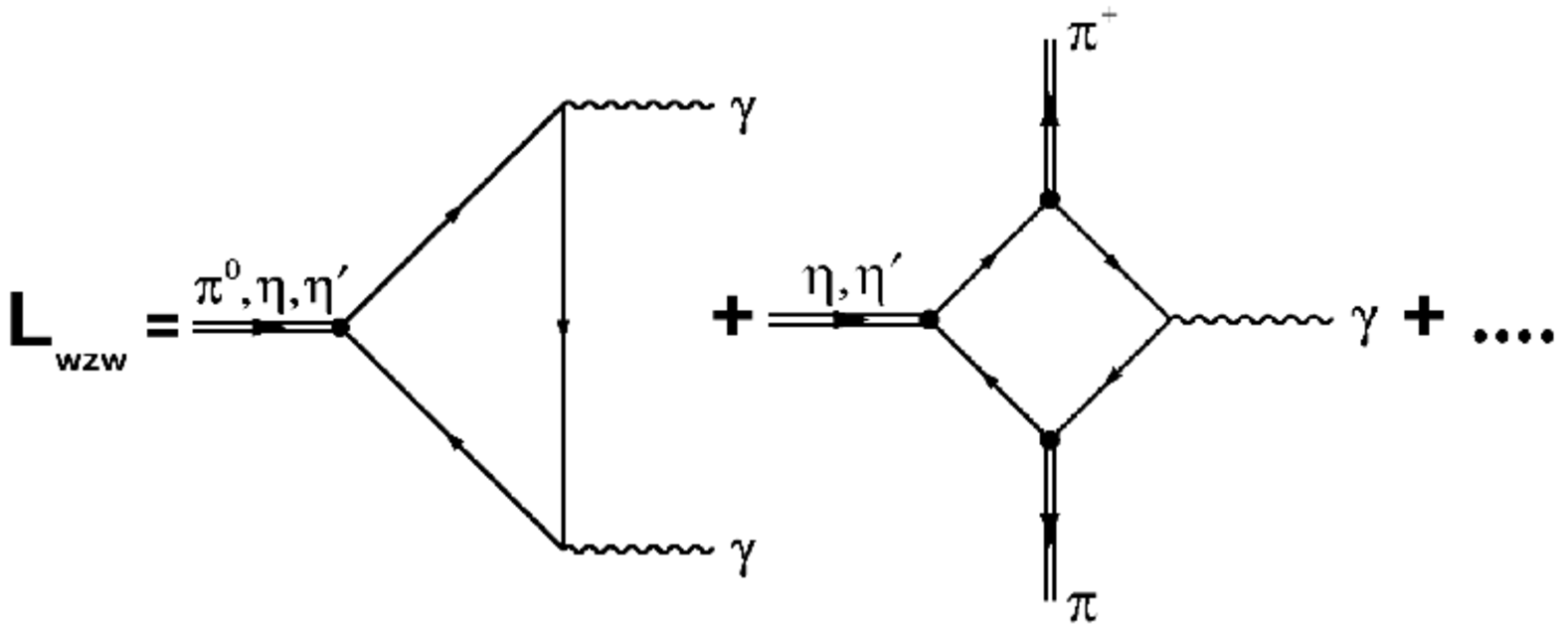}  
  \end{minipage}\hfill%
  \begin{minipage}[c]{0.45\textwidth}
    \includegraphics[width=\textwidth]{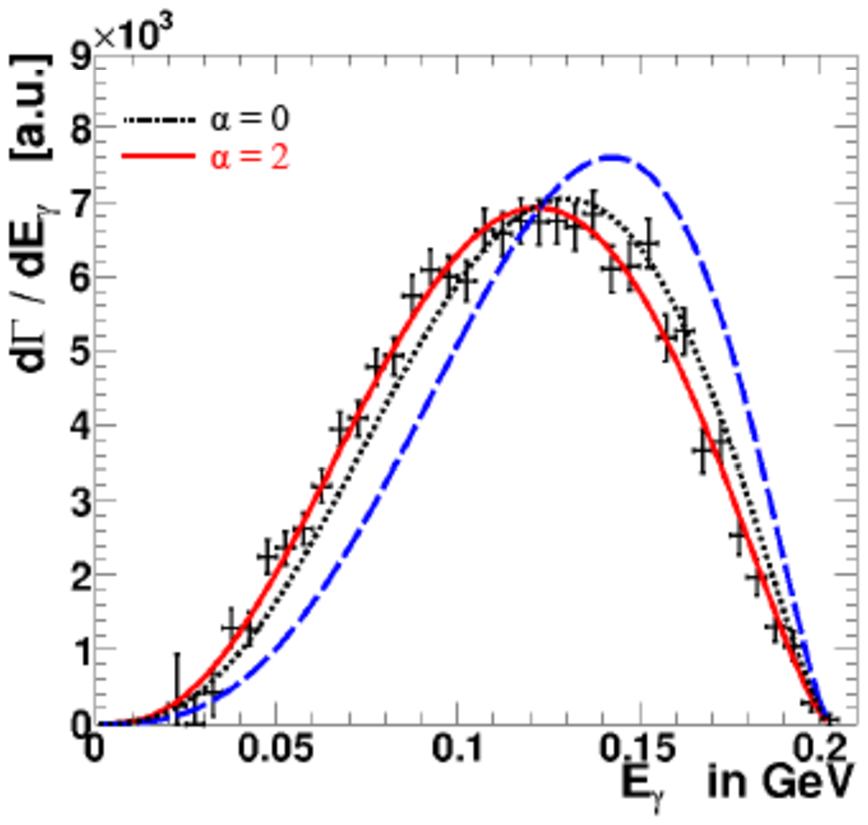}
  \end{minipage}
 \caption{\label{WZW}Left: Wess-Zumino-Witten-Lagrangian ($L_{WZW}$) including the triangle anomaly (first)
and box anomaly (second) term~\cite{ex:1}. Right: Measured $E_{\gamma}$ distribution for the reaction:
$pd\rightarrow \mathrm{^3He}[\eta\rightarrow \pi^{+} \pi^{-} \gamma]$~\cite{ex:4}. The measurement has been
performed with high statistics and was fully efficiency corrected. The measured $E_{\gamma}$ distribution is
described by multiplying the simplest gauge invariant matrix element (blue curve) with a form factor (dashed
and red curve) which depends on a single parameter $\alpha$~\cite{ex:5}.}
\end{figure}
The decay width and the shape of the $E_{\gamma}$-distribution of this channel are sensitive to the box
anomly term which is part of the Wess-Zumino-Witten-Lagrangian (see Fig.~\ref{WZW}). However, the
theoretically predicted decay width and $E_{\gamma}$-distribution do not agree with the experimental results,
if final state interactions are not included by unitarized extensions of the $L_{WZW}$. The experimental
observables for testing these extensions are (i) the branching ratio~\cite{ex:2,ex:3} or (ii) the
distribution of the single photon energy~\cite{ex:3}. A recent measurement of the photon energy (see
Fig.~\ref{WZW}) with the WASA detector can be found in~\cite{ex:4}.

The aim of this work is to measure the branching ratio and the single photon energy distribution in one
experiment using the reaction: $pp\rightarrow pp[\eta\rightarrow \pi^{+}\pi^{-}\gamma]$. The data have been
acquired during an 8 week experiment in spring 2010. At least $10^6$ $\eta\rightarrow \pi^{+}\pi^{-}\gamma$
events are expected.

The WASA detector (see poster by F. Goldenbaum) is devided into two parts: the forward detector and the
central detector.

Scattered projectiles or charged recoil particles are identified in the forward detector, a tracking
detector with a set of range hodoscopes, which uses the $\frac{dE}{E}$-method for particle identification and
reconstruction.

Decay products are measured with the central detector. Charged particles are identified by
inspecting the deposited energy as function of the momentum as measured with a tracking device in a magnetic
field. Neutral particles are detected in the calorimeter.

\Section{Data analysis}
As a first step, the reaction $\eta\rightarrow \pi^{+}\pi^{-}\pi^0$ has been analysed. On the one hand this
reaction is important for the determination of $\frac{\Gamma(\eta\rightarrow
\pi^{+}\pi^{-}\gamma)}{\Gamma(\eta\rightarrow \pi^{+}\pi^{-}\pi^0)}$, on the other hand it contributes to the
background. The $\pi^0$ is also used for monitoring the calorimeter calibration. $5\%$ of the available
preselected data have been analysed with the recent analysis chain.

The missing mass spectrum deduced from two protons is shown in Fig.~\ref{post_cuts}. The $\eta$-peak is
visible, along with background, from direct pion production. The following cuts on the invariant and missing
mass spectra of the decay products were employed to extract the $\eta$-signal from the direct pion background:
(i) The invariant mass of two photons had to be within a $0.11\,\rm{GeV}/{c^2}$ window around the $\pi^0$-mass
(ii) The missing mass deduced from two photons had to be larger than the rest mass of the two chraged pions
(iii) The missing mass squared deduced from the two charged pions had to be positive.

The energy and momentum balance (Fig.~\ref{post_cuts} right) is used to check the effect of each cut and
investigate the background contributions.
 \begin{figure}[htb]
 \centerline{\includegraphics[width=0.49\textwidth]{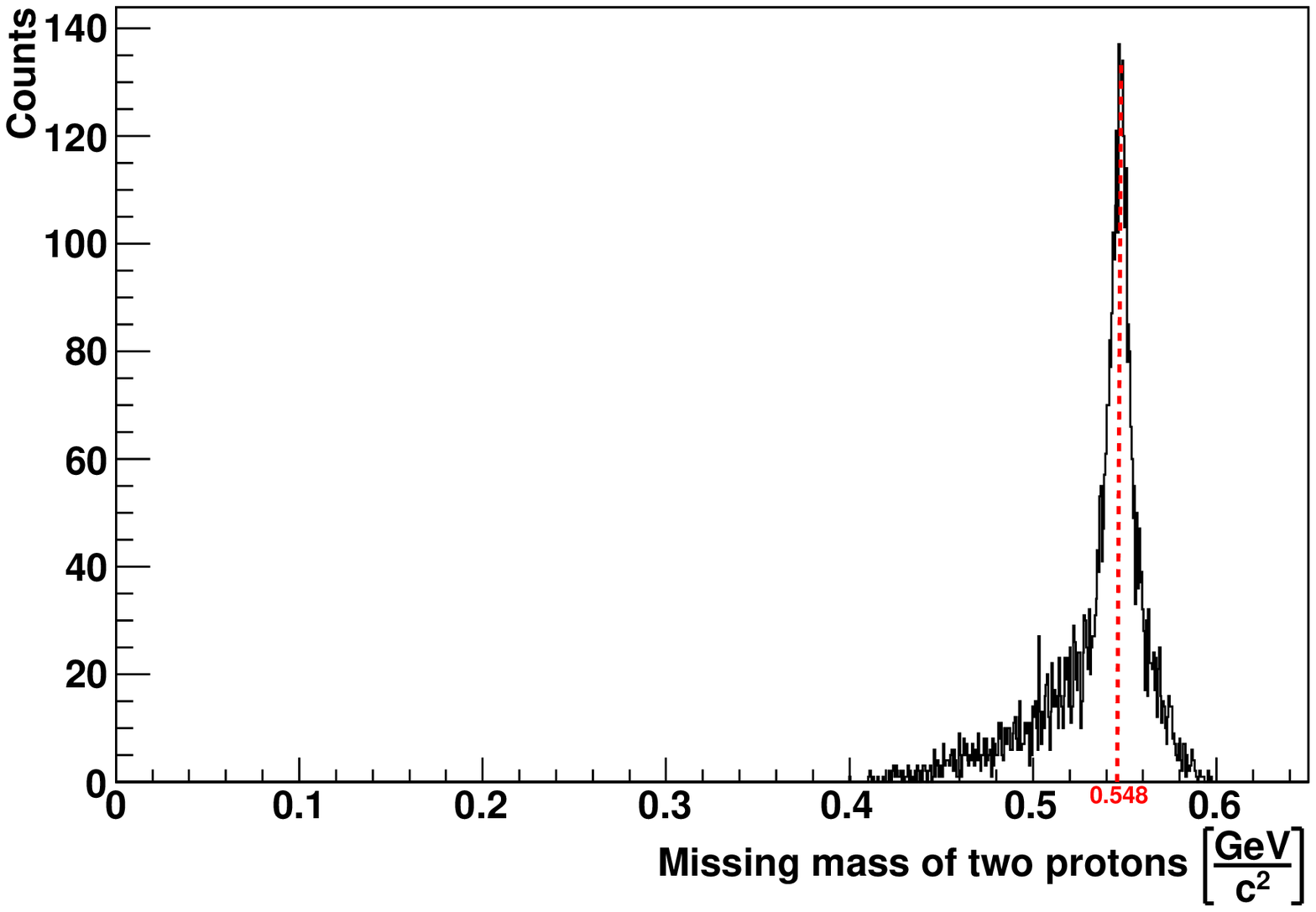}\hfill%
             \includegraphics[width=0.49\textwidth]{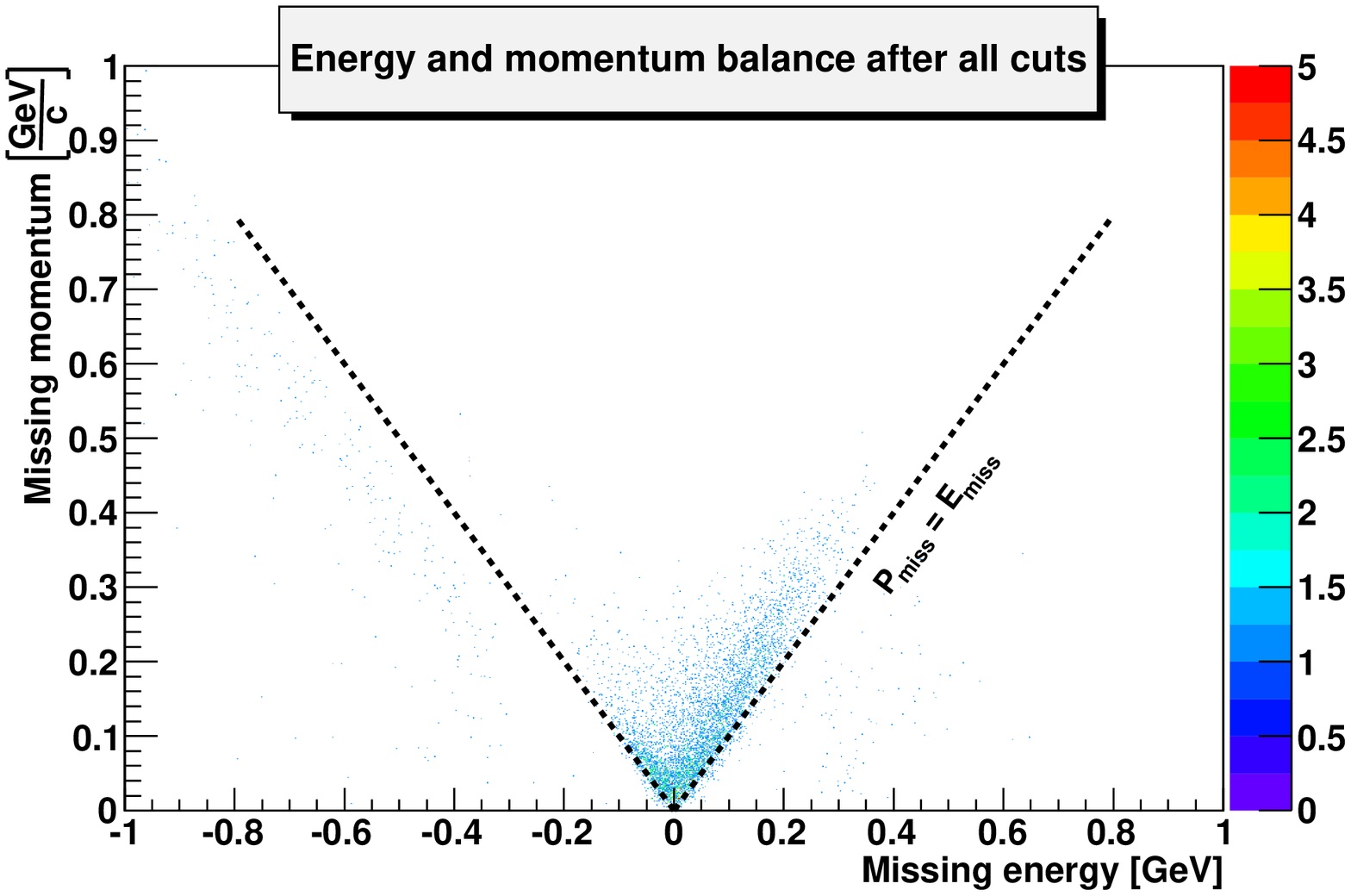}}
 \caption{\label{post_cuts}Left: Missing mass deduced from the initial state and the two final state protons.
The red line indicates the $\eta$-mass at $0.548\,\rm{GeV/c^2}$. This spectrum is obtained after proton
reconstruction and applying the cuts mentioned above. Right: Plot of the missing momentum as function of the
missing energy (derived by the momentum/energy of all initial and final particles in the reaction).}
\end{figure}

\Section{Summary \& outlook}
The $\eta$-peak is clearly visible after applying cuts on invariant and missing mass related to the decay
products. Background channels shall be identified in a second step using Monte Carlo simulations and analysing
more data runs.

The goal is to investigate the reaction $\eta\rightarrow \pi^{+}\pi^{-}\gamma$ and to determine the branching
ratio.

\section*{Acknowledgements}
This publication has been supported by the European Commission under the 7th Framework Programme through the
`Research Infrastructures' action of the `Capacities' Programme. Call: FP7-INFRASTRUCTURES-2008-1, Grant
Agreement N. 227431.

\coltoctitle{Radiative Decays of Pseudoscalars and Vectors}
\authortochead{C.~Terschl\"{u}sen}
\label{TCT}

\setcounter{chapter}{1}
\setcounter{section}{0}
\maketitle

\Section{Introduction}
Due to the running coupling constant in QCD, one cannot use perturbation theory for the low-energy regime.
Instead one can use effective theories with hadrons instead of quarks as relevant degrees of freedom. We use a
new counting scheme \cite{CS} which treats both vector mesons ($V$) and pseudoscalar mesons ($P$) as soft,
i.e. of the order of a typical momentum $q$:
\begin{eqnarray}
 m_V, \, m_P, \, \partial_\mu \sim q. \label{CS}
\end{eqnarray}
In addition, we use large-$N_c$ arguments (with $N_c$ denoting the number of colors) to suppress additional
flavor traces. We will use this counting scheme to describe radiative decays into dileptons ($l^+ l^-$). A
more detailed introduction is given in these proceedings in \cite{proc}.

\Section{\boldmath Decays $V \rightarrow P l^+ l^-$ and $P \rightarrow V l^+ l^-$}
Using the counting scheme (\ref{CS}), the leading-order Lagrangian for the decays $V \rightarrow P l^+ l^-$
and $P \rightarrow V l^+ l^-$ can be determined as
\begin{eqnarray}
 \mathcal{L}_{\rm vec.} &=& -\, \frac{1}{16 f} \, h_A \, \varepsilon^{\mu\nu\alpha\beta} \, {\rm tr} \left\{
\left[ V_{\mu\nu}, \partial^\tau V_{\tau\alpha} \right]_+ \, \partial_\beta \Phi \right\} \, - \, \frac{1}{16
f} \, b_A \, \varepsilon^{\mu\nu\alpha\beta} \, {\rm tr} \left\{ \left[V_{\mu\nu}, V_{\alpha\beta} \right]_+
\, \left[\Phi, \chi_0 \right]_+ \right\} \nonumber \\
 && - \, \frac{m_V^2}{4 f} \, n_A \, \varepsilon^{\mu\nu\alpha\beta} \, {\rm tr} \left\{ V_{\mu\nu}
V_{\alpha\beta} \right\} \eta_1 \, - \, \frac{e_V m_V}{4} \, {\rm tr} \left\{ V^{\mu\nu} Q \right\}
\partial_\mu A_\nu. \label{Lvec}
\end{eqnarray}
Thereby, $\chi_0 = {\rm diag} \left\{m_\pi^2, m_\pi^2, m_K^2 \right\}$, $Q = {\rm diag} \left\{2/3, -1/3,
-1/3 \right\}$ and the matrices $V_{\mu\nu}$ and $\Phi$ describe the vector mesons (represented by
antisymmetric tensor fields) and the pseudoscalar mesons, respectively, i.e.
\begin{eqnarray}
V_{\mu\nu} &:=&  \left(\begin{array}{ccc}
                \rho^0_{\mu\nu} + \omega_{\mu\nu} & \sqrt{2} \rho^+_{\mu\nu} & \sqrt{2} K^+_{\mu\nu} \\
                \sqrt{2} \rho^-_{\mu\nu} & - \rho^0_{\mu\nu} + \omega_{\mu\nu} & \sqrt{2} K^0_{\mu\nu} \\
		\sqrt{2} K^-_{\mu\nu} & \sqrt{2} \bar{K}^0_{\mu\nu} & \sqrt{2} \phi_{\mu\nu}
               \end{array} \right) , \\
 \Phi &:=& \left(\begin{array}{ccc}
           \pi^0 + \frac{1}{\sqrt{3}} \eta_8 & \sqrt{2} \pi^+ & \sqrt{2} K^+ \\
	   \sqrt{2} \pi^- & - \pi^0 + \frac{1}{\sqrt{3}} \eta_8 & \sqrt{2} K^0 \\
	   \sqrt{2} K^- & \sqrt{2} \bar{K}^0 & - \frac{2}{\sqrt{3}} \eta_8 
          \end{array}\right) + \sqrt{\frac{2}{3}} \,  \eta_1 \, I_{3 \times 3} \,.
\end{eqnarray}
All three terms proportional to $h_A$, $b_A$ and $n_A$ are of order $q^2$: The $h_A$ term because of the two
derivatives, the $b_A$ term because of the quark-mass insertion proportional to $\chi_0$ and the $n_A$ term
because of the separate traces (note that $\eta_1$ is proportional to ${\rm tr} \{\Phi\}$).
Obviously, this leading-order Lagrangian (\ref{Lvec}) allows only for an indirect decay via a virtual vector
meson. Furthermore, for the description of $\eta$ and $\eta'$ mesons one has to take $\eta$-$\eta'$ mixing
into account,
\begin{eqnarray}
 \eta &=& \cos\theta \, \eta_8 - \sin\theta \, \eta_1, \\
 \eta' &=& \sin\theta \, \eta_8 + \cos\theta \, \eta_1 \ \ \ \ {\rm  with } \ \ \theta \approx -20^\circ.
\end{eqnarray}
The decay of the virtual photon into a dilepton is described by usual QED.

\Subsection{Parameter determination}
To determine the parameters $h_A$, $b_A$ and $n_A$, the partial-decay widths for the two-body decays
\begin{eqnarray}
 \omega \rightarrow \pi^0 \gamma, \ \omega \rightarrow \eta \gamma, \ \phi \rightarrow \eta \gamma, \ \phi
\rightarrow \eta' \gamma, \ \eta' \rightarrow \omega \gamma
\end{eqnarray}
are compared to experimental data. First, the value $h_A$ is mainly determined via the decay $\omega
\rightarrow \pi^0 \gamma$ and, therefore, is fixed as 
\begin{eqnarray}
 h_A = 2.32
\end{eqnarray}
as in previous calculations \cite{paper}. The parameters $b_A$ and $n_A$ are then determined such that all
considered two-body widths are approximately equally well described. It turns out that one does not find fixed
values for this parameters but intervals
\begin{eqnarray}
 b_A \in [0, 0.19], \ n_A \in [0, 0.05]. \label{pr}
\end{eqnarray}
For decays into dileptons no additional parameters are needed and we have predictive power.

\Subsection{Results}
In Fig.\ \ref{Ff1}, the form factors for the transition $\omega \rightarrow \pi^0$ and $\eta' \rightarrow
\omega$ are plotted (solid lines). Both are compared to the standard vector meson dominance (VMD) form factors
(dashed lines). For both transitions, there is a deviation from VMD visible. Whereas the $\omega \rightarrow
\pi^0$ form factor is mainly determined by the term proportional to $h_A$, there is a sizable influence of
$b_A$ and $n_A$ on the $\eta' \rightarrow \omega$ form factor yielding a larger uncertainty compared to
$\omega \rightarrow \pi^0$. Obviously, data taken by the NA60 collaboration for the $\omega \rightarrow \pi^0$
transition form factor \cite{NA60} is much better described with our approach than with standard VMD.
Additionally, the calculated results for the integrated rates are within acceptable intervals and agree well
with the available experimental data \cite{PDG} if existent:
\begin{eqnarray}
 \Gamma_{\omega \rightarrow \pi^0 \mu^+ \mu^-}\, / \, \Gamma_{\omega \rightarrow \pi^0 \gamma} \in [1.36,
1.37] \cdot 10^{-3},&& \Gamma_{\omega \rightarrow \pi^0 \mu^+ \mu^-}^{\rm exp}\, / \, \Gamma_{\omega
\rightarrow \pi^0 \gamma}^{\rm exp} = (1.57 \pm 0.54) \cdot 10^{-3}, \nonumber \\
 \Gamma_{\omega \rightarrow \pi^0 e^+ e^-}\, / \, \Gamma_{\omega \rightarrow \pi^0 \gamma} \in [9.68, 9.70]
\cdot 10^{-3},&& \Gamma_{\omega \rightarrow \pi^0 e^+ e^-}^{\rm exp}\, / \, \Gamma_{\omega \rightarrow \pi^0
\gamma}^{\rm exp} = (9.30 \pm 1.04) \cdot 10^{-3},\nonumber \\
 \Gamma_{\eta' \rightarrow \omega e^+ e^-} \, / \, \Gamma_{\eta' \rightarrow \omega \gamma} \in [6.82, 6.84]
\cdot 10^{-3}.
\end{eqnarray}

\begin{figure}[htb]
 \centering
 \includegraphics[width = 0.37\textwidth, trim = 0 50 0 15]{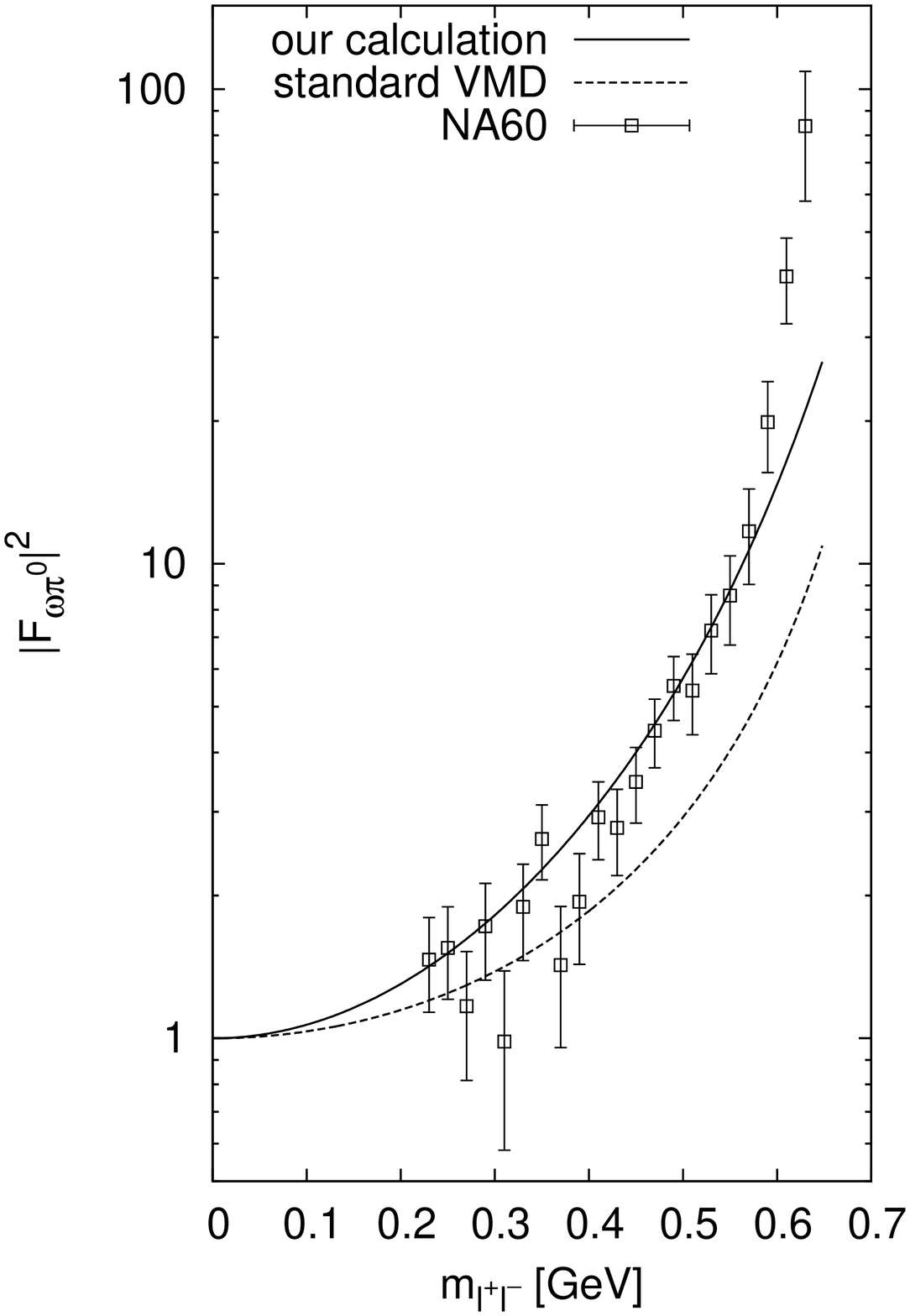}
 \includegraphics[width = 0.37\textwidth, trim = 0 50 0 15]{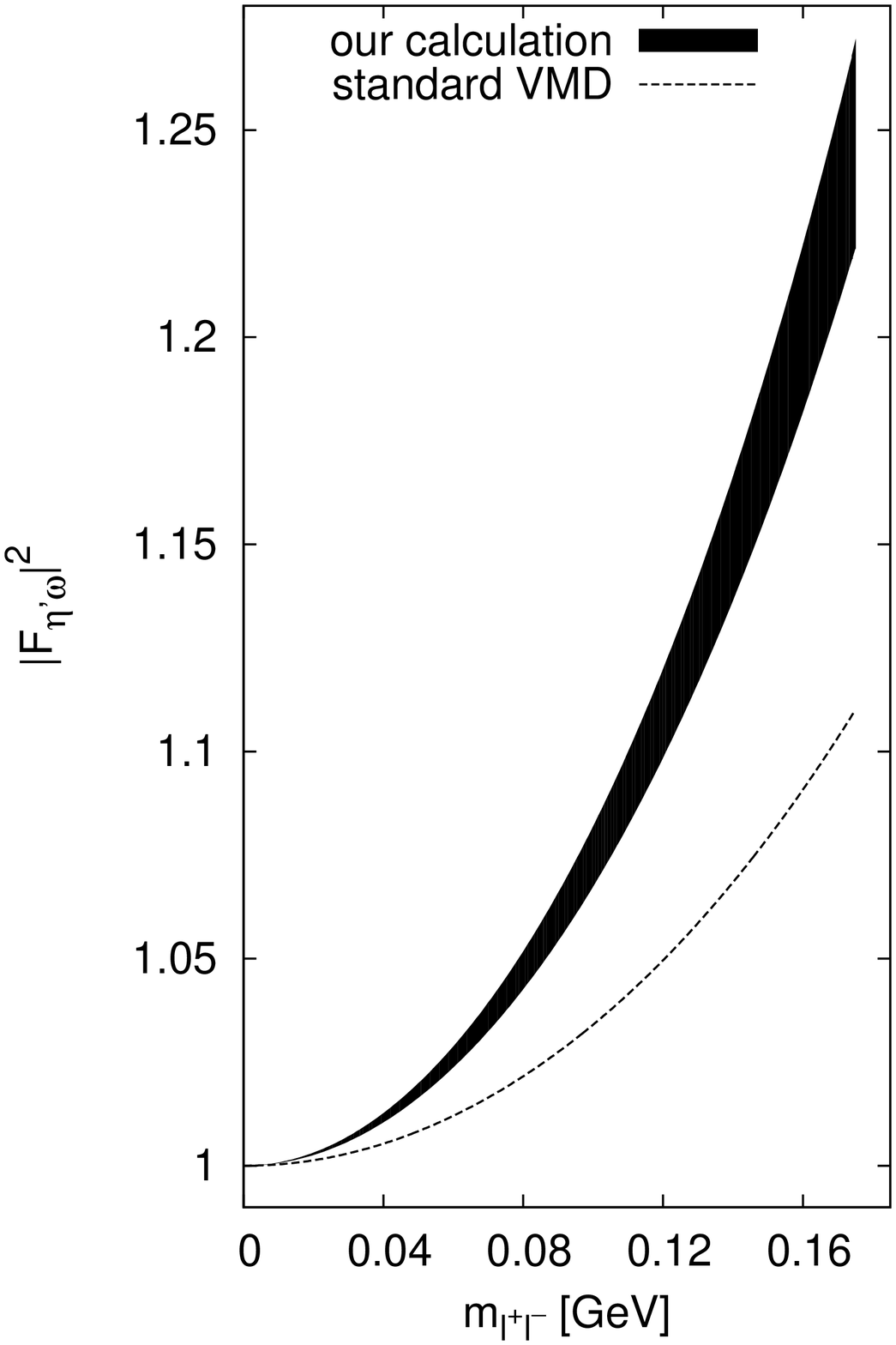} 
 \caption{\label{Ff1}\textbf{Left-hand side}: $\omega \rightarrow \pi^0$ form factor compared to dimuon data
taken by the NA60 collaboration \cite{NA60}. \textbf{Right-hand side}: $\eta' \rightarrow \omega$ transition
form factor.}
 
\end{figure}

\Section{\boldmath Decay $P \rightarrow \gamma \, l^+ l^-$}
For the transition of a pseudoscalar meson into a (real) photon, we use a more phenomenological approach by
including the Wess-Zumino-Witten action \cite{WZW}. On the one hand, a pseudoscalar meson can decay into two
(real or virtual) photons via two virtual vector mesons. In the counting scheme (\ref{CS}), this is in leading
order described by the Lagrangian $\mathcal{L}_{\rm vec.}$ (\ref{Lvec}). On the other hand, the leading-order
ChPT Lagrangian, the effective Wess-Zumino-Witten Lagrangian \cite{WZW}
\begin{eqnarray}
 \mathcal{L}_{\rm WZW} = \frac{3 e^2}{8 \pi^2 f} \, \varepsilon^{\mu\nu\alpha\beta} \, {\rm tr} \left\{ Q^2
\Phi \right\} \, \partial_\mu A_\alpha \, \partial_\nu A_\beta + \mathcal{O}(\Phi^2) \label{LWZW},
\end{eqnarray}
allows for direct decays. Formally, this term is of order $q^4$ in our counting scheme (\ref{CS}) but
nevertheless important as the low-energy leading-order term. Furthermore, the relative sign between both
Lagrangians was fixed as negative by comparing the calculated $\eta \rightarrow \gamma$ transition form factor
to experimental data.

The $\eta \rightarrow \gamma$ transition form factor is plotted on the left-hand side in Fig.\ \ref{Ff2}
(solid line) in comparison to the standard VMD calculation (dashed line) and data taken by the NA60
collaboration for the decay into a dimuon \cite{NA60}. There is nearly no deviation between the two form
factors visible, both describe the data very well. On the right-hand side in Fig.\ \ref{Ff2}, the $\eta'
\rightarrow \gamma$ transition form factor is plotted (solid line). Again, it is compared to the standard VMD
calculation (dashed line). Here, a small deviation between our calculation and standard VMD is visible. Note
that both $\rho^0$ and $\omega$ are visible in this form factor.

\begin{figure}[htb]
 \centering
 \includegraphics[width = 0.37\textwidth, trim = 0 50 0 0]{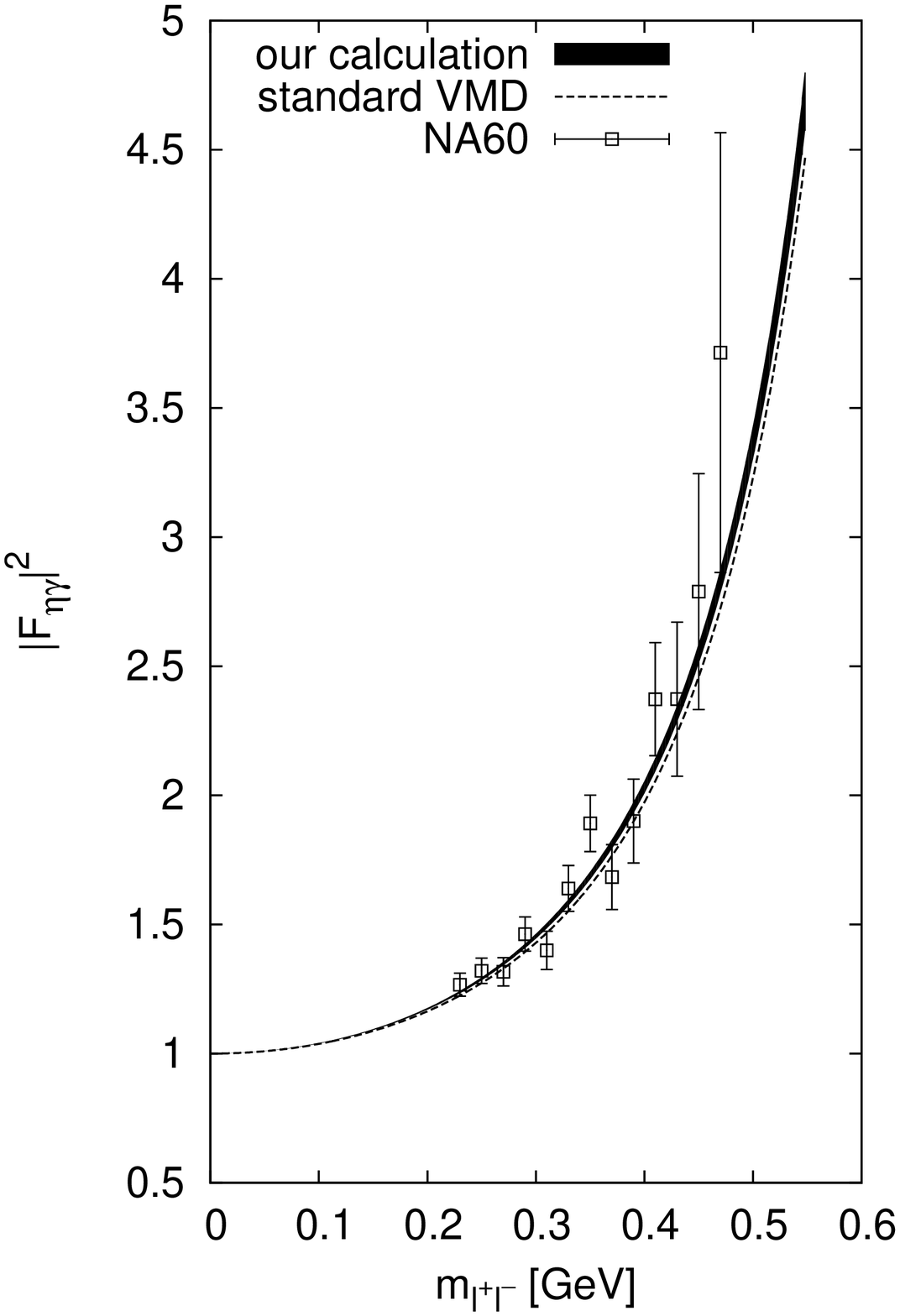}
 \includegraphics[width = 0.37\textwidth, trim = 0 50 0 0]{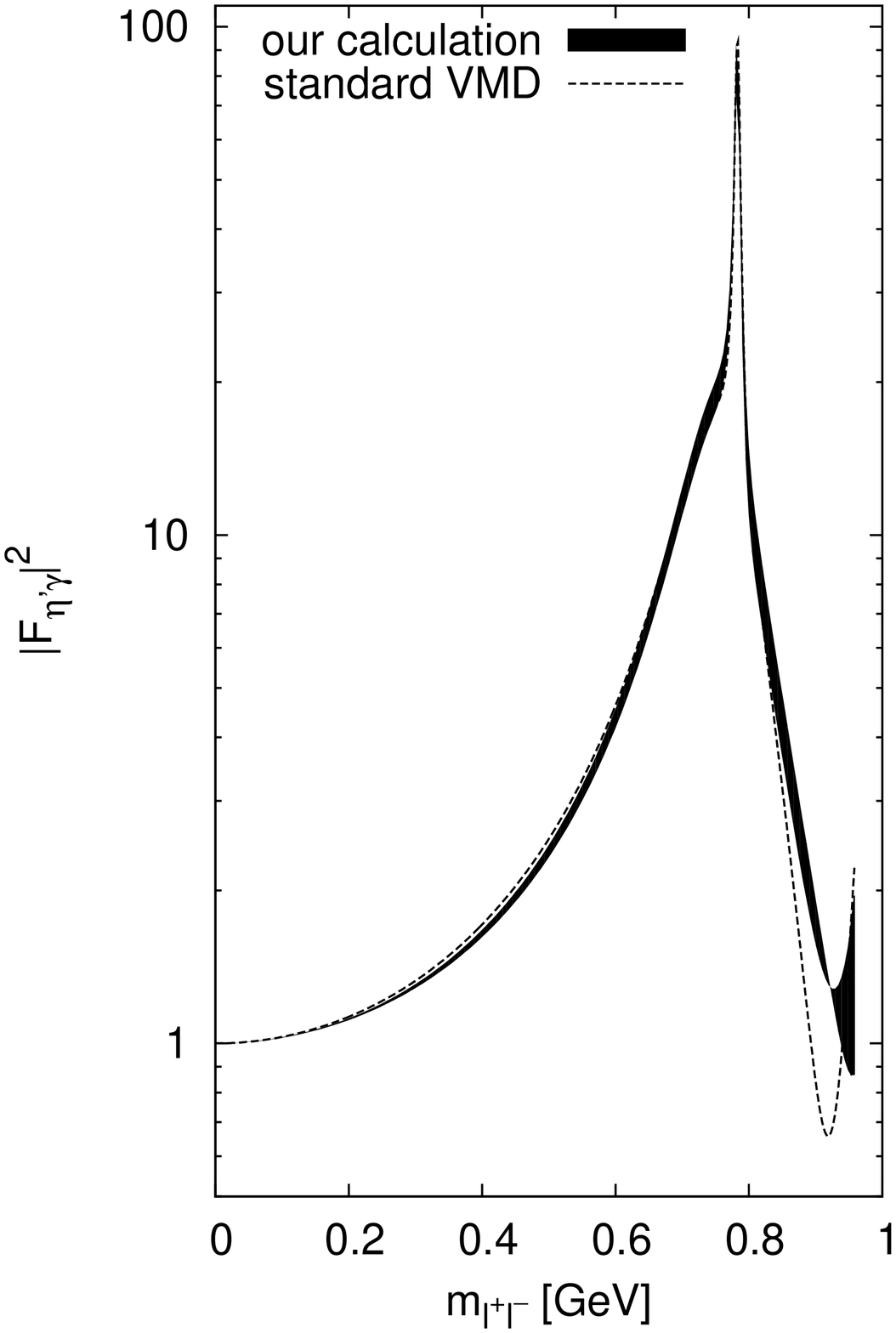}
 \caption{\label{Ff2}\textbf{Left-hand side}: $\eta \rightarrow \gamma$ form factor in comparison to dimuon
data taken by the NA60 collaboration \cite{NA60}. \textbf{Right-hand side}: $\eta' \rightarrow \gamma$ form
factor.}
\end{figure}

\Section{Summary and outlook}
The counting scheme (\ref{CS}) was used to calculate transition form factors in leading order. Thereby, the
$\omega \rightarrow \pi^0$ form-factor data was much better described than with standard VMD, the $\eta
\rightarrow \gamma$ form-factor data as good as with standard VMD. The presented partial decay widths are
within acceptable intervals and agree well with the available experimental data. As a next step,
next-to-leading order calculations have to be performed.

\coltoctitle{\boldmath$\pi^{0}$ Decays Measured with WASA-at-COSY}
\authortochead{C.-O.~Gullstr\"{o}m}
\label{GCO}

\setcounter{chapter}{1}
\setcounter{section}{0}
\maketitle

\Section{Introduction}
$\pi^{0}$ is the lightest known hadron and can only decay via the electromagnetic interaction. The most common
decay is into 2$\gamma$ (BR 98.82 $\%$) but the interesting physics occurs when decays include leptons. The
aim for WASA-at-COSY is to measure the decays \mbox{$\pi^{0}\rightarrow e^{+}e^{-}\gamma$},
\mbox{$\pi^{0}\rightarrow e^{+}e^{-}e^{+}e^{-}$} and \mbox{$\pi^{0}\rightarrow e^{+}e^{-}$}. In
\mbox{$\pi^{0}\rightarrow e^{+}e^{-}\gamma $ (BR $1.2\%$)} and \mbox{$\pi^{0}\rightarrow
e^{+}e^{-}e^{+}e^{-}$ (BR $3\cdot10^{-5}$)} the transition form factor reveals the electromagnetic
structure of the decaying meson. The $e^{+}e^{-}$ mass distribution probe the existence of a new hypothetical
boson that couples to the virtual photon~\cite{ex:1}. The current upper limit for $\pi^{0}\rightarrow
e^{+}e^{-}\gamma $ is set by the SINDRUM collaboration~\cite{ex:2} based on 100.000 events above 25 MeV
$e^{+}e^{-}$ invariant mass. For $\pi^{0}\rightarrow e^{+}e^{-}$ the KTeV Collaboration found 795 events in
$K_{L}\rightarrow 3\pi^{0}$~\cite{ex:3}. They measured a BR of $\left(7.49\pm0.29\right)\times10^{-8}$  which
exceeds the standard model prediction $\left(6.23\pm0.09\times10^{-8}\right)$~\cite{ex:4}. This leads to some
speculation about new physics. One promising teory is a new boson U that couples both to quarks and
leptons~\cite{ex:5}. Excess positrons from the U boson could also explain the intensity and shape of the
511 keV line from the galactic center~\cite{ex:6}.

\Section{Analysis}
During spring 2010 a one week test measurement has been performed with WASA-at-COSY in J\"ulich. In the
experiment proton-proton collisions were used as $\pi^{0}$ sources. The kinetic beam energy was set to 550~MeV
in order to maximize the $\pi^{0}$ cross section below threshhold for two pion production. The advantage of
excluding $\pi^{-}$ production is a clean $e^{-}$ sample for negatively charged tracks. 

\begin{figure}[htb]
 \centerline{\includegraphics[scale=0.33]{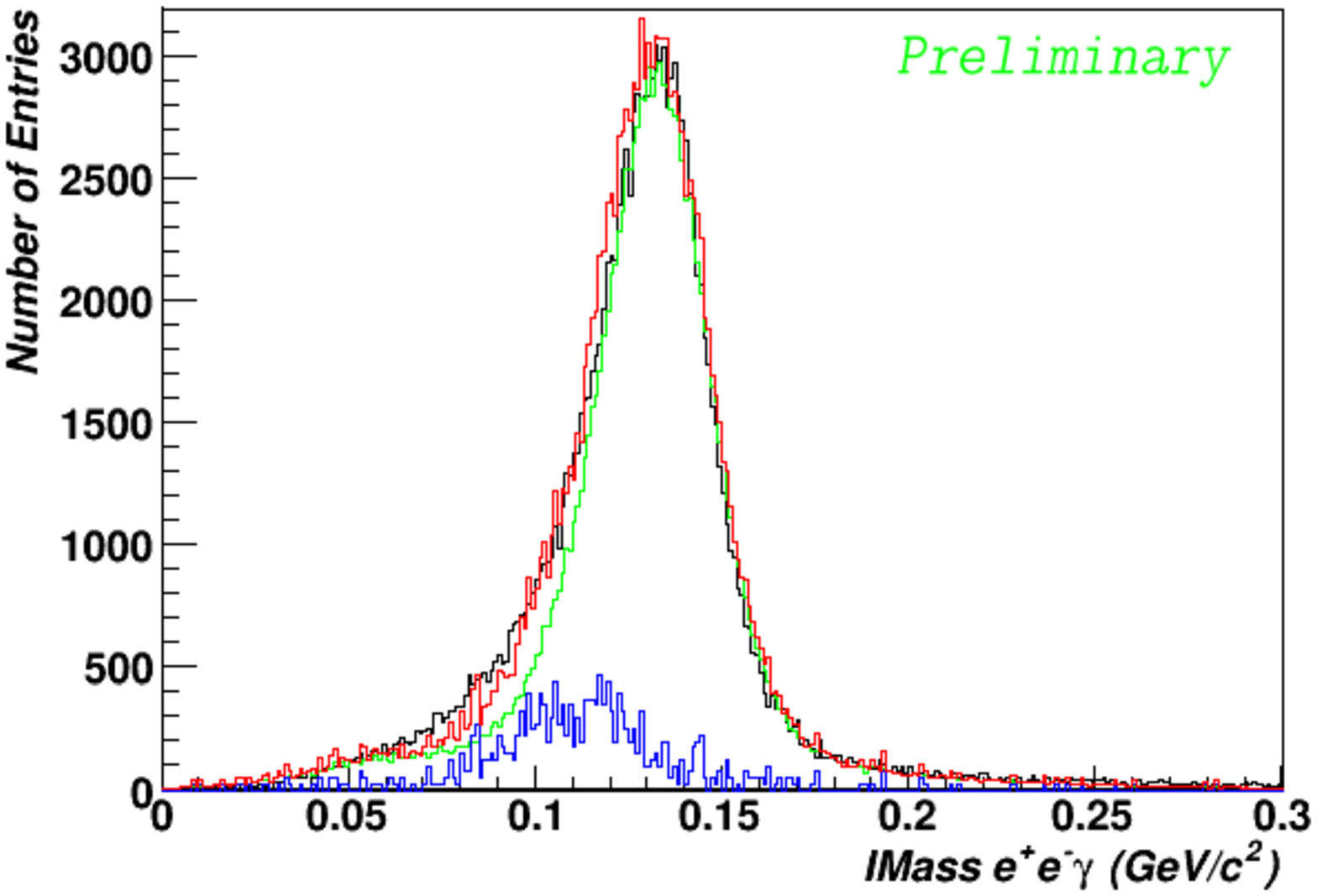}\hfill%
             \includegraphics[scale=0.33]{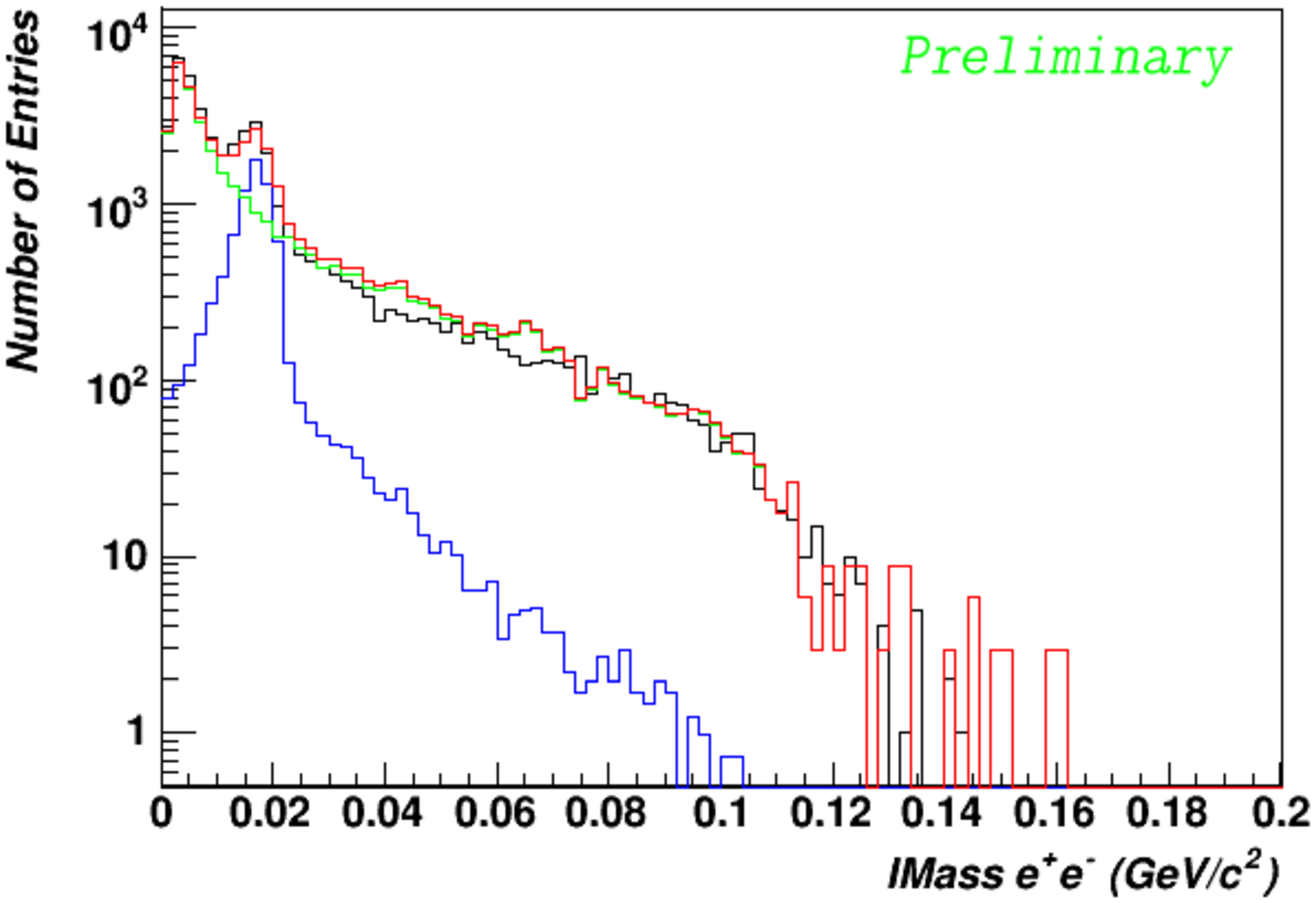}}
 \vspace{-3mm}\caption{\label{fig:IMeegee} Left: Invariant mass distribution of $e^{+}e^{-} \gamma$. Right:
Invariant mass
of $e^{+}e^{-}$ pair from the $e^{+}e^{-}\gamma$ peak. Black: Data; Green: MC $\pi^{0}\rightarrow
e^{+}e^{-}\gamma$; Blue: Background from pair production in the detector
$\pi^{0}\rightarrow\left(\gamma\rightarrow e^{+}e^{-}\right)\gamma$; Red: Total MC sample $\pi^{0}\rightarrow
e^{+}e^{-}\gamma$.}
\end{figure}

\Subsection{\boldmath$\pi^{0}\rightarrow e^{+}e^{-}\gamma$}
In $\pi^{0}\rightarrow e^{+}e^{-}\gamma$ 1.5 million reconstructed $e^{+}e^{-}\gamma$ events are available.
Out of those 50$\%$ is background from external conversion of one photon in the $\pi^{0}\rightarrow 2\gamma$
decay. The main cut for reducing such events is the vertex cuts since the main part of the photons convert at
the beamtube 30~mm from the interaction point. After cuts 40.000 $\pi^{0}\rightarrow e^{+}e^{-}\gamma$ events
are left with low background and high invariant mass resolution ( $\sim 4\%$). The Dalitz distribution in
Fig.~\ref{fig:IMeegee} shows no U boson peak and follows the existing $\pi^{0}\rightarrow e^{+}e^{-}\gamma$
transition formfactor. Tighter vertex cuts at the expense of efficiency are possible to exlude the pair
production peak at 20~MeV, so that an upper limit of U-boson exclusion could be set in this region.

\Subsection{\boldmath$\pi^{0}\rightarrow e^{+}e^{-}$}
In order to identify the $\pi^{0}\rightarrow e^{+}e^{-}$ candidates in the high-energy tail of the
$\pi^{0}\rightarrow e^{+}e^{-}\gamma $ events one needs good momentum resolution. The vertex points serve as a
check how well momentum was reconstructed by the central tracker. If the curvature is bad also the point of
closest approach the the center is bad. To reach the desired resolution even tighter cuts are set than in the
$\pi^{0}\rightarrow e^{+}e^{-}\gamma $ case. Fig.~\ref{fig:MMIM} shows a peak of 20 $\pi^{0}\rightarrow
e^{+}e^{-}$ event candidates. MC simulations show that there should be about 25 $\%$ background from
$\pi^{0}\rightarrow e^{+}e^{-}\gamma $ with a lost photon. The external conversion is completely removed in
this region by the vertex cut. 
\begin{figure}[htb]
 \centerline{\includegraphics[scale=0.33]{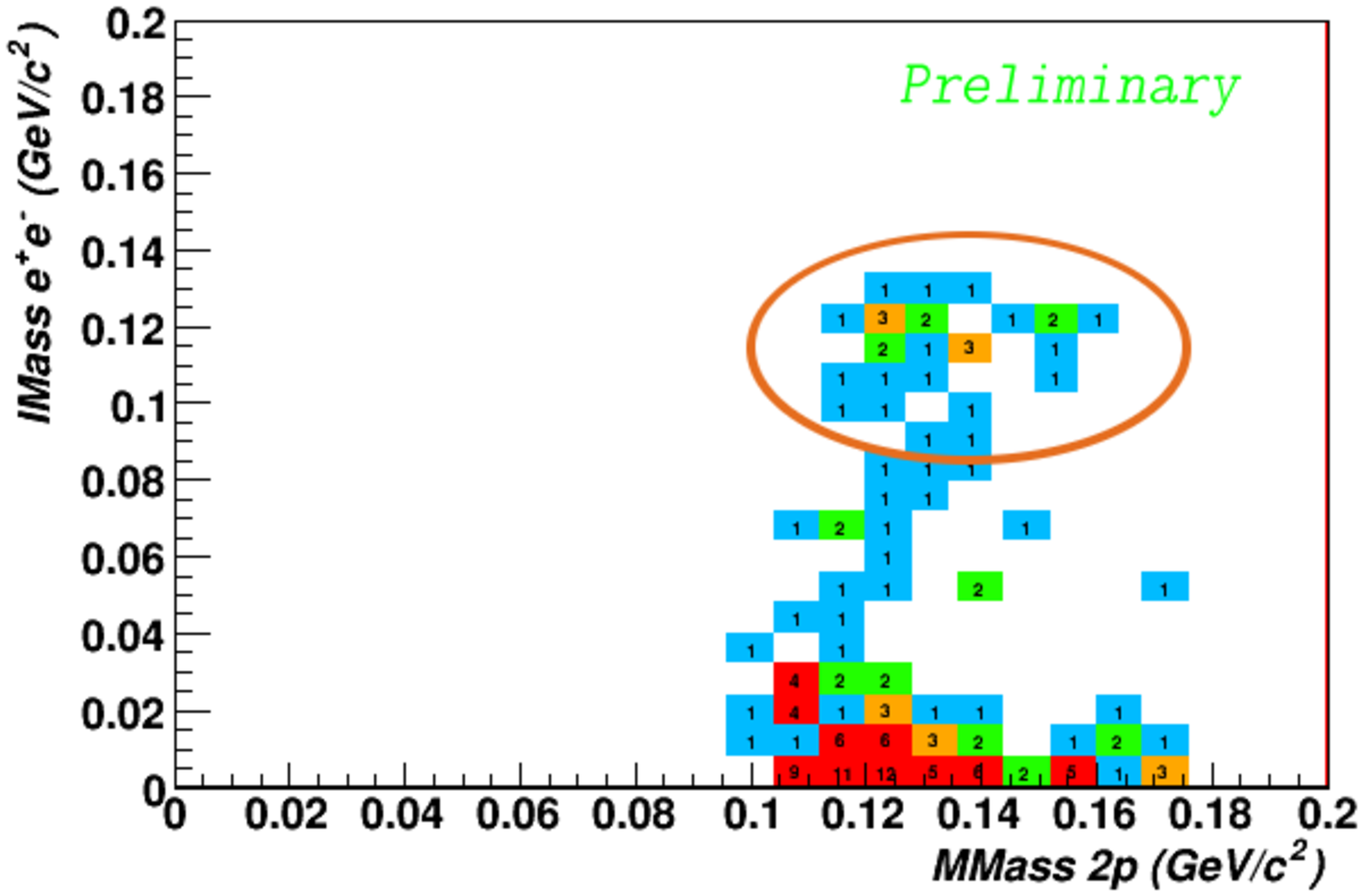}\hfill%
             \includegraphics[scale=0.33]{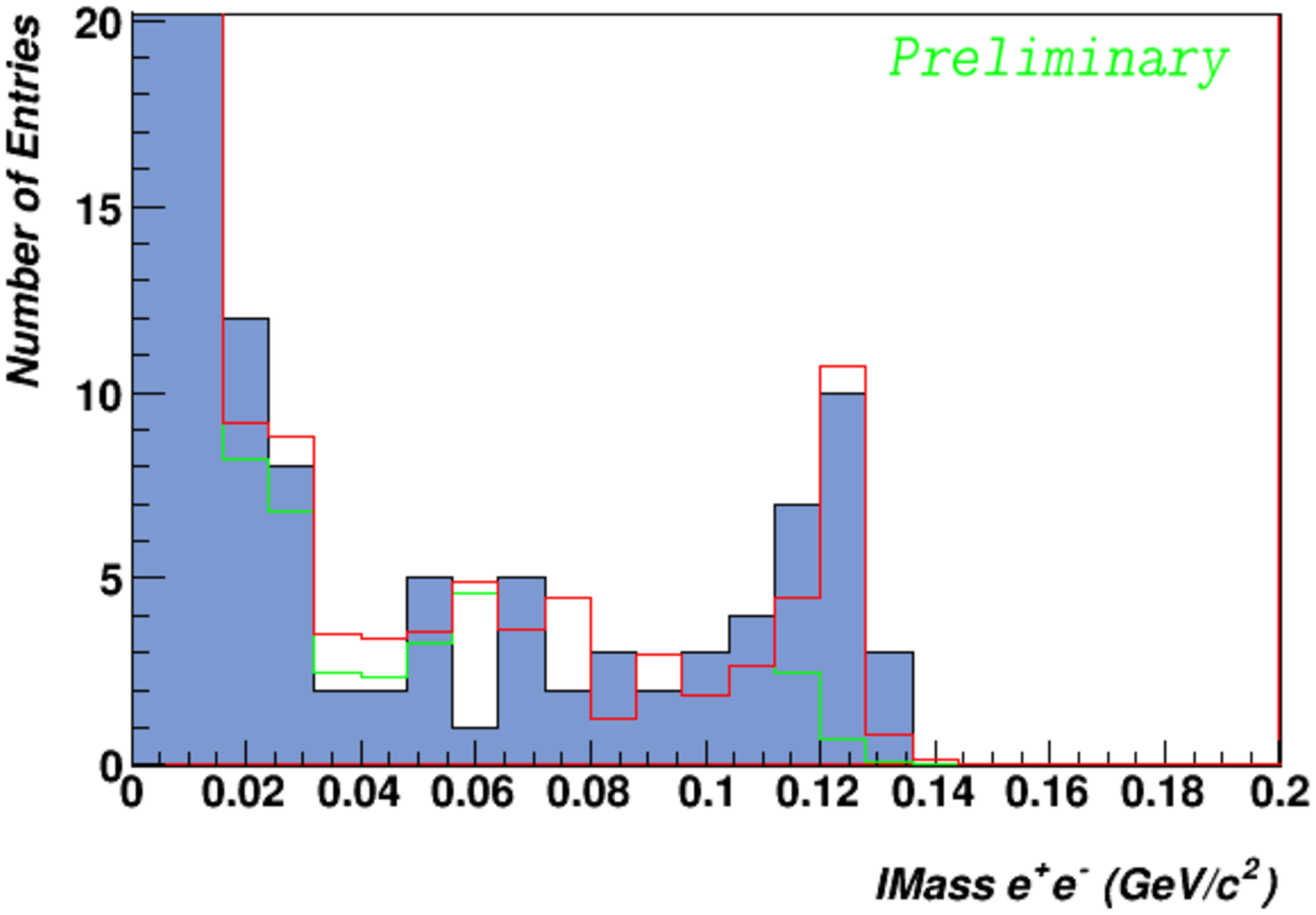}}
 \vspace{-3mm}\caption{\label{fig:MMIM}Left: Missing mass of 2 protons in the Forward Detector {\it vs.}
invariant mass of $e^{+}e^{-}$ pair. Right: invariant mass of $e^{+}e^{-}$ pair in the central detector. Blue
filled: Data; Green: MC $e^{+}e^{-}\gamma$ background;  Red: MC simulation of $\pi^{0}\rightarrow e^{+}e^{-}$
plus $e^{+}e^{-}\gamma$ and pair production background.} 
\end{figure}

\Section{Conclusion and Outlook}
For the one week test run we reconstructed 40.000 $\pi ^{0}\rightarrow e^{+}e^{-}\gamma $. This data could be
used to extend the upper limit on U Boson searches to lower masses. No deviation from the $\pi^{0}$ transition
form factor has been found. The 15 event candidates of $\pi ^{0}\rightarrow e^{+}e^{-}$ after background
subtraction motivate a longer production run in the future. The test run has not been optimized with respect
to trigger and calibration and hence it is possible to reach 50 $\pi ^{0}\rightarrow e^{+}e^{-}$ events per
week.

\addtocontents{toc}{\protect\pagebreak}
\coltoctitle{\boldmath Study of the $\eta \rightarrow e^{+}e^{-}\gamma$ Decay}
\authortochead{M.~Hodana}
\label{HM}

\setcounter{chapter}{1}
\setcounter{section}{0}
\maketitle

\Section{Introduction}
Since the $\eta$ meson is a short-lived, neutral particle, it is not possible to investigate its structure
via the classical method of particle scattering. To learn about its quark wave function, one studies the decay
processes of this meson, in which a pair of photons is produced, at least one of them being virtual. The
virtual photons have a non-zero mass and convert into lepton-antilepton pairs. The squared four-momentum
transferred by the virtual photon corresponds to the squared invariant mass of the created lepton-antilepton
pair. Therefore, information about the quarks' spatial distribution inside the meson can be achieved from the
lepton-antilepton invariant mass distributions by comparison of empirical results with predictions, based on
the assumption that the meson is a point-like particle. The latter can be obtained from the theory of Quantum
Electrodynamics. The deviation from the expected behaviour in the leptonic mass spectrum expose the inner
structure of the meson. This deviation is characterized by a form factor. It is currently not possible to
precisely predict the dependence of the form factor on the four-momentum transferred by the virtual photon in
the framework of Quantum Chromodynamics. Therefore, to perform calculations, assumptions about the dynamics of
the investigated decay are needed.

The knowledge of the form factors is also important in studies of the muon anomalous magnetic moment,
\mbox{${a_{\mu} = (g_{\mu}-2)/2}$}, which is the most precise test of the Standard Model and, as well, may be
an excellent probe of new physics. The theoretical error of calculation of \mbox{$a_{\mu}$} is dominated by
hadronic corrections and therefore limited by the accuracy of their determination. At present, the discrepancy
between the  \mbox{$a_{\mu}$} prediction based on the Standard Model~\cite{Bennett:2006fi} and its
experimental value  is ${(28.7 \pm 8.0) \cdot 10^{-10}}$  ($3.6\sigma$)~\cite{Davier:2010nc}. 

The aim of the data analysis is the investigation of the electromagnetic structure of the $\eta$ meson by
determining the transition form factor using the \mbox{$\displaystyle \eta \rightarrow e^+ e^- \gamma $} decay
mode. The probability of creation of a dielectron pair in considered decay is proportional to the probability
of emission of a virtual photon with a time-like four-momentum. The square of this four-momentum vector is
equal to the square of the  mass of created $e^+ e^-$ pair. By studying the probability of given decay as a
function of the dilepton pair mass, one obtains information about the hadron-photon transition and hence
about the electromagnetic structure of decaying neutral meson~\cite{Landsberg:1986fd}. 

\section{Current status of the analysis}
Data were collected using the \mbox{$\displaystyle pd \rightarrow {}^{3}He\, \eta$} reaction at proton beam
momentum of 1.69 GeV/c. The experiment was performed using the WASA-at-COSY detector~\cite{Adam:2004ch} in
November 2008. Data collected during 4 weeks, yielded approximately 10 million $\eta$ mesons tagged by the
\mbox{${}^{3}He$} ions measured in the Forward Detector.

The event selection in the Central Detector aims at choosing the decay channel of interest. We demand (i)
that two tracks corresponding to oppositely charged particles are reconstructed and, (ii) that at least one
neutral particle was registered and, (iii) that the signals are correlated in time with signals observed in
the Forward Detector within a 12 ns and a 31 ns window for charged and neutral tracks, respectively. The
energy deposited by a neutral particle ($E_{\gamma}$) is demanded to drop linearly with increasing opening
angle with the nearest charged particle ($\Omega$) starting from $100$~MeV for $\Omega=0^{\circ}$ and having
$E_{\gamma}=0$~MeV for $\Omega=180^{\circ}$. Additionally, in the $\eta$ centre of mass system, there has to
be only one such neutral candidate, forming with a lepton pair a $\Delta\phi$ angle in the range from
$60^{\circ}$ to $300^{\circ}$. The identification of electrons is achieved using energy-momentum plot as shown
in Fig.~\ref{fig:EdepCDBP} (right). An additional restriction is imposed on the missing mass of the \mbox{$ pd
\rightarrow X e^{+}e^{-}\gamma$} in the range of $2.66~-~2.87~GeV/c^{2}$, futher reducing background coming
from $\eta$ decay channels with pions. The efficiency of supression of the $\eta\to\pi^+\pi^-\gamma$ and
$\eta\to\pi^+\pi^-\pi^0$ can be seen in Fig.~\ref{fig:dNdM}.

Despite of asking in the analysis for events with two oppositely charged particles, there is a large
background coming from the $\eta \rightarrow \gamma\gamma$ decay channel (the one with the highest branching
ratio). Since photons may undergo conversion at the beam pipe, the $\eta$ meson decaying to two photons, may
give the same signature at the end as the decay of interest. One can see that in Fig.~\ref{fig:dNdM} (left)
which shows the \mbox{$M_{e^{+}e^{-}}$} spectra from experiment and simulations before applying cut on
photon's conversion. In order to suppress background from $\eta \rightarrow \gamma\gamma$ decay, where one of
photons undergoes conversion into $e^{+}e^{-}$ pair in the beam pipe, the radius of point of closest approach
($R_{CA}$) is used as presented in Fig.~\ref{fig:EdepCDBP} (left). 

\begin{figure}[htb]
 \centerline{\includegraphics[width=0.4\textwidth, height=5.5cm]{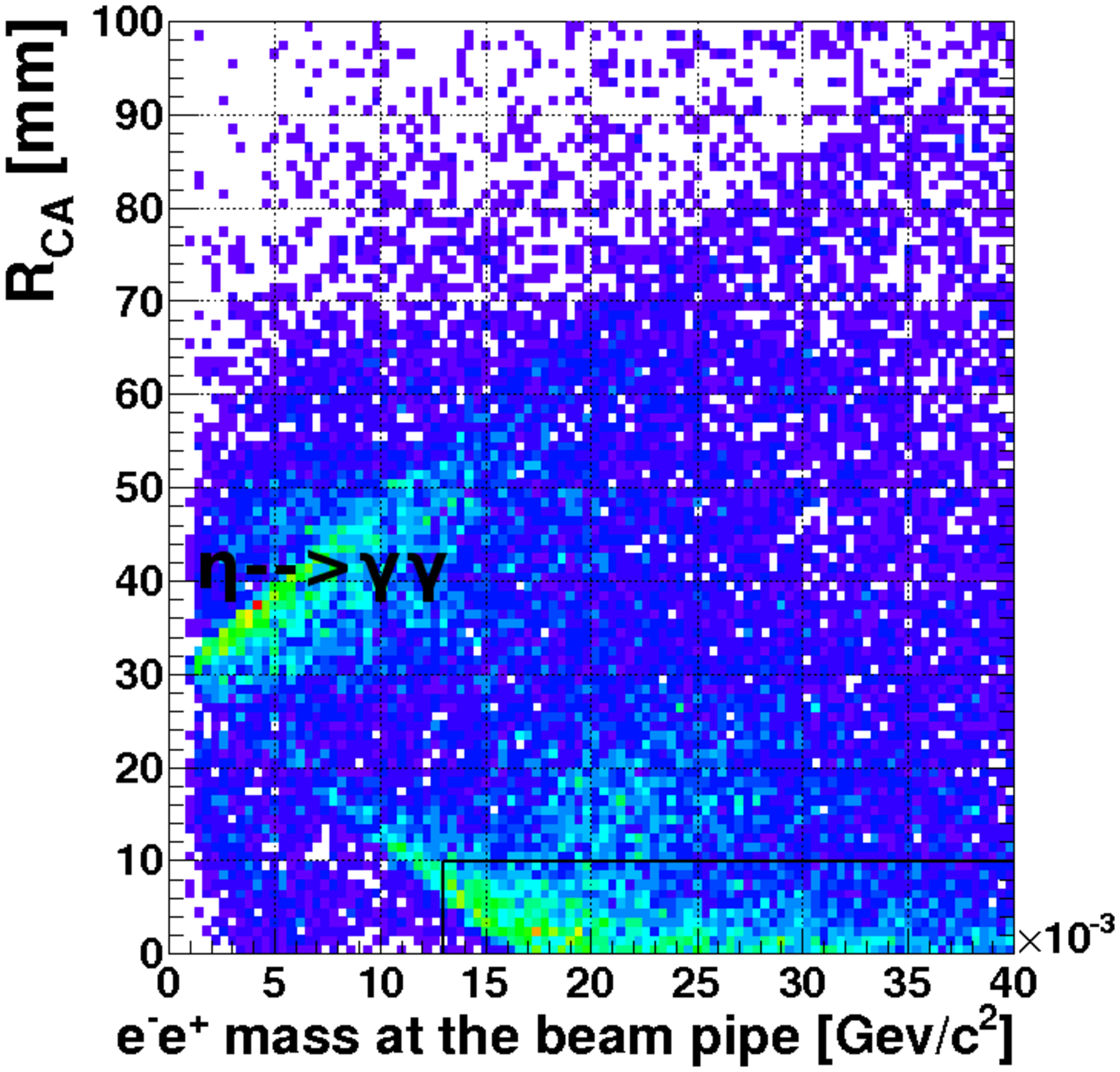}\hspace{1cm}%
             \includegraphics[width=0.4\textwidth, height=5.5cm]{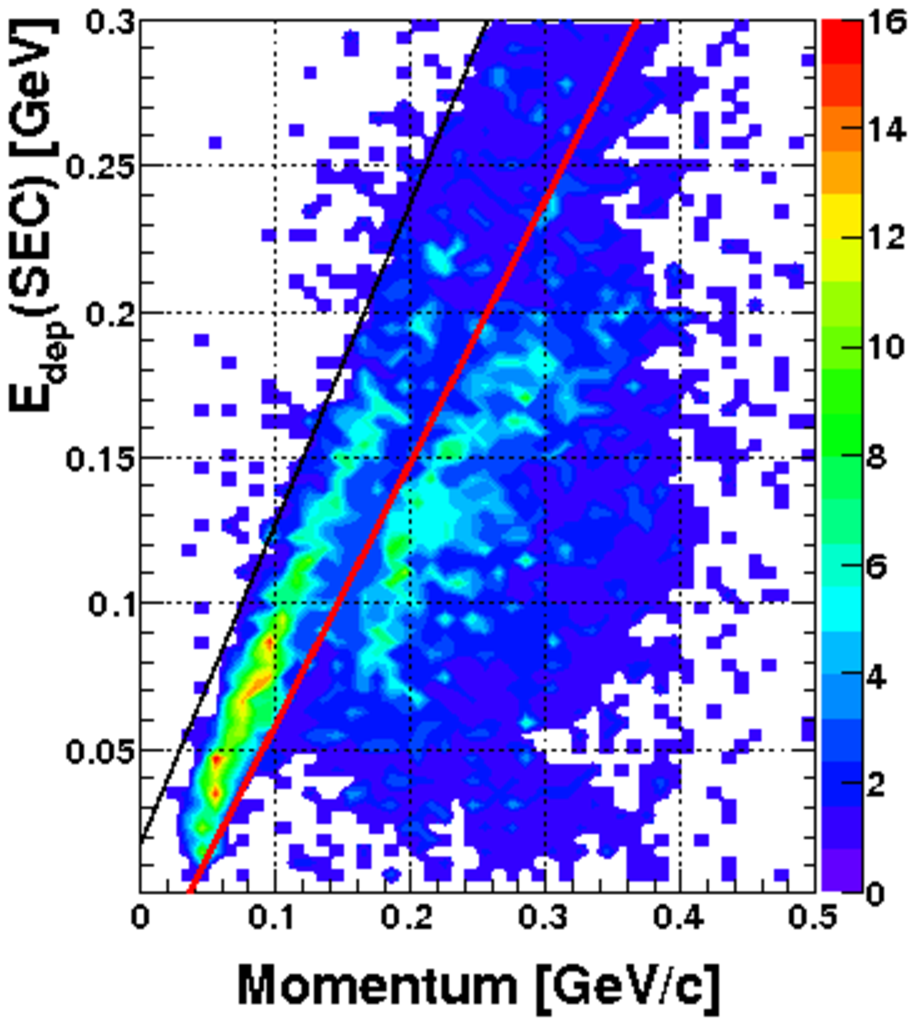}}
 \caption{\label{fig:EdepCDBP} Left: experimental distribution of the radius of the point of the closest
approach between two helices as a function of the $e^{+}e^{-}$ mass (taken from the leptons' three momenta
calculated at the beam pipe). The x-axis is zoomed in to see events from the $pd\rightarrow\gamma\gamma$
reaction. The region outside the black lines is discarded from further analysis. Right: experimental spectrum
of the energy deposited in the Electromagnetic Calorimeter as a function of particles' momenta. Electrons
arrange themselves on the  band, along the momentum-\mbox{$E_{dep}$} diagonal (between black and red lines).}
\end{figure}

\begin{figure}[htb]
 \centerline{\includegraphics[width=0.45\textwidth, height=5.5cm]{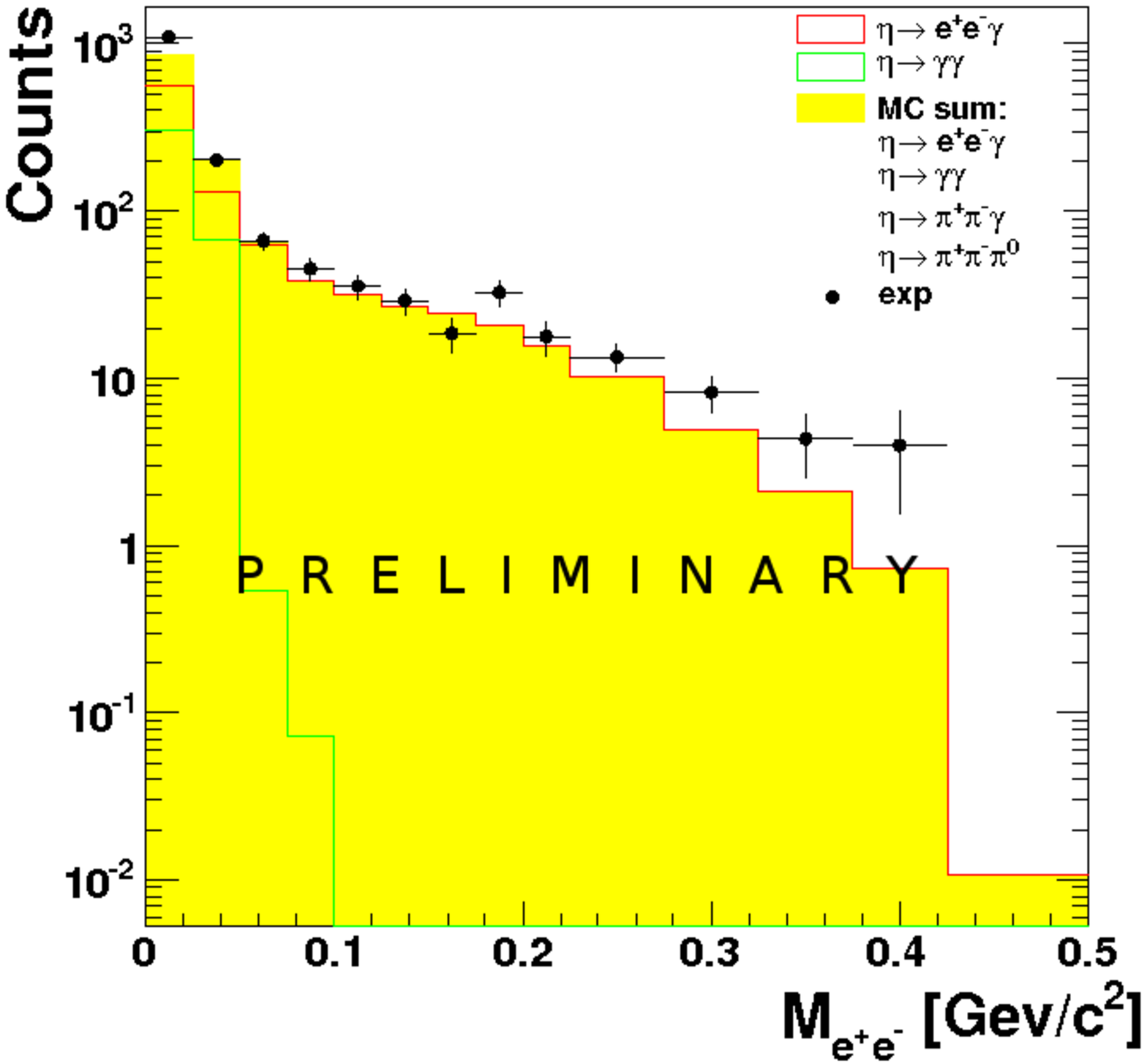}\hspace{1cm}%
             \includegraphics[width=0.45\textwidth, height=5.5cm]{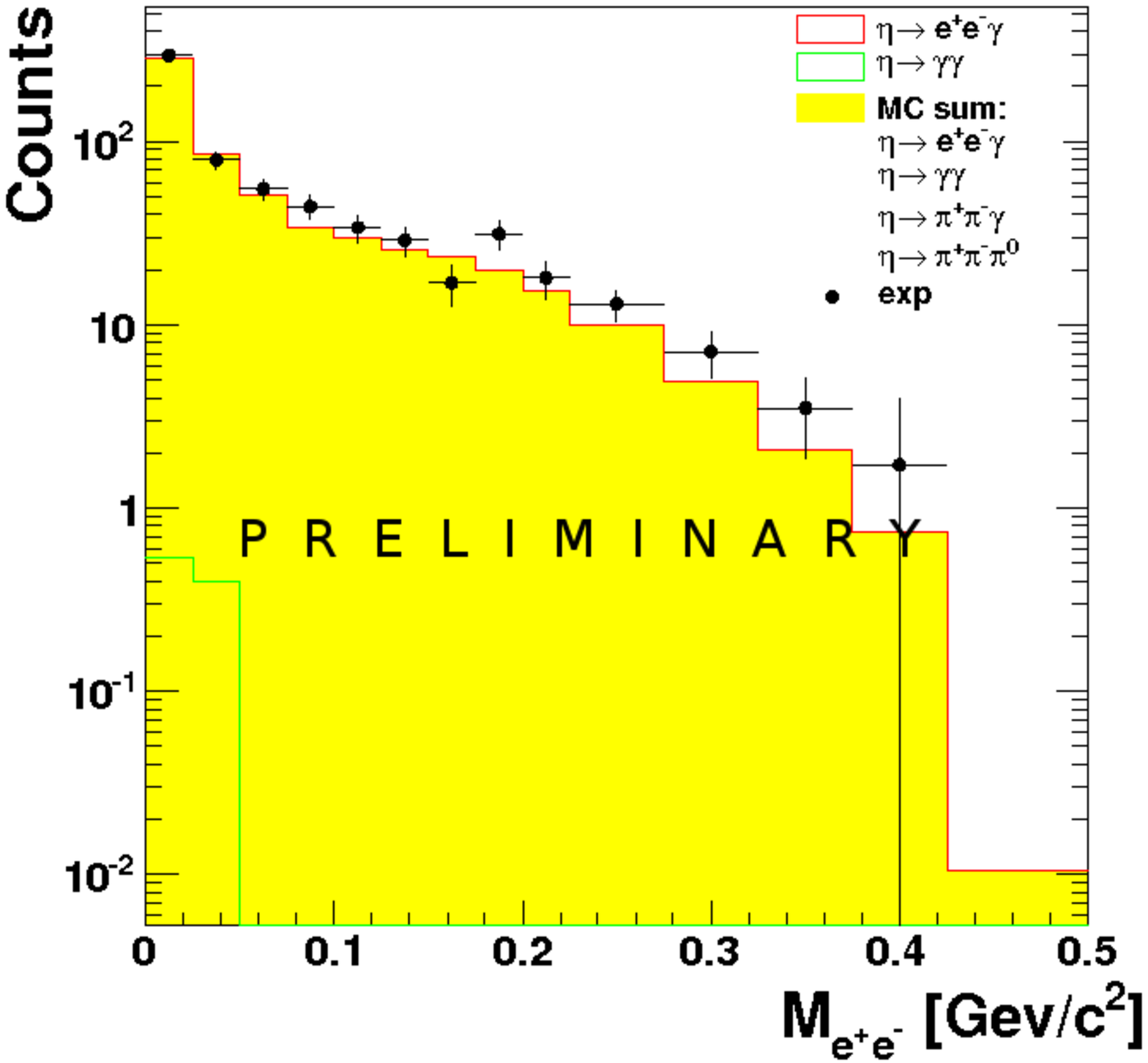}}
  \caption{\label{fig:dNdM} The \mbox{$M_{e^{+}e^{-}}$} spectra from the simulations, and the experiment
(after prompt pion background subtraction), before applying the cut on the photon conversion (left) and with
this cut (right). The spectra are not acceptance corrected. }
\end{figure}

The \mbox{$M_{e^{+}e^{-}}$} spectrum after selection of \mbox{$\displaystyle \eta \rightarrow e^+ e^- \gamma
$} decay channel as described above is shown in Fig.~\ref{fig:dNdM} (right). Approximately 700 events of
\mbox{$\displaystyle \eta \rightarrow e^+ e^- \gamma $} has been reconstructed. The extraction of the
transition form factor and the systematic studies are in progress.  

\section*{Acknowledgements}
This publication has been supported by COSY-FFE and by the European Commission under the 7th Framework
Programme through the ’Research Infrastructures’ action of the ’Capacities’ Programme, Call:
FP7-INFRASTRUCTURES-2008-1, Grant Agreement N. 227431 and by the Polish Ministry of Science and Higher
Education under grants No. 2861/B/H03/2010/38 and 0320/B/H03/2011/40.

\coltoctitle{\boldmath Analysis of the Double Dalitz Decay $\eta \rightarrow e^+ e^- e^+ e^-$}
\authortochead{P.~Wurm}
\label{WP}

\setcounter{chapter}{1}
\setcounter{section}{0}
\maketitle

\vspace{2ex}
\Section{Introduction}
The Wide Angle Shower Apparatus (WASA) is a large-acceptance detector to study the decay channels of light
mesons~\cite{Adam,Bargholtz}. It is operated at the Cooler Synchrotron COSY J\"ulich.

A large number of $\eta$ mesons is being produced in proton-deuteron and proton-proton collisions. The data
permit the study of very rare $\eta$-decay channels, like the double Dalitz decay, where the $\eta$ meson
decays via two virtual photons into two electron-positron pairs.

So far, the only published value for the branching ratio is $\mathcal{BR} = (2.4 \pm 0.2_{\mathrm{stat}} \pm
0.1_{\mathrm{syst}}) \times 10^{-5}$~\cite{KLOE}. One objective of the WASA-at-COSY experiment is to determine
a value for this branching ratio. A first analysis is based on a $p \, d \rightarrow {}^3_{}\!\mathrm{He} \,
\eta$ beam time from 2008, which comprises approximately $10^7$ $\eta$ mesons~\cite{Yurev}. The analysis
presented here is based on a total amount of $3 \times 10^7$ $\eta$ mesons, which includes additional $2
\times 10^7$ $\eta$ mesons, which have been produced using the same reaction in 2009.

\vspace{3ex}%
\Section{Analysis chain}
In the production reaction $p \, d \rightarrow {}^3_{}\!\mathrm{He} \, \eta$ the $\eta$ meson is tagged by the
${}^3_{}\!\mathrm{He}$ nuclei, measured in forward tracking detectors and an arrangement of hodoscopes.
${}^3_{}\!\mathrm{He}$ can be identified with the $\Delta E - E$ method, where the energy deposit in a thin
plastic scintillator is plotted versus the energy deposit in the stopping layer of the hodoscope. The
${}^3_{}\!\mathrm{He}$-missing mass distribution is used as a monitoring spectrum to count the number of
$\eta$ events after every cut.

The central part of the detector is used to reconstruct and identify the $\eta$-decay products. First, it is
checked if the event has two or more positively and negatively charged particles. The charge of a particle is
determined by the curvature of the track helix. The helix is measured by a Mini Drift Chamber, embedded in a
solenoid magnet.

Due to the large amount of pions, an effective method to distinguish electrons (positrons) and pions must be
found. WASA provides particle identification with the momentum information, which is measured by the Mini
Drift Chamber, combined with the energy deposit in a Plastic Scintillator Barrel and in an Electromagnetic
Calorimeter. Fig.~\ref{fig:bands} shows the electron (positron) and pion bands.
\begin{figure}[htb]
  \begin{minipage}[c]{0.49\textwidth}
    \centering
    \includegraphics*[width=1.00\textwidth]{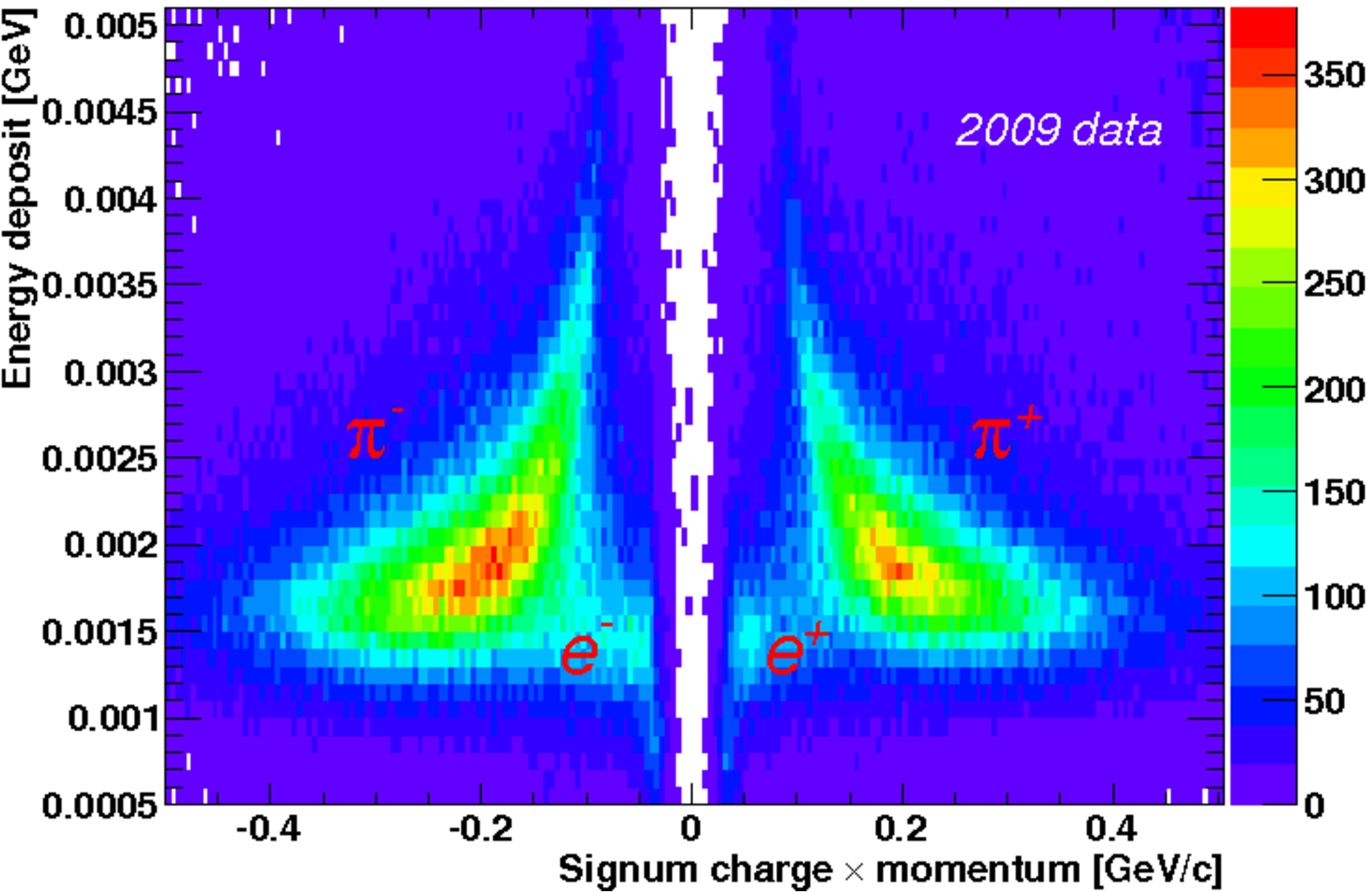}
  \end{minipage}
\hspace{0.02\textwidth}
  \begin{minipage}[c]{0.49\textwidth}
    \centering
    \includegraphics*[width=1.00\textwidth]{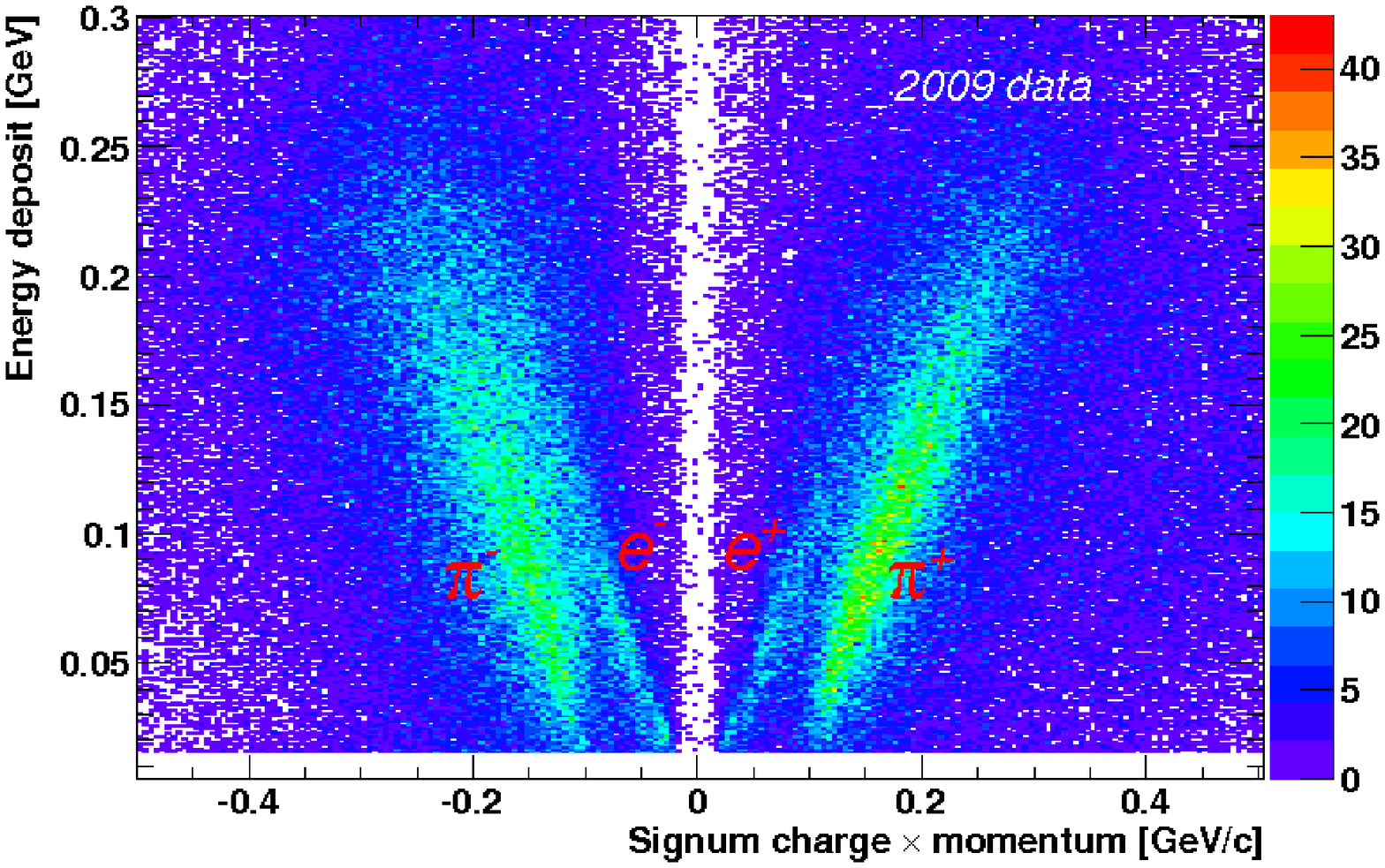}
  \end{minipage}
  \caption{\label{fig:bands}Energy deposit in the central part of WASA against the momentum of a particle
times its charge. Left: Plastic Scintillator Barrel. Right: Electromagnetic Calorimeter.}
\end{figure}

The decision, whether the particle is an electron (positron) or a pion, is based on an artificial neural
network implemented in ROOT~\cite{Root}. After the particle identification, an event candidate must have at
least two electrons and two positrons. Out of these tracks all possible pair-combinations are build, which can
be the decay product of one of the two virtual photons. In the following, each combination is evaluated to be
a double Dalitz decay event.

To get rid of pions which have been misidentified as electrons (positrons), the opening angle between the
supposed electron and the positron from the same virtual photon is checked. In case of an electron-positron
pair it has to peak at small values.

Another source of background are events, where a photon converts into an electron-positron pair. In case of
the single Dalitz decay $\eta \rightarrow e^+ e^- \gamma$, there is the same number of electrons and positrons
in the final state as in case of the double Dalitz decay. To suppress such events, the distance between the
point of closest approach of the positively and negatively charged particle and the beam center is analyzed as
a function of the invariant mass of the lepton-pair, calculated from their four-vectors at the beam pipe. For
conversion events, the invariant mass is zero, since the lepton-pair comes from a real photon. The radius of
the closest approach peaks around $30 \, \mathrm{mm}$, since this is the radius of the beam pipe. The left
panel of Fig.~\ref{fig:conversion} shows the distribution for simulated conversion events and for signal
events, and the right panel shows the distribution from the data.
\begin{figure}[htb]
  \begin{minipage}[c]{0.49\textwidth}
    \centering
    \includegraphics*[angle=90,width=1.00\textwidth]{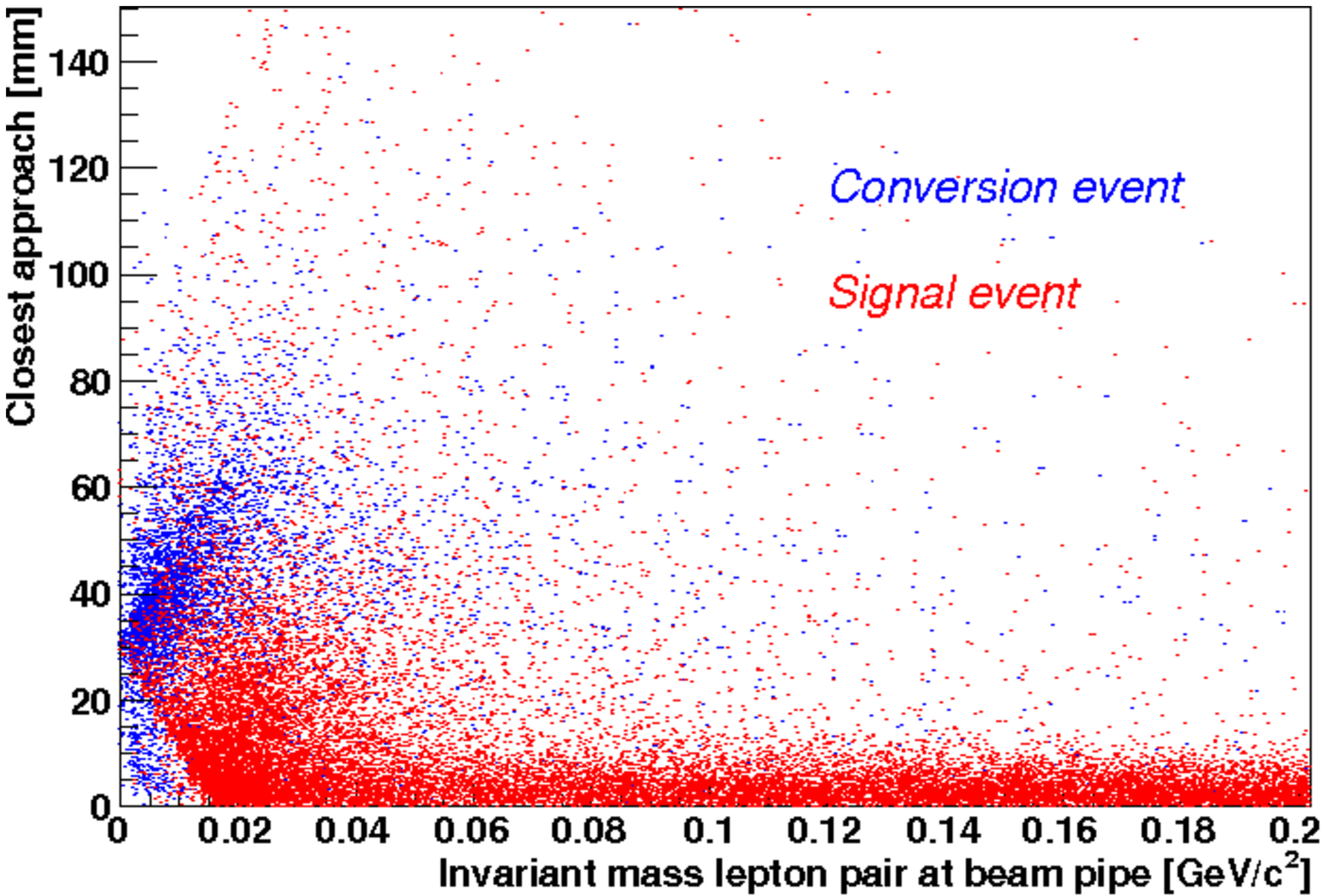}
  \end{minipage}
\hspace{0.02\textwidth}
  \begin{minipage}[c]{0.49\textwidth}
    \centering
    \includegraphics*[width=1.00\textwidth]{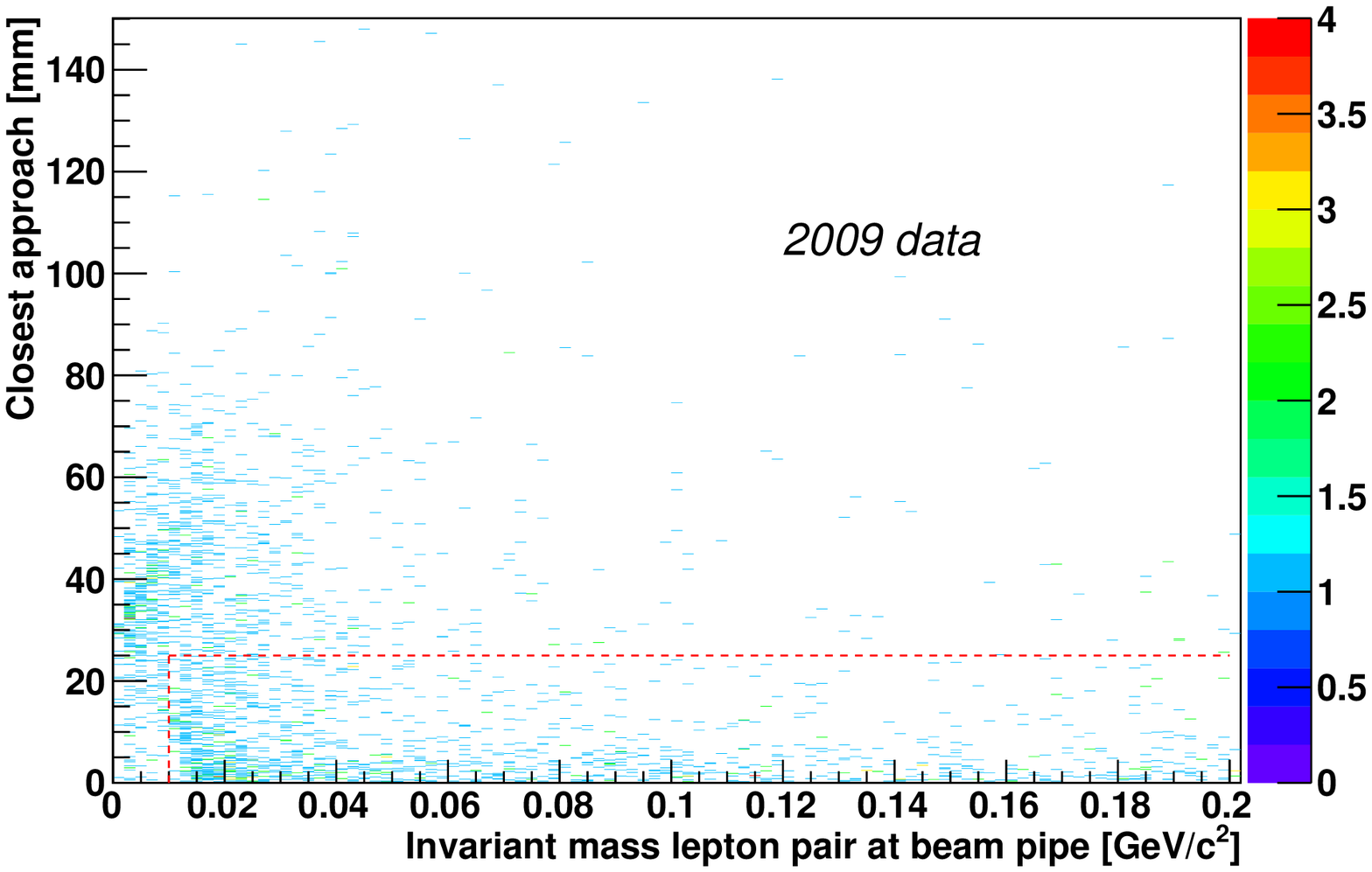}
  \end{minipage}
  \caption{\label{fig:conversion}Radius of the closest approach of the positively and negatively charged
particle against the invariant mass calculated from their four-vectors at the beam pipe. Left: Simulation.
Right: 2009 data. The dashed line in denotes the used cut.}
\end{figure}

After all these cuts the signal to background ratio for $\eta$ events is $\approx 2/1$, where the main
background channels are $\eta \rightarrow \pi^0 \pi^0 \pi^0$ and $\eta \rightarrow \pi^+ \pi^- \pi^0$.

\vspace{3ex}%
\Section{Results}
Fig.~\ref{fig:mmDevelopment} shows the development of the ${}^3_{}\!\mathrm{He}$-missing mass distribution for
the 2009 data after each cut. The remaining background events are mainly from direct pion production. To get
rid of those events the final ${}^3_{}\!\mathrm{He}$-missing mass distribution is fitted with a polynomial
function for the direct pion background and a Gaussian function for the $\eta$ signal. Fig.~\ref{fig:results}
shows the ${}^3_{}\!\mathrm{He}$-missing mass distribution from the 2008 and 2009 data after subtracting the
direct pion background. After subtracting the background from $\eta$ decays besides the double Dalitz decay,
$52 \pm 13$ events remain in the distribution.
\begin{figure}[htb]
  \begin{minipage}[c]{0.49\textwidth}
    \centering
    \includegraphics*[width=1.00\textwidth]{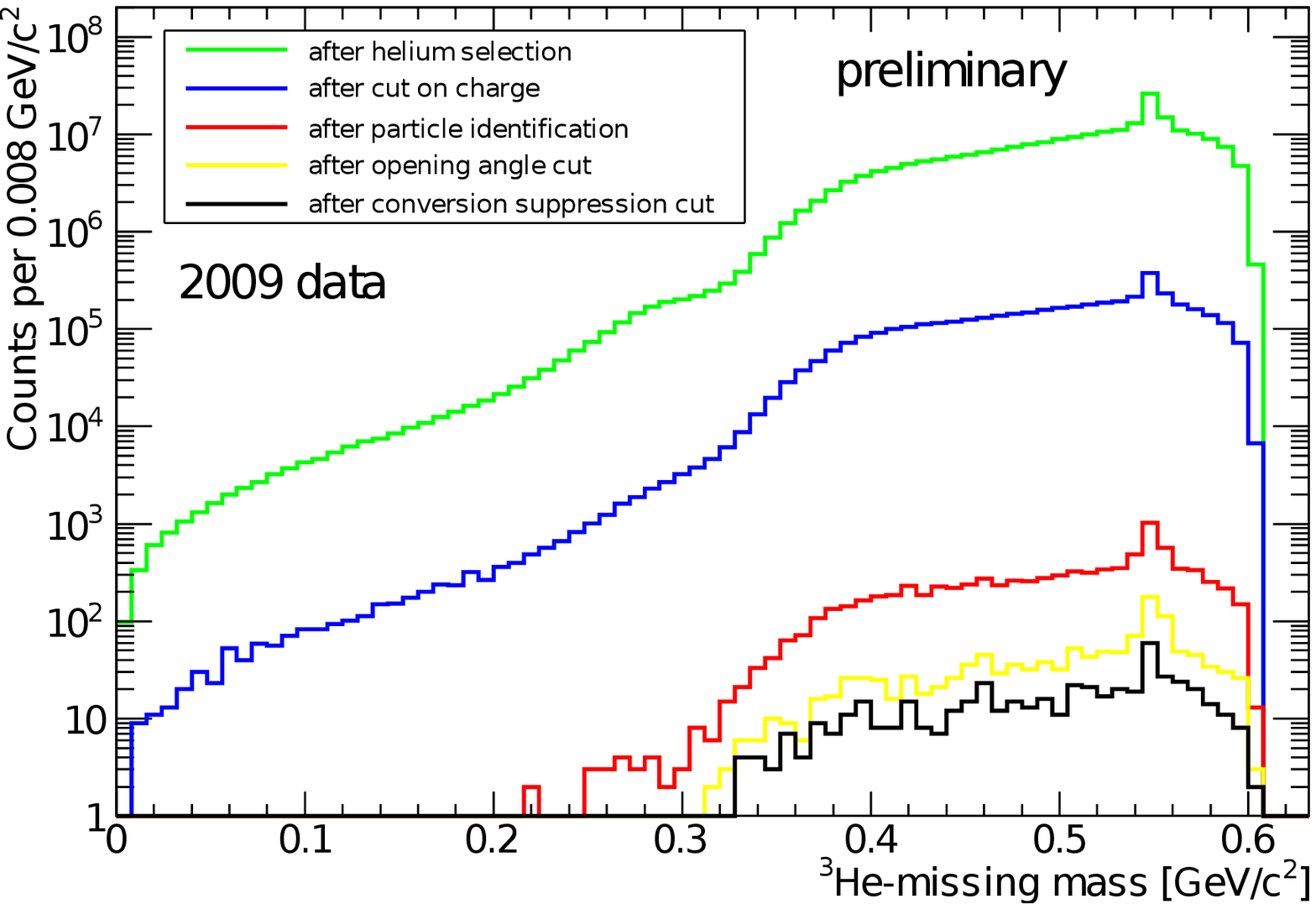}
  \caption{\label{fig:mmDevelopment}Development of the ${}^3_{}\!\mathrm{He}$-missing mass distribution (2009
data).}
  \end{minipage}
\hspace{0.02\textwidth}
  \begin{minipage}[c]{0.49\textwidth}
    \centering
    \includegraphics*[width=1.00\textwidth]{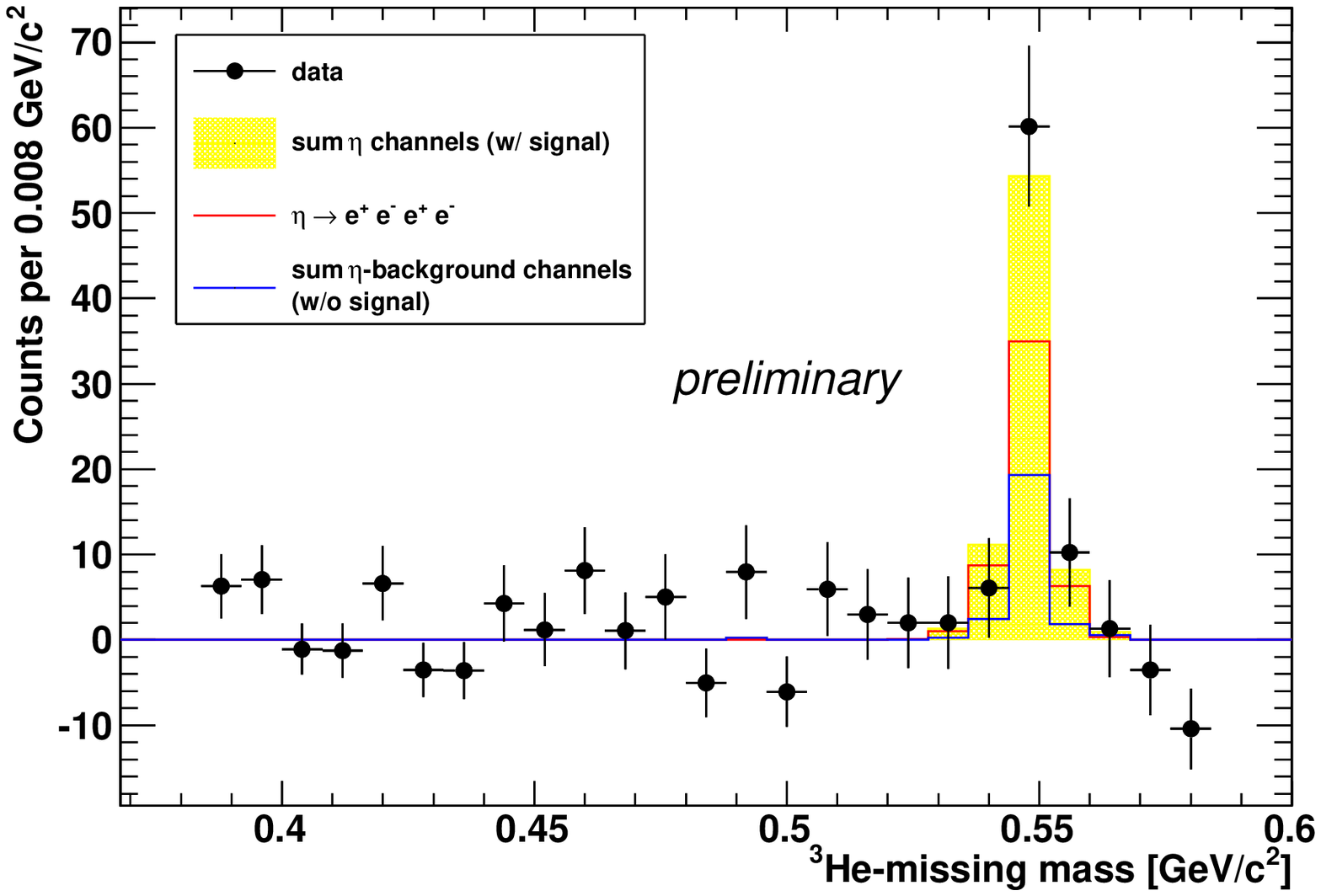}
    \caption{\label{fig:results}Final ${}^3_{}\!\mathrm{He}$-missing mass distribution after direct pion
background subtraction (2008 and 2009 data).}
  \end{minipage}	
\end{figure}

The combined result from the analysis presented in Ref.~\cite{Yurev} and from the analysis, which is
presented here, is $\mathcal{BR} = (3.0 \pm 0.8_{\mathrm{stat}} \pm 0.7_{\mathrm{syst (norm.)}}) \times
10^{-5}$ (preliminary), which is in agreement with the value from Ref.~\cite{KLOE}. The systematic error comes
from the branching ratio calculation based on two different normalization channels, e.g. $\eta \rightarrow e^+
e^- \gamma$ and $\eta \rightarrow \pi^+ \pi^- e^+ e^-$.

\Section{Outlook}
As a next step the systematic error, which is approximately on the same order of magnitude as the statistical
uncertainty, has to be studied in more detail.

The statistics can be improved with $p \, p \rightarrow p \, p \, \eta$ reactions, where WASA-at-COSY has
already measured $> 10^8$ $\eta$ mesons. On the one hand the production cross section is in this reaction
around twenty times larger compared with the $p \, d \rightarrow {}^3_{}\!\mathrm{He} \, \eta$ production
reaction, but on the other hand there is also a larger contribution from direct pions.

\coltoctitle{\boldmath Determination of the Electromagnetic $\eta$ Transition Form Factor}
\authortochead{V.~Metag}
\label{MV}
\import{./MetagVolker/Metag}

\coltoctitle{\boldmath Dalitz Decay of the $\omega$ Meson}
\authortochead{F.~A.~Khan}
\label{KFA}

\setcounter{chapter}{1}
\setcounter{section}{0}
\maketitle

\Section{Introduction}
Our motivation is to measure the electromagnetic transition form factor of the $\omega$ meson Dalitz decay
$\omega\to\pi^{0}$e$^{+}$e$^{-}$. The differential decay width at the vertex of the $\omega$-pion-transition
can be described as a product of the differential decay width for point like - particles (a fuction of
invariant mass of dilepton) and the transition form factor, where the form factor is the description of the
electromagnetic structure arising at the transition point. The form factor is a function of q$^{2}$ where q is
the transferred four momentum of the dileptons equivalant to their invariant mass. In this Dalitz decay, the
$\omega$ vector meson decays into a pseudoscalar meson $\pi^{0}$ and a dilepton. According to the VMD
(Vector Meson Dominance) assumption this decay occurs via a virtual $\rho$ meson, \mbox{m$_{\omega}$ $\approx$
m$_{\rho}$}~=~0.77~GeV, thus a resonance at m$_{\gamma}^{*}$ = $\sqrt(q^2)$ = m$_{\rho}$. Experimentally, the
form factor is determined by comparing the dilepton invariant mass spectrum with the point-like Quantum
Electrodynamics prediction.

The issue is that the transition form factor for the $\omega$ meson does not agree with VMD predictions
compared to other meson decays involving dileptons; for example, $\pi^{0}\to\gamma^{0}$e$^{+}$e$^{-}$ and
$\eta\to\gamma^{0}$l$^{+}$l$^{-}$ are consistent with VMD predictions. However, even extensions of VMD do not
seem to work for the $\omega$ meson. There is agreement when the decays of narrow light vector mesons into
pseudoscalar mesons and dileptons are calculated to leading order by treating pseudoscalar and vector mesons
on equal footing. These theoretical efforts can attempt to go beyond VMD in a systematic way and are in need
of further experimental input~\cite{ex:1}. Experiments have been performed and results for the $\omega$ meson
form factor measured from $\mu^{+}\mu^{-}$ are found in the 1986 review by L.G.Landsberg~\cite{ex:2} which
shows disagreement with calculations. Recently, the NA60 collaboration has claimed to have confirmed the old
data~\cite{ex:3}. It would be  useful to have data with e$^{+}$e$^{-}$ as an additional experimental approach
giving access to smaller virtual photon masses including the full reconstruction of the decaying meson.

Two sets of experiments have been performed with the WASA detector at COSY using two different reaction
mechanism. The intention is to compare the quality of the data between pd and pp reactions, in the sense of a
feasibility and background study for $\omega\to$e$^{+}$e$^{-}\pi^{0}$ decays. For \mbox{p + d $\to$ $^{3}$He +
$\omega$} 20~TB of data have been recorded during 12~days of beamtime at two different beam kinetic energies
1.5~GeV and 1.45~GeV. The excess energies are 88~MeV and 63~MeV, respectively. This reaction has a smaller
cross section and smaller background than the p+p reaction. The cross section at 1.45~GeV is
\mbox{83.6$\pm$1.5$\pm$2.2 nb}~\cite{ex:4}. The two beam energies have been selected to improve the background
subtraction. The pilot experiment for the \mbox{p + p $\to$ p +p + $\omega$} reaction has been performed for
11~days at 2.063~GeV beam kinetic energy, 60~MeV excess energy and 18TB of data have been recorded. This
reaction has a bigger cross section and significantly more background than p+d. The cross section for this
reaction at 2.063~GeV of beam kinetic energy is \mbox{5.7$\pm$0.6$\pm$0.8$\pm$0.9$\mu$b}~\cite{ex:5}. The
bigger background makes the online and offline event selection  more challenging.

\Section{Preliminary analysis}
The first analysis is being performed using the p d induced reaction. A preselection, i.e. a subset of
acquired data is used for the analysis. The preselection of the p+d reaction was based on the $^{3}$He
selection by extracting those events where $^{3}$He was formed. Particles emitted in the forward direction
i.e. $^{3}$He, can be detected in the forward detection system that includes tracking detectors and range
hodoscopes using the $\Delta$E-E method. The decay particles i.e. e$^{+}$,e$^{-}$,$\gamma$ can be identified
in the central part of the detector. The central detector consists of a mini drift chamber operated in the
field of a superconducting solenoid, a plastic scintillator barrel and an electromagnetic calorimeter. We aim
to analyse $\omega\to\gamma^{*}\pi^{0}\to$e$^{+}$e$^{-}\pi^{0}$ but start with the real photon case
$\omega\to\gamma\pi^{0}$ which is one of our reference channels. For
$\omega\to\gamma\pi^{0}\to\gamma\gamma\gamma$, the $\pi^{0}$ is reconstructed via the two decay photons.
$\omega$ mesons can be tagged via the missing mass deduced from the $^{3}$He detected in forward direction
M$_{\omega}$ = $\sqrt{(E_{p}+E_{d}-E_{He})^{2}-(P_{p}-P_{He})^{2}}$. The $\omega$ is also fully reconstructed
with the invariant mass M$_{\pi^{0}\gamma}$ =
$\sqrt{(E_{\pi^{0}}+E_{\gamma})^{2}-(P_{\pi^{0}}+P_{\gamma})^{2}}$. A peak is expected at the $\omega$ mass
0.782 GeV/c$^{2}$ in both spectra. A two-dimensional spectrum is shown on the left panel of Figure 1. In this
histogram one can see an enhancement at the $\omega$ mass indicated by the black lines. The background comes
from multi pion production. The projection of the spectra is shown on the right panel of Figure 1, one can see
a peak at the $\omega$ mass (red line) for both energies of the p+d data set.In this histogram one can see an
enhancement at the $\omega$ mass indicated by black lines. The black dashed line corresponds to
1.5~GeV/c$^{2}$ and the blue line correspond 1.45 GeV/c$^{2}$ beam kinetic energy. One can see the $\omega$
peak at the same point for both energies but with different background features. This will allow to have a
better understanding of the background.

\begin{figure}[!hbt]
  \centerline{\includegraphics[width=6.5cm, height=5.0cm]{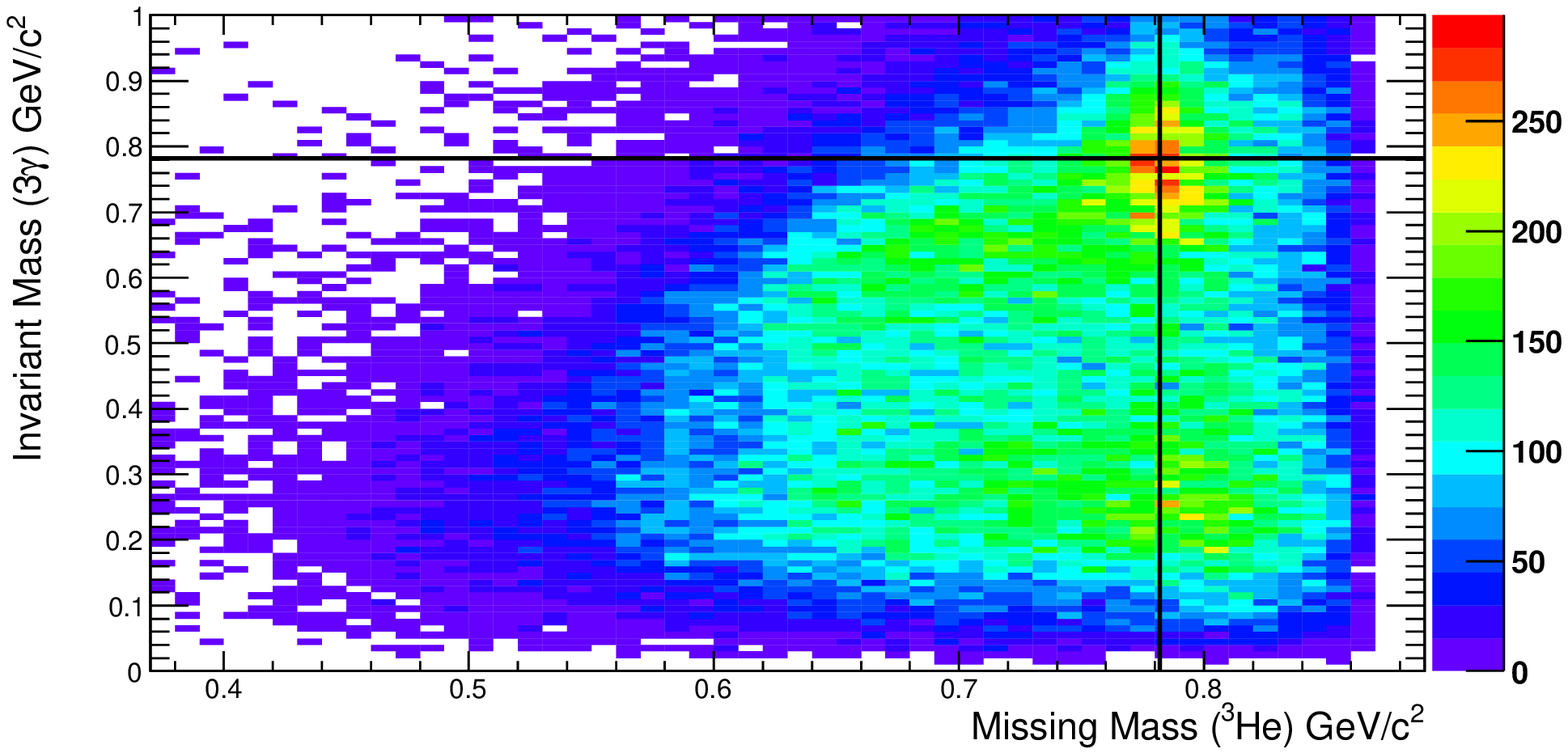}%
              \includegraphics[width=6.5cm, height=5.0cm]{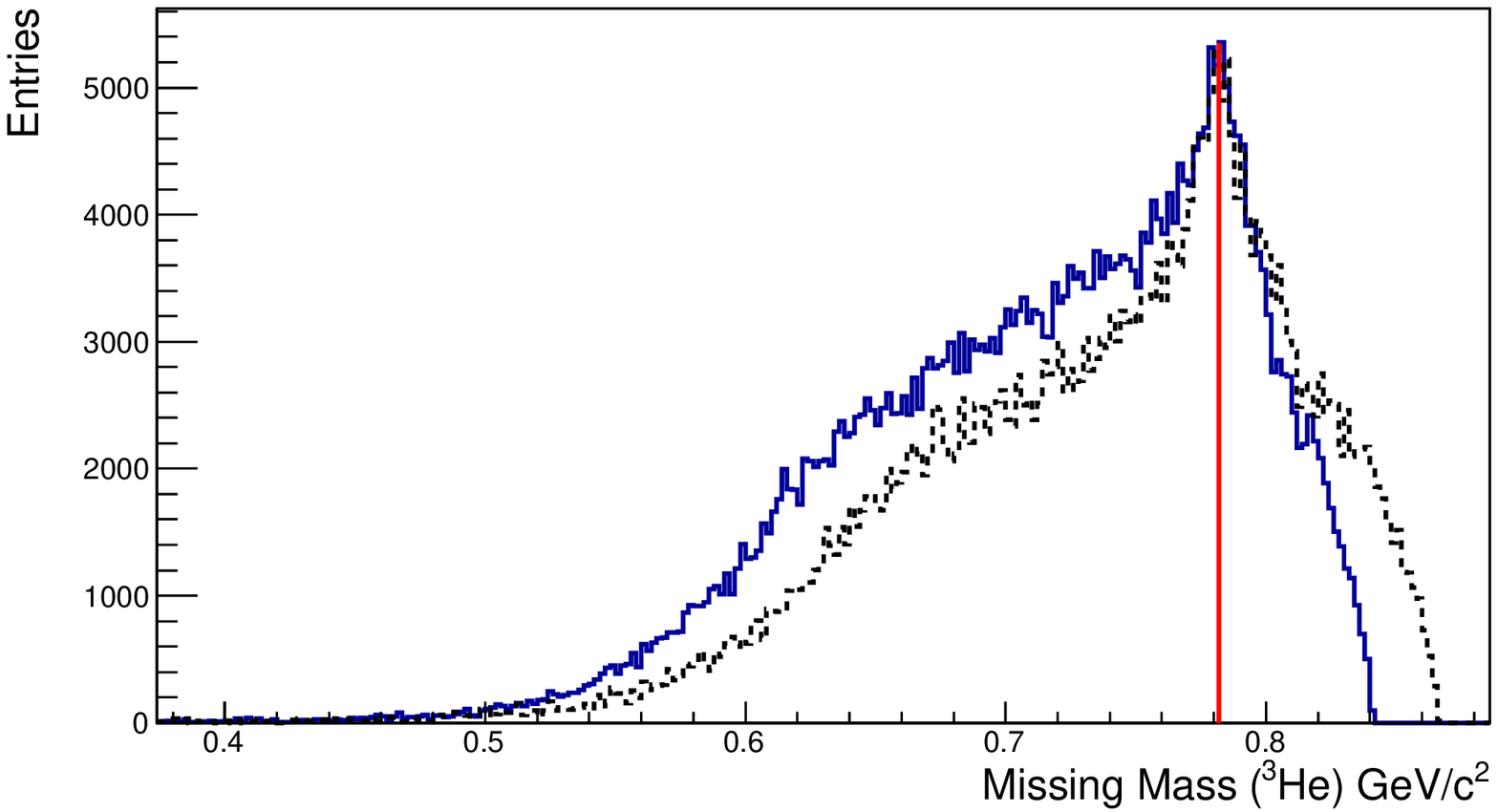}}
  \caption{\label{Fig}\small The left panel shows the invariant mass vs missing mass plot. The intersection of
the two black lines indicate the location of $\omega$ peak. The figure on the right panel is the projection
for invariant masses above 0.5~GeV/c$^{2}$. This figure shows the peak at 0.782~GeV/c$^{2}$ (red line) for
both energies. The black dashes correspond to 1.5~GeV/c$^{2}$ and the blue histogram corresponds to the
1.45~GeV/c$^{2}$ data set, showing the two phase space regions and the different position of the peak relative
to the background.}
\end{figure}

For fast analysis we need preselected data for p+p reaction as well. The preselection has been started and
data is saved into four different streams, using dedicated experiment triggers and extracting those events
where at least two proton tracks were found. Central detector conditions for the four streams are:
\begin{description}
 \item [\it stream1:] at least 3 neutral tracks in central detector ($\gamma\pi^{0}$) 
 \item [\it stream2:] at least 2 charged tracks and at least 2 neutral tracks in central
detector \\ (e$^{+}$e$^{-}\pi^{0}$, and other decays with similar topology) 
 \item [\it stream3:] at least 2 charged in central detector
 \item [\it stream4:] exactly 2 neutral tracks in central detector (background studies)
\end{description}

The goal of the analysis is to find the event candidates in the p+d reaction and compare with the p+p
reaction and then decide which reaction can be used for the production beamtime.

\coltoctitle{\boldmath Study of the $\phi\to\eta e^{+}e^{-}$ Decay at KLOE}
\authortochead{J.~Zdebik}
\label{ZJ}

\setcounter{chapter}{1}
\setcounter{section}{0}
\maketitle

\Section{Introduction}
The analysis of $\phi\to\eta\,e^+e^-$ decay is interesting from several points of view. The structure of
$\phi$ and $\eta$ mesons and underlying quark dynamics in the transition region can be extracted from $e^+e^-$
invariant mass spectrum. By comparing the experimentally measured spectrum of the lepton pair with QED
calculations for pointlike particles,  it is possible to determine transition form factor in the time-like
region of momentum transfer~\cite{landsberg}. 

The only one measurement of the form factor comes from SND  collaboration is not in good agreement with
predictions from the Vector Meson Dominance (VMD) framework~\cite{landsberg,vmdtheory}. The form factor is
often parametrized in one-pole approximation: 
\begin{equation}
  F_{\phi\eta}(q^2) = \frac{1}{1-q^2/\Lambda^2} \, .
 \label{Eq:FF}
\end{equation}
where $q=M_{ee}$ and $\Lambda$ is a free parameter. 

The theoretical calculation for $\Lambda$ is 1.0 GeV (VMD), which the value measured by SND is
0.5$\pm$0.1~GeV~\cite{snd}. Also is interesting to compare experimental results with another theoretical
models~\cite{leupold}. In this paper preliminary results of the investigation of the $\phi\to\eta\,e^+e^-$
decay at KLOE are presented. 

Additional goal of this work is also to understand the signature of this channel in the KLOE detector, since
it is the main and nonreductive background for dark matter particles searches~\cite{fayetU,simonadark}. 

\Section{The KLOE detector}
The KLOE (\textbf{Klo}ng \textbf{E}xperiment) detector is installed at the interaction point of the electron
and positron beams of the DA$\Phi$NE (\textbf{D}ouble \textbf{A}nnular \textbf{$\phi$}-factory for
\textbf{N}ice \textbf{E}xperiments) collider operating in the Laboratori Nazionali di Frascati (LNF).

DA$\Phi$NE is an $e^+e^-$ collider running at a center of mass energy of $\sim 1020$~MeV (the mass of the
$\phi$ meson). The positron and electron beams collide at an angle of $\pi$-(25 mrad), producing $\phi$ mesons
nearly at rest. At the interaction point (IP) the beam pipe has the shape of a sphere which is made of a
beryllium-aluminium alloy with 10~cm diameter and 50~$\mu$m thickness~\cite{simonadark}. 

This detector was fully constructed by the end of the year 1998~\cite{AFlavorofKLOE}. It consists of two main
subsystems: an electromagnetic calorimeter and a large drift chamber. The drift chamber and the calorimeter
are inside a superconducting coil which produces a 0.57~T magnetic field parallel to the beam axis. 

Energy and time resolutions for calorimeter are $\sigma_E/E = 5.7\%/\sqrt{E\ {\rm(GeV)}}$ and  \newline
$\sigma_t = 57\ {\rm ps}/\sqrt{E\ {\rm(GeV)}} \oplus100\ {\rm ps}$, respectively. For the drift chamber, the
spatial resolutions are $\sigma_{xy} \sim 150\ \mu$m and $\sigma_z \sim$~2~mm. The momentum resolution is
$\sigma(p_{\perp})/p_{\perp}\approx 0.4\%$. Vertices are reconstructed with a spatial resolution of $\sim$
3~mm. 

\newpage
\Section{Event Selection}
The analysis of the $\phi\to\eta\,e^+e^-$ decay with subsequent, $\eta\to\pi^+\pi^-\pi^0$, has been performed
on 1.52~fb$^{-1}$ of the KLOE dataset. The signal Monte Carlo (MC) simulation has been produced with
$d\Gamma(\phi\to\eta\,e^+e^-)/dq$ weighted according to Vector Meson Dominance model~\cite{landsberg}, using
the form factor parametrization from the SND experiment~\cite{snd}. Data-MC corrections for cluster energies
and tracking efficiency, evaluated with radiative Bhabha events and $\phi\to\rho\pi$ samples respectively,
have been applied~\cite{simonadark}. 

The first step of the analysis was preselection of events, that have to satisfy the following criteria:
\begin{enumerate}
\item two positive and two negative tracks with point of closest approach to the beam line inside a cylinder
around the interaction point (IP), with transverse radius R=4~cm and length Z=20~cm;
\item two energy clusters in calorimeter with $E > 7$~MeV not associated to any track, in an angular
acceptance $|\cos\theta_\gamma|<0.92$ and in the expected time window for a photon  
($|T_\gamma-R_\gamma/c|<{\rm MIN}(5\sigma_t,2\,{\rm ns})$); 
\item best $\pi^+\pi^-\gamma\gamma$ match to the $\eta$ mass with the pion hypothesis to assign $\pi^\pm$
tracks; the other two tracks are then assigned to $e^\pm$; 
\item loose cuts on $\eta$ and $\pi^0$ invariant masses (\mbox{$495 < M_{\pi^+\pi^-\gamma\gamma} < 600$ MeV}, 
  \mbox{$70 < M_{\gamma\gamma} < 200$ MeV}).
\end{enumerate}
After this preselection, a clear peak corresponding to $\phi\to\eta\,e^+e^-$ events is observed in the
distribution of the recoil mass to the $e^+e^-$ pair (Fig.~\ref{Fig:Mmiss}). The second peak at $\sim 590$ MeV
is due to $\phi\to K_S K_L$, $K_S\to\pi^+\pi^-$ events with a wrong mass assignment. Events in the \mbox{$535
< M_{\rm recoil}(ee) < 560$~MeV} window are retained for further analysis.
\begin{figure}[htb]
 \centerline{\includegraphics[angle=0,width=0.49\textwidth]{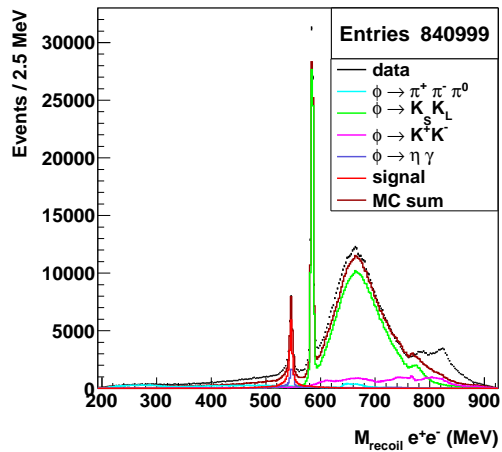}}
 \caption{\label{Fig:Mmiss}Recoiling mass against the $e^+e^-$ pair for the data sample and MC events after
preselection.} 
\end{figure}

The next selection level was performed taking into account the information from the signal channel MC
simulation. After the preselection, the remaining background contamination mainly comes from 
$\phi\to\eta\gamma$ reaction and events which have more than two charged pions in the final state. 

The background, due to $\phi\to\eta\gamma$ events with photon conversion on beam pipe (BP) or drift chamber
walls (DCW), is rejected by tracing back the tracks of the $e^+$, $e^-$ candidates and reconstructing their
invariant mass ($M_{ee}$) and distance ($D_{ee}$) at the BP/DCW surfaces. As both quantities are small in case
of photon conversions, $\phi\to\eta\gamma$ background is removed by rejecting events with: \mbox{$M_{ee}(BP) <
10$ MeV} and \mbox{$D_{ee}(BP) < 2$ cm}, \mbox{$M_{ee}(DCW)< 80$ MeV} and \mbox{$D_{ee}(DCW)< 10$ cm}. The
second relevant background, originated from $\phi\to K \bar{K}$ decays surviving analysis cuts, has more than
two charged pions in the final state and is suppressed using time-of-flight (ToF) to the calorimeter. When an
energy cluster is connected to a track, the arrival time to the calorimeter is evaluated both using the
calorimeter timing ($T_{\rm cluster}$) and the track trajectory ($T_{\rm track}=L_{\rm track}/\beta c$). The
$\Delta T = T_{\rm track}-T_{\rm cluster}$ variable is then evaluated for both electron ($\Delta T_e$) and
pion ($\Delta T_\pi$) mass hypotheses. Events with an $e^+$, $e^-$ candidate outside a 3\,$\sigma$'s window on
the $\Delta T_e$ variables are rejected. The distribution of invariant mass of $e^-e^+$ pair after all
analysis cuts is reported in Fig.~\ref{Fig:Mee}. 

\begin{figure}[htb]
 \centerline{\includegraphics[angle=0,width=0.49\textwidth]{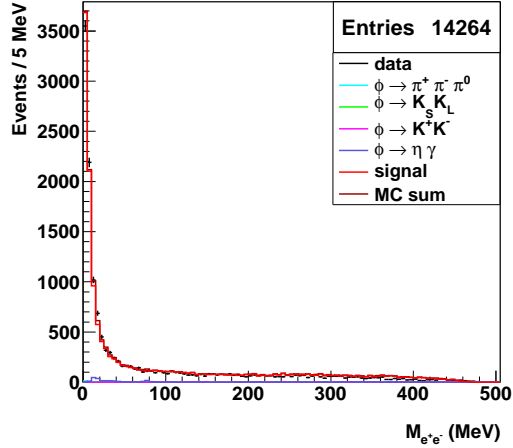}}
 \caption{\label{Fig:Mee}Distribution of the $e^+e^-$ invariant mass spectra for the process
$\phi\to\eta\,e^+e^-$ with $\eta\to\pi^+\pi^-\pi^0$ decay chain. }
\end{figure}

After all selection cuts the contamination coming from background events is less than 5\% in the final sample 
and more than 14000 events of $\phi\to\eta\,e^{+}e^{-}$ reaction we reconstructed. The statistics is almost
two orders of magnitude larger than in any previous measurment. 

\Section{Preliminary Results}
A fit to the invariant mass distribution corrected with acceptance and background subtracted was done using 
parametrizantion from Ref.~\cite{landsberg}. 

\begin{eqnarray}
  \frac{d\Gamma(\phi\to\eta\,e^+e^-)}{dq^2}
  = \ \ \ \ \ \ \ \ \ \ \ \ \ \ \ \ \ \ \ \ \ \ \ \ 
  \ \ \ \ \ \ \ \ \ \ \ \ \ \ \ \ \ \ \ \ \ \ \ \ \ 
   \ \ \ \ \ \ \ \ \ \ \ \ \ \ \ \ \ \ \ \ \ \ \ \ \
  \ \ \ \ \ \ \ \ \nonumber \\
  \frac{\alpha}{3\pi}\frac{|F_{\phi\eta}(q^2)|^2}{q^2} 
  \sqrt{1-\frac{4m^2}{q^2}} \left(1+\frac{2m^2}{q^2}\right) 
  \times \left[ \left( 1 + \frac{q^{2}}{m_{\phi}^{2}-m_{\eta}^{2}}    \right)^{2} - \frac{4m_{\phi}q^{2}}{\left(m_{\phi}^{2}-m_{\eta}^{2}\right)^{2}}  \right]^{\frac{3}{2}}
\end{eqnarray}

Free parameters of the fit are $\Lambda$ (reported in Eq.~1) and an overall normalization factor. The fit
result is shown in Fig.~\ref{Fig:Fit}. 

\newpage

\begin{figure}[htb]
 \centerline{\includegraphics[width=0.59\textwidth]{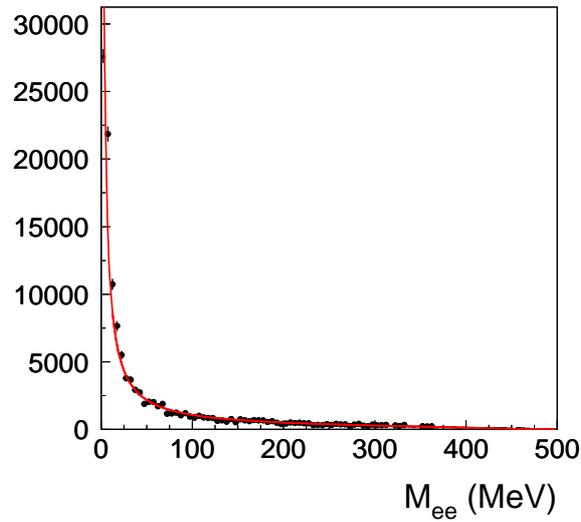}}
 \caption{\label{Fig:Fit}Fit to the $e^+e^-$ invariant mass distribution corrected  with acceptance and
background subtracted.}
\end{figure}

The achieved statistical error from the fit for $\Lambda$ is about 0.005 GeV. The evaluation of systematic
errors and implementation of the smearing matrix into fit procedure is in progress.

\coltoctitle{Electromagnetic Transition Form Factors of Pseudoscalar and Vector Mesons}
\authortochead{C.~Terschl\"{u}sen}
\label{TCP}

\setcounter{chapter}{1}
\setcounter{section}{0}
\maketitle

\Section{Introduction}
The aim of this contribution is the description of two kinds of decays,
\begin{itemize}
 \item light vector meson ($\rho$, $\omega$, $K^\ast$, $\phi$) into a light pseudoscalar meson ($\pi$, $K$, $\eta$, $\eta'$) and a dilepton,
 \item pseudoscalar meson into a photon and a dilepton.
\end{itemize}
Both processes might contain intermediate vector mesons.

The most general transition form factor for a decay including a virtual vector meson with mass $m$ is given as
\begin{eqnarray}
 F_{\rm gen.}(q) = g_0 \, \frac{m^2}{m^2-q^2} \, + \, \left( 1- g_0 \right) \, + \, g_1 \, \frac{q^2}{m^2} \, + g_2 \, \frac{q^4}{m^4} \,+ \,\ldots
\end{eqnarray}
with $q^2$ denoting the square of the invariant mass of the dilepton. In general, this form factor is an infinite series. Therefore, the parameters cannot be fixed. Progress can be made if it is possible to introduce a power counting scheme which orders the coupling constants $g_0$, $g_1$, $g_2$, $\ldots$ according to their importance. Therewith, it is possible to perform calculations up to a given order with a finite number of parameters. Additionally, one will be able to improve the results systematically by adding the next order. 

Due to the running coupling constant in QCD, perturbation theory can only be used for high but not for low energies. A possible solution are effective theories which take hadrons instead of quarks as relevant degrees of freedom. The effective theory called Chiral perturbation theory (ChPT) \cite{WZW} takes the light pseudoscalar mesons as relevant degrees of freedom and treats all other mesons, in particular the vector mesons as heavy. Therefore, it is not applicable for the energy range of the hadronic resonances. \\
We are using a new counting scheme \cite{CS} which treats the masses of both light vector and pseudoscalar mesons as soft, i.e. of the order of a typical momentum $q$. For decays, all involved momenta are smaller than the mass of the decaying particle and, hence, also of the order of $q$, 
\begin{eqnarray}
 m_V^{\vphantom{-1}}, \, m_P^{\vphantom{-1}}, \, \partial_{\mu} \sim q. \label{CS}
\end{eqnarray}

In ChPT, the range of applicability, i.e. the range for $q$ is limited (on tree level) by the not-considered mesons, in practice by $m_V^{\vphantom{-1}}$, and (for loops) by the scale $4 \pi f$, where $f$ denotes the pion decay constant. In the scheme of \cite{CS}, where vector mesons are included and where two-particle reducible diagrams (rescattering processes) are resummed, it is suggestive that the range of applicability can be pushed to larger energies. Next-to-leading-order calculations are necessary to assess this proposition. At present, phenomenological consequences are worked out in leading order and compared to data.

\Section{Decay of a light vector meson into a pseudoscalar meson and a dilepton}

Using the counting scheme (\ref{CS}), one can determine the leading-order Lagrangian for the decay of a vector meson into a pseudoscalar meson and a (real or virtual) photon, \newpage
\begin{eqnarray}
 \mathcal{L}_{\rm vec.} &=& -\, \frac{1}{16 f} \, h_A \, \varepsilon^{\mu\nu\alpha\beta} \, {\rm tr} \left\{ \left[ V_{\mu\nu}, \partial^\tau V_{\tau\alpha} \right]_+ \, \partial_\beta \Phi \right\} \, - \, \frac{1}{16 f} \, b_A \, \varepsilon^{\mu\nu\alpha\beta} \, {\rm tr} \left\{ \left[V_{\mu\nu}, V_{\alpha\beta} \right]_+ \, \left[\Phi, \chi_0 \right]_+ \right\} \nonumber \\
 && - \, \frac{e_V m_V}{4} \, {\rm tr} \left\{ V^{\mu\nu} Q \right\} \partial_\mu A_\nu. \label{Lvec}
\end{eqnarray}
Thereby, $V_{\mu\nu}$ describes the vector mesons in antisymmetric tensor representation and $\Phi$ the pseudoscalar mesons,
\begin{eqnarray}
V_{\mu\nu} &:=&  \left(\begin{array}{ccc}
                \rho^0_{\mu\nu} + \omega_{\mu\nu} & \sqrt{2} \rho^+_{\mu\nu} & \sqrt{2} K^+_{\mu\nu} \\
                \sqrt{2} \rho^-_{\mu\nu} & - \rho^0_{\mu\nu} + \omega_{\mu\nu} & \sqrt{2} K^0_{\mu\nu} \\
		\sqrt{2} K^-_{\mu\nu} & \sqrt{2} \bar{K}^0_{\mu\nu} & \sqrt{2} \phi_{\mu\nu}
               \end{array} \right) , \\
 \Phi &:=& \left(\begin{array}{ccc}
           \pi^0 + \frac{1}{\sqrt{3}} \eta_8 & \sqrt{2} \pi^+ & \sqrt{2} K^+ \\
	   \sqrt{2} \pi^- & - \pi^0 + \frac{1}{\sqrt{3}} \eta_8 & \sqrt{2} K^0 \\
	   \sqrt{2} K^- & \sqrt{2} \bar{K}^0 & - \frac{2}{\sqrt{3}} \eta_8 
          \end{array}\right) + \sqrt{\frac{2}{3}} \,  \eta_1 \, I_{3 \times 3} \,. 
\end{eqnarray}
Furthermore, $\chi_0 = {\rm diag} \left\{m_\pi^2, m_\pi^2, m_K^2 \right\}$, $Q = {\rm diag} \left\{2/3, -1/3, -1/3 \right\}$ and $A_\nu$ denotes the photon field. Note that strictly speaking the non-Goldstone boson $\eta_1$ was not included in the scheme of \cite{CS}.\\
The first two terms in (\ref{Lvec}) describe the decay of a vector meson into a pseudoscalar meson and a virtual vector meson, the last term the decay of the virtual vector meson into a photon. The decay of the photon into a dilepton is described by usual QED. Obviously, the counting scheme (\ref{CS}) allows in leading order only for decays via virtual mesons. Additionally, the parameters $h_A$ and $b_A$ can be fixed by comparing the calculated results of two-body decays into real photons to experimental data. There are no additional parameters needed to describe the decays into dileptons so that our calculations for such decays a real predictions.

If the decay can happen only via one type of virtual vector meson with mass $m$, the standard vector-meson dominance (VMD) form factor equals
\begin{eqnarray}
 F_{\rm VMD}(q) = \frac{m^2}{m^2-q^2}.
\end{eqnarray}
Our form factor calculated with (\ref{Lvec}) gets an additional non-VMD term,
\begin{eqnarray}
 F(q) = g_0 \, \frac{m^2}{m^2-q^2} \, + \left( 1- g_0 \right).
\end{eqnarray}

On the left-hand side of Fig.\ \ref{fig:VMff}, the $\omega \rightarrow \pi^0$ transition form factor is plotted \cite{paper} (solid line) in comparison to the standard VMD form factor (dashed line) and data taken by the NA60 collaboration for the decay into a dimuon \cite{NA60}. Obviously, the standard VMD form factor fails to describe the data whereas our calculation misses only the last three data points. In the middle of Fig.\ \ref{fig:VMff}, the $\phi \rightarrow \eta$ transition form factor is plotted \cite{paper} (solid line). Hereby, the physical $\eta$ meson is approximated by the octet state $\eta_8$. Again, it is compared to the standard VMD form factor (dashed line) and data taken at the VEPP-2M detector for the decay into a dielectron \cite{VEPP}. Due to the relatively large error bars there is no assessment possible which form factor describes the data better. In the near future better data are expected form KLOE\footnote{See the contribution by Jaroslaw Zdebik in these proceedings.}.

\Section{Decay of a light pseudoscalar meson into a photon and a dilepton}

As a more phenomenological approach for the decay of a light pseudoscalar meson into two (real or virtual) photons, we take both our Lagrangian (\ref{Lvec}) describing the decay via two virtual vector mesons and the Wess-Zumino-Witten Lagrangian \cite{WZW}
\begin{eqnarray}
 \mathcal{L}_{\rm WZW} = \frac{3 e^2}{8 \pi^2 f} \, \varepsilon^{\mu\nu\alpha\beta} \, {\rm tr}  \left\{ Q^2 \Phi \right\} \partial_\mu A_\alpha \, \partial_\nu A_\beta + \mathcal{O}(\Phi^2)
\end{eqnarray}
which describes the direct decay into two photons. Additionally, the $\eta$-$\eta'$ mixing is given by
\begin{eqnarray}
 \eta = \cos\theta \, \eta_8 - \sin\theta \, \eta_1 \, \wedge \, \eta' = \sin\theta \, \eta_8 + \cos\theta \, \eta_1
\end{eqnarray}
with the mixing angle $\theta \approx - 19.5^\circ$.

The $\eta \rightarrow \gamma$ transition form factor is plotted on the right-hand side of Fig.\ \ref{fig:VMff} (solid line) in comparison to the standard VMD form factor\footnote{The decay $\eta \rightarrow \gamma \, l^+ l^-$ can happen via all three neutral light vector mesons and, therefore, the standard VMD form factor is more complicated than for the case of only one type of possible virtual meson.} (dashed line) and data taken by the NA60 collaboration for the decay into a dimuon \cite{NA60}. Here, the form factors are on top of each other and describe the data equally well.

\begin{figure}[htb]
 \includegraphics[width = 0.2\textheight, trim = 50 0 0 0]{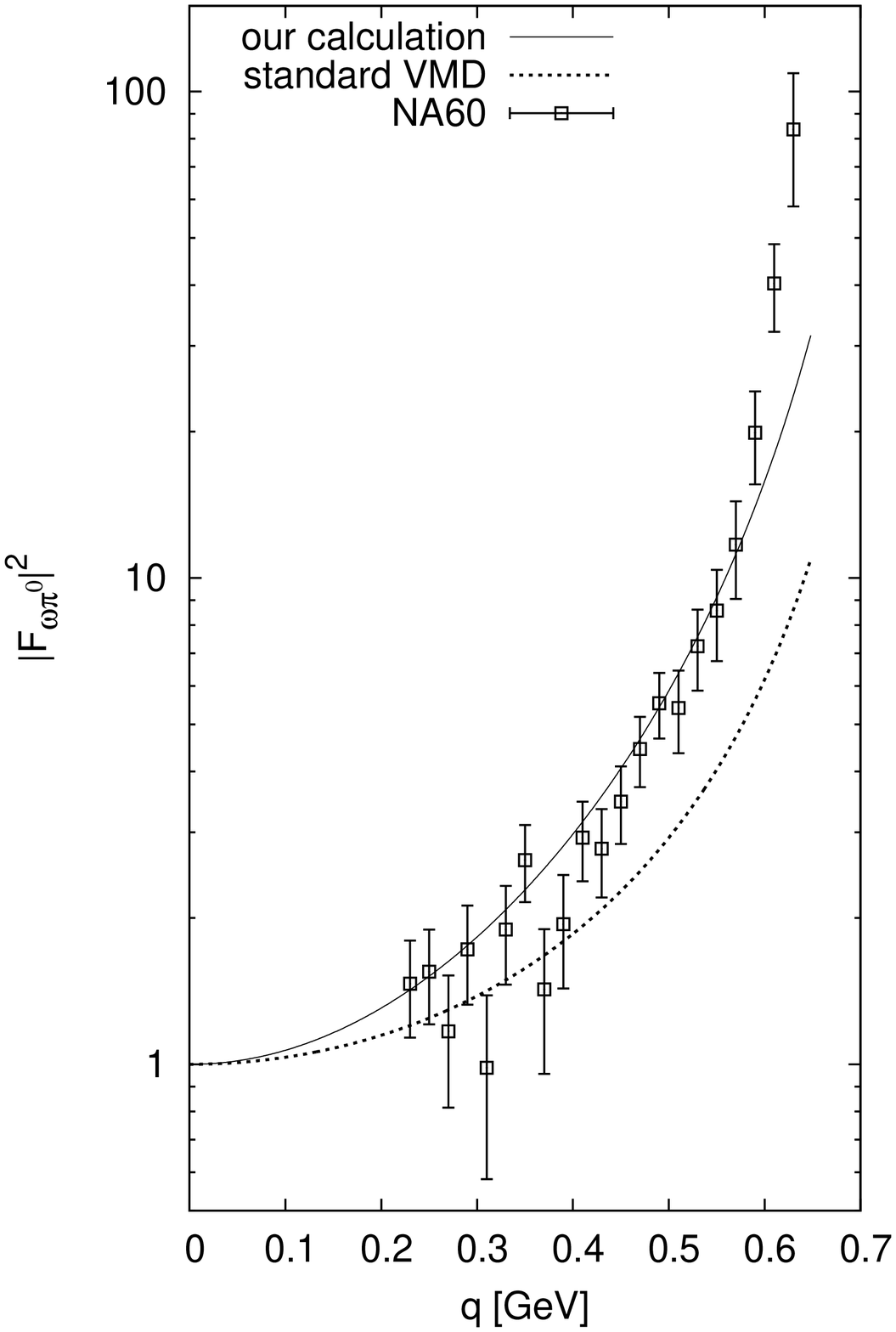}
 \includegraphics[width = 0.2\textheight, trim = 50 0 0 0]{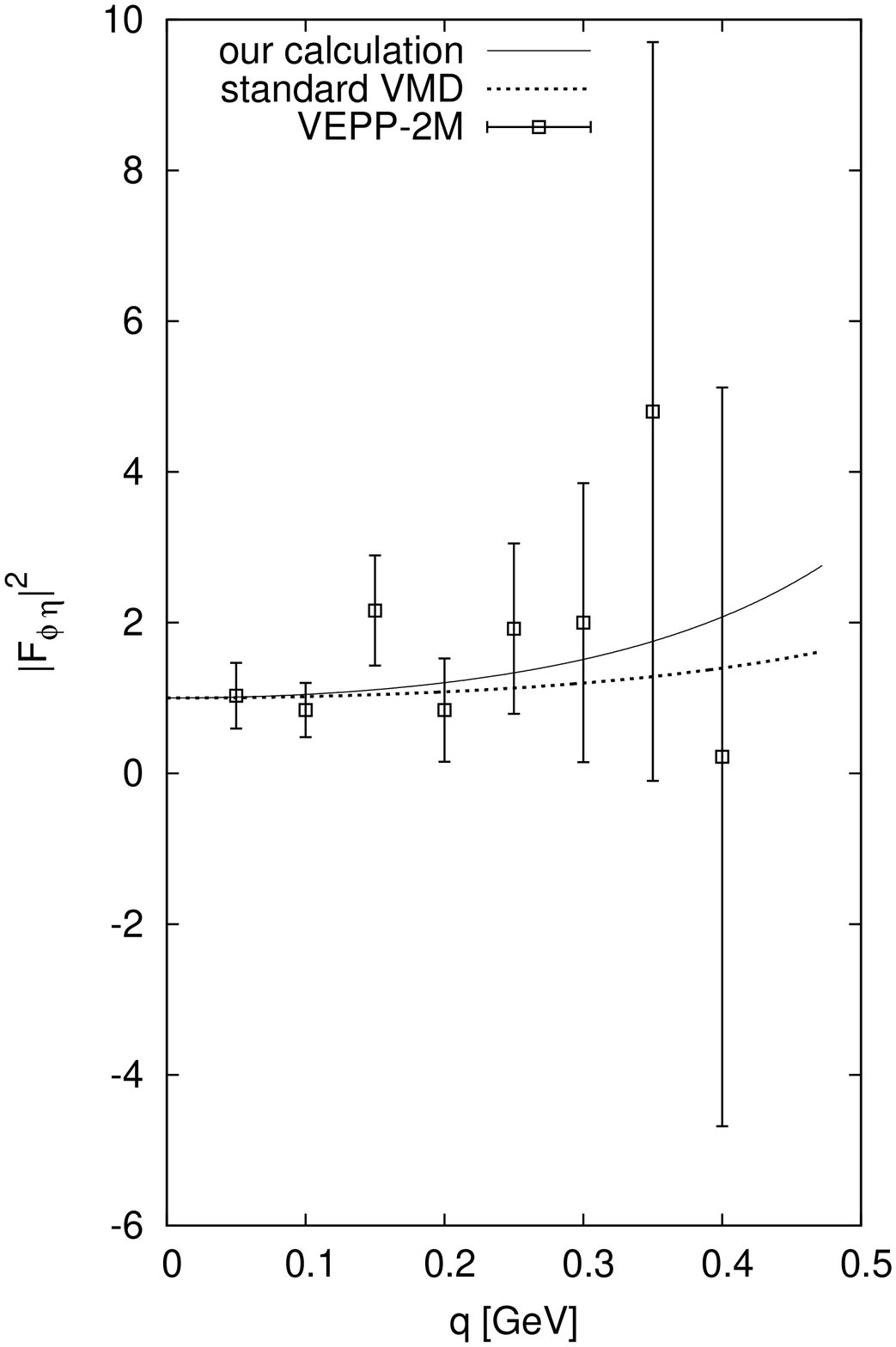}
 \includegraphics[width = 0.2\textheight, trim = 50 0 0 0]{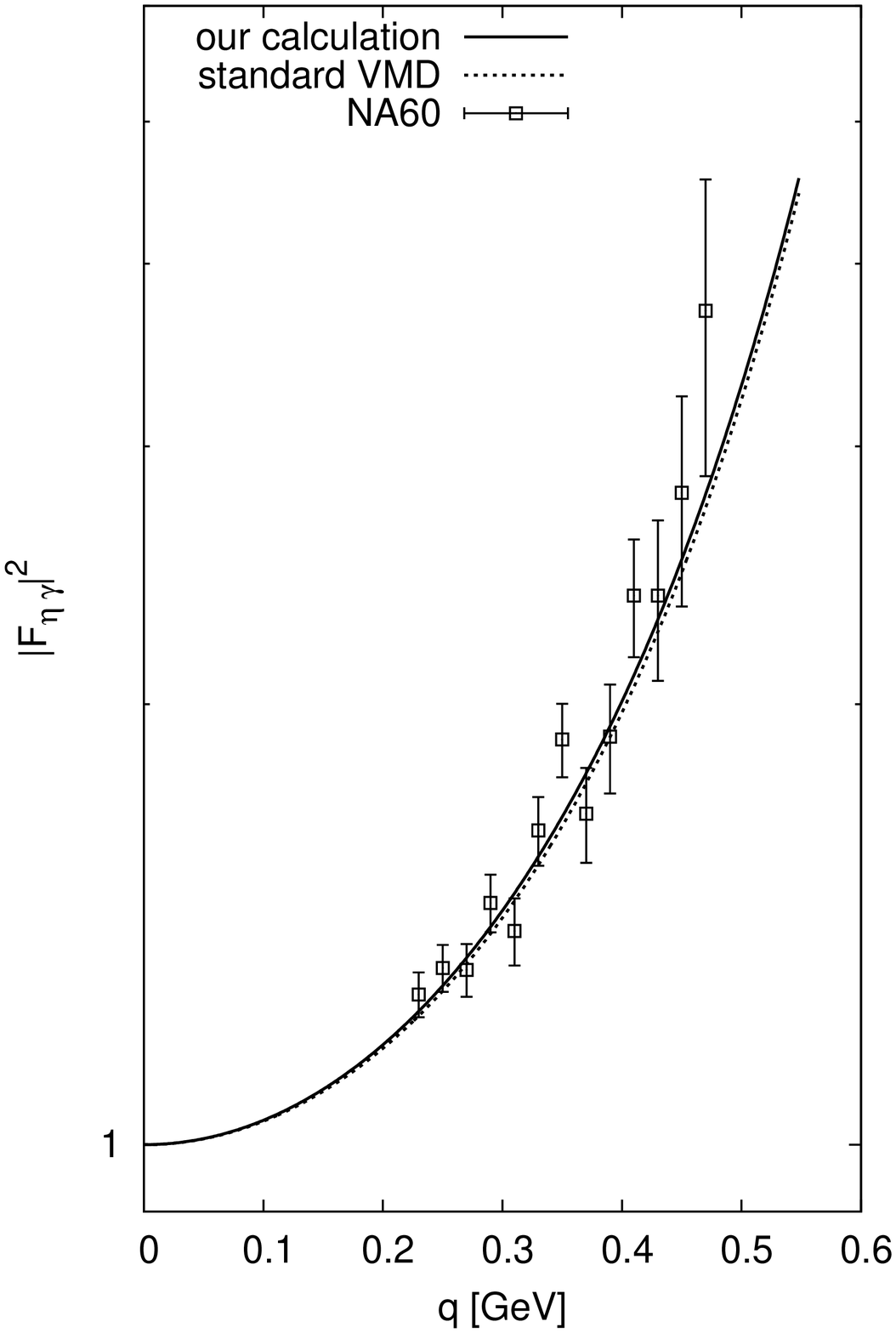}
 \caption{\textbf{Left-hand side}: $\omega \rightarrow \pi^0$ form factor compared to dimuon data taken by the NA60 collaboration \cite{NA60}. \textbf{Middle}: $\phi \rightarrow \eta$ form factor compared to dielectron data taken at the VEPP-2M detector \cite{VEPP}. \textbf{Right-hand side}: $\eta \rightarrow \gamma$ form factor compared to dimuon data taken by the NA60 collaboration \cite{NA60}. The first two results are published in \cite{paper}.}
 \label{fig:VMff}
\end{figure}

\Section{Summary and outlook}

Using a new counting scheme (\ref{CS}) we were able to determine the leading-order Lagrangian for the decays $V \rightarrow P l^+ l^-$ and for the vector part of the decay $P \rightarrow \gamma \, l^+ l^-$. The form factors for the transitions $\omega \rightarrow \pi^0$, $\phi \rightarrow \eta$ and $\eta \rightarrow \gamma$ describe the experimental data very well, especially the $\omega \rightarrow \pi^0$ transition is much better described than with standard VMD. Additionally, we were able to describe the partial decay widths very well \cite{paper}. As a next step, next-to-leading order calculations have to be performed.

\coltoctitle{\boldmath Investigation of $\omega\ \to\ \pi^+\pi^-\pi^0$ with WASA-AT-COSY}
\authortochead{L.~Heijkenskj\"{o}ld and S.~Sawant}
\label{HLSS}

\setcounter{chapter}{1}
\setcounter{section}{0}
\maketitle

\Section{Introduction}
The WASA-at-COSY collaboration has recently performed experiments dedicated to studies of the $\omega$ meson
decays. One of the decay channels that will be under detailed investigations is the most probable decay of the
$\omega$ meson into $\pi^+ \pi^- \pi^0$. The goal of this experiment is the investigation of the $\omega$
decay mechanism to three pions. This information may be gained from a comparison of the density distribution
in Dalitz plot with appropriate theoretical predictions. To reach this goal a high statistics experimental
Dalitz plot distribution is necessary. Measurements of the Dalitz plot for $\omega \to \pi^+ \pi^- \pi^0$
decay have been done before in experiments on the determination of the spin and parity of the $\omega$
meson~\cite{intro0}. However, the statistics of existing measurements is limited to about 4600 $\omega\to3\pi$
events. The WASA-at-COSY experiment has produced a higher amount of fully reconstructed $\omega\to3\pi$ events
using proton-proton and proton-deuteron collisions.

The decay mechanism was considered within Vector Meson Dominance model as proceeding via the $\rho\pi$
intermediate state~\cite{motiv0}. More recent predictions base on a counting scheme for flavor-SU(3) systems
of Goldstone bosons and light vector mesons~\cite{motiv1}. The importance of the $\pi\pi$ final state
interaction is also considered~\cite{motiv2}. However, the expected deviations from Vector Meson Dominance
model are small.

\Section{WASA-at-COSY}
The experiments have been performed with the WASA detector setup at COSY~\cite{detector}. COSY (COoler
SYnchrotron) is a synchrotron and storage ring which can provide a beam of (un)polarized protons or deuterons
with a momentum range of 600-3700~MeV/c.

WASA (Wide Angle Shower Apparatus) is a nearly $4\pi$ detector optimised for studying production and decay of
light mesons. A unique pellet generator provides frozen pellets of hydrogen and deuterium as a fixed target.
The central part of WASA can detect both charged and neutral particles while the forward detector is designed
to detect scattered projectiles and charged recoil particles like protons, deuterons and He nuclei.

\Section{The current analyses}
In spring 2011, WASA-at-COSY collected data of $\omega$ production in proton beam on proton target (p-p)
collisions ($\mathrm{p}+\mathrm{p}\to \mathrm{p}+\mathrm{p}+ \omega$) at T$_p$=2.063~GeV and in proton beam on
deuteron target (p-d) collisions ($\mathrm{p}+\mathrm{d}\to {}^3\mathrm{He}+\omega$) at T$_p$=1.45~GeV and
T$_p$=1.5~GeV. The corresponding excess energies are 60~MeV for the p-p reaction and 63~MeV and 88~MeV for the
p-d reaction, respectively. These two reaction have complementary advantages and drawbacks, which are
summarised in Tab.~\ref{table:prodreac}.
\begin{table}[!h]
\caption{The table displays the advantages and drawbacks of the two production reactions.
\label{table:prodreac}}
\begin{tabular}{|p{0.1\linewidth}|p{0.4\linewidth}|p{0.4\linewidth}|}
\hline
Reaction & Advantage & Drawback \\
\hline
p-p & Higher production cross section, & Fast protons, majority punches\\
 & at T$_{p}$ = 2.063 GeV $\sigma_{\mathrm{tot}} = 5.7\ \mu$b \cite{xsec1}. & through forward part of detector.\\
 &  & Requires elaborate trigger. \\
\hline
p-d & Simple triggering on ${}^3$He. & Lower production cross section,\\
 & ${}^3$He stops in forward part of detector. Advantageous for tagging. &  at T$_{p}$ = 1.45 GeV $\sigma_{\mathrm{tot}} = 83.6$ nb \cite{xsec2}.\\
\hline
\end{tabular}
\end{table}

\Subsection{First preliminary results}
Preliminary analyses have been performed on both data sets. In these analyses the protons or the ${}^3$He
particles are identified in the forward part of WASA using the $\Delta$E/E technique. The decay products of
the $\omega$ are detected in the central part of WASA. For charged pions, the charge and momentum are
reconstructed using information from the central mini drift chamber placed in the solenoid magnetic field. The
$\pi^0$ is reconstructed using the energy deposit and angle in the central calorimeter of the two photons.

In presented analysis 11\% for the p-d data and 15\% for the p-p data of the detected events have been
analysed. On the left panel of Fig.~\ref{fig:IMMM}, which shows the missing mass of the ${}^3\mathrm{He}$
detected for $\mathrm{p}+\mathrm{d}\to {}^3\mathrm{He}+ \pi^+ + \pi^- + \pi^0$, a peak at the value of
$\omega$ mass is clearly visible. In the middle plot of Fig.~\ref{fig:IMMM} the three pion invariant mass
versus the missing mass of two protons detected for $\mathrm{p}+\mathrm{p}\to \mathrm{p}+\mathrm{p}+ \pi^+ +
\pi^- + \pi^0$ reaction is presented. By putting a cut shown in the figure on the invariant mass of three
pions around the $\omega$ meson mass, the missing mass of two protons (right plot in Fig.~\ref{fig:IMMM})
does not show a clear peak but only a shoulder at the value of $\omega$ mass. This suggests a small $\omega$
meson signal hidden under the three pion background.

\begin{figure}[!h]
\centering
\includegraphics[scale=0.25]{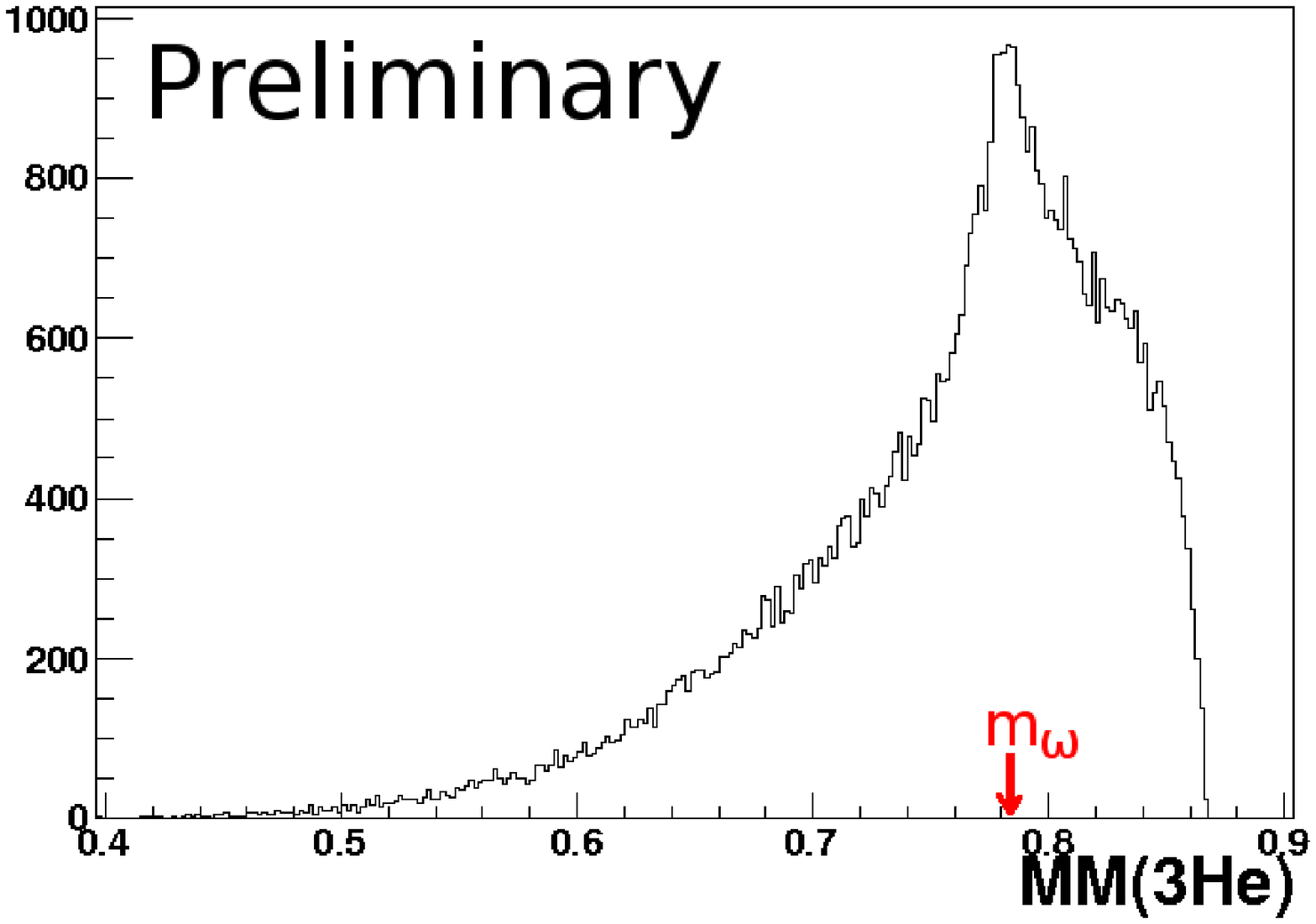}
\includegraphics[scale=0.23]{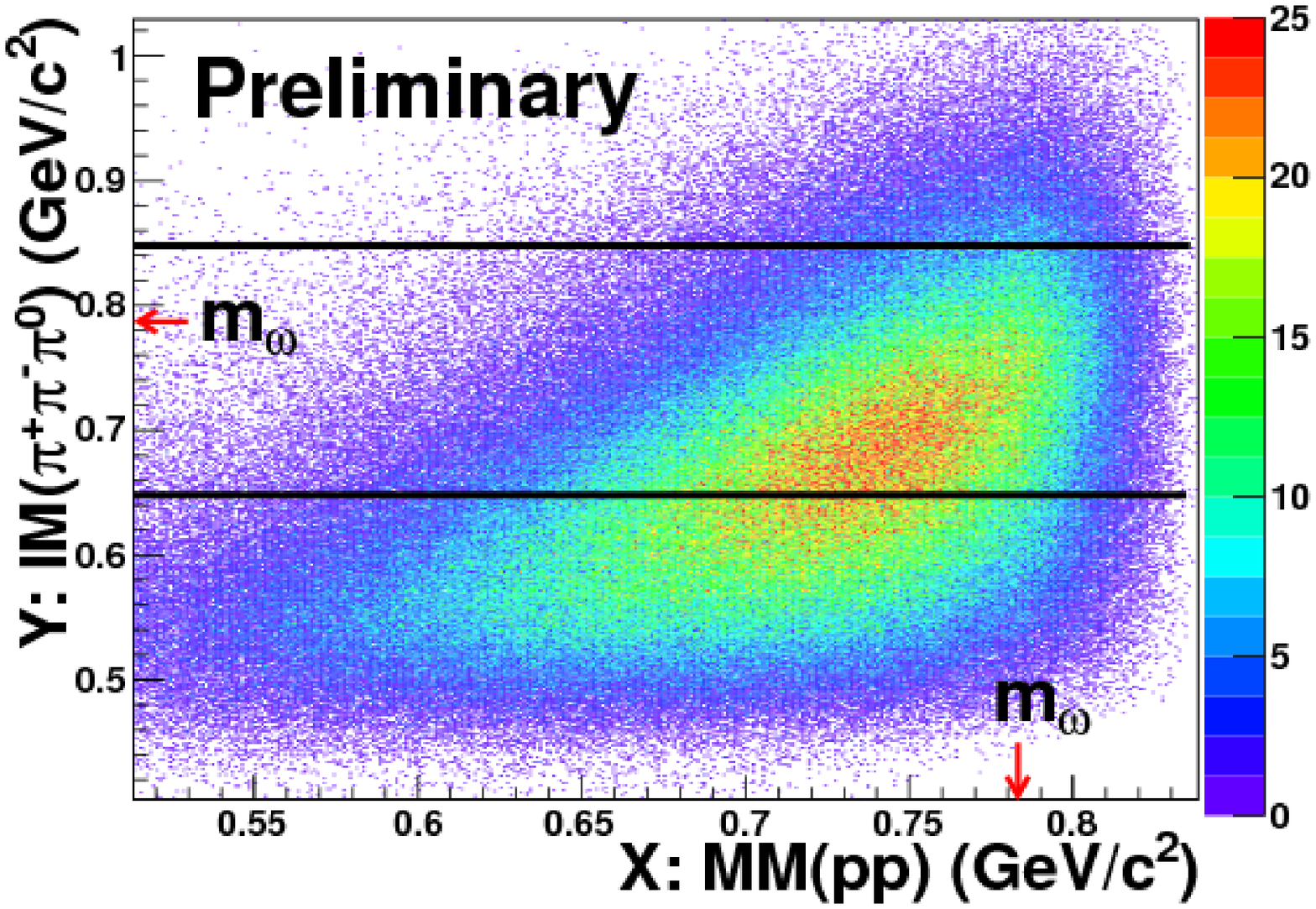}
\includegraphics[scale=0.23]{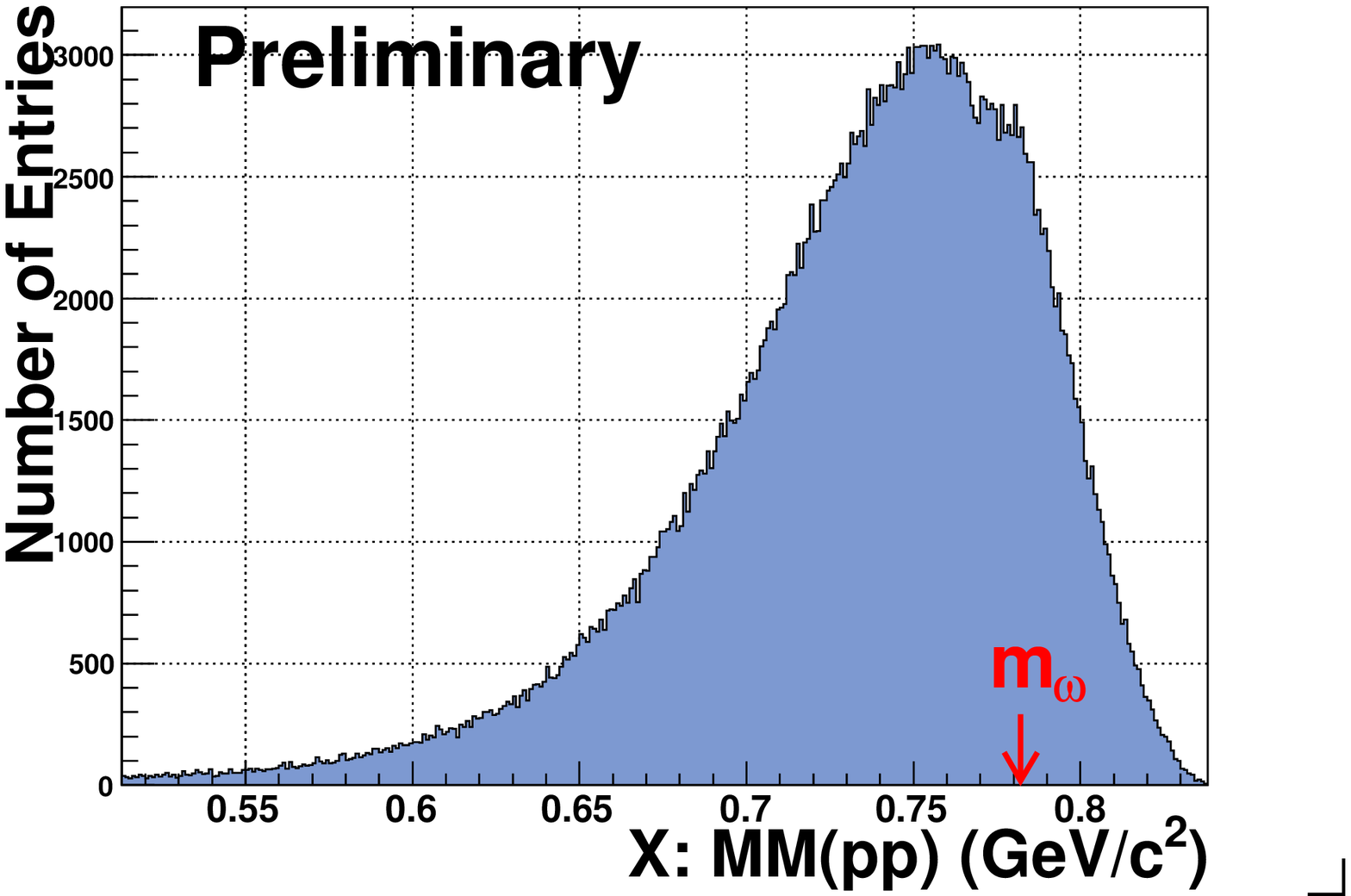}
\caption{\label{fig:IMMM}{\bf Left}: The missing mass deduced from the ${}^3$He in the p-d data at
T$_{p}$~=~1.5~GeV. {\bf Middle}: The invariant mass of three pions versus the missing mass from the protons in
the p-p data. Black lines show the cut on the invariant mass of $\pi^+\pi^-\pi^0$. {\bf Right}: The missing
mass from the protons in the p-p data after the cut on the invariant mass of the three pions.}
\end{figure}

\Subsection{Outlook}
In order to extract the $\omega$ signal, the next steps in the analyses would be fine tuning of energy
calibration of all detectors. Then the kinematic fitting method will be applied to improve overall resolution.
These steps should allow to obtain the best signal to background ratio and properly select the $\omega$
signal.

\coltoctitle{\boldmath Dispersive Analyses of $\omega/\phi\to3\pi$ and $\eta'\to\eta\pi\pi$}
\authortochead{S.~P.~Schneider and F.~Niecknig}
\label{SSNF}

\setcounter{chapter}{1}
\setcounter{section}{0}
\maketitle

\Section{Introduction}
The advent of high-statistics measurements of hadronic three-body decays calls for equally accurate
theoretical descriptions. In this context rescattering effects between decay products play a very important
role if one aspires to perform precision amplitude analyses. While perturbative approaches, such as chiral
perturbation theory or non-relativistic approaches, implement final-state interactions up to a certain order
in a small power-counting parameter and are restricted to a low-energy regime, dispersive approaches resum
hadronic rescattering up to all orders and thus allow for an extension to higher energies. While the
groundwork for dispersive analyses of hadronic three-body decays has been laid long ago~\cite{KT},
high-accuracy parameterizations of phase-shifts and sufficient computational power, which are necessary for
such an enterprise, have only recently become available.

We apply the dispersive approach to two specific examples. $\omega/\phi\to3\pi$ is the simplest imaginable
system in which three-body effects in hadronic rescattering can be studied, due to the fact that the decaying
particles are of vector type. If one neglects contributions from F- and higher partial waves, the system is
completely determined by P-wave interactions. In the case of $\phi\to3\pi$ the phase space is sufficiently
large to study three-body effects on the $\rho$-resonance. With its large existing ($\phi$:
KLOE/CMD-2~\cite{KLOE,CMD}) and upcoming ($\omega$: WASA-at-COSY) data base $\omega/\phi\to3\pi$ provides an
ideal testing ground for the approach.

$\eta'\to\eta\pi\pi$ on the other hand is one of the very few processes in which effects from $\pi\eta$
scattering might be studied, in the hope of gathering information on $\pi\eta$ phase shifts. Additionally, the
neutral channel $\eta'\to\eta\pi^0\pi^0$ exhibits a cusp effect that may allow for a determination of $\pi\pi$
S-wave scattering lengths~\cite{Cuspetap}. Finally, recent and upcoming experimental efforts from various
collaborations (BES-III, WASA-at-COSY, ELSA, CB-AT-MAMI-C) call for a renewed analysis.

\Section{The framework}
The goal of the dispersive approach is to relate the decay amplitudes to a set of integral equations derived
from the fundamental principles of unitarity and maximal analyticity, which are then solved numerically. We
follow a Khuri--Treiman type dispersion relation ansatz~\cite{KT}, where the dispersive result for two-body
scattering is linked to the three-body decay by analytic continuation in the mass of the decaying particle and
the center-of-mass energy.

\Subsection{Decay amplitude for \boldmath{\(\omega/\phi\to\pi^{0}\pi^{+}\pi^{-}\)} }
In the following we resort to the standard Mandelstam representation for three-body decays,
\mbox{$s=(p_{\pi^{-}}+p_{\pi^{+}})$}, \mbox{$t=(p_{\pi^{-}}+p_{\pi^{0}})^{2}$},
\mbox{$u=(p_{\pi^{+}}+p_{\pi^{0}})^{2}$} with \(s+t+u=3s_{0}\equiv M_{V}^{2}+3M_{\pi}^{2}\), where
$M_V=M_{\omega/\phi}$ is the mass of the decaying vector meson. The process is of odd intrinsic parity and we
may decompose the amplitude \(\mathcal{M}(s,t,u)\) according to
\begin{equation}
\mathcal{M}(s,t,u)=\epsilon_{\mu\nu\alpha\beta}n^{\mu}p_{\pi^{+}}^{\nu}p_{\pi^{-}}^{\alpha}p_{\pi^{0}}^{\beta}
\,\mathcal{F}(s,t, u)~,
\end{equation}
where $n^\mu$ is the polarization vector of the decaying $\omega/\phi$. Due to Bose symmetry only odd angular
momenta contribute in the partial-wave expansion of \(\mathcal{F}(s,t,u)\). We restrict our analysis to P-wave
contributions only, since the effects of F- and higher partial waves are expected to be significantly
suppressed. The P-wave part of the amplitude can be decomposed into a sum of single-variable amplitudes
\(\mathcal{F}(s)\) (compare~\cite{Stern}),
\begin{equation}
\mathcal{F}(s,t,u)=\mathcal{F}(s)+\mathcal{F}(t)+\mathcal{F}(u)~. 
\end{equation}
Imposing unitarity and elastic final-state interactions provides us with the discontinuity of
$\mathcal{F}(s)$~\cite{Watson},
\begin{equation}
 \mathrm{disc}\,\mathcal{F}(s)=2i\big[\mathcal{F}(s)+\hat{\mathcal{F}}(s)\big]e^{-i\delta_1(s)}
\sin(\delta_1(s))~, 
\end{equation}
where \(\mathcal{F}(s)+\hat{\mathcal{F}}(s)\) is the P-wave projection of \(\mathcal{F}(s,t,u)\) and
\(\delta_1(s)\) the \(\pi\pi\) P-wave phase shift taken from phenomenology~\cite{Caprini,KMP}. The
inhomogeneity $\hat{\mathcal{F}}(s)$ is given as
\begin{align}
\hat{\mathcal{F}}(s) &= 3\langle (1 - z^{2})\mathcal{F}\rangle ~,\quad \langle z^{n} f \rangle=
\frac{1}{2}\int_{-1}^{1}\mathrm{d}z\,z^{n}f\left(\frac{3s_{0}-s+z\,\kappa(s)}{2}\right)~, \nonumber\\
\kappa(s)
&=\sqrt{1-\frac{4M_{\pi}^{2}}{s}}\sqrt{\left((M_{V}+M_{\pi})^2-s\right)\left((M_{V}-M_{\pi})^2-s\right)}~.
\end{align}
To solve for the amplitude we use a product ansatz \(\mathcal{F}(s)=\Omega(s)\phi(s)\) where \(\Omega(s)\) is
the well-known Muskhelishvili--Omn\`es (MO) solution to the homogeneous equation
\(\hat{\mathcal{F}}(s)=0\)~\cite{Omnes}.
For $\mathcal{F}(s)$ we thus obtain
\begin{equation}
\mathcal{F}(s)=\Omega(s)\bigg\{\alpha+\frac{s}{\pi}\int_{4M_{\pi}^{2}}^{\infty}\frac{\mathrm{d}s'}{s'}\frac{
\hat{ \mathcal{F}} (s)\sin(\delta(s'))}{|\Omega(s')|(s'-s)}\bigg\}~, \quad \Omega(s)=\exp
\bigg\{\frac{s}{\pi}\int_{4M_{\pi}^{2}}^{\infty}\frac{\mathrm{d}s'}{s'}\frac{\delta(s')}{(s'-s)}\bigg\}~. 
\end{equation}
The subtraction constant \(\alpha\) is fixed by the partial decay width. The number of subtractions is chosen
such that the dispersion integral is guaranteed to converge. The issue of performing the double integration
correctly in view of the analytic continuation in \(M_{V}\) and \(s\) will not be discussed here. We refer
to~\cite{Bronzan} and references therein.

\Subsection{\boldmath{\(\eta'\to\eta\pi\pi\)} }
Similarly to the \(\omega/\phi\to3\pi\) case we can decompose the full amplitude into isospin components,
\begin{equation}
\mathcal{M}(s,t,u)=\mathcal{M}_{0}^{\pi\pi}(s)+\mathcal{M}_{0}^{\pi\eta}(t)+\mathcal{M}_{0}^{\pi\eta}(u)~,
\end{equation}
with the amplitudes of the $\pi\pi$ S-wave $\mathcal{M}_{0}^{\pi\pi}$ and the $\pi\eta$ S-wave
$\mathcal{M}_0^{\pi\eta}$. We neglect the exotic $\pi\eta$ P-wave throughout. The set of integral equations
can be obtained
analogously as before,
\begin{align}
\mathcal{M}_{0}^{\pi\pi}(s)&=\Omega_{0}^{\pi\pi}(s)\bigg\{a_{0}+b_{0}s+c_{0}s^{2}+\frac{s^{3}}{\pi}\int_{4M_{
\pi}^{2}}^{\infty}\frac{\mathrm{d}s'}{s'^{3}}\frac{\sin\delta_{0}^{\pi\pi}(s')\hat{\mathcal{M}}_{0}^{\pi\pi}
(s')}{ |\Omega_{0}^{\pi\pi}(s')|(s'-s-i\epsilon)}\bigg\}~,\nonumber\\
\mathcal{M}_{0}^{\pi\eta}(t)&=\Omega_{0}^{\pi\eta}(t)\bigg\{a_{1}+\frac{t^{2}}{\pi}\int_{t_{\eta}}^{\infty}
\frac{\mathrm{d}t'}{t'^{2}}\frac{\sin\delta_{0}^{\pi\eta}(t')\hat{\mathcal{M}}_{0}^{\pi\eta}(t')}{|\Omega_{0}^
{\pi\eta} (t')|(t'-t-i\epsilon)}\bigg\},\quad t_{\eta}=(M_{\eta}+\mathcal{M}_{\pi})^{2}~.
\end{align}
We fix the $\pi\eta$ phase shift $\delta_{\pi\eta}$ using a simplified unitarization model. The subtraction
constants are matched to Resonance Chiral Theory (RChT, see~\cite{RChT}), using a narrow-width-resonance
approximation for the Omne\`es function $\Omega(s)\to{M_S^2}/{(M_S^2-s)}$. This is by no means a perfect
method to fix the subtraction constants, since obviously the $\pi\pi$ S-wave is not well-described by a
resonance approximation. An alternative would be to match to large-$N_C$ chiral perturbation theory. In this
case, however, it is not clear how to properly treat resonance contributions in the low-energy constants.

\Section{Preliminary numerical results}
The integral equations are solved numerically by an iterative procedure. We start out with an arbitrary input
function  and calculate the inhomogeneity \(\hat{\mathcal{F}}(s)\). The single-variable amplitude is then
calculated from this inhomogeneity, which is in turn used as input for the following iteration step. We repeat
the procedure until convergence is reached up to satisfactory accuracy. 

\Subsection{\boldmath{\(\phi/\omega\to 3\pi\)} amplitudes and Dalitz plot}
\begin{figure}[t]
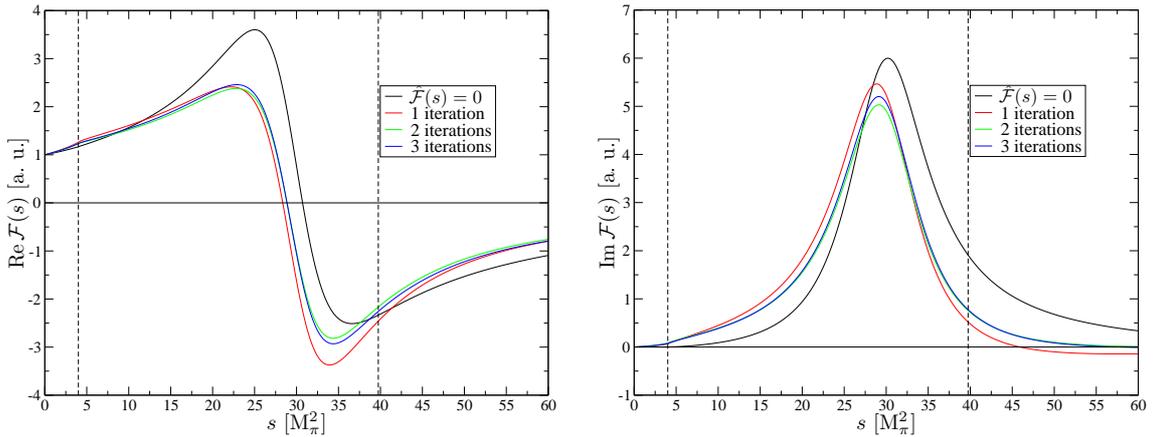

 \centering 
\psfrag{xlabel}[c][b][0.8]{$s~[{\rm M_{\pi}^2}]$}
\psfrag{Mhat=0}[c][][0.7]{~~$\hat{\mathcal{F}}(s)=0$}
\psfrag{1. iteration}[c][][0.7]{1 iteration~~}
\psfrag{2. iteration}[c][][0.7]{2 iterations}
\psfrag{3. iteration}[c][][0.7]{3 iterations}
\psfrag{RealM(s)}[c][][0.8]{$\rm{Re}\mathcal{F}(s)$}
\psfrag{ImagM(s)}[c][][0.8]{$\rm{Im}\mathcal{F}(s)$}
\psfrag{ylabel}[c][][0.8]{${\rm Re}\,\mathcal{F}(s)~[\rm{a.~u.}]$}
\includegraphics[width= 0.48\linewidth]{./SchneiderNiecknig/schneider1.eps} \hfill
\psfrag{ylabel}[c][][0.8]{${\rm Im}\,\mathcal{F}(s)~[\rm{a.~u.}]$}
\includegraphics[width= 0.48\linewidth]{./SchneiderNiecknig/schneider2.eps}
 \caption{Successive iterations steps of real (left) and imaginary (right) part of the amplitude
$\mathcal{F}(s)$ for
$\phi\to 3\pi$. The dashed lines denote the physical region of the decay.}
\label{fig:amplitudephi}
\end{figure}
To illustrate the convergence of the iteration procedure the \(\phi\to3\pi\) amplitude $\mathcal{F}(s)$ is
plotted after each iteration step in Fig.~\ref{fig:amplitudephi}. We have chosen the MO function as the
starting point for the iteration. It should be mentioned that the final result of the iteration is independent
of the specific choice of the starting point; however, choosing the MO function allows us to quantify
crossed-channel rescattering effects in a plausible way. We find that these effects are sizable and from the
point of view of precision analyses certainly not negligible. As seen in Fig.~\ref{fig:amplitudephi},
convergence of the amplitude is reached after three iterations; for \(\omega\to 3\pi\) (not shown here), this
is the case after two iterations.

For a three-body decay, the Dalitz plot distribution is of particular interest. The Dalitz plot is a
two-dimensional scatter plot in terms of two kinematic variables; in the case of $\phi/\omega\to3\pi$ it is
common practice to use $x={(t-u)}/{(2M_{V})}$ and $y ={[(M_{V}-M_{\pi})^2-s]}/{(2M_{V})}$. The
\emph{normalized} Dalitz plot is shown in Fig.~\ref{fig:Dalitzomega}; we note that it is independent of the
subtraction constant, so that Fig.~\ref{fig:Dalitzomega} is free from any input aside from the $\pi\pi$ P-wave
phase shift, which is well-established up to roughly $1.2$ GeV. 

\begin{figure}[b]
\centering
\includegraphics[width = 0.49\linewidth]{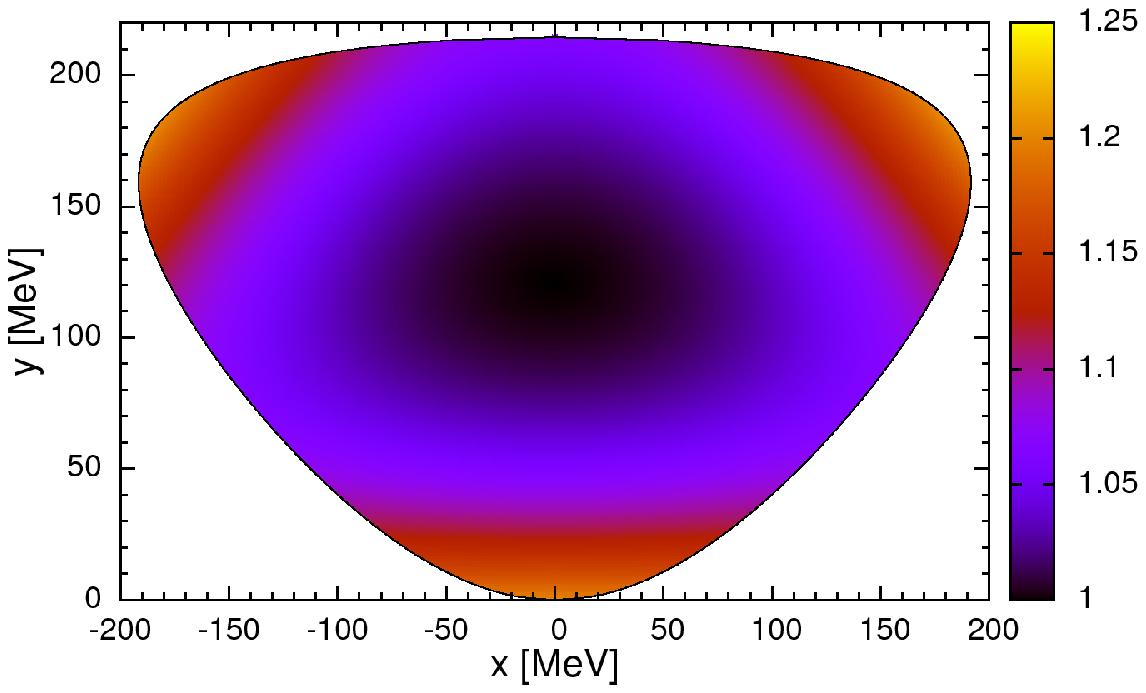} \hfill
\includegraphics[width = 0.49\linewidth]{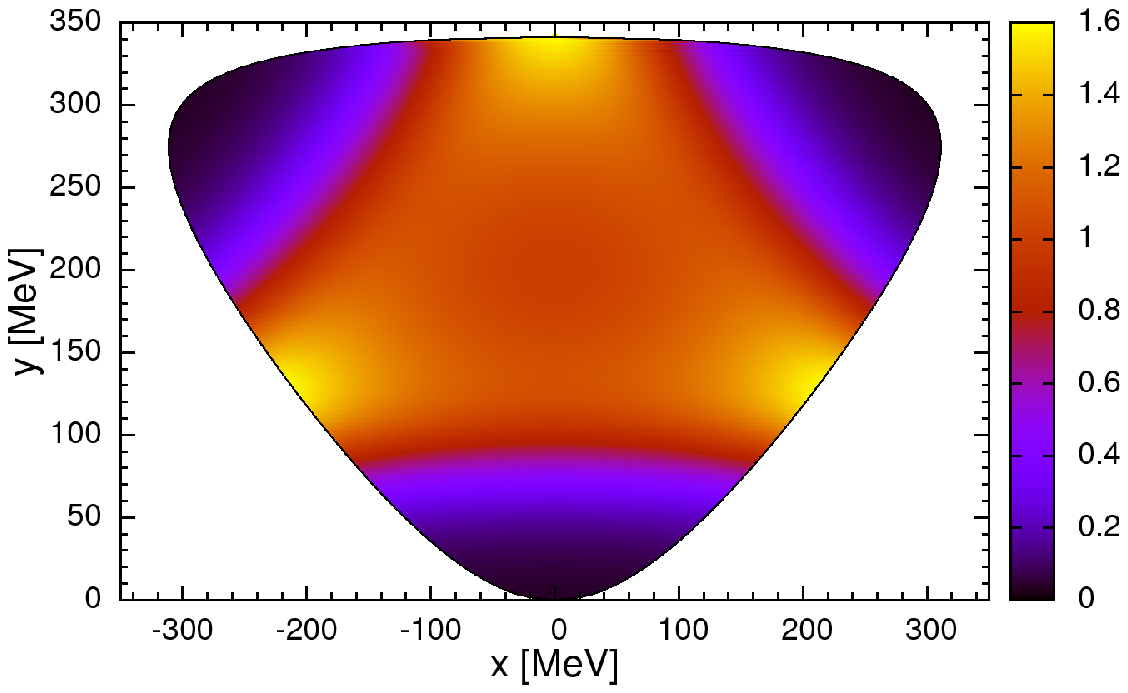}
\caption{Normalized Dalitz plot for $\omega\to3\pi$ (left) and $\phi\to3\pi$ (right).}
\label{fig:Dalitzomega}
\end{figure}

The $\omega\to3\pi$ Dalitz plot is a relatively smooth distribution, which rises from the center to its
border. The outer corners show a maximum increase of roughly $20\%$ with respect to the center. In contrast,
the $\phi\to3\pi$ Dalitz plot  exhibits significantly more structure. From its center the Dalitz plot rises
towards the bands created by the $\rho$ resonance, and then steeply falls off towards the outer corners. The
profile of the $\phi\to3\pi$ Dalitz plot seems to agree qualitatively with the one shown in~\cite{KLOE}; for
an in-depth analysis, closer comparison to the data is certainly required.

Finally we wish to study the effects of crossed-channel rescattering on the Dalitz plot distribution. For that
purpose, we follow two different approaches to fix the subtraction constant. First we assume that the
subtraction constant is given by some independent method, and we are interested in what bearings the
crossed-channel effects have on both the overall shape of the Dalitz plot and the partial decay width. We fix
the subtraction constant from the experimental decay width for the case $\hat{\mathcal{F}}=0$. The effect on
the decay width in both processes amounts to roughly $20\%$. However, while the partial width of
$\omega\to3\pi$ sees an increase, it suffers a decrease in $\phi\to3\pi$. The same qualitative behaviour shows
up in the Dalitz plot distribution.

In our second approach, we match the subtraction constant to reproduce the experimental decay width in both
cases, that is with and without crossed-channel rescattering effects included. We then study the effects on
the profile of the Dalitz plot. The result of this approach is shown for both $\omega\to3\pi$ and
$\phi\to3\pi$ in Fig.~\ref{fig:DalitzfixedGamma}.  One observers that the effects of crossed-channel
rescattering are to a large degree absorbed in the partial width. The overall behavior is very similar: in
both processes one observes an increase from the outer corners to the center of the Dalitz plot. It is
noteworthy that the $\rho$ resonance bands in $\phi\to3\pi$ are left relatively untouched.

\begin{figure}[t]
\centering
\includegraphics[width = 0.49\linewidth]{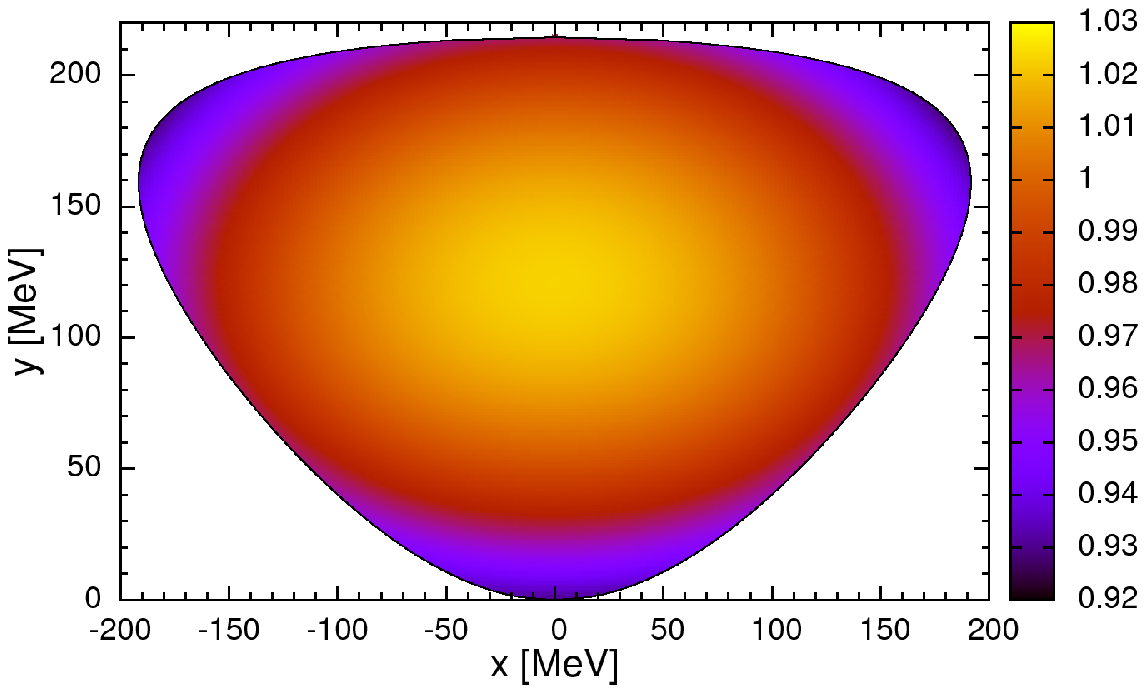}\hfill
\includegraphics[width = 0.49\linewidth]{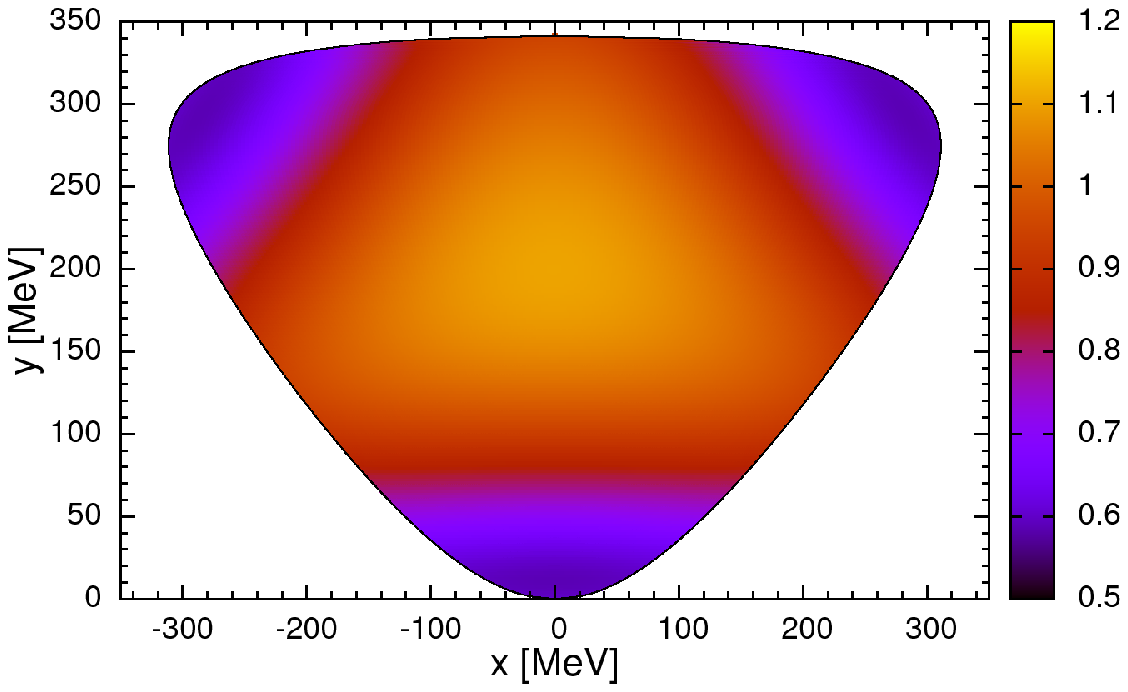}
\caption{
$|\mathcal{F}_{\rm full}|^2/|\mathcal{F}_{\hat{\mathcal{F}}=0}|^2$ for $\omega\to3\pi$ (left) and
$\phi\to3\pi$ (right). The subtraction constant is fixed so that it reproduces the decay width before and
after the iteration.}
\label{fig:DalitzfixedGamma}
\end{figure}

For a rough error estimate we have varied the cutoff in both the integral of the MO function and the
dispersion integral. Further we have varied between two sets of phenomenological representations of the
$\pi\pi$ P-wave phase shifts~\cite{Caprini, KMP}. We found that these effects are small and certainly not
comparable to the sizable crossed-channel effects.

\Subsection{\boldmath{$\eta'\to\eta\pi\pi$} Dalitz plot}
The $\eta'\to\eta\pi\pi$ amplitude exhibits a similar convergence behavior as $\phi\to3\pi$. Crossed-channel
effects are sizable and convergence is reached only after three iterations. It is also interesting to study
the effects of crossed-channel rescattering on the Dalitz plot, more specifically the Dalitz plot parameters.
Slightly different conventions for the Dalitz plot variables in $\eta'\to\eta\pi\pi$ as opposed to
$\omega/\phi\to3\pi$ are used, namely
\begin{equation}
y=\frac{(M_\eta+2M_{\pi})(M_{\eta'}-M_\eta)^2-s}{2M_{\eta'}M_{\pi}(M_{\eta'}-M_\eta-2M_{\pi})}-1~,\quad
x=\frac{\sqrt{3}(t-u)}{2M_{\eta'}(M_{\eta'}-M_\eta-2M_{\pi})}~,
\end{equation}
where $s$ and $t$ are defined analogously as before. The amplitude squared can then be expanded around the
center of the Dalitz plot in terms of $x$ and $y$ according to
\begin{equation}
|\mathcal{M}(x,y)|^2=|\mathcal{N}|^2\,\{1+a\,y+b\,y^2+d\,x^2\}~,
\end{equation}
neglecting terms of cubic order and higher, where $\mathcal{N}$ is the overall normalization and $a$, $b$ and
$d$ are the Dalitz plot parameters. A term $\propto x$ is forbidden by C-parity. Table~\ref{tab:Dalitzparam}
shows the parameters as determined by our procedure.
\begin{table}[hb]
\centering
\begin{tabular}{c c c c c}
\toprule
  	    &  ~~~RChT~\cite{RChT} ~~~  &$~~~\hat{\mathcal{M}}=0~~~$	& ~~~full~~~&	~~~VES
Collaboration~\cite{VES}~~~		\\
\midrule
$a$ &$~~~-0.119~~~$&$~~~-0.239~~~$	&$~~~-0.478~~~$ & $-0.127\pm0.016\pm0.008$	\\
$b$ &$~~~+0.001~~~$&$~~~-0.025~~~$	&$~~~+0.053~~~$	& $-0.106\pm0.028\pm0.014$	\\
$d$ &$~~~-0.056~~~$&$~~~-0.023~~~$	&$~~~-0.043~~~$	& $-0.082\pm0.017\pm0.008$	\\
\toprule
\end{tabular}
\caption{Determination of the $\eta'\to\eta\pi\pi$ Dalitz plot parameters by RChT, the pure MO solution
($\hat{\mathcal{M}}=0$), the full dispersive result and the experimental measurement by the VES
collaboration.\label{tab:Dalitzparam}
}
\end{table}
The following comments are in order: we matched the dispersive representation to RChT in the limit
$\Omega(s)\to{M_S^2}/{(M_S^2-s)}$. This is not fulfilled well by the $\pi\pi$ S-wave, which manifests itself
in large modifications of the parameter $a$ from the RChT result to the MO solution. From that vantage point,
if the RChT result is already in marginal agreement with experiment, one cannot expect the outcome of the full
analysis to be close to the measured parameters, especially if the crossed-channel rescattering corrections
are as large as evidenced in the table. Certainly the matching procedure and thus the determination of the
subtraction constants is still the weakest link in our analysis and requires further improvement. 

Finally, while we do observe effects of $\pi\eta$ phase shifts on the Dalitz plot parameter $d$, these are
not large enough to extract precision information about $\pi\eta$ scattering, especially if the subtraction
constants are not under control.

\coltoctitle{\boldmath Measurement of $\eta^\prime\to\eta\pi^+\pi^-$ with KLOE and KLOE-2}
\authortochead{C.~F.~Redmer}
\label{RCF}
\import{./RedmerChristoph/Redmer}

\coltoctitle{\boldmath Hard Pion Chiral Perturbation Theory:\\ What is it and is it relevant for $\eta^\prime$
decays?}
\authortochead{J.~Bijnens}
\label{BJ}

\setcounter{chapter}{1}
\setcounter{section}{0}
\maketitle

\vspace{-3ex}\Section{Introduction}
In this talk I will try to convince you that we can give predictions from chiral symmetry also for cases where
not all pions are soft. This is something I called hard pion Chiral Perturbation Theory (HPChPT) and there
have been a few recent papers using this~\cite{FS,BC,BJ1,BJ2,BJ3}.

I will first give a short introduction to effective field theory (EFT) and remind you of the underlying
principles of Chiral Perturbation Theory (ChPT). I will remind you of the fact that in ChPT with baryons and
other heavy particles a power-counting has been achieved by consistently absorbing the heavy mass dependence
into the low-energy-constants (LECs).

The arguments will then be generalized to the case of processes with high energy or hard pions. The arguments
also apply to cases where we can treat the strange quark mass as small as well.

After that I will show applications to $K\to\pi\pi$, to semileptonic decays of pseudo-scalar mesons or to more
general vector form-factors and to charmonium decays to two pseudo-scalars.

Unfortunately, there seem to be no $\eta^\prime$ decays where the present method are applicable.

\Section{Effective field theory and ChPT}
The underlying idea of EFT is a general theme in science, restrict yourself to the relevant degrees of
freedom. So, in cases where there exists an energy or mass gap we keep only the lower degrees of freedom.
Lorentz-invariance and quantum mechanics imply that we are restricted to a field theory but we should build
the most general one with our chosen degrees of freedom. We have no predictability left since the most general
Lagrangian will have an infinite number of parameters. This can be cured if we find an ordering principle,
power-counting, for the importance of terms.

ChPT is ``exploring the consequences of the chiral symmetry of QCD and its spontaneous breaking using
effective field theory techniques'' and was introduced as an EFT in~\cite{Weinberg0,GL1}. The degrees of
freedom are the Goldstone bosons from the spontaneous breakdown of the global chiral $SU(n)_L\times SU(n)_R$
to the diagonal vector subgroup $SU(n)_V$. That the Goldstone boson interactions vanish at zero momentum
allows to construct a consistent power-counting~\cite{Weinberg0}.

The basic form of ChPT has since been extended to baryons, mesons and baryons containing a heavy quark, vector
mesons, structure functions and related quantities as well as beyond the pure strong interaction by including
weak and electromagnetic internal interactions. Many models of alternative Higgs sectors also use the same
technology.

\Section{Power-counting and one large scale}
In purely mesonic ChPT the power-counting is essentially dimensional counting and this works since all the
lines in all diagrams have ``small'' momenta. Already when discussing baryons, this lead to problems
because now there is a large scale, the baryon mass. However, by setting the baryon momentum $p_B= M_B v+k$
with $v$ the baryon four-velocity, a consistent power-counting can be achieved. This works ``obviously'' since
the heavy line goes through the entire diagram and all momenta apart from $M_B v$  are soft as indicated by
the thick line in Fig.~\ref{figbaryonChPT}(a).
\begin{figure}
\centerline{\includegraphics[width=0.7\textwidth]{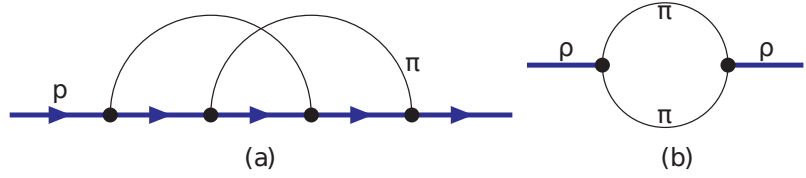}}
\caption{\label{figbaryonChPT}
(a) A typical baryon ChPT diagram with the baryon going through
the entire diagram. (b) An example of a diagram in vector meson ChPT
with no continuous vector meson line.}
\end{figure}
The same arguments apply to ChPT for mesons containing a heavy quark. It works because the ``soft'' stuff can
be expanded in ``soft$/M_B$'' and the remaining $M_B$ dependence can be absorbed in LECs. For vector meson
ChPT there is a problem since they can decay. A typical diagram is shown in Fig.~\ref{figbaryonChPT}(b).
However it was argued that the non-analytic dependence on the light quark mass could still be obtained, see
the discussions in \cite{BGT}. Again, the underlying idea is that the large $M_V$ allows to expand in
``soft/$M_V$'' and the remaining $M_V$ dependence is absorbed in the LECs. 

\Section{Several large scales or HPChPT}
In~\cite{FS} the authors applied heavy Kaon ChPT to $K_{\ell3}$ at the endpoint, i.e. the pion is soft, this
works as in usual ChPT. They also applied it to the region for small $q^2$ where the pion has a large momentum
and gave arguments based on partial integrations why this would give a correct chiral logarithm. The argument
was generalized in~\cite{BC,BJ1,BJ2,BJ3}. The underlying idea is similar to the previous section. The
``heavy/fast/hard'' dependence on the soft stuff can always be expanded and the remaining dependence goes into
the LECs. That this might be possible follows also from current algebra. Non-analyticities in the light masses
come from the soft lines and soft pion couplings are restricted by current algebra via
$
\lim_{q\to 0}\langle\pi^k(q)\alpha|O|\beta\rangle
 = -\frac{i}{F_\pi}\langle\alpha|\left[Q_5^k,O\right]|\beta\rangle\,.
$
Nothing prevents hard pions to be in the states $\alpha$ or $\beta$, so by heavily using current algebra one
can get the light quark mass non-analytic dependence

A field theoretic argument is: (1) Take a diagram with a given external and internal momentum configuration.
(2) Identify the soft lines and cut them. (3) The resulting part is analytic in the soft stuff, so it can be
described by an effective Lagrangian with coupling constants dependent on the external given momenta
(Weinberg's folklore theorem~\cite{Weinberg0}). (4) The non-analytic dependence on the soft stuff is
reproduced by loops in the latter Lagrangian.
\begin{figure}
\centerline{\includegraphics[width=0.7\textwidth]{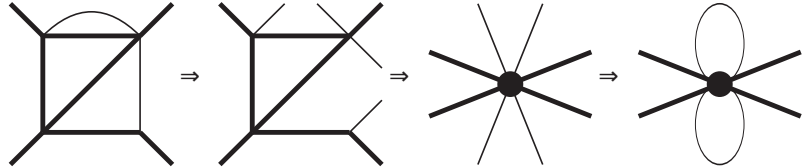}}
\caption{\label{figcuthardlines} The process of cutting the soft lines and reproducing the non-analytic
dependence by the diagram on the right. The hard lines are thick, soft lines are shown thin.}
\end{figure}
The process is depicted in Fig.~\ref{figcuthardlines}.

The remaining problem is that we have no power-counting. The Lagrangian that reproduces the non-analyticities
is fully general. In the HPChPT papers it was shown that for the processes at hand, all higher order terms can
be reduced to those with the fewest derivatives thus allowing the light quark mass chiral logarithm to be
predicted. The underlying arguments were tested by comparing to a two-loop calculation~\cite{BJ2} and by
explicitly keeping some higher order terms~\cite{BC,BJ1,BJ2,BJ3}.

\Section{Applications}
We have applied the method to $K\to\pi\pi$ decays~\cite{BC} where we treat the Kaon as heavy and look for the
dependence on the pion mass $M^2$. The result is, up to linear in $M^2$ and higher order:
$$
A_0^{NLO}/A_0^{LO} =1+({3}/{8})\hat A\,,
\quad\!\!
A_2^{NLO}/A_2^{LO}=1+({15}/{8})\hat A\,,
\quad\!\!
\hat A = -{M^2}\ln({M^2}/{\mu^2})/({16\pi^2 F^2}).
$$

The scalar and vector form-factors of the pion are known to two-loops in ChPT~\cite{BCT}. HPChPT predicts at
large $t$~\cite{BJ2} with $F_V(t,0)$ and $F_S(t,0)$, the form-factors at large $t$ in the chiral limit,
completely free:

$$
F_V(t,M^2) = F_V(t,0)\,
(1+\hat A)\,,\quad
F_S(t,M^2) = F_S(t,0)\,
(1+({5}/{2})\hat A)\,.
$$
The full two-loop result expanded for large $t$ should have this form and it does with for e.g. $F_V$
$$
F_V(t,0) = 1+ ({t}/({16\pi^2F^2}))
\left({5}/{18}-16\pi^2 l_6^r+{i\pi}/{6}-({1}/{6})\ln({t}/{\mu^2})
 \right)\,.
$$

The first application was semileptonic form-factors in $K_{\ell3}$. We extended this to $B,D\to
D,\pi,K,\eta$-decays in~\cite{BJ1,BJ2}. This allowed to test our results experimentally. The form factor
$f_+(t)$ as measured by CLEO~\cite{CLEO} in $D\to\pi$ and $D\to K$ decays are different by about the amount
expected from the chiral logarithms as shown in Fig.~\ref{figfvpCLEO}.
\begin{figure}
\begin{minipage}{0.49\textwidth}
\includegraphics[angle=270, width=0.99\textwidth]{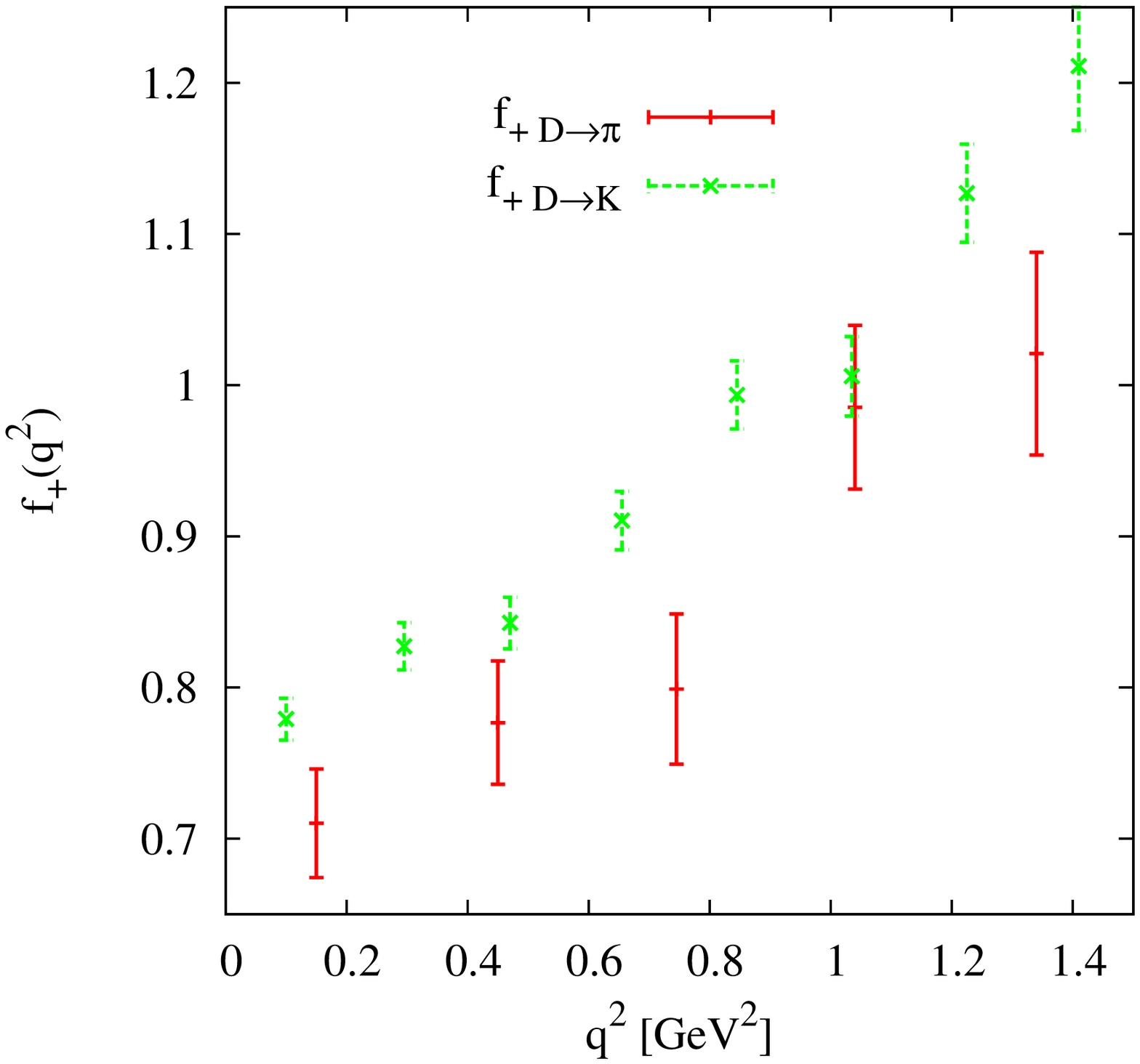}
\end{minipage}
\begin{minipage}{0.49\textwidth}
\includegraphics[angle=270, width=0.99\textwidth]{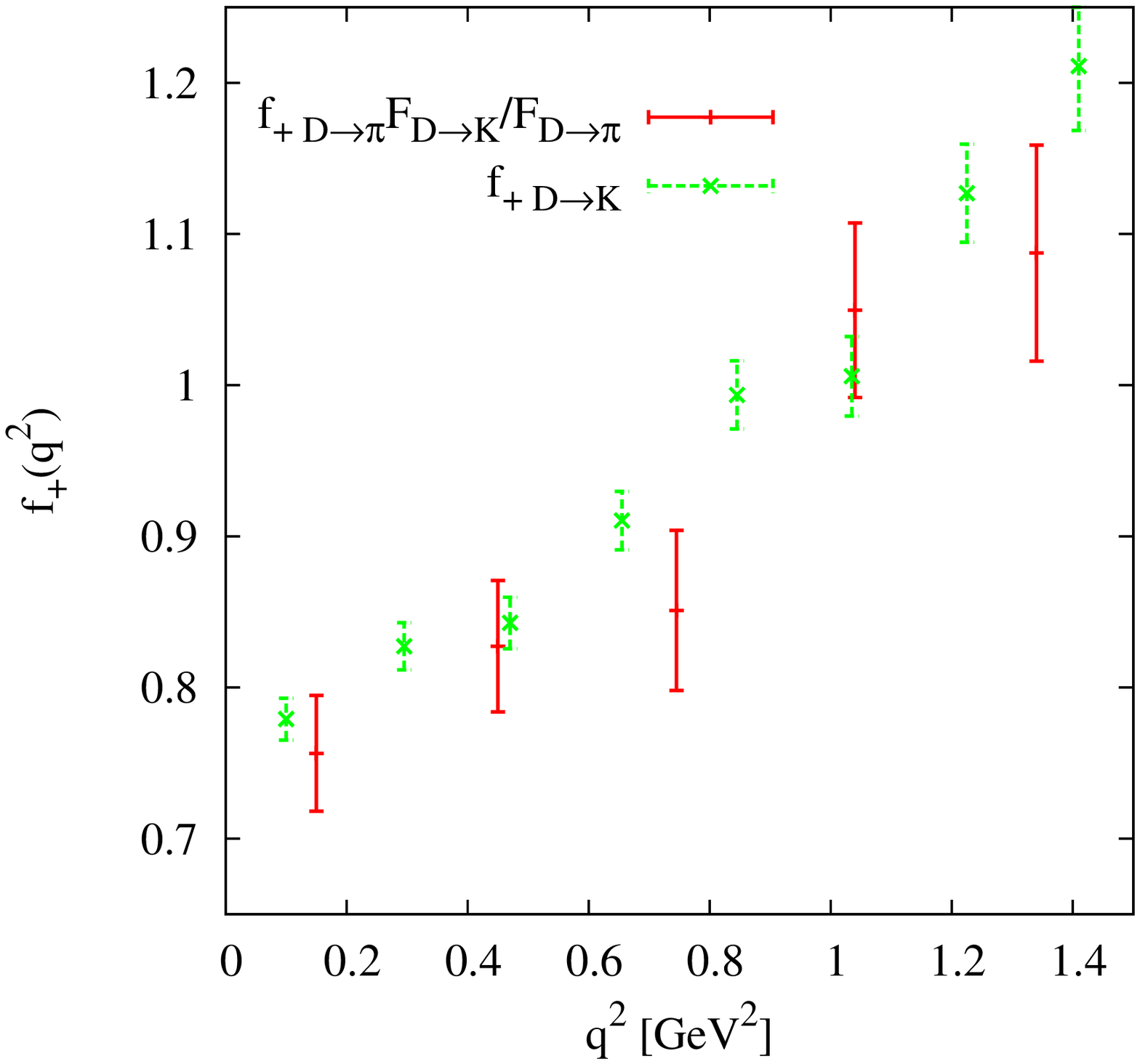}
\end{minipage}
\caption{\label{figfvpCLEO} The CLEO data on $D\to\pi$ and $D\to K$. Form-factors as measured to the the left
and corrected with the chiral logarithms to the right. Note the improved agreement.}
\end{figure}
One puzzling observation~\cite{BJ2} was that in the limit of a hard pseudo-scalar in the final state the
correction was always the same for $f_+$ and $f_-$. This is in fact due to the LEET relation~\cite{LEET} which
shows that there is only one form-factor in this limit and it is nice to see that our calculation respects
this without it being used as input.

A last application was to $\chi_{c0,2}\to\pi,KK,\eta\eta$~\cite{BJ3}. Here it was found that there were no
chiral logarithms to the order considered. Comparing with the known experimental results indeed shows $SU(3)$
breaking to be somewhat smaller than in e.g. $F_K/F_\pi$. Details can be found in~\cite{BJ3}.

\section*{Acknowledgments}\vspace{-8pt}
This work is supported in part by the EU, HadronPhysics2 Grant Agreement n. 227431, and the Swedish Research
Council, grants 621-2008-4074 and 621-2010-3326.

\coltoctitle{Hunting Resonance Poles}
\authortochead{P.~Masjuan}
\label{MP}

\setcounter{chapter}{1}
\setcounter{section}{0}
\maketitle

\Section{Introduction}
The non-perturbative regime of QCD is characterized by the presence of physical resonances, complex poles of
the amplitude in the transferred energy at higher (instead of the physical one) complex Riemann sheets. From
the experimental point of view, one can obtain information about the spectral function of the amplitude
through the Minkowsky region ($q^2>0$) and also about its low energy region through the experimental data on
the Euclidean region ($q^2<0$).

In reference~\cite{MJJP-VFF}, the particular case of the $\pi\pi$ Vector Form Factor (VFF) was analyzed using
the available Euclidean data and its first and the second derivatives were determined at $q^2=0$ with Pad\'{e}
Approximants (PA) centered at the origin trough a fit procedure to that data \cite{MJJP-VFF}. In such a way,
the vector quadratic radius $\langle r^2\rangle_V^{\pi}$ and the curvature $c_V^{\pi}$ were extracted from the
fit and, as a consequence, a value for the low-energy constant $L_9=(6.84\pm0.07)\cdot 10^{-3}$ was obtained,
\cite{MJJP-VFF,Masjuan}.
Similar ideas are applied in reference~\cite{TFF} to study the pion Transition Form Factor (TFF) and its
first ($a_{\pi}$) and second ($b_{\pi}$) derivatives with better precision compared to the current world
average result~\cite{PDG}. The implications of these parameters on the $\pi^0$ contribution to the hadronic
light-by-light scattering to the muon anomalous magnetic moment are also studied showing the relevance of this
kind of model-independent methods.

Despite the nice convergence and the systematical treatment of the errors, this procedure does not allow us
to obtain properties of the amplitude above the threshold, such as in the case of the $\pi\pi$ VFF, the
$\rho$-meson pole position, or trough the $\pi\pi$  phase shift the $\sigma$-pole. The reason is simple: the
convergence of a sequence of Pad\'{e} Approximants centered at the origin of energies ($q^2=0$) is limited by
the presence of the $\pi-\pi$ production brunch cut: the PA sequence converge everywhere except on the
cut~\cite{Masjuan}. Still, the mathematical Pad\'{e} Theory, through the Montessus de Ballore theorem, allow
us to produce a model independent determination of resonance poles when certain conditions are fulfilled. The
most important one is to center our PA sequence above the branch-cut singularity (beyond the first production
threshold) instead of at origin of energies ($q^2=0$). This small modification also provides the opportunity
to use Minkowskian data in our study instead of the Euclidean one. The relevance of this model-independent
method to extract resonance poles is clear since does not depend on a particular lagrangian realization or
modelization on how to extrapolate from the data on the real energy axis into the complex plane.

The Montessus de Ballore's theorem states that the sequence of one-pole Pad\'{e} Approximants $P_1^N$ around $x_0$,

\begin{equation}\label{PAeq}
P_1^N(x,x_0)=\sum_{k=0}^{N-1}a_k(x-x_0)^k+\frac{a_N(x-x_0)^N}{1-\frac{a_{N+1}}{a_N}(x-x_0)}\, ,
\end{equation}

converges to $F(x)$ in any compact subset of the disk excluding the pole $x_p$, i.e, $\lim_{N\rightarrow
\infty} P_1^N (x,x_0)\, =\, F(x)$.

As an extra consequence of this theorem, one finds that the PA pole $x_{PA}=x_0+\frac{a_N}{a_{N+1}}$ converges
to $x_p$ for $N\rightarrow \infty$. Since experiments provide us with values of $F_j$ at different $x_j$
instead of the derivatives of our function, we can use the rational functions $P_1^N$ as fitting functions to
the data in a similar way as in Ref.~\cite{MJJP-VFF, TFF} In this way, as $N$ grows $P_1^N$ gives us an
estimation of the series of derivatives and the $x_p$ pole position.

For a physical amplitude, the function $F(x)$ (without a right-hand cut) is analytic from $x=-\infty$ up to
the first production threshold $x_{th}$ and within the disk $B_{x_{th}}(0)$. In the $\pi \pi$ VFF case, the
threshold is found to be at $x_{th}=4m_{\pi}^2$, where $m_{\pi}$ is the mass of the pion. One can use Pad\'{e}
Approximants in a safe way by using the Montessus' theorem and center the approximants at $x_0+i0^+$ over the
brunch cut between the first and the second production threshold. In the $\pi\pi$ VFF that would correspond to
the range between $x_{th}=4m_{\pi}^2$ and $\tilde{x}_{th}=4m_K^2$ (assuming small multipion channels). In such
a way we unfold the Second Riemann sheet due to the analytical extension of the function $F(x)$ from the first
Riemann sheet at $x+i0^+$ into the second one.

In the case of resonant amplitudes, a single pole appears in the neighborhood of the real $x$ axis in the
second Riemann sheet which is related to a hadronic state, a resonance. Once we have unfold the Second Riemann
sheet by locating our approximants over the brunch cut, the application of the Montessus' theorem in the disk
of convergence (which is the region defined between the thresholds) is straightforward and allows us to locate
the position of the resonant pole if lies inside that region. If that is the case, our $P_1^N$ Pad\'{e}
approximant sequence determine systematically its position in a model independent way.

\Section{Application of the method in a physical case}
We would like to use the method to analyze the final compilation of ALEPH $\pi\pi$ VFF data for the squared
modulus $|F_{\pi\pi}(q^2)|^2$, \cite{ALEPH}, and the $I=J=1$ $\pi\pi$ scattering phase-shift
$\delta_{\pi\pi}$, identical to the $\pi\pi$ VFF phase-shift in the elastic region $4m_{\pi}^2<q^2<4m_K^2$ (if
multipion channels are neglected), which conforms the range of applicability of our $P_1^N$ sequence (more
details can be found in Ref. \cite{ProcJJ}). For $N\geq3$ the fit $\chi^2$ already lies within the 68$\%$
confidence level (CL) and becomes statistically acceptable.At this point one needs to reach the typical
fitting compromise. On one hand the experimental errors have an statistical origin and the contribution of
this error increase as one considers higher order $P_1^N$, with larger number of parameters. On the other
hand, the systematic theoretical error decreases as $N$ increases and the PA converges to the actual VFF and
the phase-shift. For the $\rho$-mass determination we have taken $N=6$ as our best estimate as the new
parameters of PA with $N\geq7$ turn out to be all compatible with zero, introducing no more information with
respect to the previous $P_1^6$. Furthermore, the model studied in Ref.\cite{ProcJJ} shows that the
theoretical errors for mass and width results are smaller than $10^{-2}-10^{-3}$MeV for $N\geq6$, being
negligible compared to the ${\cal O}$(1 MeV) experimental errors. This yields to the determination:

\begin{equation}\label{result}
\centering
M_{\rho}=763.7\pm1.2 \mathrm{MeV}, \quad \Gamma_{\rho}=144\pm3 \mathrm{MeV}\, ,
\end{equation}

which is found to be in good agreement with previous determinations using more elaborated and complex
procedures and with similar size of uncertainties as shown in Table \ref{tab:results}.

{\footnotesize
\begin{table}
\centering
\begin{tabular}{|c|c|c|}
 \hline
  Ref. & $M_{\rho}(\mathrm{MeV})$ & $\Gamma_{\rho}(\mathrm{MeV})$\\
  \hline
  \cite{Ananthanarayan}  & $762.5\pm2  $&$ 142\pm7$ \\
  \cite{Pelaez}                             & $754  \pm18  $&$ 148\pm20$ \\
 \cite{Zhou}            & $763.0\pm0.2  $&$ 139.0\pm0.5$ \\
 \cite{PichCillero}        & $764.1\pm2.7^{+4.0}_{-2.5}  $&$ 148.2\pm1.9^{+1.7}_{-5.9} $ \\
 \hline
  this work                       & $763.7\pm1.2  $&$ 144\pm 3 $\\
  \hline
\end{tabular}
\caption{\label{tab:results}Comparison of different results for the determination of the mass and width of the
$\rho$-meson.}
\end{table}
}

Similar results are found when analyzing the $\pi-\pi$ phase-shift data for the $\sigma$-pole position,
yelding in this case

\begin{equation}\label{result2}
\centering
M_{\sigma}=451\pm21 \mathrm{MeV}, \quad \Gamma_{\sigma}=619\pm18 \mathrm{MeV}\, ,
\end{equation}

again in good agreement with previous determinations with more complex methods~\cite{PDG}.

\Section{Conclusions}
We have develop a new model-independent method for extracting resonance poles (mass and width) from physical
amplitudes based on the well defined mathematical theory of Pad\'{e} Approximants which makes use of the
Montessus' theorem to unfold de Second Riemann sheet of the desired amplitud to locate the resonant pole in
the complex plane. This method systematize the algorithm of extracting the desired resonance pole (or low
energy properties of these amplitudes as commented in the introduction) by building up a convergence sequence
of those PA which at the very same time provides with a simple way to estimate the systematic errors. Our
method has, however, a larger application since does not rely on a particular lagrangian or in a modelization
on how to extrapolate from the data on the real energy axis into the complex plane (such as other methods
discussed in Ref.~\cite{MSCV}).

In the particular case presented here we have analyzed on one hand the experimental Euclidean data  to obtain
the first and the second derivatives at $q^2=0$ for both the $\pi$-VFF and for the $\pi^0$ TFF. And on the
other hand, also Minkowskian data have been analyzed for $\pi\pi$-VFF and also $\pi\pi$-scattering data by a
$P_1^N$ Pad\'{e} Approximant sequence centered between the first and the second production thresholds (in such
a way that we can unfold the second Riemann sheet of our amplitude). We have obtained a prediction of the
$\rho$-meson and $\sigma$-meson pole positions with compatible accuracy compared to other determinations  but
with a simpler and systematic method.

\section*{Acknowledgments}\vspace{-8pt}
This work has been supported by MICINN, Spain (FPA2006-05294), the Spanish Consolider-Ingenio 2010 Programme
CPAN (CSD2007-00042) and by Junta de Andaluc\'ia (Grants P07-FQM 03048 and P08-FQM 101).

\coltoctitle{Exploring Low-Energy QCD Using a Generalized Linear Sigma Model}
\authortochead{A.~H.~Fariborz}
\label{FA}

\setcounter{chapter}{1}
\setcounter{section}{0}
\maketitle

The decays of $\eta'$ can provide very useful information on the physics of scalar mesons. For example, the
decay $\eta'\rightarrow \eta \pi \pi$ involves intermediate scalar states $f_0(600)$ (or sigma), $f_0(980)$
and $a_0(980)$ as shown in Fig.~\ref{F_etap_decays}. Therefore, at least in principle, such decays can shed
light on various aspects of scalar meson. However, in practice, this objective is overshadowed by various
uncertainties such as different final state interactions as well as various underlying mixing among scalar
states. Nevertheless, one would hope to find useful approximations that could provide some insights into the
physics of scalar mesons. The present work summarizes one such approach within the context of a generalized
linear sigma model.
\begin{figure}[h]
\centering
 \epsfig{file=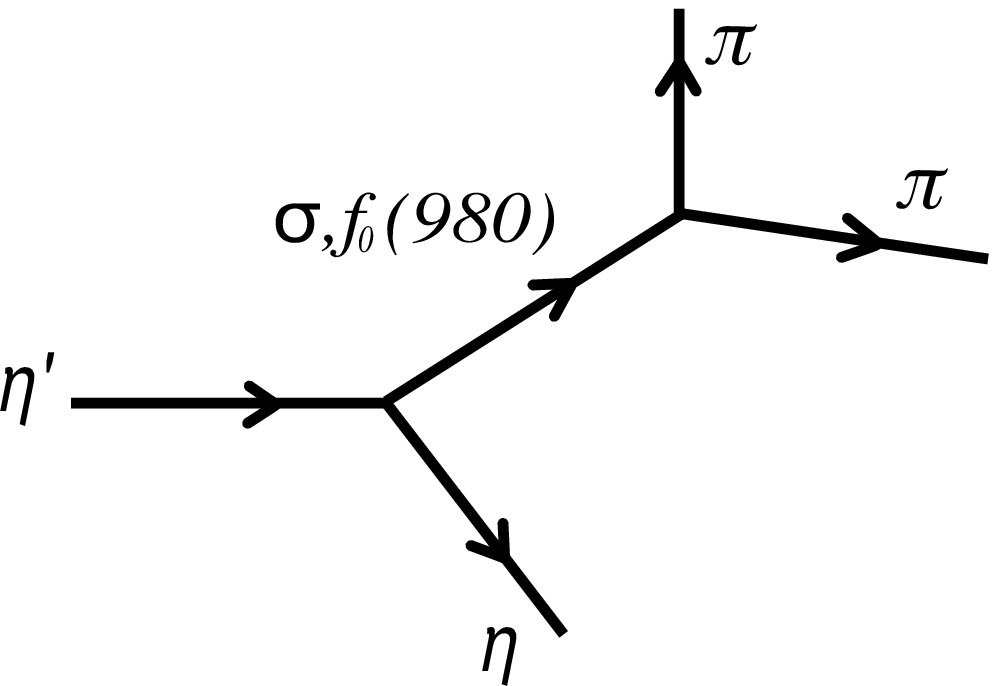,width=1.5in}
 \epsfig{file=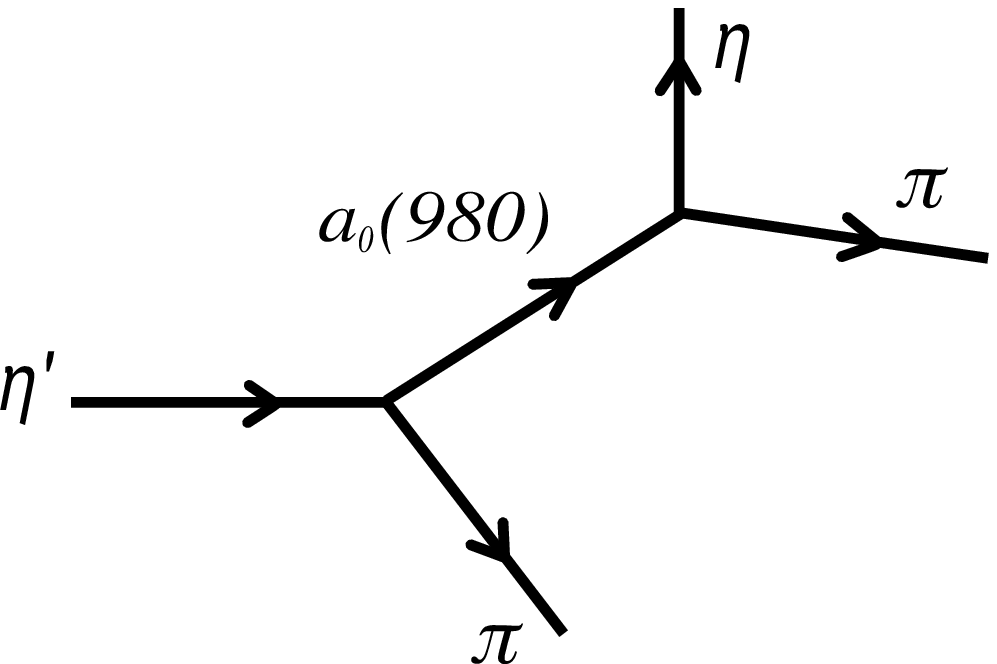,width=1.5in}
 \caption{\label{F_etap_decays}Contribution of scalar mesons to decay $\eta'\to\eta\pi\pi$.}
\end{figure}\\
Below 1 GeV, the known scalar states are listed in Fig.~\ref{F_j0_states}~\cite{pdg}: the light and broad
isosinglet $f_0(600)$ or sigma, followed by isodoublet $K^*(800)$ or kappa meson and the two nearly degenerate
states, isosinglet $f_0(980)$ and isotriplet $a_0(980)$. Above 1 GeV, the known scalar states are also listed
in Fig. \ref{F_j0_states}: The isosinglet state $f_0(1370)$ followed by the isodoublet $K_0^*(1430)$, the
isotriplet $a_0(1450)$ and the two isosinglet states $f_0(1500)$ and $f_0(1710)$. There have been numerous
works addressing the physics of these illusive states and at times conflicting point of views have surfaced.  
Of course, getting into the  history of these states and various approaches by different investigators is
beyond the scope of this short summary and the reader is encouraged to refer to the available literature for
details. However, in a nutshell, the scalar states below 1 GeV seem to be close to four-quark
states~\cite{j77} (with some deviations) and those above 1 GeV seem be close to two-quark states (with some
deviations). To generate such deviations from pure four-quark and pure two-quark pictures, it seems natural
to investigate a possible underlying mixing among four- and two-quark states. Such a mixing was first
investigated within a nonlinear chiral Lagrangian framework in ref. \cite{Mec} and it was shown that it
provides a natural understanding of some of the puzzling aspects of $K_0^*(1430)$ and $a_0(1450)$. The same
type of mixing was further extended to the case of isosinglet scalar states below and above 1 GeV (which can
also mix with scalar glueballs)  in \cite{Mec_f}, in which again consistent results were obtained and
estimates on the quark and glue substructure of the isosinglet scalars were made.
\begin{figure}[h]
\centering
 \epsfig{file=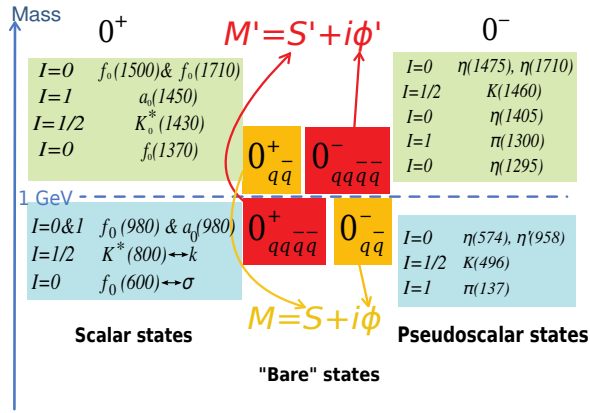,width=3in}
 \caption{\label{F_j0_states}Scalars (left) and pseudoscalars (right) below 2 GeV together with their ``bare''
(unmixed) underlying two- and four-quark mixing nonets (middle).  }
\end{figure}
Such mixing patterns were further studied within the framework of a generalized linear sigma model which is
the topic of the present summary. In this approach, the pseudoscalar mesons below and above 1 GeV (also given
in Fig. \ref{F_j0_states}) are brought into the picture. The main motivations for using a generalized linear
sigma model are: First, it provides an independent check of the results obtained for scalar mixings in the
nonlinear chiral Lagrangian approach mentioned above, and tests whether the results obtained are robust, or
are artifacts of the framework used.   Second, as scalar mesons are probed in various pseudoscalar
interactions (such as, for example, $\pi\pi$, $\pi K$, and $\pi \eta$ scatterings and decays such as
$\eta'\rightarrow \eta\pi\pi$, $\cdots$) it seems important to see what roles (if any) two- and four-quark
mixings play in the physics of pseudoscalars mesons, for which, at least for pseudoscalars below 1 GeV, the
phenomenology is fairly established, and as a result, allowing pseudoscalars to participate in these mixing
patterns can provide a reliable probe into the understanding of the mixing.  Third, some of the heavier
pseudoscalar states above 1 GeV are known to be dubious and speculated to be of a non quark-antiquark nature,
therefore, extending the mixing mechanism to the pseudoscalar sector may provide useful insight into the
nature of some of the heavier pseudoscalars. A generalized linear sigma model [which is formulated in terms of
two scalar meson nonets (a two-quark nonet and a four-quark nonet) and two pseudoscalar meson nonets (a
two-quark nonet and a four-quark nonet)] seems to be an appropriate framework for exploring the mixing
patterns in both scalar as well as pseudoscalar sectors. The idea of mixing within this framework is
investigated in Ref.~\cite{global} (and references therein) and the outcome (within the approximation used
in~\cite{global}) is symbolically summarized in Fig.~\ref{F_j0_states}. In this approach, it is found that (as
expected) the underlying four-quark scalar nonet is lighter than the two-quark scalar nonet, and the situation
in reversed for the two pseudoscalar nonets (i.e. the two-quark pseudoscalar nonet tends to become lighter
than the four-quark pseudoscalar nonet). This of course agrees with the conventional phenomenology for light
pseudoscalars where their properties are understood based on a quark-antiquark picture. Further details can
be seen in Ref.~\cite{global}. The notation is clarified in Fig.~\ref{F_j0_states} in which the chiral nonet
$M$ is formed out of quark-antiquark scalar nonet $S$ and quark-antiquark pseudoscalar nonet $\phi$
(similarly, the chiral nonet $M'$ is formed out of four-quark nonets $S'$ and $\phi'$). The general structure
of the Lagrangian in terms of $M$ and $M'$ is:
\begin{equation}
{\cal L} = - \frac{1}{2} {\rm Tr}
\left( \partial_\mu M \partial_\mu M^\dagger
\right) - \frac{1}{2} {\rm Tr}
\left( \partial_\mu M^\prime \partial_\mu M^{\prime \dagger} \right)
- V_0 \left( M, M^\prime \right) - V_{SB},
\label{mixingLsMLag}
\end{equation}
where $V_0(M,M^\prime) $ stands for a function made from SU(3)$_{\rm L} \times$ SU(3)$_{\rm R}$ (but not
necessarily U(1)$_{\rm A}$) invariants formed out of $M$ and $M^\prime$ and $V_{SB}$ is the flavor symmetry
breaking term. In dealing with this Lagrangian,  two approaches have been considered: Approach 1 is based on
the underlying chiral symmetry and the resulting generating equations~\cite{toy_model}. In this approach no
specific choice for the chiral invariant part of $V_0$ is made,  and only the axial anomaly, the $V_{SB}$ and
the condensates are modeled. The details of this approach can be found in Ref.~\cite{toy_model} and are not
discussed here. The main conclusions are: (i) The light pseudoscalars, as expected from conventional
phenomenology, remain dominantly close to quark-antiquark states; (ii) The kappa meson tends to become a
dominantly four-quark state; and (iii) This approach  does not provide any information on $a_0$ and $f_0$
systems. To study  $a_0$ and $f_0$ systems, specific choices for $V_0$ should be made, and that leads to the
second approach which is summarized here. In approach 2, a specific choice for the chiral invariant part of
$V_0$ is made~\cite{global} (in addition to modeling the axial anomaly, the $V_{SB}$ and the condensates).  
The main conclusions are: (i) Again, light pseudoscalars have the tendency of becoming quark-antiquark states;
(ii) Light scalars tend to become mainly four-quark states; and (iii) Predictions for various low-energy
processes such as $\pi\pi$, $\pi K$, $\pi \eta$ scatterings, and decays such as $\eta'$ decays, semileptonic
decays of $D_s$, $\cdots$ can be made.  Here, we only give a summary of approach 2. Obviously, there are
infinite number of terms that can be written down for $V_0$. Up to dimension 4, there are twenty one
SU(3)$_{\rm L}\times$ SU(3)$_{\rm R}$ invariant terms in $V_0$ which can be made out of $M$ and $M'$. To work
with such a high number of terms,  it seems reasonable to consider an approximation scheme in which $V_0$ is
organized in terms of the number of quark and antiquark lines.   In the first approximation attempt:
 1) We only consider terms with eight or fewer quark or antiquark lines;
 2) We mock up the $U(1)_{\rm A}$ anomaly exactly;
 3) We ignore terms that are not consistent with OZI rule; and
 4) We consider the leading flavor symmetry breaking term that mocks up the quark mass term in the
fundamental QCD Lagrangian.  These approximations lead to the potential:
 \begin{eqnarray}
V =&-&c_2 \, {\rm Tr} (MM^{\dagger}) +
 c_4^a \, {\rm Tr} (MM^{\dagger}MM^{\dagger})
 + d_2 \,
{\rm Tr} (M^{\prime}M^{\prime\dagger})
 + e_3^a(\epsilon_{abc}\epsilon^{def}M^a_dM^b_eM'^c_f +
{\rm h.c.})
\nonumber \\
&+& c_3\left[\gamma_1 {\rm ln}\, \left(\frac{{\rm det}\,
M}{{\rm det}
M^{\dagger}}\right)
 +(1-\gamma_1)
{\rm ln} \,
\left(\frac{{\rm Tr}\,
(MM'^{\dagger})}{{\rm Tr}\, (M'M^{\dagger})}\right)\right]^2
+ k_1[ {\rm Tr} (AM) +{\rm h.c.}].
\label{V_N8}
\end{eqnarray}
where  matrix $A=$ diag. $(A_1, A_2, A_3)$ induces the flavor symmetry breaking and the two- and four-quark
condensates of fields $S$ and $S'$ obey
$\langle S_a^b \rangle = \delta_a^b \alpha_a$
and $\langle {S'}_a^b \rangle = \delta_a^b \beta_a$.
We assume isospin limit which implies:
$A_1 = A_2 \ne A_3$ and $\alpha_1$ = $\alpha_2$ $\ne$ $\alpha_3$ and
$\beta_1 = \beta_2 \ne \beta_3$.
At the level of the approximate potential of Eq.~(\ref{V_N8}), altogether, there are 12 unknown parameters
$c_2$, $c_4^a$, $d_2$, $e_3^a$, $c_3$, $\gamma_1$, $\alpha_1$, $\alpha_3$, $\beta_1$, $\beta_3$, $A_1$, $A_3$
that need to be determined using 12 inputs.   We take the following eight experimental inputs: $m[a_0(980)] =
984.7 \pm 1.2\, {\rm MeV}$;
$m[a_0(1450)] = 1474 \pm 19\, {\rm MeV}$;  $m[\pi(1300)] = 1300 \pm 100\, {\rm MeV}$;  $m_\pi = 137 \, {\rm
MeV}$; $F_\pi = 131 \, {\rm MeV}$;
${A_3\over A_1} = 20\rightarrow 30$; $\det(M_\eta^2) = \det(M_\eta^2)^{\rm exp.}$; and ${\rm Tr} (M_\eta^2) =
{\rm Tr} (M_\eta^2)^{\rm exp}$,
together with four minimum conditions:
$\left\langle\frac{\partial V}{\partial S_1^1}\right\rangle =
\left\langle\frac{\partial V}{\partial S_3^3}\right\rangle =
\left\langle\frac{\partial V}{\partial {S'}_1^1}\right\rangle =
\left\langle\frac{\partial V}{\partial {S'}_3^3}\right\rangle = 0.
$
These inputs and minimum conditions allow a determination of the Lagrangian parameters and the detailed
numerical analysis of this study in given in \cite{global}.  The main uncertainties are clearly on
$m[\pi(1300)]$ as well as the ratio ${A_3\over A_1}$,  and therefore,  the results have some variations that
reflect these two uncertainties. The work o~ \cite{global} shows that the underlying mixing among two- and
four-quark pseudoscalars does not change the well-established picture of the light pseudoscalars as dominantly
being quark-antiquark states (the situation is of course reversed for the heavy pseudoscalars). The present
level of this investigation predicts four etas: Two of the predicted etas are below 1 GeV and are close to
the physical states $\eta(547)$ and $\eta(958)$. The two heavier etas above 1 GeV may be identified with two
of the several physical eta states above 1 GeV. In the work of~\cite{global} various scenarios for this
identification is studied in detail and the closest agreement corresponds to identifying the two predicted
etas above 1 GeV with $\eta(1295)$ and $\eta(1760)$. The situation for scalar mesons is the opposite of
pseudoscalars: Those below 1 GeV come out close to four-quark states and those above 1 GeV tend to become more
of two-quark states. These natural tendencies for scalar mesons are consistent with those studied within
nonlinear chiral Lagrangians described above. Therefore, in summary, it seems that the generalized linear
sigma model discussed here is an appropriate framework for learning about scalar and pseudoscalar mesons, and
already has provided useful insights into their quark substructure. Further extensions will involve inclusion
of  higher order terms in the potential [with higher number (more than eight) of underlying quark and
antiquark lines]. Also inclusion of glueballs seems to be quite important. Various predictions of the model
for several low-energy processes are being currently looked at and will be reported elsewhere. Particularly,
the present approach is hoped to provide useful insights into the physics of $\eta'$ and the eta's above
1~GeV.

The author wishes to thank the organizers of PrimeNet Workshop for a very productive meeting. This work is
based on an ongoing collaboration with R. Jora, J. Schechter and M.N. Shahid, and the author thanks
collaborators for many helpful discussions. This work has been partially supported by the NSF Grant 0854863
and by a 2011 grant from the Office of the Provost, SUNYIT.

\coltoctitle{\boldmath The $\eta$ Decay Program at WASA-at-COSY}
\authortochead{D.~Coderre}
\label{CD}

\setcounter{chapter}{1}
\setcounter{section}{0}
\maketitle

\Section{Introduction}
The $\eta$ meson provides an excellent laboratory for the study of matter in the low-energy regime. Because
all strong and electromagnetic decays of the $\eta$ are forbidden in the first order, rare processes are
accessible which would otherwise be highly suppressed. The measurement of the decays of the $\eta$ can test
fundamental symmetries, probe the structure and dynamics of hadrons, and investigate possible new physics. 

A major goal of the WASA-at-COSY experiment is to make precision measurements of $\eta$ decays. $\eta$ mesons
are produced by impinging a proton beam onto a proton or deuteron target. With the WASA detector, both
forward-scattered hadrons and wide-angle decay products of the $\eta$ can be measured. 

WASA has recorded $3 \times 10^{7}$ $\eta$ mesons in the $pd \rightarrow {^{3}He}\eta$ reaction. In the $pp
\rightarrow pp\eta$ reaction, over $10^{8}$ $\eta$ mesons have been produced with a trigger biased towards
charged decays. Proton-deuteron reactions feature a favorable ratio between $\eta$ production and background
processes, but a lower rate of $\eta$ production  of about $10 \eta s^{-1}$. Proton-proton reactions have a
factor of ten higher rate of $\eta$ production but significantly higher background. The two production methods
are complementary.

\Section{Current Work on Selected Decay Channels}
In the following, the current status of the analysis of selected decay channels will be described.

The decay $\eta \rightarrow \pi^{+}\pi^{-}\gamma$ provides the possibility to search for a contribution of
the box-anomaly term from the Wess-Zumino-Witten (WZW) Lagrangian~\cite{Wess_1975,Witten_1983}. The predicted
decay width calculated with chiral perturbation theory is significantly lower than the measured value,
partially because the phase space available for this decay is constrained by the mass of the $\pi$ and is far
from the chiral limit. Attempts have been made to include final-state interactions as unitarized extensions to
the WZW Lagrangian. In order to test these predictions, both the decay rate and the dynamics of the final
state particles can be investigated.

In the WASA measurement, a total of $13960 \pm 140$ event candidates for this reaction were recontructed from
$pd \rightarrow {^{3}He}\eta$ data and the shape of the relevant kinematic distributions have been
determined~\cite{WASA_pipig}. The data deviates from calculations using the simplest gauge-invariant matrix
element. For a more realistic description, a form factor as described in~\cite{pipig_Theo} has been applied to
the prediction of the simplest matrix element. This allows for a model independent description of the
distribution with a single free parameter, $\alpha$. This parameter can be compared to theoretical
predictions.

Another aspect of this decay is the measurement of the branching ratio. A value about 3-$\sigma$ lower than
the former experimental average was recently measured by the CLEO collaboration and confirmed by the KLOE
collaboration~\cite{CLEO_pipig,KLOE_pipig}. As a next step for this channel, WASA will measure both the
branching ratio and the photon-energy dependence in one analysis from high-statistics proton-proton data.

The decay $\eta \rightarrow \pi^{+}\pi^{-}e^{+}e^{-}$ is a rare decay with a branching ratio of $(2.68 \pm
0.11) \times 10^{-4}$~\cite{PDG}. One goal of the analysis is to confirm the experimental value of the
branching ratio, which has only been measured once with high statistics~\cite{KLOE_ppee}. This decay is
interesting because it may provide a novel test for CP-violation outside of the standard model. It has been
shown that the asymmetry of the angle between the electron and $\pi$ decay planes is sensitive to a
contribution of the CP-violating E-1 transition to the decay amplitude~\cite{Gao,Geng}. Unlike the reaction
$\eta\rightarrow\pi^{+}\pi^{-}$, CP-violation in this system is not suppressed by experimental limit on the
electric dipole moment of the neutron. The theoretical upper limit for this asymmetry is about 1\%. 

So far $(230 \pm 22)$ signal event candidates have been identified in proton-deuteron data, shown in
Figure~\ref{fig:eta_fig}a. This number is consistent with a branching ratio near the measured and
predicted values. Systematic studies are currently being performed in order to extract the branching ratio
with the lowest possible error. The decay is also being analyzed in proton-proton data where a larger sample
of event candidates is visible. The goal is to measure the branching ratio with proton-deuteron data and to
subsequently use the proton-proton data to confirm the branching ratio and measure the decay plane asymmetry.

Because it proceeds via two virtual photons, the decay $\eta \rightarrow e^{+}e^{-}e^{+}e^{-}$ has a very
small branching ratio, with a QED prediction of $2.6 \times 10^{-5}$. It has been measured only once and the
experimental value was found to be within errors of the QED calculation~\cite{KLOE_eeee}. 

WASA has measured this decay in the $pd \rightarrow ^{3}He\eta$ reaction. Two independent analyses report
$(50 \pm 14)$ event candidates. The event sample from one analysis is shown in Figure~\ref{fig:eta_fig}b.
The branching ratio returned by these two analyses is consistent, allowing WASA to quote a preliminary value
of $(3.0 \pm 0.8_{stat} \pm 0.7_{sys. (norm)}) \times 10^{-5}$ which is in agreement with theoretical and
experimental values. Systematic studies as well as consistency checks between the two analyses are currently
being performed. 

\begin{figure}[hbt]
 \begin{minipage}[t]{0.5\textwidth}
  \centerline{\includegraphics[width=0.9\textwidth]{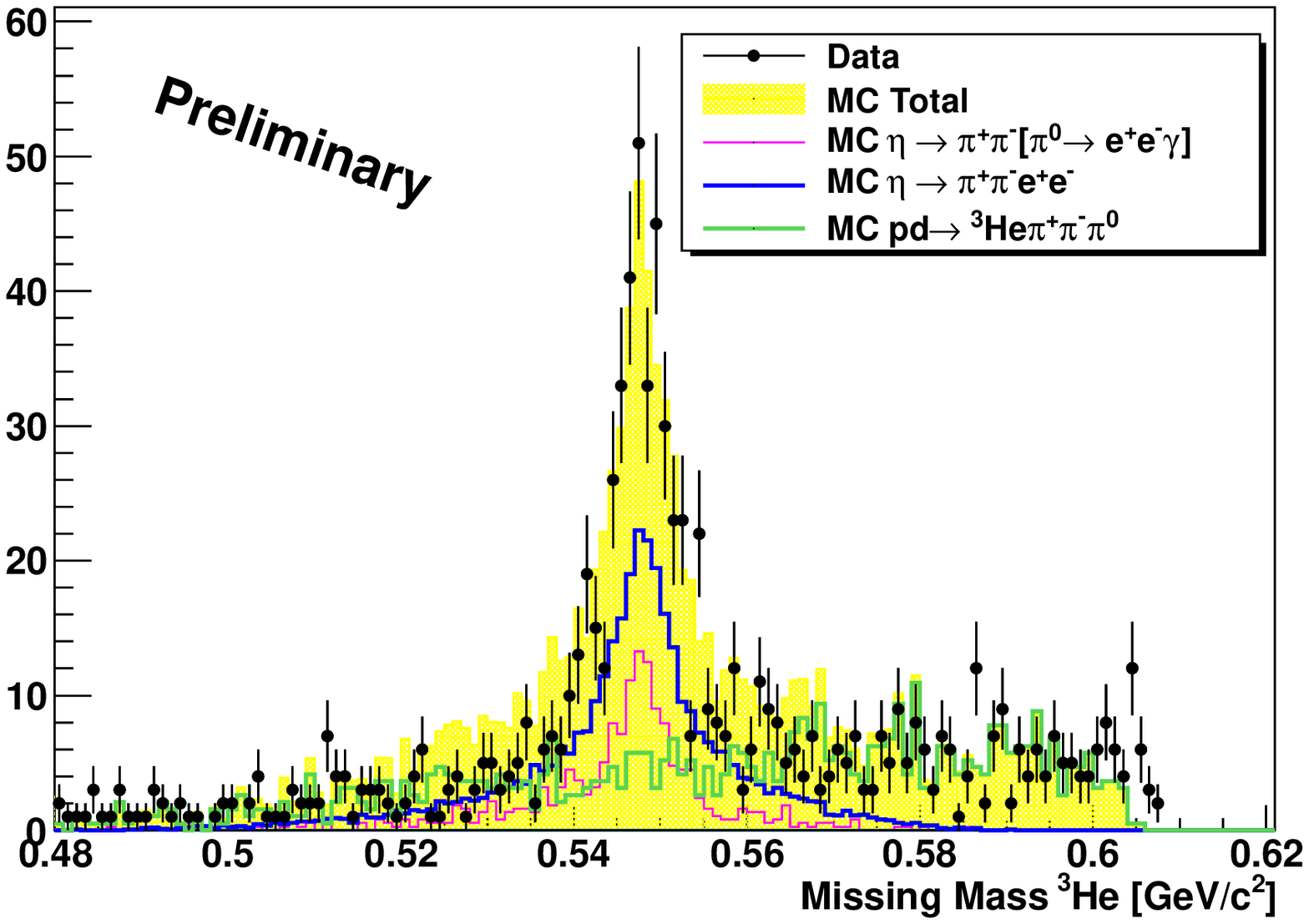}}
  \centering\parbox{0.9\textwidth}{\centering{\footnotesize (a)}}
 \end{minipage}%
 \begin{minipage}[t]{0.5\textwidth}
  \centerline{\includegraphics[width=0.82\textwidth]{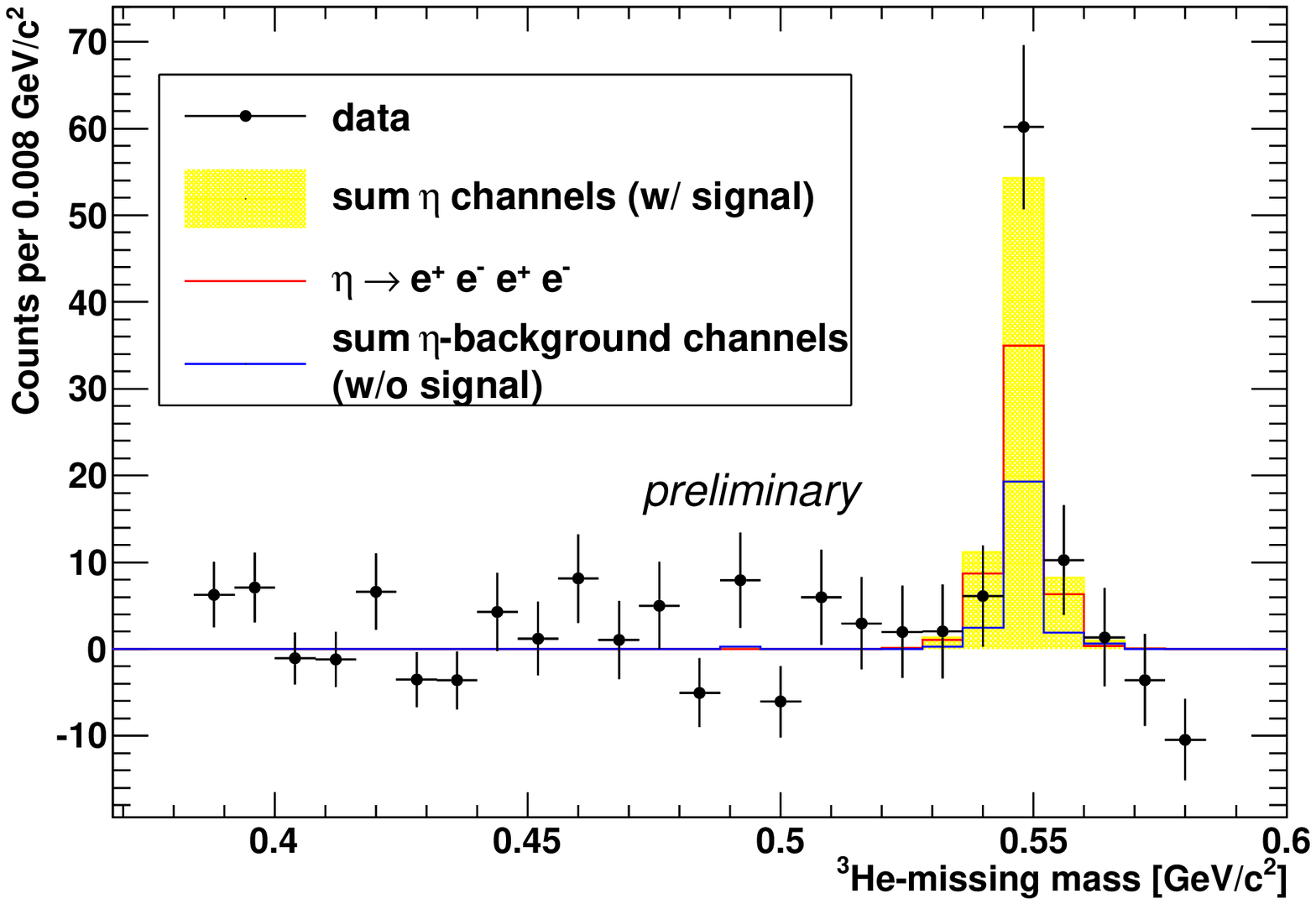}}
  \centering\parbox{0.9\textwidth}{\centering{\footnotesize (b)}}
 \end{minipage}
 \caption{\label{fig:eta_fig}Missing Mass of $^{3}He$ for events satisfying conditions for (a)
$\eta\rightarrow\pi^{+}\pi^{-}e^{+}e^{-}$ and (b) $\eta\rightarrow e^{+}e^{-}e^{+}e^{-}$.}
\end{figure}

The decay of $\eta$ into two electrons is a fourth order electromagnetic process proceeding through two
virtual photons and additionally suppressed due to helicity conservation to the order of $10^{-9}$. This
extremely low branching ratio makes the reaction very sensitive to any processes that could enhance the decay
width, including possible effects outside the standard model~\cite{ee_1,ee_2,ee_3}. Measurements of the
closely related $\pi^{0} \rightarrow e^{+}e^{-}$ decay have found branching ratios above theoretical
predictions. Measurements in the $\eta$ system should be even more sensitive to such effects due to the lower
branching ratio predicted by the standard model.

The current experimental upper limit for this decay is $7.7 \times 10^{-5}$~\cite{PDG} and was measured by the
CELSIUS-WASA collaboration. Using two weeks of data in proton-proton interactions from WASA-at-COSY, a value
one order of magnitude lower has been calculated. When considering the full statistics available in the $pp
\rightarrow pp\eta$ reaction, it is expected that additional sensitivity can be gained corresponding to at
least another order of magnitude improvement. Additionally, a more intelligent trigger has been developed and
is being tuned to enable data collection at higher luminosities. 

\section{Conclusion}
The WASA-at-COSY collaboration is in the processes of analyzing the large amount of data taken on $\eta$
decays while simultaneously gearing up for extended production runs to increase statistics for rare processes.
Results for some decay channels have been released while several more are at the stage of careful systematic
checks.

Additionally, the meson program at WASA-at-COSY is expanding. A pilot run intended to produce large numbers
of $\pi^{0}$ mesons in order to measure the very rare $\pi^{0}\rightarrow e^{+}e^{-}$ decay is currently being
analyzed. Also, the first data on $\omega$ decays was taken in proton-deuteron and proton-proton reactions in
Spring of 2011. 

\section*{Acknowledgements}
This publication has been supported by COSY-FFE and by the European Commission under the 7th Framework
Programme through the 'Research Infrastructures' action of the 'Capacities' Programme. Call:
FP7-INFRASTRUCTURES-2008-1, Grant Agreement N. 227431.

\coltoctitle{Decays of Light Mesons in CLAS}
\authortochead{M.~Amaryan}
\label{AM}

\setcounter{chapter}{1}
\setcounter{section}{0}
\maketitle

\Section{Introduction}
At low energies the coupling constant of Quantum Chromodynamics (QCD) becomes large and renders perturbation
theory useless. The phenomenology of the strong interactions in this domain remains one of the major
challenges of modern particle physics. Experimental study of decays of light mesons in general and
$\pi^0,\eta$ and $\eta^{\prime}$ pseudoscalars in particular is a unique way to explore low energy QCD. By
comparing data obtained at different facilities like KLOE, CLEO, BES, MAMI, COSY, and CLAS with a predictions
of effective field theory at low energy, Chiral Perturbation Theory (ChPT), one can get an insight into the
non-perturbative regime of strong interactions and provide important information for a firmer foundation of
hadronic physics rooted in the standard model. Until recently the quest for high statistics accurate
experimental data was not fulfilled. 

In this talk we  give a brief overview of experimental data on a decay of light mesons collected with the
CLAS setup at JLAB either  with unprecedented high statistics or comparable with the worlds highest statistics
measurements depending on the decay mode. Below we outline the quality and statistics of available data at
CLAS  in a different decay channels of light mesons, rather than present results of finalized analyses that
are in progress.

\Section{Light mesons in CLAS}
The tagged beam of photons produced from 6 GeV CEBAF electron accelerator is used to study photo-production of
light mesons from liquid hydrogen target. Until now the main focus of the experiments in CLAS related to light
mesons was the study of different aspects of photo-production cross sections aiming to understand mechanism of
their production. However as it is shown below the wealth of data allows to make a studies exploiting
different decay modes detached from the production vertex and considering decaying mesons as a laboratory by
themselves.

Experimental data are presented by accenting on photo-production reactions $\gamma+p\rightarrow
p\pi^0(\eta,\eta^{\prime})$ collected in the following decay modes:
\begin{itemize}
\item Dalitz decays $\pi^0(\eta,\eta^{\prime})\rightarrow e^+e^-\gamma$
\item Radiative decays $\eta(\eta^{\prime})\rightarrow \pi^+\pi^-\gamma$
\item Hadronic decays $\eta(\eta^{\prime})\rightarrow \pi^+\pi^-\pi^0$
\item Hadronic decay $\eta^{\prime}\rightarrow \pi^+\pi^-\eta$
\end{itemize}

\Subsection{Dalitz decays}
The branching ratios of radiative decay of pseudoscalar mesons $\pi^0$ and $\eta$ were measured and are
recorded in PDG, however there is only an upper limit quoted for $\eta^{\prime}$.  Below in
Fig.~\ref{fig:dalitz_decay} (left panel) we present distribution of the invariant mass of $e^+e^-\gamma$
measured in photoproduction reaction with CLAS. As one can there are clear peaks of all three light
pseudoscalar mesons $\pi^0,\eta$ and $eta^{\prime}$. One of the less studied channel of the $\eta\rightarrow
e^+e^-\gamma$ decay allows to extract form factor of $\eta$ in the timelike region by studying distribution of
the invariant mass of $e^+e^-$ pairs presented in Fig.~\ref{fig:dalitz_decay} (right panel). In both panels in
Fig.~\ref{fig:dalitz_decay} number of events are presented without corrections due to the CLAS acceptance and
efficiency.
\begin{figure}[htb]
  \centerline{\includegraphics[width=0.48\textwidth,clip,trim = 0mm 70mm 5mm
70mm]{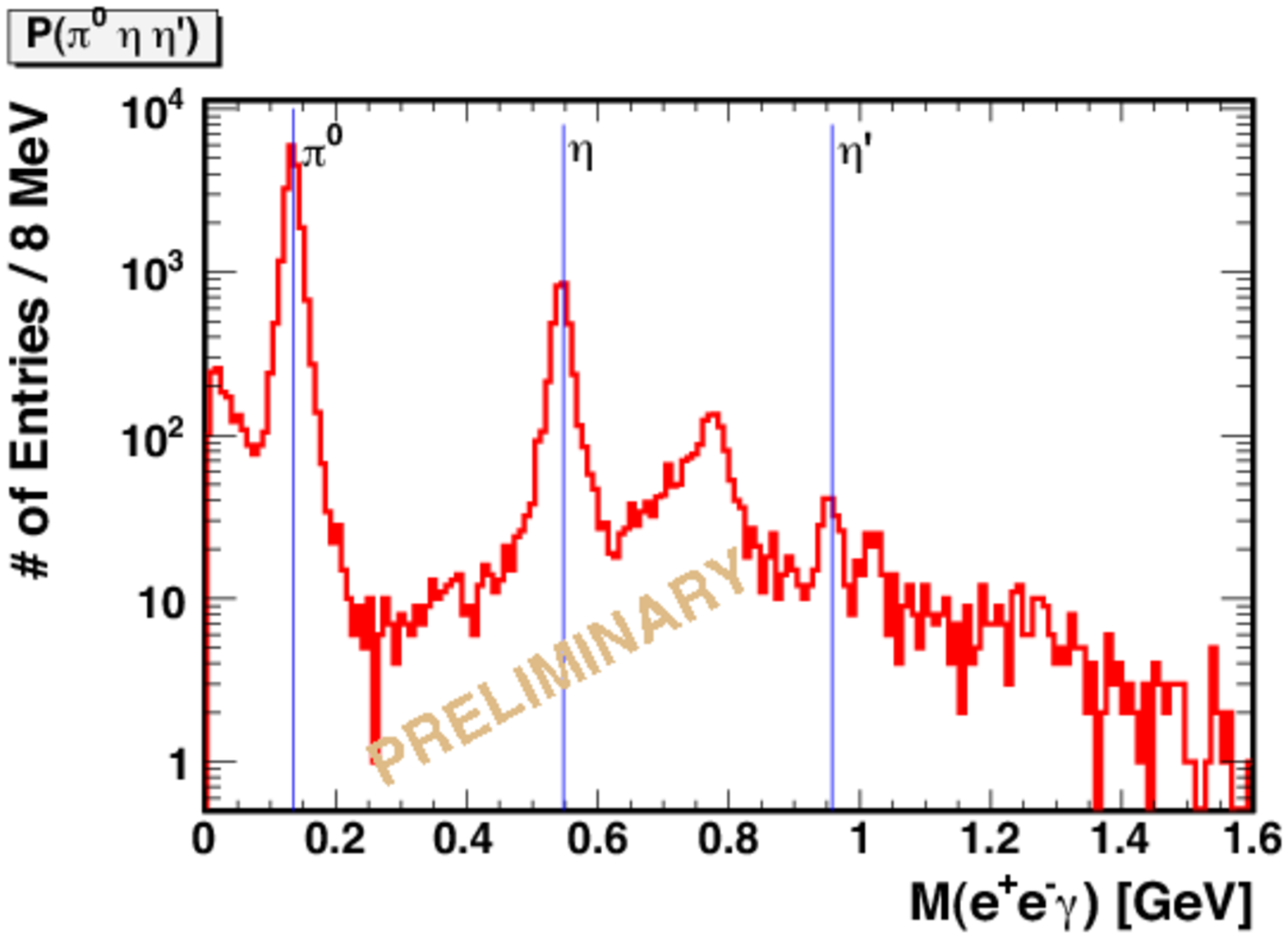}\hfill%
              \includegraphics[width=0.48\textwidth,clip,trim = 0mm 70mm 5mm
70mm]{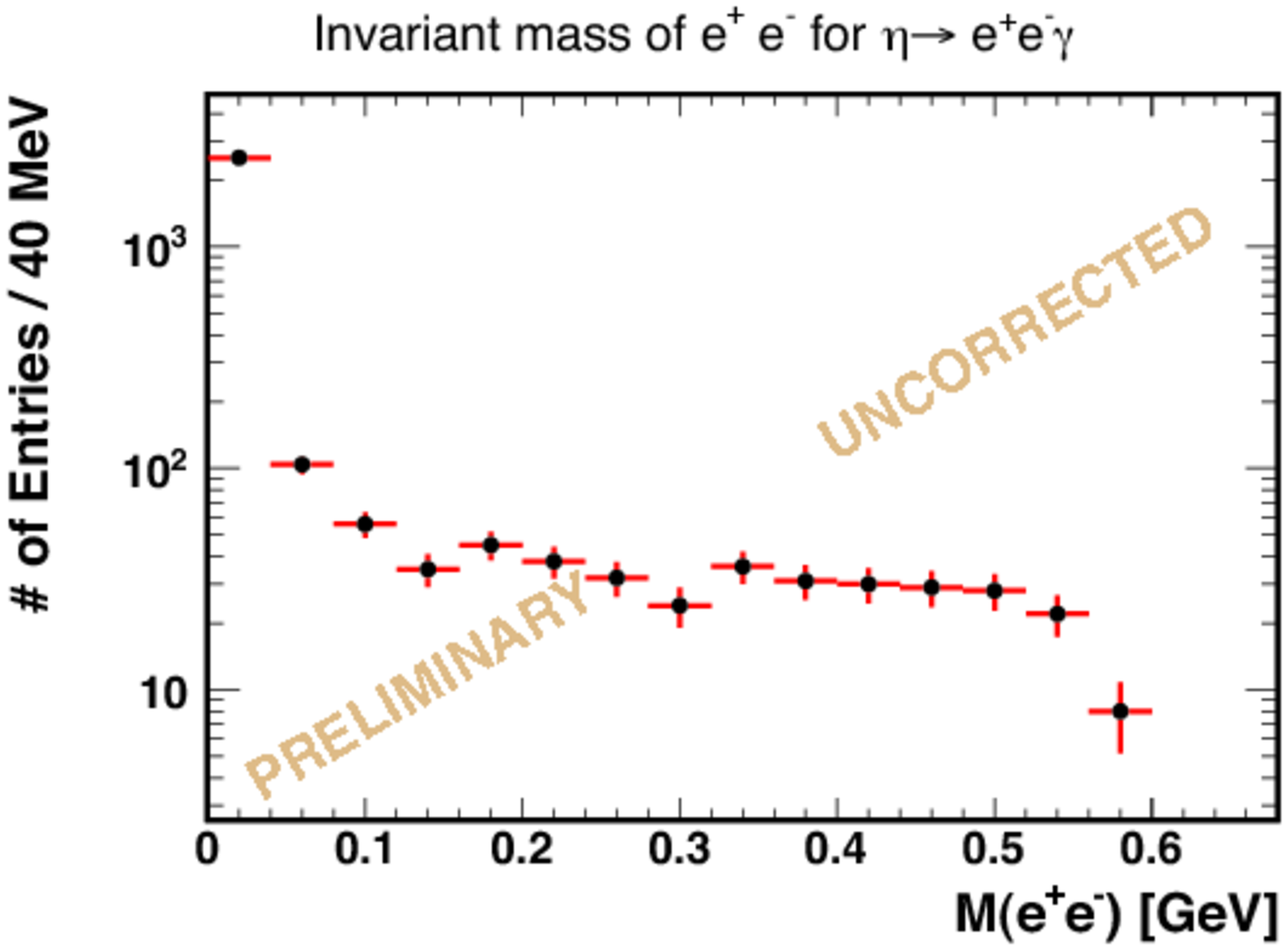}}
  \caption{\label{fig:dalitz_decay}(Color online). Left panel: the $e^+e^-\gamma$ invariant mass distribution
in the reaction $\gamma+p\rightarrow pe^+e^-\gamma$,  the vertical lines show positions of pseudoscalar mesons
$\pi^0,\eta,\eta^{\prime}$.  Right panel: invariant mass of $e^+e^-$ lepton pairs for events under the $\eta$
peak $M(e^+e^-\gamma)=0.548\pm0.01$~GeV.}
\end{figure}

\Subsection{Radiative decays}
Another very interesting decay of $\eta$ and $\eta^{\prime}$ mesons is their radiative decay  
$\eta(\eta^{\prime})\rightarrow \pi^+\pi^-\gamma$. Recent theoretical paper ~\cite{meissner} demonstrates
that the photon energy distribution from radiative decay of $\eta$ and $\eta^{\prime}$ mesons provides very
important test of theoretical models. In Fig.~\ref{fig:pipi_g} (left panel) missing mass of the proton is
presented for exclusive reaction $\gamma+p\rightarrow p\pi^+\pi^-\gamma$.  In both $\eta$ and $\eta^{\prime}$
peaks a statistics is by more than an order of magnitude higher than existing world data (see ~\cite{meissner}
and references therein). In Fig.~\ref{fig:pipi_g} (middle (right) panel) distribution of the energy of the
photon from the decay of $\eta(\eta^{\prime}$ in the center-of-mass frame of the parent is are presented.

\begin{figure}[htb]
  \centerline{\includegraphics[height=1.80in,width=2.0in]{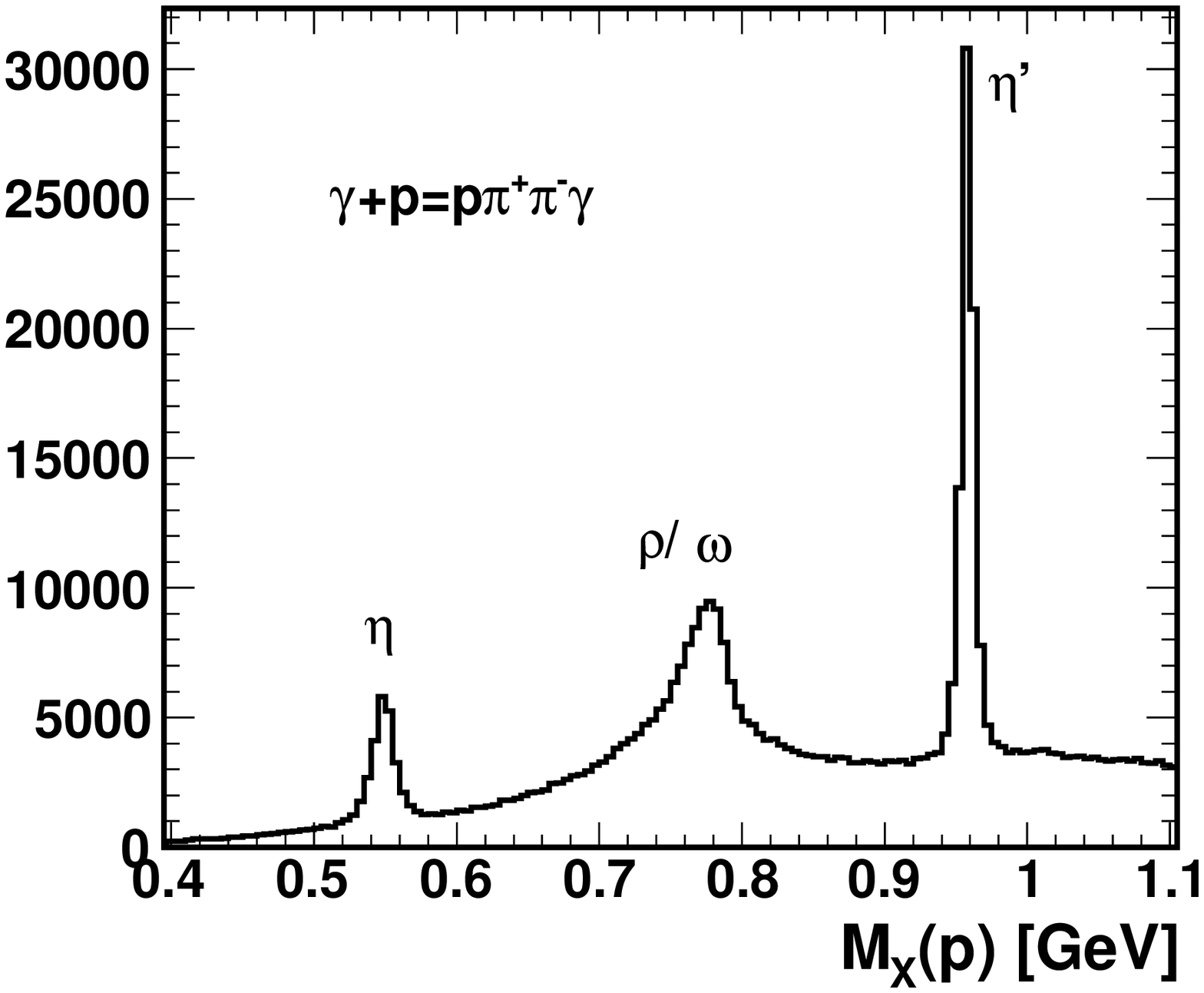}%
              \includegraphics[height=1.80in,width=2.0in]{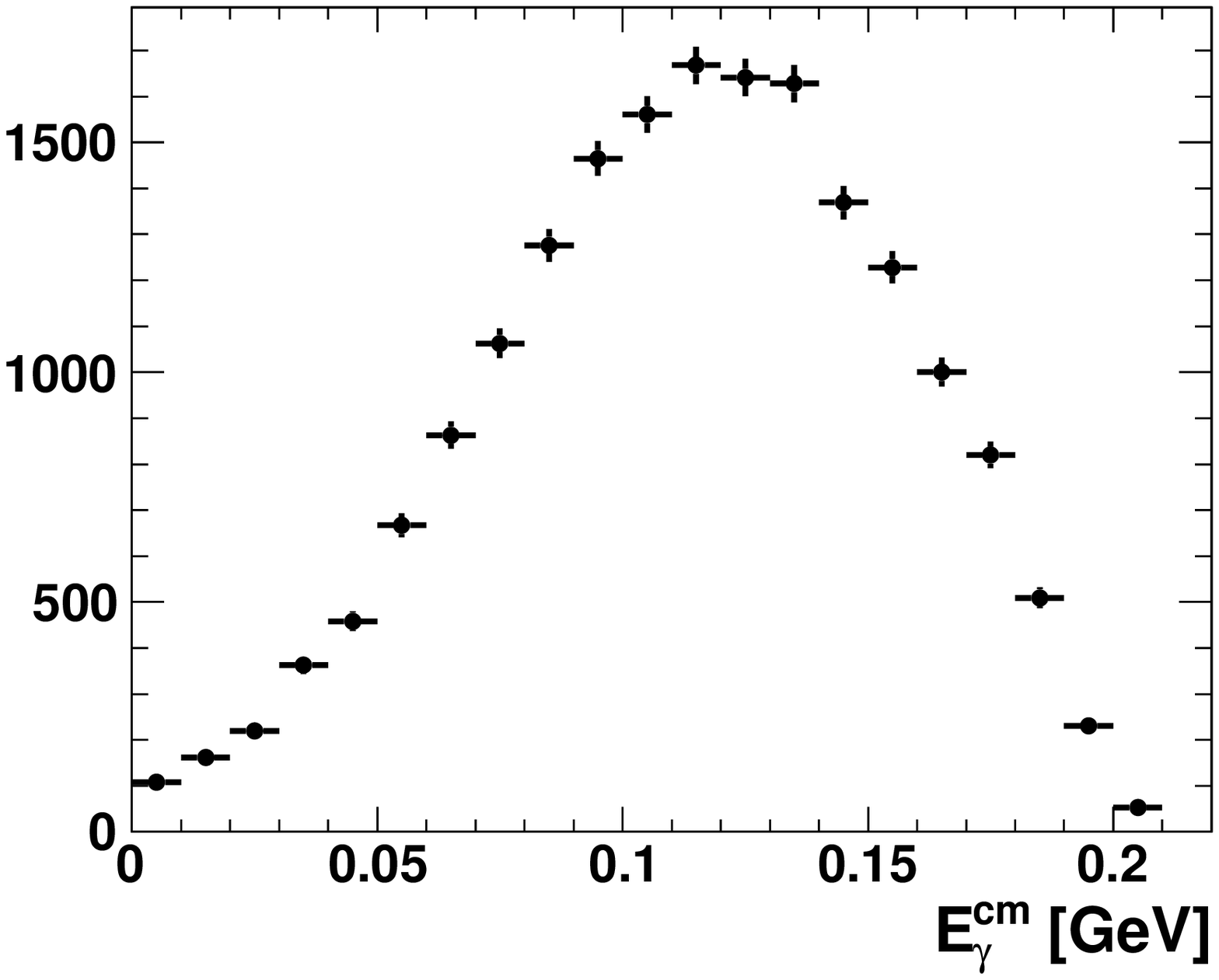}%
              \includegraphics[height=1.80in,width=2.0in]{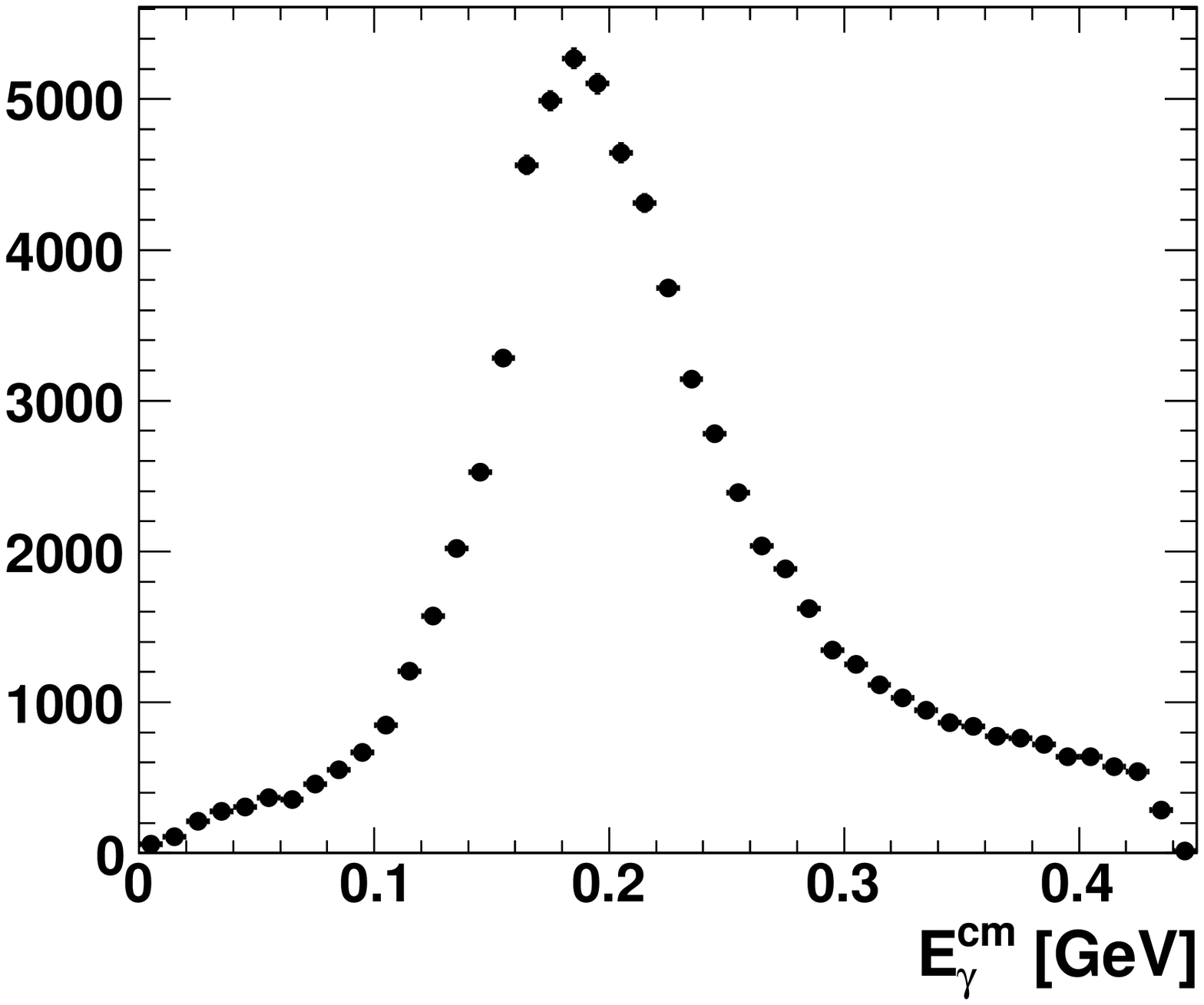}}
 \caption{\label{fig:pipi_g}Left panel: distribution of missing mass of the proton in exclusive reaction
$\gamma+p\rightarrow p\pi^+\pi^-\gamma$. Middle panel: the distribution of the $\gamma$ energy in the
center-of-mass of $\eta$. Right panel: same as the middle panel plotted for the $\eta^{\prime}$ decay. All
three plots show raw number of events before the acceptance and efficiency corrections.}
\end{figure}

\Subsection{Hadronic decays}
In this section we present experimental data for the reaction $\gamma+p\rightarrow p\pi^+\pi^-X(\pi^0(\eta))$,
where $\pi^0(\eta)$ are identified via missing mass of proton, $\pi^+$ and $\pi^-$. In Fig.~\ref{fig:three_pi}
(left panel) a distribution of missing mass of the proton is presented showing clear peaks of $\eta$ and
$\omega$ mesons with $\sim$2M and $\sim$20M events respectively as well as hints of $\eta^{\prime}$ and $\phi$
mesons. To see peaks of $\eta^{\prime}$ and $\phi$ better in Fig.~\ref{fig:three_pi} (right panel) we plot the
same distribution for the range above $\omega$ meson. As one can see there we observe one of rare decays
$\eta^{\prime}\rightarrow\pi^+\pi^-\pi^0$ (Br=$3.6\times10^{-3}$) and OZI violating decay
$\phi\rightarrow\pi^+\pi^-\pi^0$ (Br=$15.3\%$) observed for the first time in the photo-production
experiments. The photo-production of the latter can be studied  to check the influence of meson-baryon
interference that might occur in the charge and neutral $K\bar K$ decay modes and is absent in this channel.

\begin{figure}[htb]
  \centerline{\includegraphics[clip,trim = 70mm 0mm 0mm 0mm,
width=0.37\textwidth]{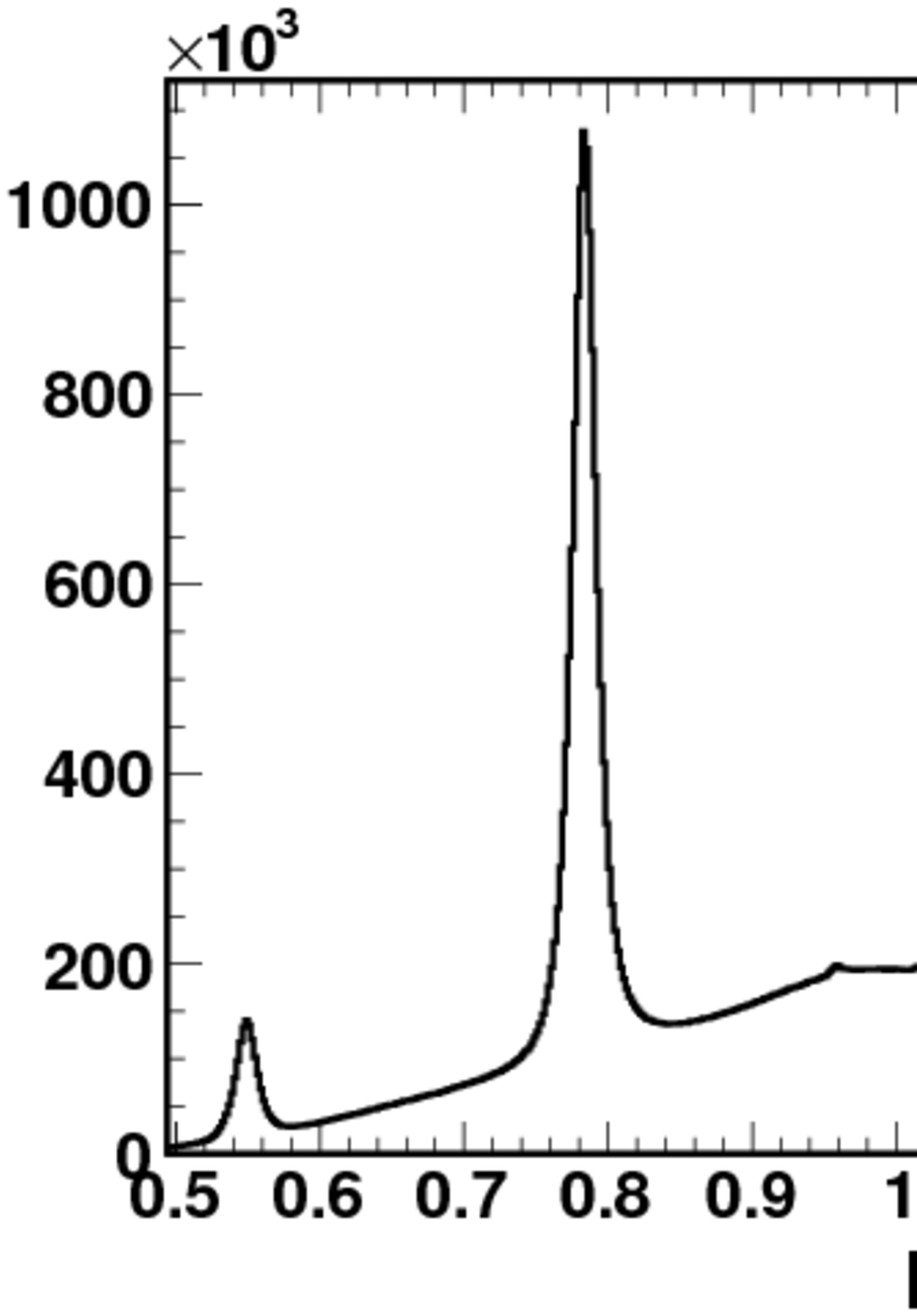}\hspace{0.1\textwidth}%
              \includegraphics[clip,trim = 70mm 0mm 0mm 0mm,
width=0.37\textwidth]{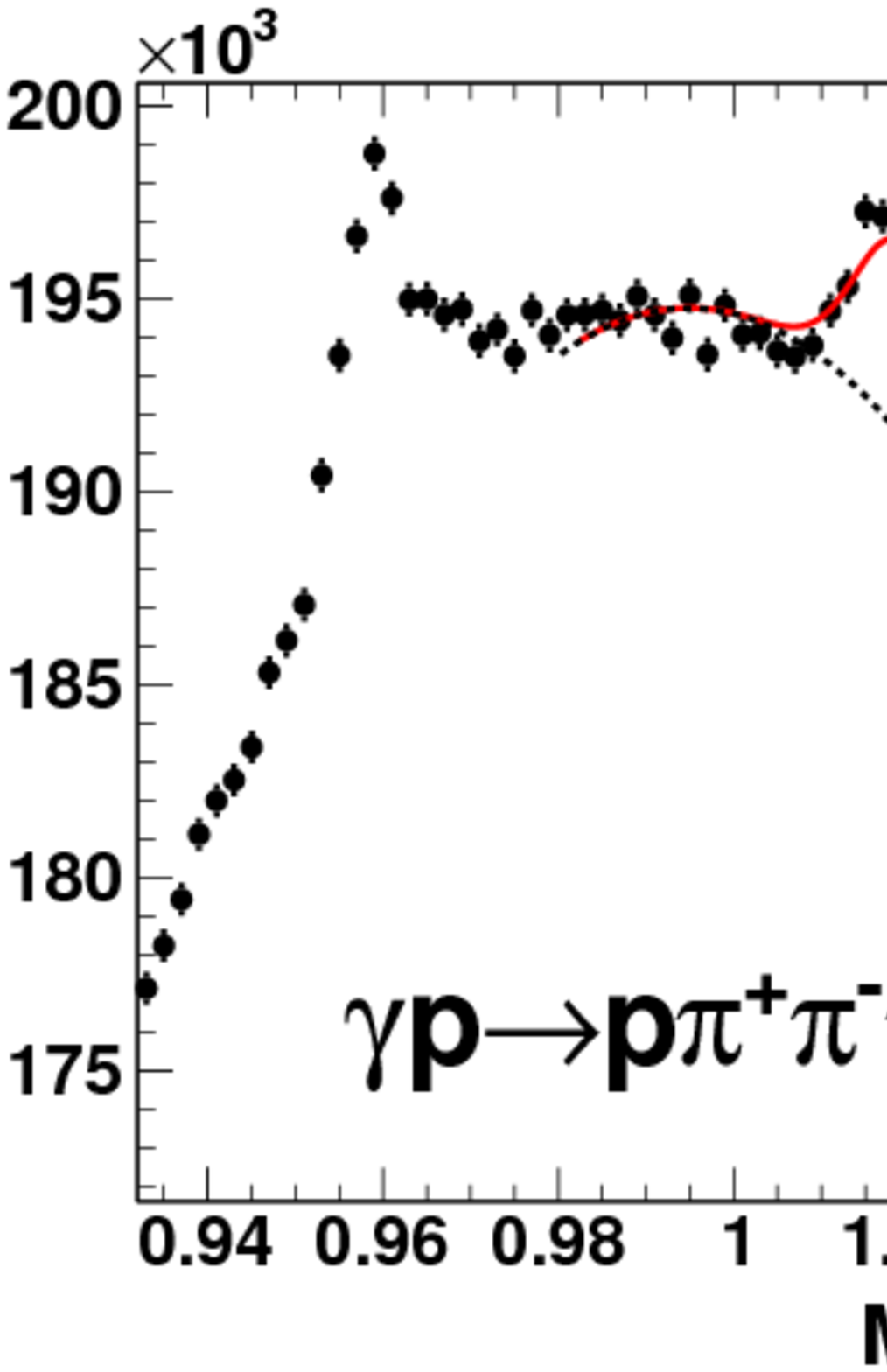}} 
 \caption{\label{fig:three_pi}Left panel: distribution of missing mass of the proton in the reaction
$\gamma+p\rightarrow p\pi^+\pi^-\pi^0$.  Right panel: the same for the range of invariant mass above $\omega$
meson production.}
\end{figure}
In Fig.~\ref{fig:eta_pipi} (left panel) we present distribution of missing mass of the proton in the reaction
$\gamma+p\rightarrow p\pi^+\pi^-\eta$, where $\eta$ is reconstructed in the missing mass of proton, $\pi^+$
and $\pi^-$. As one can see there is a clear peak of $\eta^{\prime}$ with $\sim$340K events which is almost an
order of magnitude higher than recent BES~\cite{BES} data. This will allow to extract coefficients of
polynomial distributions of Dalitz plot projections with much higher statistical precision. On the other hand
there appears a hint of G-parity decay mode of $\phi\rightarrow\pi^+\pi^-\eta$. In  In Fig.~\ref{fig:eta_pipi}
(right panel) the same distribution is plotted above the $\eta^{\prime}$ peak. Here we see clear peak of
$\phi$ meson. This decay mode has never been observed and PDG quotes only an upper limit
Br$<1.8\times10^{-5}$. Theoretical model~\cite{achasov} predicts even smaller upper limit on the order of
$\sim3\times10^{-7}$.

\begin{figure}[htb]
  \centerline{\includegraphics[width=0.54\textwidth]{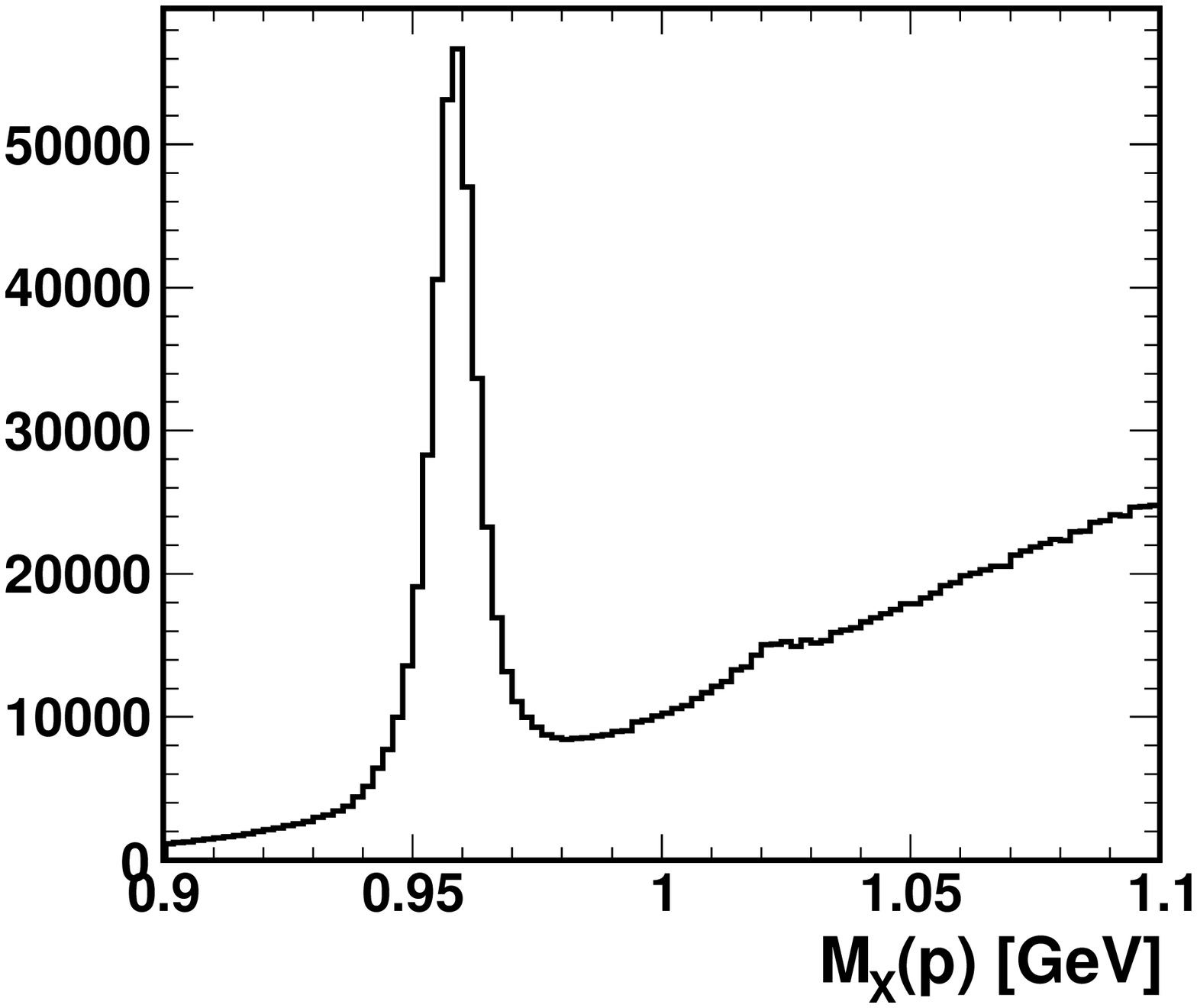}\hfill%
              \includegraphics[width=0.44\textwidth,clip,trim = 0mm 5mm 5mm
80mm]{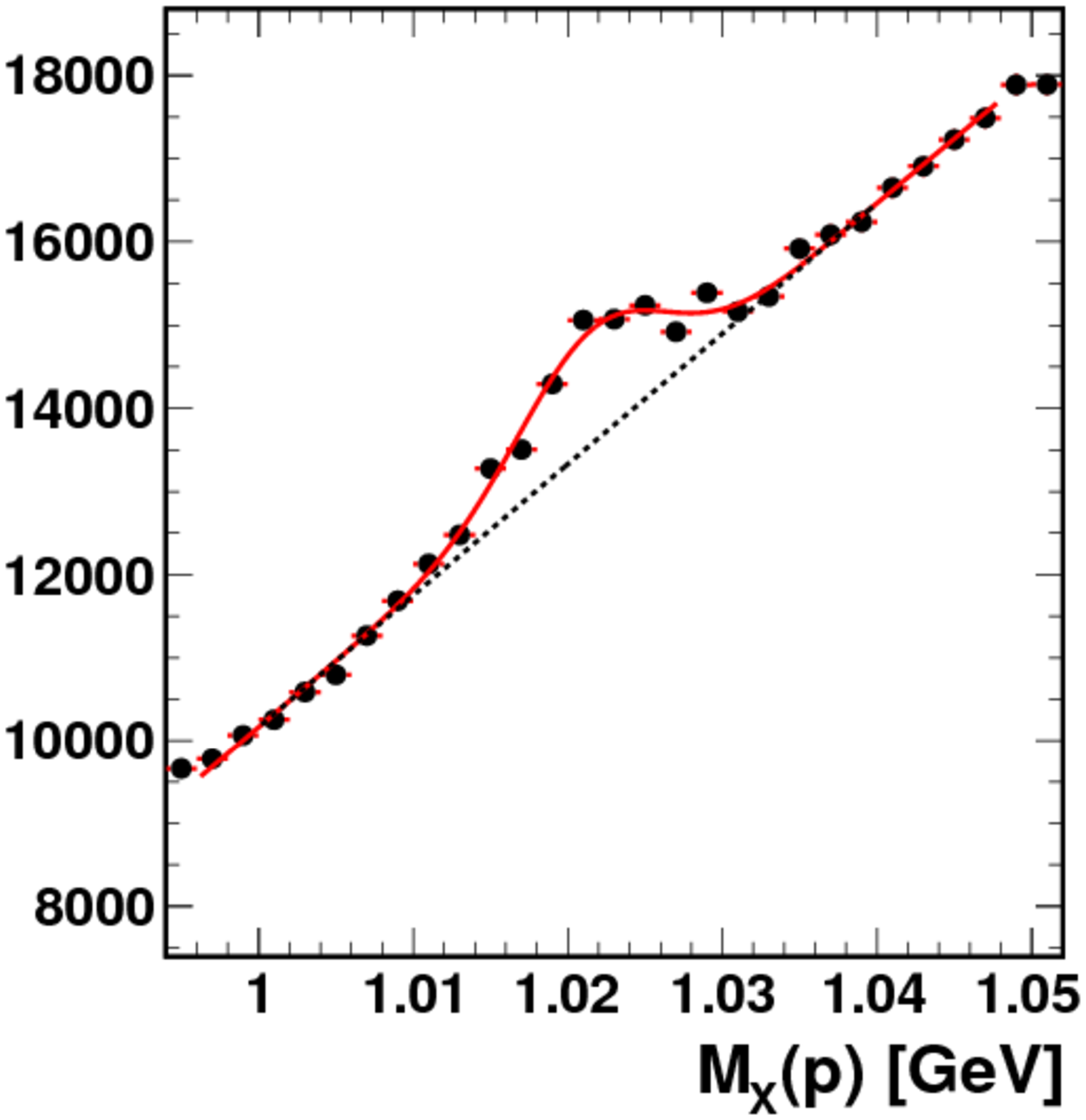}}
  \caption{\label{fig:eta_pipi}Left panel: distribution of missing mass of the proton for the reaction
$\gamma+p\rightarrow p\pi^+\pi^-\eta$. Right panel: the same for the range of invariant mass above
$\eta^{\prime}$ meson production.}
\end{figure}

\vfill%
\Section{Aknowledgements}
We take this opportunity to express our gratitude to organizers for the invitation to give a talk at very
fruitful PrimeNet workshop in J{\"{u}}lich and for their kind hospitality. This work is supported by the DOE
Grant No. 350314.

\title{Pseudoscalar Meson Studies with Charm}
\author{J.~Messchendorp}
\label{MJ}
\import{./MesschendorpJohan/Messchendorp}
\end{papers}
\cleardoublepage

\pagestyle{plain}
\begin{center}
 \vspace*{0.3\textheight} {\Huge\textbf{\boldmath Meson Production}}\\
 \vspace{2cm} {\Large\textbf{in photon-induced reactions and from NN collisions \\ as well as
e$^+$e$^-$ production from pp collisions}}
 \addcontentsline{toc}{part}{\sffamily Meson Production in photon-induced reactions and from NN collisions}
\end{center}
\cleardoublepage
\pagestyle{fancy}

\begin{papers}
\coltoctitle{\boldmath Study of $\eta$ Meson Production with a Polarized Proton Beam}
\authortochead{I.~Ozerianska}
\label{OI}

\setcounter{chapter}{1}
\setcounter{section}{0}
\maketitle

\Section{Introduction}
In the low energy regime of Quantum Chromodynamics, the interaction between quarks and gluons cannot be
treated perturbatively and so far the  understanding  of the processes governed by the strong forces is
unsatisfactory. Therefore, it is essential to carry out measurements involving the production and decay of
hadrons and to interpret them in the framework of effective field theories experiencing recently an enormous
development in applications to the description of meson decays and production. In this contribution we
concentrate on the $\eta$ meson. The progress in understanding of the production processes of the $\eta$ meson
will strongly rely on the precise determination of spin and isospin observables. So far these observables have
been determined only for few excess energies and with low statistics~[1-5].

\Section{Partial waves}
For an unambiguous understanding of the production process relative magnitudes from the partial waves
contributions must be well established. This may be achieved by the measurement of the analysing power which
would enable to perform the partial wave decomposition  with an accuracy by far better than resulting from the
measurements of the distributions of the spin averaged cross sections~\cite{ex:6,ex:7}. Analysing power $A_y$
may be understood as a measure of the relative deviation between the differential cross section for the
experiment with and without polarized beam ($\sigma$ and $\sigma_0$ respectively):
\begin{equation}
\sigma(\theta,\varphi )=\sigma_{0}(\theta,\varphi )\cdot[1+A_{y}(\theta)\cdot{P}\cdot\cos(\varphi)]
\label{Eq:xdef111}
\end{equation}
where $P$ denotes the beam polarization.

\Section{ Studies of $A_y$ with the WASA detector at COSY}
Using the WASA-at-COSY detector~\cite{ex:8} we intend to determine the energy and the angular dependence of
A$_y$(Q,$\theta$) and the total and differential cross sections for the $\vec{p}p\to pp\eta$ reaction in the
excess energy range from the threshold up to 100~MeV. In November 2010 first measurements for Q~=~15~MeV 
and Q~=~72~MeV have been conducted \cite{ex:9}. From Table~\ref{table} we can see the beam parameters and the
expected number of events for each excess energy.

\begin{table}[h!]
 \begin{small}
  \begin{center}
   \begin{tabular}{|c|c|c|c|c|c|c|}\hline
Q [MeV/c] &P [MeV/c] & $\sigma_{tot}$[$\mu$b]  & Acceptance & $N_{\eta \rightarrow\gamma \gamma}$ &  $N_{\eta\rightarrow}$3$\pi^{0}$  \\
\hline
15 &2026 &10$^{3}$  &0.55 &99770 &81861  \\
\hline
72 &2188 &$5\cdot10^{3}$ & 0.63&447739 &375580 \\
\hline
   \end{tabular}
  \end{center}
  \vspace{-0.7cm}
  \begin{center}
   \caption{ Estimate of the number of produced $\eta$ mesons \label{table}}
  \end{center}
 \end{small}
\end{table}

Protons from the $\vec{p}p\to pp\eta$ reaction are registered in the forward part of the detector and photons
from the $\eta$ meson decay are detected in the Electromagnetic Calorimeter of the central part.
Simultaneously to the $\vec{p}p\to pp\eta$ reaction elastically scattered protons were registered. The
$\vec{p}p\to pp$ reaction will be used for monitoring of the polarization degree, luminosity and the detector
performance. In the case of the $\vec{p}p\to pp$ reaction one proton is registered in the Forward Detector and
the other in the Central Detector.

Within the geometrical acceptance of the Forward Detector (from $3^{\circ}$ to $18^{\circ}$), the Central
Detector covers proton scattering angle from $60^{\circ}$ to $84^{\circ}$. In the center of mass system that
corresponds to a scattering angle in the range of ~30${}^{\circ}$ to ~46${}^{\circ}$ as seen in
Fig.~\ref{fig:dp}a.

In Fig.~\ref{fig:dp}b the scattering angle of one proton is depicted as a function of the scattering angle
of the second proton. The first proton is registered in the Central Detector, the second one in the Forward
Detector. Colored lines are predictions for different beam momenta. The blue line corresponds to a beam
momentum of 2~GeV/c,  close to the experimental value.

\begin{figure}[hbt]
 \begin{minipage}[t]{0.5\textwidth}
  \centerline{\includegraphics[width=0.8\textwidth]{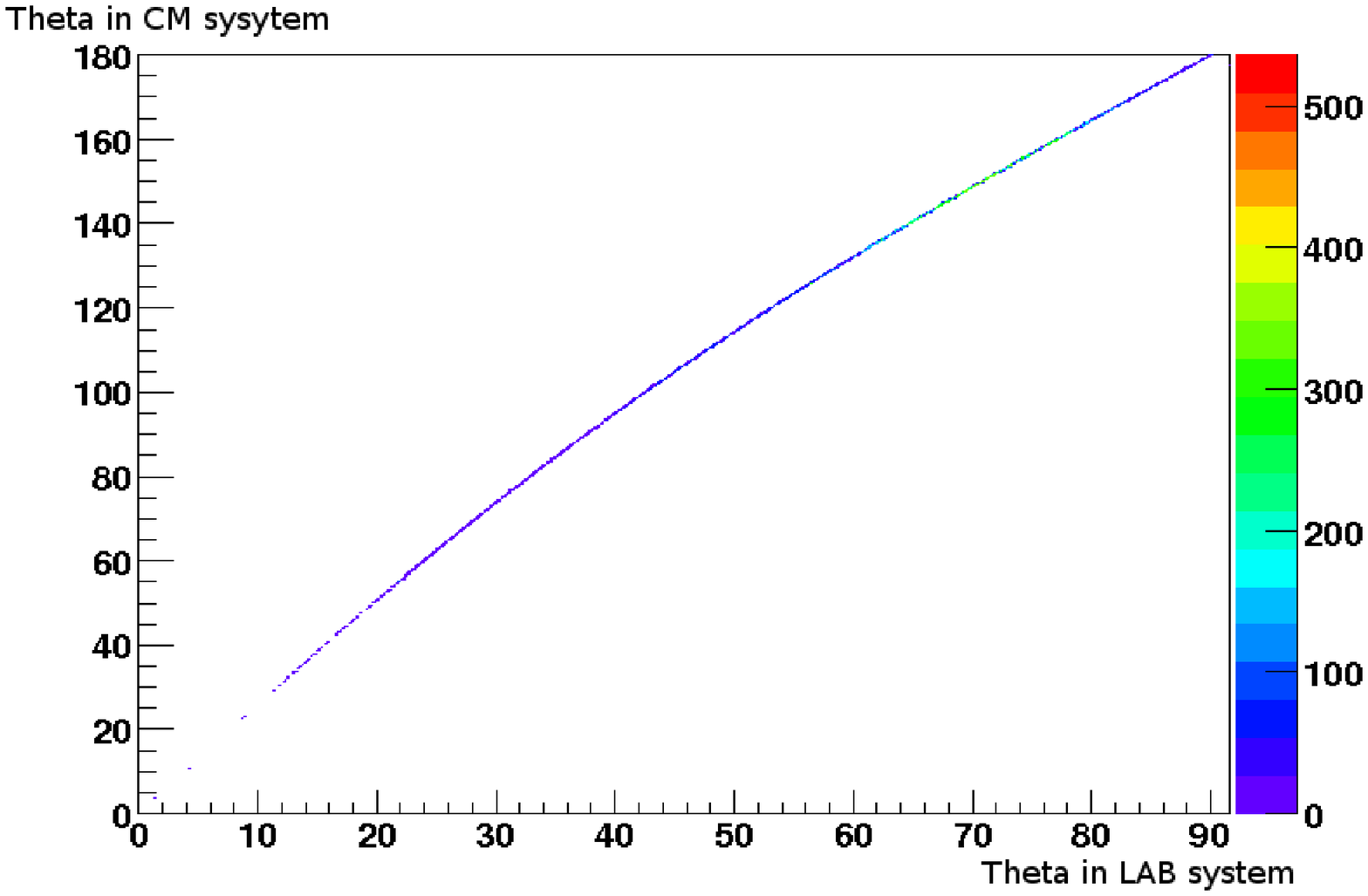}}
  \centering\parbox{0.9\textwidth}{{\small (a) The scattering angle in the centre of mass system is presented
as a function of the scattering angle in the laboratory frame.}}
 \end{minipage}%
 \begin{minipage}[t]{0.5\textwidth}
  \centerline{\includegraphics[width=\textwidth]{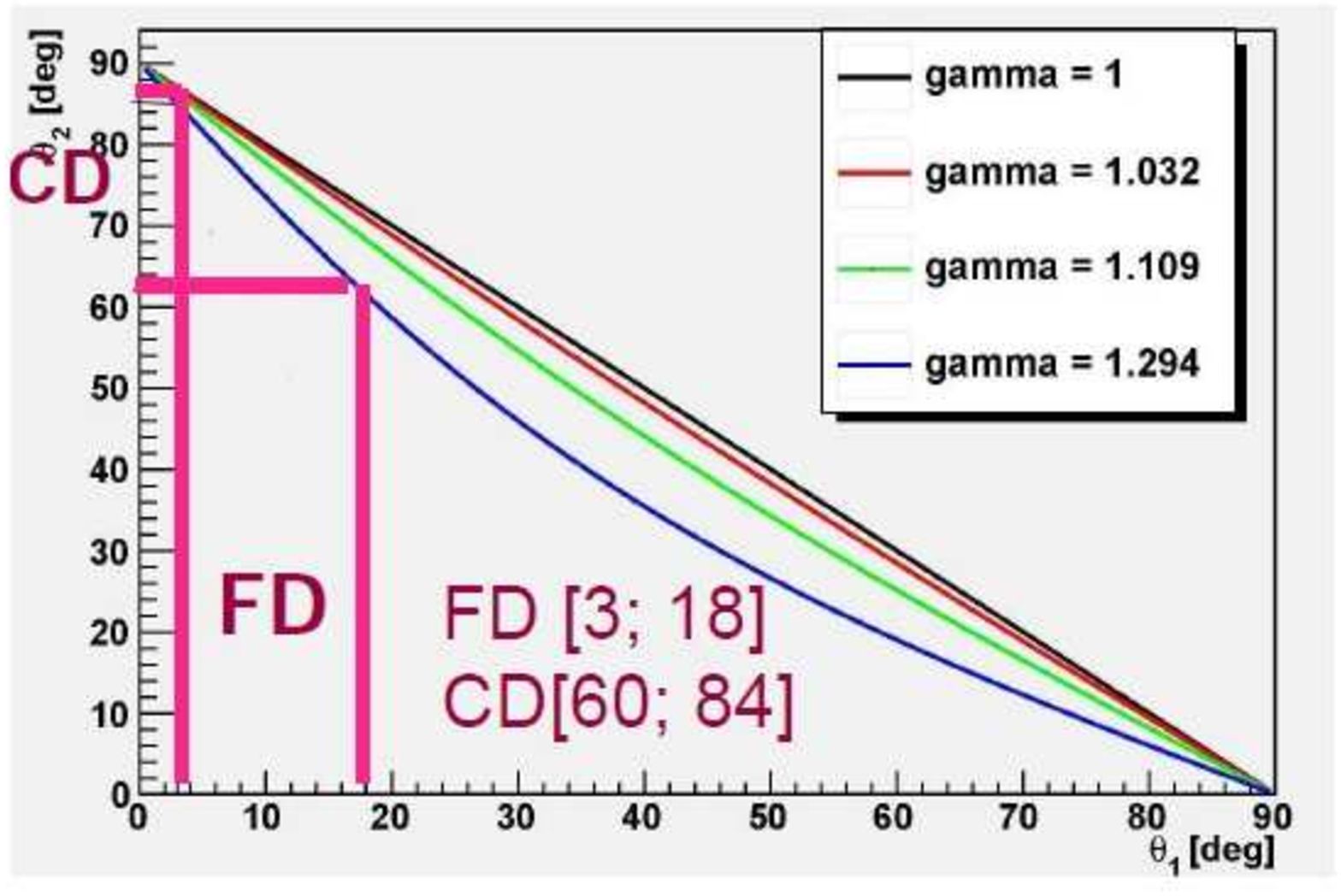}}
  \centering\parbox{0.9\textwidth}{{\small (b) The distribution of the scattering angle of proton registered
in the Central Detector as a function of the scattering angle of the proton registered in the Forward
Detector. For more details see the text.}}
 \end{minipage}
 \caption{\label{fig:dp}The protons angular distributions in the Monte Carlo simulations of the $pp\to pp$
reaction.}
\end{figure}

\vspace{-4ex}\Section{ Extraction of $A_y$ from the experiment}
In the first step we can obtain the value of the polarization, P, from the $\vec{p}p\to pp$ reaction using the
analysing power already measured by the EDDA experiment~\cite{ex:10}. In the second step, we will calculate
the geometrical averages, $N_{\pm}$, of the number of $\eta$ meson produced in direction ($\theta$, $\varphi$)
during the spin down $N^{\downarrow}$ and spin up $N^{\uparrow}$ modes, defined as follows:

\begin{eqnarray}\nonumber
N_{-}&\equiv\sqrt{N_{-}^{\uparrow}N_{-}^{\downarrow}}=\sqrt{\frac{N_{R}^{\uparrow}}{\epsilon_{R}{L^{\uparrow}}}\cdot\frac{N_{L}^{\downarrow}}{\epsilon_{L}{L^{\downarrow}}}}\\
N_{+}&\equiv\sqrt{N_{+}^{\uparrow}N_{+}^{\downarrow}}=\sqrt{\frac{N_{L}^{\uparrow}}{\epsilon_{L}{L^{\uparrow}}}\cdot\frac{N_{R}^{\downarrow}}{\epsilon_{R}{L^{\downarrow}}}}
\label{eq:xdef2}
\end{eqnarray}
$N_{-}^{\downarrow}$ and $N_{+}^{\uparrow}$ can be determined according to the Madison convention\cite{ex:11}.
 Knowing the angles $\theta$ and $\varphi$ of the outgoing $\eta$ meson we can
calculate the analysing power for the three particle final state using the following formula:

\begin{equation}
\ 
A_{y}(\theta)=\frac{1}{Pcos\varphi}\cdot\frac{N_{+}(\theta,\varphi)-N_{-}(\theta,\varphi)}{N_{+}(\theta,\varphi)+N_{-}(\theta,\varphi)}
\label{eq:xdef3}
\end{equation}
where the polarization $P$ was calculated in the first step.

An $A_{y}$ for the proton-proton elastic scattering at 2026 MeV and 2188~MeV beam momenta, based on the
results of the EDDA experiment is shown in Fig.~\ref{fig4:dP}. We can conclude that values of $A_{y}$ for
Q=15~MeV and Q=72~MeV are within the range of [0.32-0.38]. 

\begin{figure}[hbt]
 \begin{minipage}[c]{0.5\textwidth}
  \centerline{\includegraphics[width=\textwidth]{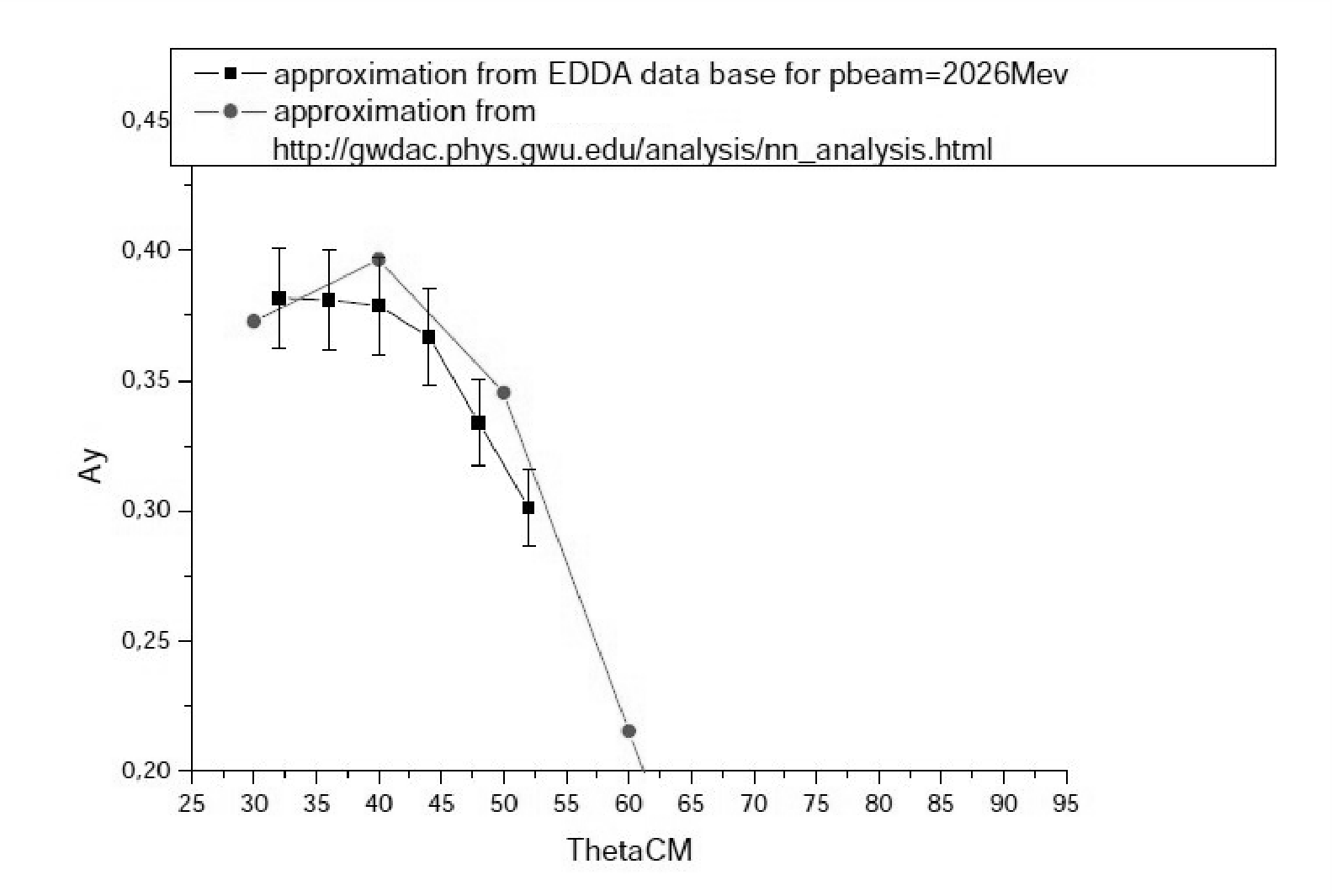}}
  \centerline{{\small $p_{beam}=2026$~MeV}}
 \end{minipage}%
 \begin{minipage}{0.5\textwidth}
  \centerline{\includegraphics[width=\textwidth]{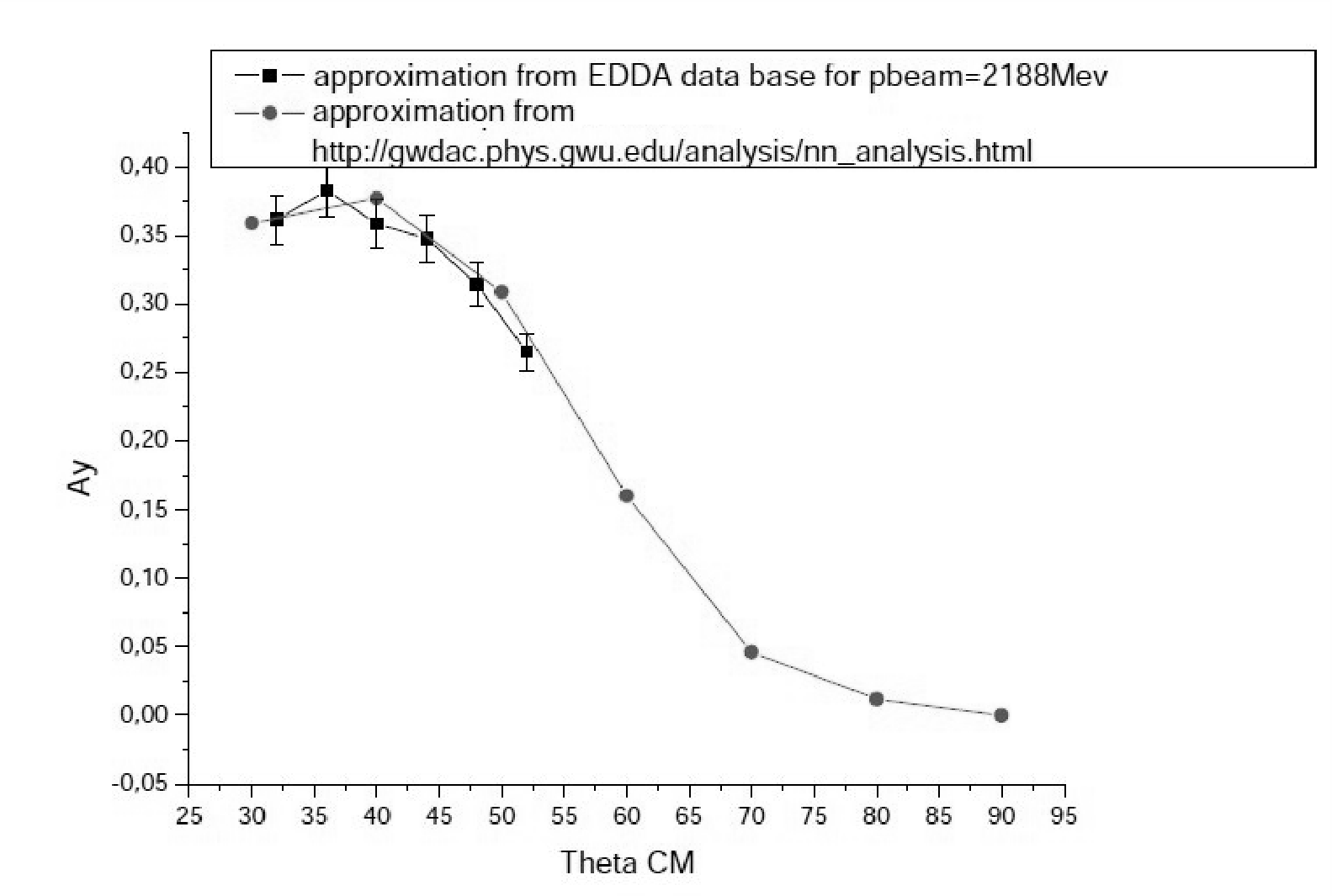}}
  \centerline{{\small $p_{beam}=2188$~MeV}}
 \end{minipage}
 \caption{\label{fig4:dP}Distribution of $A_y$ for $\vec{p}p\to pp$ reaction as a function of the protons
scattering angle, $\theta$, in the centre of mass system. Square points denotes data based on EDDA
experiment~\cite{ex:12} and superimposed line indicates results of Ref.~\cite{ex:13}.}
\end{figure}

\vspace{7.1cm}

\section*{Acknowledgements}
This publication has been supported by COSY-FFE and by the European Commission under the 7th Framework
Programme through the Research Infrastructures action of the Capacities Programme, Call:
FP7-INFRASTRUCTURES-2008-1, Grant Agreement N. 227431 and by the Polish Ministry of Science and Higher
Education under grants No. 2861/B/H03/2010/38 and 0320/B/H03/2011/40.

\coltoctitle{\boldmath Determination of the $\eta$ Mass with the Crystal Ball at MAMI-B}
\authortochead{A.~Nikolaev}
\label{NA}
\import{./NikolaevAlexander/Nikolaev}

\coltoctitle{\boldmath High Precision $\eta$ Meson Mass Determination at COSY-ANKE}
\authortochead{P.~Goslawski}
\label{GP}

\setcounter{chapter}{1}
\setcounter{section}{0}
\maketitle

\noindent Recent measurements on the $\eta$ meson mass performed at different experimental facilities (i.e.
CERN-NA48, COSY-GEM, CESR-CLEO, DA$\Phi$NE-KLOE, MAMI-Crystall Ball) resulted in very precise data but differ
by up to more than eight standard deviations, i.e. $0.5$~MeV/c$^{2}$~\cite{r1}. In order to clarify this
situation a high precision measurement using the ANKE spectrometer at the COoler SYnchrotron COSY has been
realized.

Using the two-body reaction $d\,p \rightarrow \, ^3\text{He}\,\eta$  at low excess energies the $\eta$ mass
can be determined only from pure kinematics by the determination of the production threshold. Therefore,
twelve data points at fixed excess energies in the range of $Q = 1 - 12$~MeV were investigated. The final
state momentum $p_f$ of the $^{3}$He -particles in the Center of Mass System (CMS)
\begin{equation}
\label{eq1}
p_f \left(s\right) = \frac{\sqrt{ \left[s - \left(m_{^{3}\text{He}} + m_{\eta} \right)^2 \right] \cdot
\left[s - \left(m_{^{3}\text{He}} - m_{\eta} \right)^2 \right]    }} {2 \cdot \sqrt{s}} \ ,
\end{equation}
measured with the ANKE spectrometer, is very sensitive on the $\eta$ mass and the total energy $\sqrt{s}$,
where the latter one is completely defined in a fixed target experiment by the masses of the initial particles
and the momentum of the deuteron beam $p_d$: 
\begin{equation}
\label{eq2}
\sqrt{s} = \sqrt{2m_p \sqrt{m_d^2 + p_d^2} + m_d^2 +m_p^2} \ .
\end{equation}
For a precise determination of the production threshold both quantities, the final state momenta of the
$^{3}$He -particles and the corresponding deuteron beam momenta have to be measured with highest accuracy.
Fitting the dependency of the final state momentum $p_f$ on the beam momentum $p_d$ and the $\eta$ mass, $p_f
= p_f(p_d,\, m_{\eta})$, for the twelve points the mass can be extracted as fitting parameter (see
Fig.~\ref{fig3}). \\

\noindent The beam momentum for each fixed excess energy was determined using a method developed at VEPP-2M at
Novosibirsk at an electron-positron machine~\cite{r2} using the spin dynamics of a polarized beam. Here the
spin precession frequency of a relativistic particle is disturbed by an artificial spin resonance induced by a
horizontal rf-magnetic field of a solenoid leading to a depolarization of the polarized accelerator beam. The
depolarizing resonance frequency $f_r$ depends on the kinematical $\gamma$-factor (i.e. the beam momentum $p =
m \sqrt{\gamma^2 - 1}$) and the beam revolution frequency $f_0$ via the resonance condition:
\begin{equation}
f_r = (k + \gamma G) \,  f_0 \, ,
\end{equation}
where $k$ is an integer and $G$ the gyromagnetic anomaly of the beam particle. By measuring these two
frequencies the beam momentum of a polarized beam can be determined with a precision below $\Delta p/p <
10^{-4}$. For the first time this method was used at COSY with a vector polarized deuteron beam and the
momenta in the threshold range of $3.1-3.2$~GeV/c were determined with an accuracy of $\Delta p/p <
6\cdot 10^{-5}$, i.e. with 170~keV/c~\cite{r3}. \\

\noindent The correct final state momenta for the twelve different energies of the $^{3}$He -nuclei of the
reaction $d\,p \rightarrow \, ^3\text{He}\,\eta$ can only be extracted fulfilling two conditions: a clear
identification of the reaction of interest and a precise detector calibration. At ANKE the produced $^{3}$He
-nuclei can be identified by the energy loss and time of flight information. By this, the background
consisting mainly of protons and deuterons of the dp elastic scattering and the deuteron break-up, can be
suppressed effectively. For a two-body reaction at a fixed center of mass energy $\sqrt s$ the final state
momenta in the CMS are distributed on a momentum sphere with constant radius $p_f$, which can be visualized by
plotting the transversal versus the longitudinal reconstructed momentum, as shown in Fig~\ref{fig1}a. 

\begin{figure}[htb]
  \centerline{\includegraphics[width=0.45\linewidth]{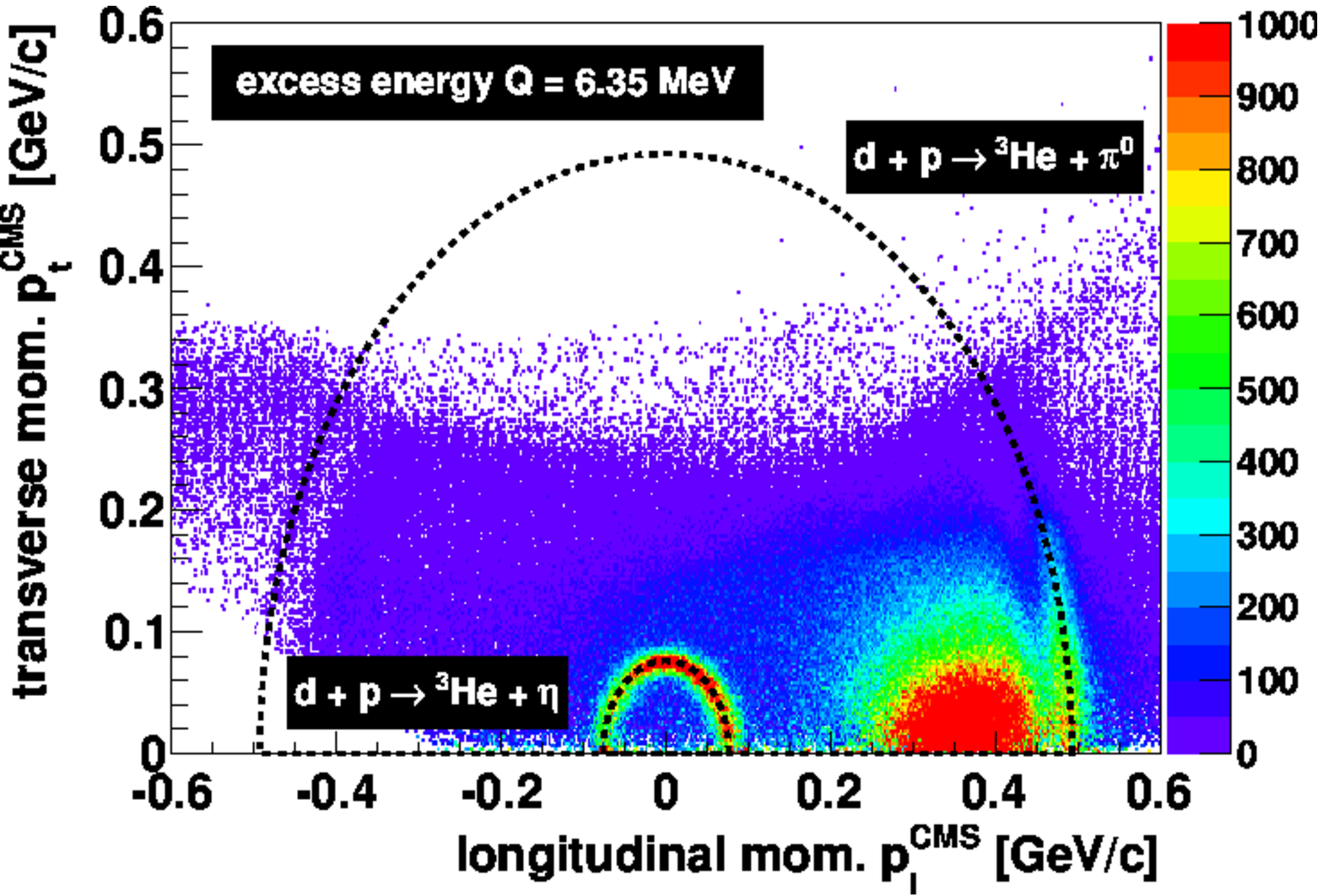}\hfill%
              \includegraphics[width=0.45\linewidth]{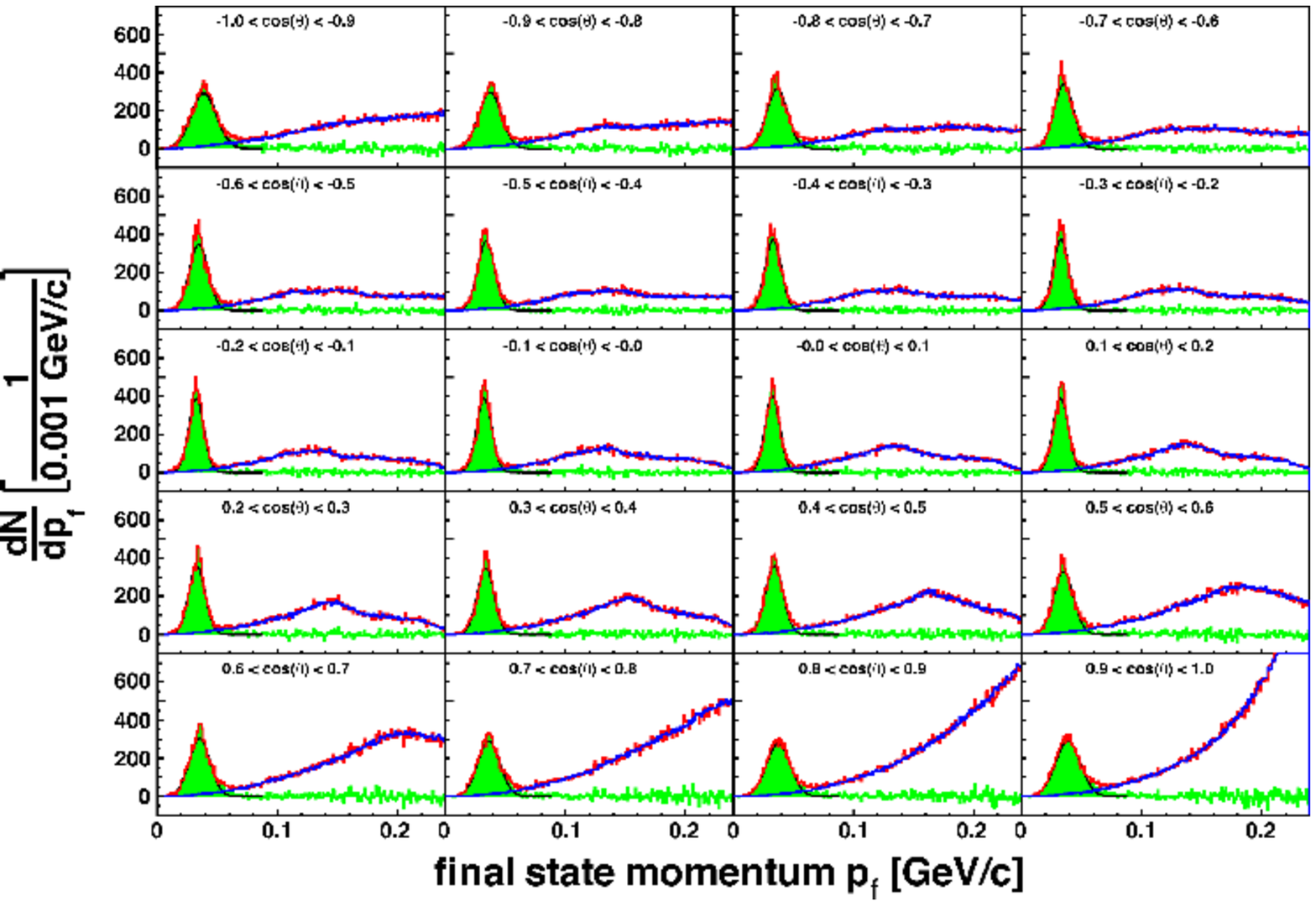}}
  \caption{\label{fig1} {\bf a)} The momentum loci for the $^3\rm{He} \, \eta \; $ and $^3\rm{He} \, \pi$
channels. For the $^3\rm{He} \, \eta \; $ channel ANKE covers near threshold the full solid angle, while for
the $^3\rm{He} \, \pi$ channel only forward scattered $^{3}$He -nuclei are detected. 
  {\bf b)} Final state momentum $p_f = p_f (\cos \vartheta)$ (red) at an excess energy of $Q \approx 1$~MeV,
the background description (blue) and the extracted $^{3}\rm{He} \, \eta$ signal (green).}
\end{figure}

\noindent According to Eq.~\ref{eq1}, one expects a centered momentum locus with a fixed radius $p_f = (p_x^2
+ p_y^2 + p_z^2)^{1/2}$, indicated in Fig.~\ref{fig1}a as dashed line. Using the feature that the ANKE
facility has full geometrical acceptance for the reaction $d\,p \rightarrow \, ^3\text{He}\,\eta$  near
threshold up to $15$~MeV the detector calibration can be verified and improved by studying the
kinematics of this two-body reaction. The main idea to verify the calibration is that the momentum sphere has
to be completely symmetric in $p_x$, $p_y$ and $p_z$ (or $\vartheta$ and $\phi$) so that the final state
momentum $p_f$ should be constant in all directions. By a careful investigation of the momentum dependency on
the cosine of the polar angle $\vartheta$ and the azimuthal angle $\phi$
\begin{eqnarray}
	p_f &=& p_f \left( \cos \vartheta \right) \\
	p_f &=& p_f \left( \phi \right)
\end{eqnarray}
the shape of the momentum sphere or locus can be studied in more detail. Deviations from this symmetric shape
will indicate the need of an improvements of the calibration. Therefore the $^{3}\rm{He} \, \eta$ signal has
to be extracted background free. The background left after cutting on $^{3}$He -nuclei (see Fig.~\ref{fig1}a
originating mainly from the multi pion production can be subtracted by data taken below the $\eta$ threshold
at an excess energy of $Q \approx -5$~MeV, but analyzed as if they were taken above. In~\cite{r4} the
successful applicability of this approach on missing mass spectra is shown, but it is also applicable to final
state momentum spectra as shown in Fig.~\ref{fig1}b for different $\cos \vartheta$ bins for a data point close
to threshold, i.e. Q $\approx$ 1~MeV. Similarly the final state momentum dependency on $\phi$ can be studied.
The background free $^3\rm{He} \, \eta \; $ distribution allows to extract the mean final state momentum
$p_f(\cos \vartheta)$ and $p_f(\phi)$ as shown in Fig.~\ref{fig2}.
\begin{figure}[htb]
  \centering
  \includegraphics[width=0.8\linewidth]{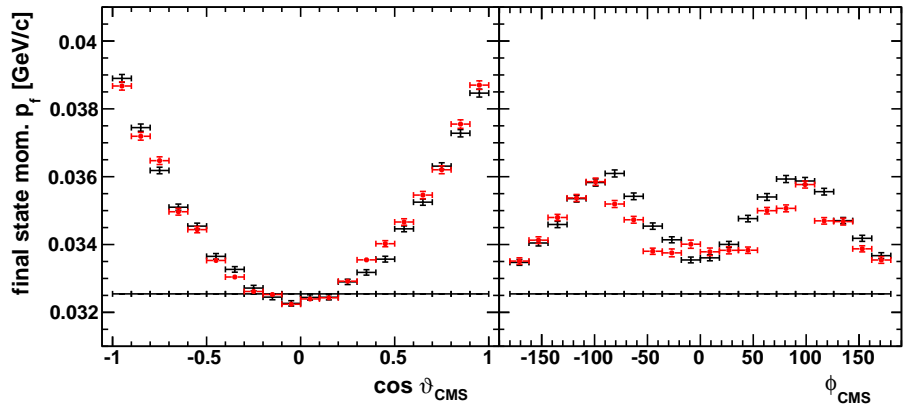}
  \caption{\label{fig2} Final state momentum dependency: $p_f = p_f (\cos \vartheta)$ and $p_f(\phi)$ for
data (red) simulations without momentum smearing (black line) and with momentum smearing (black).}
\end{figure}
In contrast to Monte-Carlo simulations on $p_f(\cos \vartheta)$ and $p_f(\phi)$ without momentum smearing
(black line) the obtained data points indicated as red crosses show a dependency of the final state momentum
in $\cos \vartheta$ and $\phi$. The shape of the momentum sphere is stretched to values $\cos \vartheta
\rightarrow \pm 1$ and shows an oscillation in $\phi$. This behavior is caused by a kinematic effect due to
different momentum resolutions of the ANKE detector for $p_x$, $p_y$ and $p_z$. Assuming that the $p_x$, $p_y$
and $p_z$ distributions are gaussian distributed with different widths~$\sigma$, it is possible to reproduce
the final state momentum dependency on $\cos \vartheta$ and $\phi$ with Monte-Carlo simulations shown as black
crosses.

For the final state momentum determination the plots shown in Fig.~\ref{fig2} are of high importance because
of following three reasons
\begin{enumerate}
 \item[i)] Improvement of the calibration. Asymmetric shapes in $\cos \vartheta$ and $\phi$ can be corrected
by minor changes of the ANKE calibration parameters. These changes are so small that typical calibration
quantities like missing masses of different reactions show no variations, because they are not sensitive
enough.
 \item[ii)] Extraction of the correct momentum resolution in $(p_x, p_y, p_z)$. Assuming gaussian
distributions with different widths we found for ANKE $(\sigma_x, \sigma_y, \sigma_z) = (3.2, 7.8,
16.4)$~MeV/c.
 \item[iii)] Correction of the reconstructed final state momentum. Because of the finite momentum resolution
the extracted momentum (red) are larger compared to the original one (black line). This has to be considered
in the determination of the twelve final state momenta.
\end{enumerate}

\noindent Currently the final state momentum analysis is still in progress but already now the twelve momenta
in the range of $30 - 100$~MeV/c can be determined with a precision below $400$~keV/c. Fitting the $p_f =
p_f(p_d,\, m_{\eta})$ dependency, as shown in Fig.~\ref{fig3} the $\eta$ mass is determined preliminary to 
\begin{equation}
	m_\eta = (547.869 \pm 0.007 \pm 0.040)\textrm{~MeV/c}^2 .
\end{equation}
The accuracy, which will be achieved at ANKE, will be comparable and competitive to the precision achieved at
other recent experiments.
\begin{figure}[htb]
  \centering
  \includegraphics[width=0.6\linewidth]{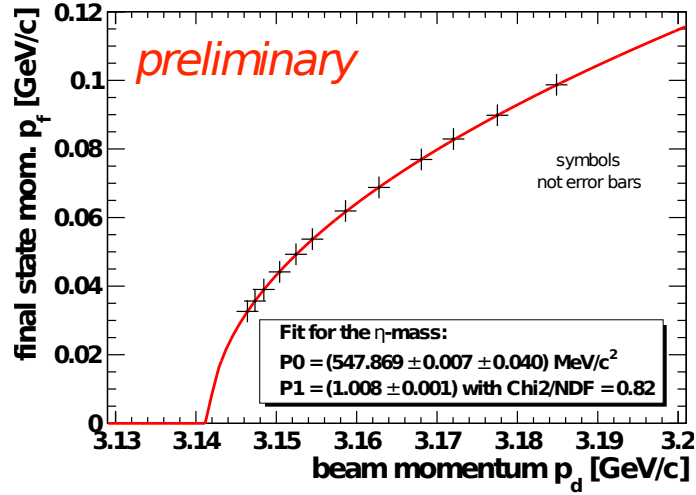}
  \caption{\label{fig3} Determination of the $\eta$ mass by fitting the dependency $p_f = p_f(p_d,\, m_{\eta})$.}
\end{figure}

\vfill This work is supported by FZ J\"ulich FFE.

\coltoctitle{\boldmath Measurement of the $\eta'$ Meson Total Width at the COSY--11 Facility}
\authortochead{E.~Czerwi\'nski}
\label{CE}

\setcounter{chapter}{1}
\setcounter{section}{0}
\maketitle

\Section{COSY--11 detection setup}
\begin{wrapfigure}{r}{0.5\textwidth}
 \begin{center}
    \vspace{-1cm}\includegraphics[width=0.5\textwidth]{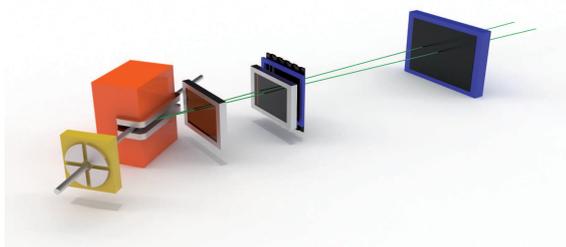}
 \end{center}
 \caption{\label{c113d}COSY--11 detector setup. From left to right: quadrupole (yellow) and dipole (orange)   
          magnets of COSY, two drift chambers (silver) and two scintillator detectors (blue-black).}
\end{wrapfigure}
The reported experiment has been performed in the Research Centre J{\"u}lich at the cooler synchrotron     
COSY~\cite{Maier} by means of the \mbox{COSY--11} detector system~\cite{Brauksiepe,Moskal5} presented in
Figure~\ref{c113d}.

The collision of a proton from the COSY beam with a proton cluster target may cause an $\eta'$ meson creation.
In that case all outgoing nucleons have been registered by the COSY--11 detectors, whereas for the
$\eta^{\prime}$ meson identification the missing mass technique was applied.

The cluster target width was reduced by a factor of 5 compared to the normal operation mode in order to
decrease the systematic uncertanities during the determination of the total width of the $\eta'$ meson. The
dimensions of the target stream were determined using a wire probe installed for this purpose.
\Section{Total width of the $\eta'$ meson}
The mass of unregistered meson was determined via the missing mass technique, while the total width was
directly derived from the mass distributions established at five different beam
momenta~\cite{ErykPRL,ErykPhD}. Based on the sample of more than 2300 reconstructed $pp\to pp\eta'$ events the
determined total width of the $\eta'$ meson amounts to
$\Gamma_{\eta'}=0.226\pm0.017(\textrm{stat.})\pm0.014(\textrm{syst.})$~MeV/c$^2$, which is the most precise
measurement until now~\cite{pdg,Wurzinger,Binnie}. On the plots in Figure~\ref{mmbcgfit} the experimental data
are presented as black points, while the histograms correspond to the sum of the Monte Carlo generated signal
for $\Gamma_{\eta'}=0.226$~MeV/c$^2$ and a normalised second order polynomial obtained as a fit to the
signal-free background region for another energy~\cite{ErykPRL,ErykPhD}. The plot in the bottom right corner
presents the $\chi^2$ distribution as a function of the $\Gamma_{\eta'}$ determined by the comparison of the
missing mass spectra from the measurement with Monte Carlo simulations.
\begin{figure}[!t]
\vspace{-1.0cm}
  \begin{center}
    \includegraphics[width=0.35\textwidth]{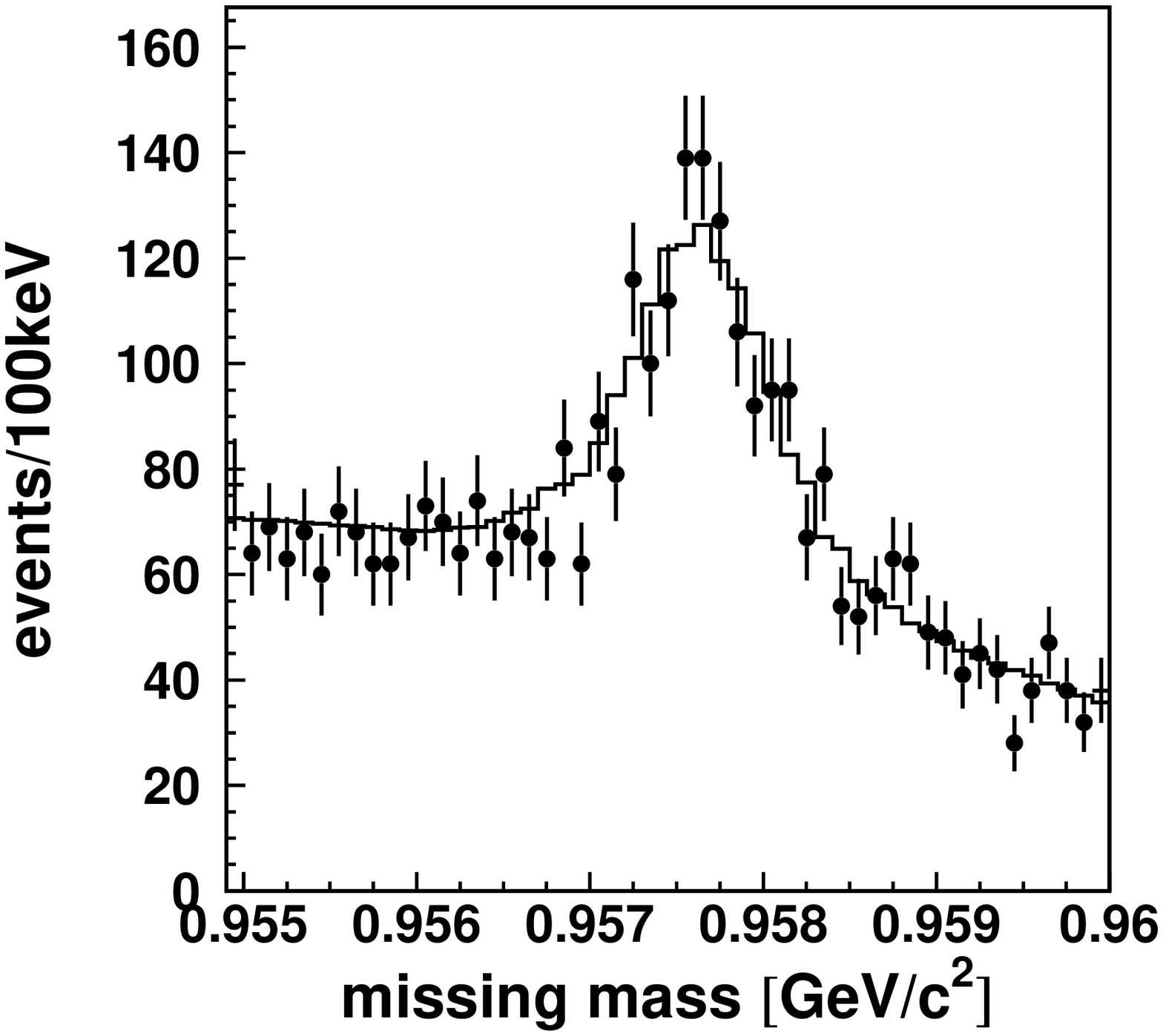}
\vspace{-1.0cm}
    \includegraphics[width=0.35\textwidth]{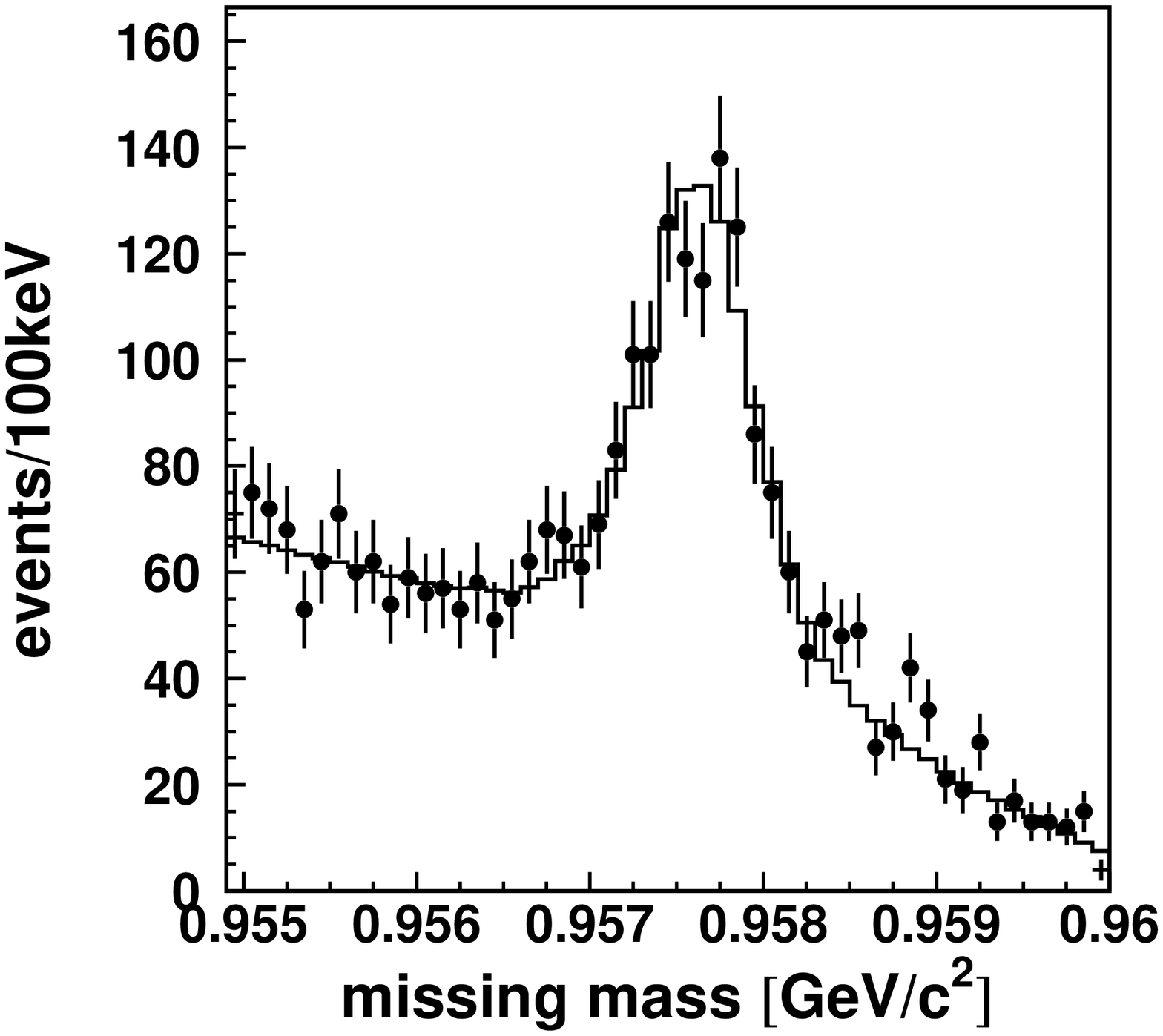}
\vspace{-1.0cm}
    \includegraphics[width=0.35\textwidth]{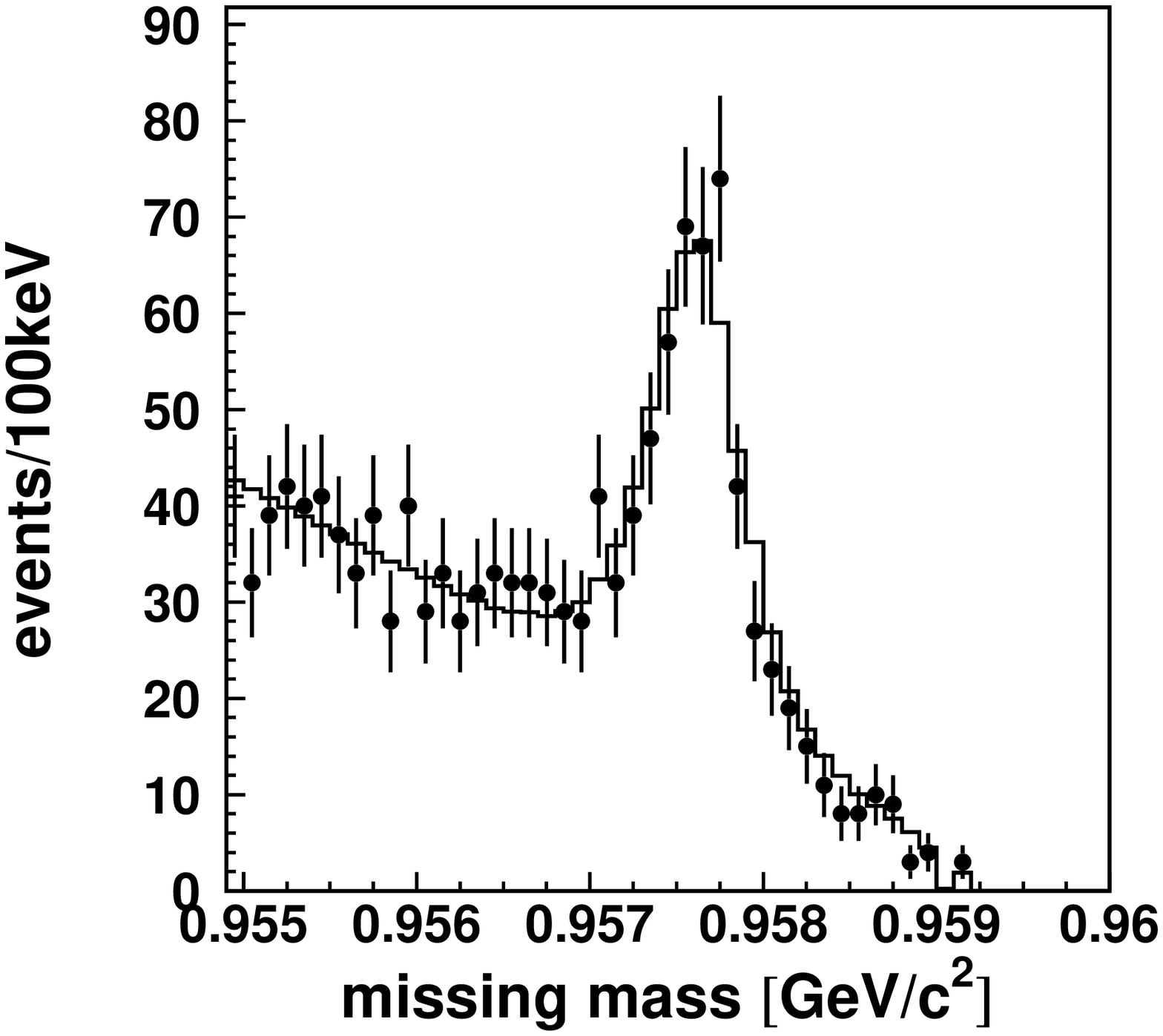}
    \includegraphics[width=0.35\textwidth]{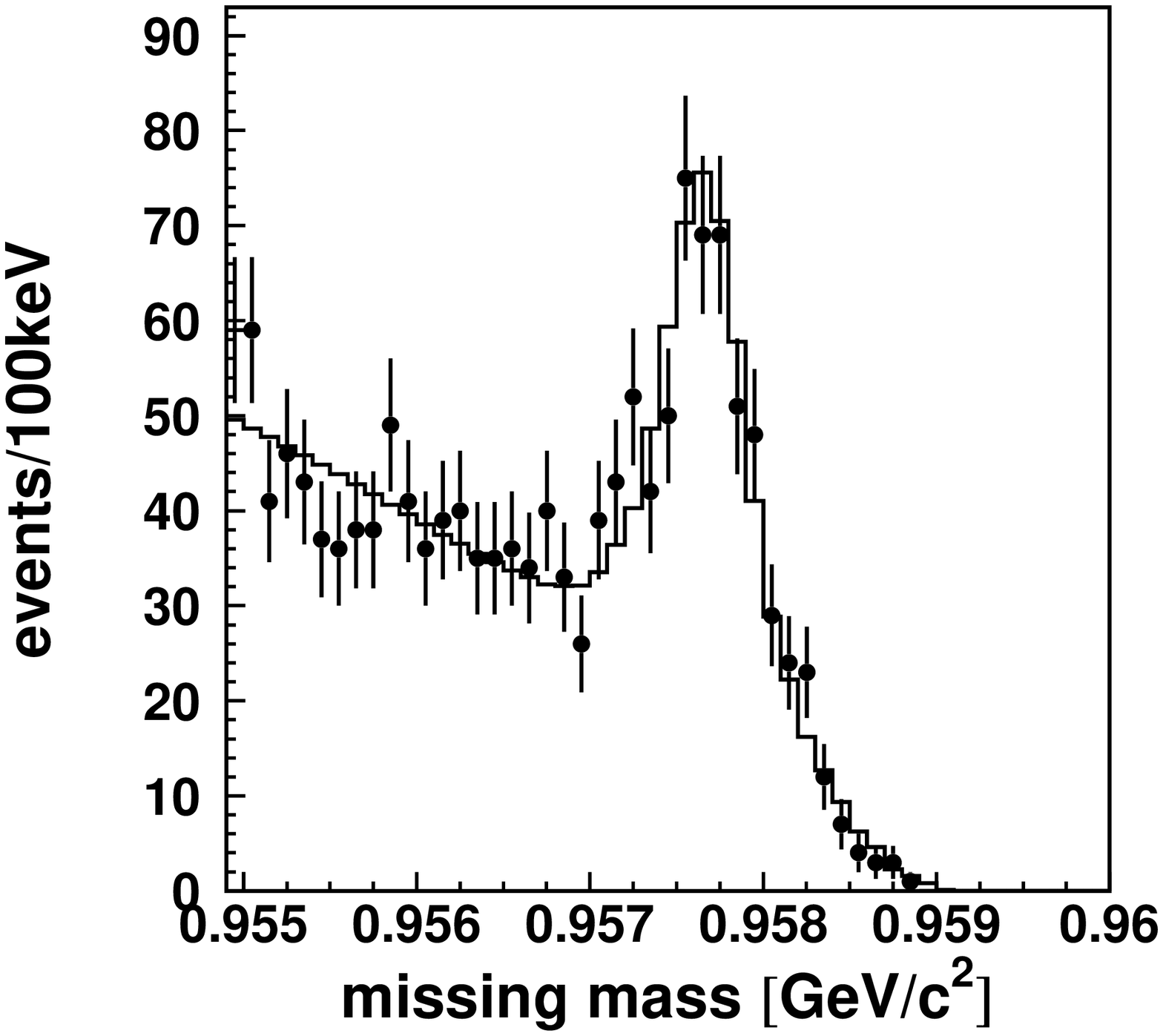}
    \includegraphics[width=0.35\textwidth]{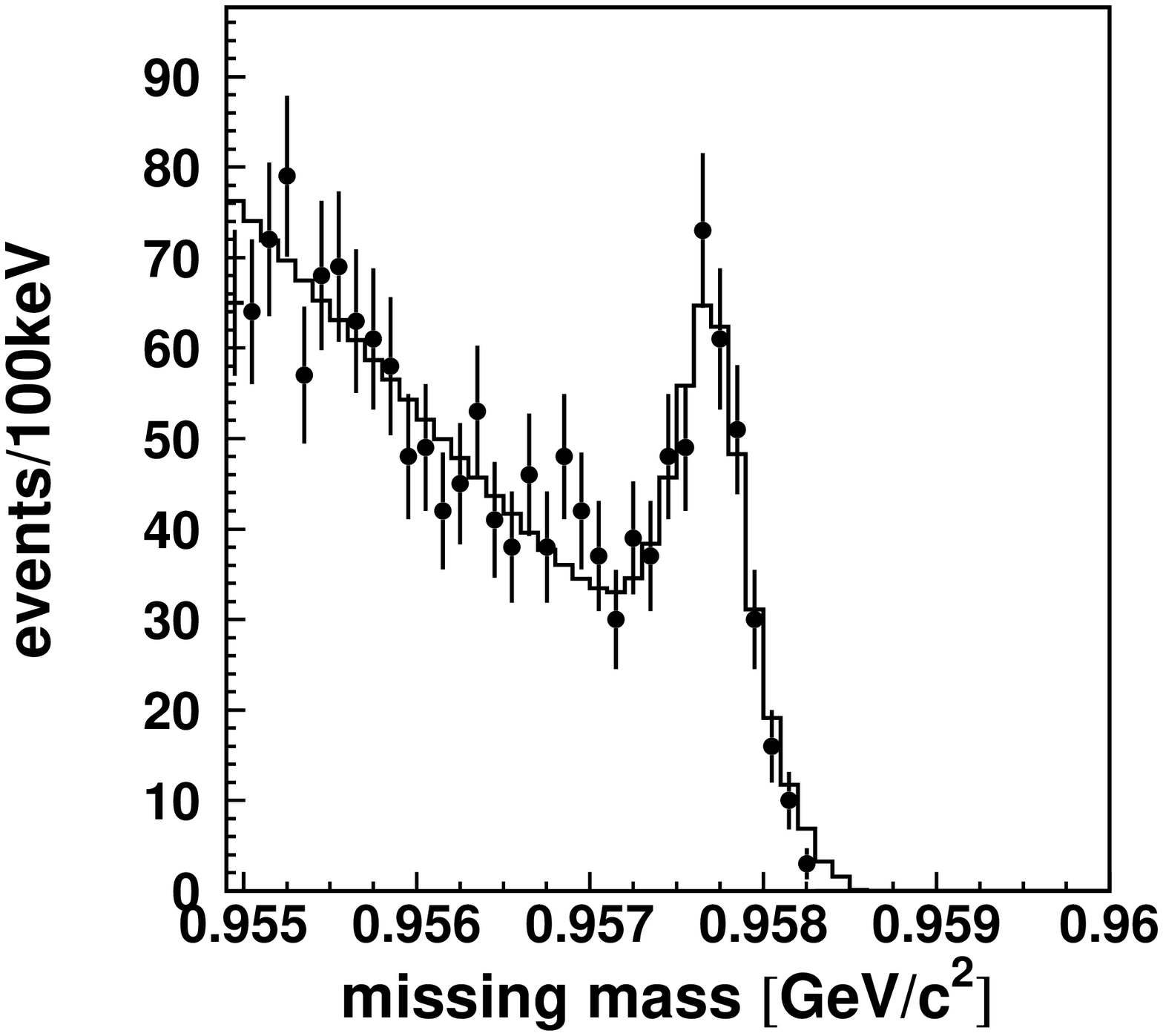}
    \includegraphics[width=0.35\textwidth]{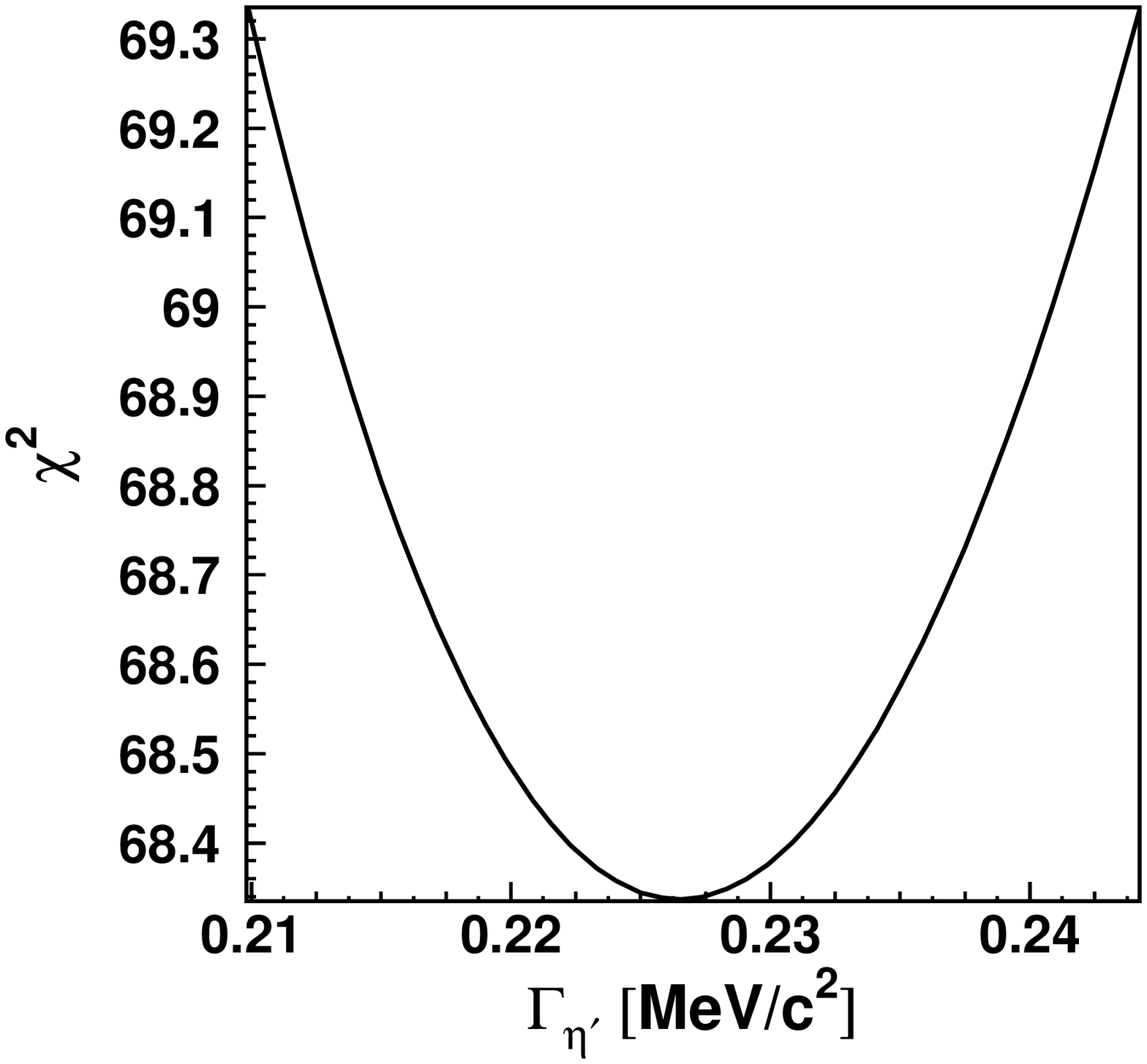}
  \end{center}
\vspace{-1.0cm}
 \caption{\label{mmbcgfit}The missing mass spectra for the $pp\to ppX$ reaction for excess energies in the CM
system equal to 4.8, 2.8, 1.7, 1.4, and 0.8~MeV (from left to right, top to bottom). The $\eta'$ meson signal
is clearly visible. The experimental data are presented as black points, while in each plot the histogram
corresponds to the sum of the Monte Carlo generated signal for $\Gamma_{\eta'}=0.226$~MeV and the shifted and
normalised second order polynomial obtained as a fit to the signal-free background region for another energy.
Bottom right: $\chi^2$ as a function of $\Gamma_{\eta'}$ in the range up to $\chi^2=\chi^2_{min}+1$. The
minimum corresponds to the determined $\Gamma_{\eta'}$ value.}
\end{figure}

\Section{Acknowledgments}
The work was partially supported by the European Commission under the 7$^{\textrm{th}}$ Framework Programme
through the \emph{Research Infrastructures} action of the \emph{Capacities} Programme. Call:
FP7\--INFRASTRUCTURES\--2008\--1, Grant Agreement N. 227431, by the PrimeNet,  by the Polish Mi\-nistry
of Science and Higher Education through grants No. 1202/\-DFG/\-2007/\-03 and 1253/\-B/\-H03/\-2009/\-36, by
the German Research Foundation (DFG) and by the FFE grants from the Research Center J{\"u}lich.

\coltoctitle{Two-Pion Production in Isoscalar NN Collisions:\\ ABC Effect and Resonance} 
\authortochead{M.~Bashkanov}
\label{BaM}

\setcounter{chapter}{1}
\setcounter{section}{0}
\maketitle

\vspace{3ex}%
\Section{Introduction}
More than fifty years ago Abashian, Booth and Crowe~\cite{abc} observed in the reaction $pd \to ^3$HeX that
the missing mass of the detected $^3$He ejectile corresponding kinematically to the production of a pion pair,
{\it i.e.} X = $\pi\pi$, contains a peculiar enhancement starting right at threshold at low $\pi\pi$-masses.
In subsequent bubble-chamber~\cite{bar,abd} and single-arm magnetic spectrometer
measurements~\cite{hom,hal,ban,ban1,plo,col,wur,cod} this enhancement was found in fusion reactions to $d$,
$^3$He and $^4$He, where an isoscalar pion pair was produced, however, not in the  fusion reaction leading to
tritium, where an isovector pion pair was produced. From these results it was concluded that this effect,
which later-on was named ABC effect after the initials of the first authors, only appears in the double-pionic
fusion in combination with the production of an isoscalar pion pair.

Since these experiments were undertaken at beam energies, which energetically allow the mutual excitation of
two colliding nucleons into their first excited state, the $\Delta(1232)P_{33}$, theoretical attempts were
undertaken to understand the ABC effect by the excitation of a $\Delta\Delta$ system via $t$-channel meson
exchange or variations of it~\cite{ris,barn,anj,gar,mos,alv}. These calculations, indeed, predicted a low-mass
enhancement, however, also a high-mass enhancement. Surprisingly enough, the inclusive single-arm measurements
seemed to support such a high-mass enhancement. And it was only recently that the first exclusive and
kinematically complete measurements of solid statistics~\cite{MB,OK,SK,AP,fzj,prl2011} over practically the
full phase space revealed that such a high-mass enhancement is not present. 

\vspace{3ex}%
\Section{Recent Experiments}
In order to understand the two-pion production process in $pp, pd$ and $dd$ collisions in more detail,
systematic experimental studies were initiated at CELSIUS and continued later-on at COSY. These exclusive and
kinematically complete high-statistics measurements have been carried out at the detector installations
PROMICE/WASA~\cite{Johanson200275,PhysRevLett.88.192301,PhysRevC.67.052202},
CELSIUS/WASA~\cite{barg,springerlink:10.1140/epja/i2008-10569-6,OK,Skorodko200930,FK,Skorodko2011115,nnpipi},
COSY-TOF~\cite{springerlink:10.1140/epja/i2008-10637-y,evd}, WASA-at-COSY~\cite{WASApro2004,prl2011,tt} and
COSY-ANKE~\cite{PhysRevLett.102.192301} covering the energy range from the two-pion production threshold up to
$\sqrt s$~=~2.5~GeV (corresponding to a nucleon beam energy of 1.4~GeV). The bulk of these experiments has
been performed with the WASA detector setup, which has an angular coverage of nearly $4\pi$ and features
windowless hydrogen and deuterium pellet target systems~\cite{barg,WASApro2004} as well as a very reliable
particle identification for $\gamma, \pi, p, n, d, ^3$He and $^4$He. Measurements at WASA generally are
performed with detection of all ejectiles of an event (with the exception of spectator nucleons) leading thus
to kinematic fits with overconstraints in the data analysis. For the investigation of the $pn$ initiated
reactions the quasifree $pd$ process with a spectator proton resulting from the target deuteron is utilized.
By exploiting the Fermi motion of the active nucleon in the deuteron target that way the energy dependence of
$pn$ and $pp$ initiated reactions can be measured over a range of more than 100 MeV by use of a single
beam energy.

\begin{figure}[htb]
 \centerline{\includegraphics[width=0.99\columnwidth]{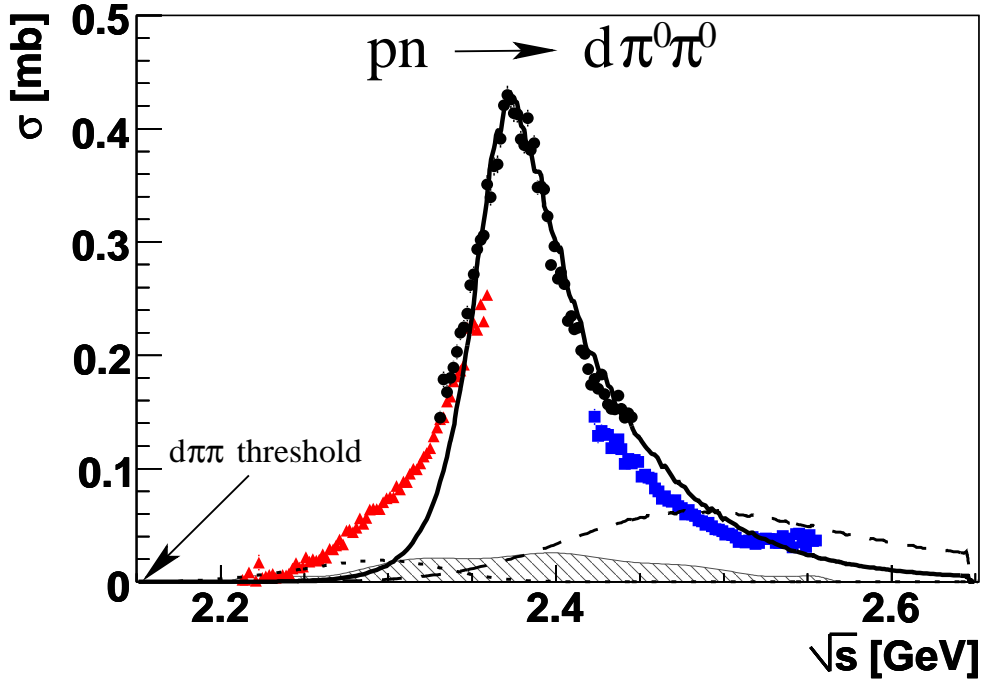}}
 \caption{Total cross sections obtained from this experiment on $pd\to d\pi^0\pi^0 + p_{spectator}$ for the
beam energies $T_p$~=~1.0~GeV (triangles), 1.2~GeV (dots) and 1.4 GeV (squares) normalized independently.
Shown are the total cross section data after acceptance, efficiency and Fermi motion corrections. The hatched
area indicates systematic uncertainties. The drawn lines represent the expected cross sections for the Roper
excitation process (dotted) and the $t$-channel $\Delta\Delta$ contribution (dashed) as well as a calculation
for a $s$-channel resonance with m~=~2.37~GeV and $\Gamma$~=~68~MeV (solid). From Ref.~\cite{prl2011}}
\end{figure}

\Section{Results}
From the overwhelming amount of high-quality data for total (integral) and differential cross sections in the
various two-pion production channels the following features emerge for the energy region from threshold up to
$\sqrt s$~=~2.6~GeV:
\begin {itemize}
 \item The {\bf isovector} two-pion production~\cite{TS} initiated by $pp$ collisions can be quantitatively
understood by $t$-channel meson exchange with 
\begin{itemize}
 \item excitation of one of the colliding nucleons into the Roper resonance,   $N^*(1440)P_{11}$ and its
subsequent decay into the $N\pi\pi$ system either   directly ($N\sigma$ channel) or via the $\Delta\pi$
intermediate state,
 \item mutual excitation of the two colliding nucleons into the $\Delta$ state each with subsequent decay
into $N\pi$, or
 \item excitation of a higher lying $\Delta$ state, favorably the $\Delta(1600)P_{33}$ state with subsequent
decay into $N\pi\pi$ via $\Delta\pi$. Such an excitation, however, plays a dominant role only in the 
$nn\pi^+\pi^+$ channel possessing an isotensor pion pair.
\end{itemize}

Whereas the Roper process dominates at lower energies close to threshold, the $\Delta\Delta$ process dominates
at higher energies close to $\sqrt s = 2 m_{\Delta}$ = 2.46 GeV. In particular the isovector double-pionic
fusion reaction $pp \to d\pi^+\pi^0$ is well described in its energy dependence up to at least 3 GeV by the
latter process~\cite{FK}.  

\item The {\bf isoscalar} two-pion production reaction $pn \to d\pi^0\pi^0$ -- measured as quasifree process
in $pd$ collisions~\cite{OK,prl2011} -- exhibits a pronounced low-mass enhancement in the $\pi\pi$-invariant
mass distribution as expected in view of the ABC history on double-pionic fusion reactions. However, what was
not expected is a strict correlation of this ABC effect with the appearance of a resonance-like structure in
the total cross section -- see Fig.~1 -- with m~=~2.37~GeV and $\Gamma$~=~70 MeV, {\i.e.} roughly 100~MeV
below $2 m_{\Delta}$ and three times narrower than expected from the $t$-channel $\Delta\Delta$ process. From
the angular distributions the spin-parity of this structure is derived with the result that $I(J^P) = 0(3^+)$.

Note that this reaction constitutes the only purely isoscalar two-pion production channel in $NN$ collisions.
Hence exotic processes in the isoscalar channel are likely to be seen best in this particular reaction without
being diluted by isovector contributions.
\item The {\bf isoscalar} double-pionic fusion reaction $dd \to ^4$He$\pi\pi$ also exhibits the ABC effect in
strict correlation with a resonance-like structure in the total cross section at the same energy above the
two-pion threshold~\cite{AP,fzj}. The increased width of this structure can be explained by the Fermi motion
of the active  {\bf isoscalar} $pn$ pair in target, projectile and fused nucleus. The energy dependence of the
reaction $pd \to ^3$He$\pi\pi$ is under investigation. Preliminary results~\cite{EP} indicate a similar trend
as observed for the $^4$He. If true than this {\bf isoscalar} $pn$ resonance structure is obviously strong
enough to survive even in helium nuclei. 
\end {itemize}

\Section{Acknowledgments}
We acknowledge valuable discussions with L. Alvarez-Ruso, C. Hanhart, E. Oset, A. Sibirtsev and C. Wilkin on
this issue. This work has been supported by BMBF(06TU9193), Forschungszentrum J\"ulich (COSY-FFE) and DFG
(Europ. Grad. School 683).

\addtocontents{toc}{\protect\pagebreak}
\coltoctitle{\boldmath ABC Effect in Double-Pionic Fusion to $^3$He} 
\authortochead{E.~Perez~del~Rio}
\label{PRE}

\setcounter{chapter}{1}
\setcounter{section}{0}
\maketitle

\vspace{-3ex}\Section{Introduction}
The ABC effect named after Abashian, Booth and Crowe~\cite{abc} denotes a low-mass enhancement in the
$\pi\pi$-invariant mass distribution resulting from two-pion production processes. More than fifty years ago
Abashian, Booth and Crowe were the first to observe this enhancement in the reaction $pd \to ^3$HeX by
measuring the momentum spectrum of the detected $^3$He ejectile. This spectrum translates into a
$^3$He-missing mass spectrum, which corresponds to the $\pi\pi$-invariant mass spectrum for X = $\pi\pi$. In
subsequent bubble-chamber~\cite{bar,abd} and single-arm magnetic spectrometer
measurements~\cite{hom,hal,ban,ban1,plo,col,wur,cod} this enhancement was observed in double-pionic fusion
reactions leading to $d$, $^3$He and $^4$He, if an isoscalar pion pair was produced. However, such an
enhancement was not seen in fusion reactions leading to deuteron and triton, if an isovector pion pair was
produced. These results led to the conclusion that this effect only appears in the double-pionic fusion in
combination with the production of an isoscalar pion pair.

Recent exclusive and kinematically complete measurements~\cite{mb,OK,prl2011} of the $pn \to d\pi^0\pi^0$
reaction from threshold up to $\sqrt s$~=~2.5~GeV revealed a strict correlation between the appearance of the
low-mass enhancement in the $\pi\pi$-invariant mass spectrum (ABC effect) and a narrow resonance-like
structure in the total cross section of this reaction. This surprising structure peaks at m~=~2.37 GeV, {\it
i.e.} roughly 100~MeV below $2 m_{\Delta}$, where the conventional excitation of a \mbox{$\Delta\Delta$
system} peaks. Early attempts to theoretically understand the ABC phenomenon favored exactly this conventional
$t$-channel process for its understanding~\cite{ris,barn,anj,gar,mos,alv}. The recent high-statistics data do
not only show that the reaction peaks far below the conventional  $\Delta\Delta$ process, but that also the
narrow width $\Gamma$~=~70~MeV of the observed resonance-like structure is three times less than expected from
the conventional process. 

In fact, all $pp$-induced, {\it i.e.} {\bf isovector} two-pion production processes including the
double-pionic fusion reaction $pp \to d\pi^+\pi^0$~\cite{FK} show neither a pronounced ABC effect nor a narrow
resonance-like structure in their total cross sections~\cite{TS}. They rather are well understood by
conventional $t$-channel nucleon excitations. From this follows that any unconventional process in connection
with the ABC effect has to be associated not only with the {\bf isoscalar} $\pi\pi$ system but also with the
{\bf isoscalar} $NN$ system. This hypothesis is supported by the finding that also in the double-pionic fusion
to $^4$He the appearance of the ABC effect is strictly connected with a resonance-like structure in the total
cross section~\cite{AP,fzj}. 

\vspace{-1ex}\Section{New Experiments and Results}
Since the only other double-pionic fusion reaction exhibiting an ABC effect is the one leading to $^3$He, this
reaction is ideally suited to test this hypothesis. Hence corresponding exclusive and kinematically complete
measurements have been performed at the WASA-at-COSY facility~\cite{WASApro2004}. These measurements covering
the energy range from threshold up to the $\Delta\Delta$ region have been carried via the reactions $pd \to
^3$He$\pi^0\pi^0$ at $T_p$~=~1.0~GeV and via the quasifree process $dd \to ^3$He$\pi^0\pi^0 + n_{spectator}$,
which covers a wide range of energies due to the nucleons' Fermi motion in the target. The experiments are
complemented by the recent CELSIUS/WASA measurement at $T_p$ = 0.875~GeV~\cite{MB}. The data analysis is
still in progress. As a preliminary result we show the energy dependence of the total cross section in Fig.1,
where also the results from previous experiments at CELSIUS~\cite{MB,Anders} and COSY~\cite{Momo} are plotted.
Again we find a resonance-like behavior of the total cross section connected with the appearance of the ABC
effect in the $\pi\pi$ -invariant mass spectrum.

\begin{figure}[htb]
 \centerline{\includegraphics[width=0.99\columnwidth]{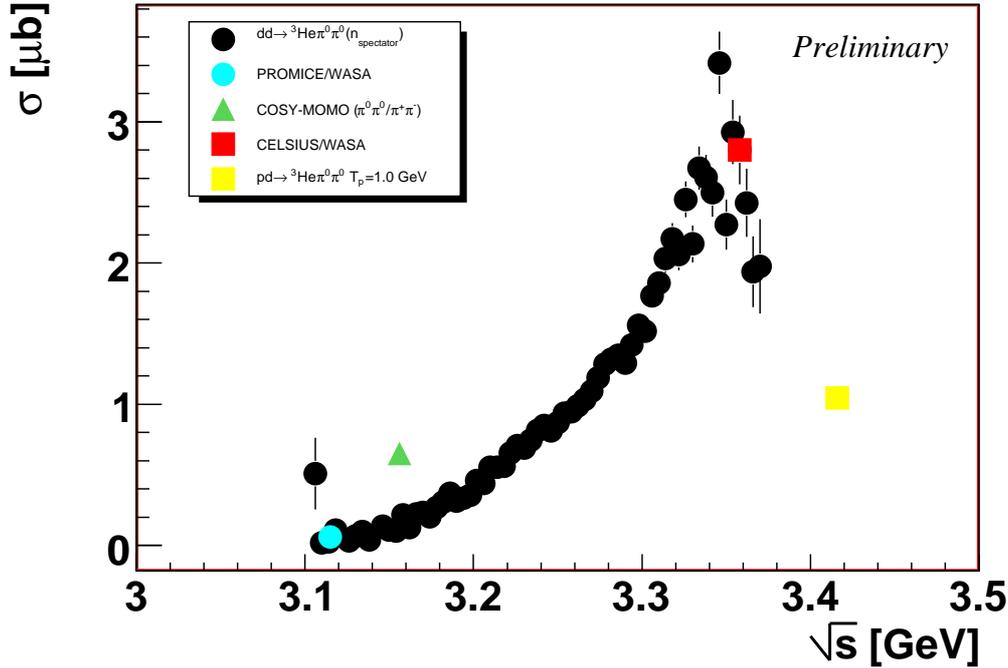}}
 \caption{Preliminary results for the total cross section for the double-pionic fusion to $^3$He obtained from
this experiment on \mbox{$pd \to ^3$He$\pi^0\pi^0$} at $T_p$~=~1.0~GeV (yellow square) and \mbox{$dd \to
^3$He$\pi^0\pi^0 + n_{spectator}$} for the beam energy $T_d$~=~1.4~GeV (black dots) in comparison with
previous results from PROMICE/WASA~\cite{Anders}, CELSIUS/WASA~\cite{MB} and COSY-MOMO~\cite{Momo}.}
\end{figure}

\Section{Acknowledgments}
We acknowledge valuable discussions with L. Alvarez-Ruso, C. Hanhart, E. Oset, A. Sibirtsev and C. Wilkin on
this issue. This work has been supported by BMBF(06TU9193), Forschungszentrum J\"ulich (COSY-FFE) and DFG
(Europ. Grad. School 683)

\coltoctitle{Two-Pion Production in Proton-Proton Collisions} 
\authortochead{T.~Skorodko}
\label{ST}

\setcounter{chapter}{1}
\setcounter{section}{0}
\maketitle

\Section{Introduction}
Two-pion production in $NN$ collisions gives access to the investigation of nucleon excitations in course of
the production process. In particular, it provides the unique possibility to study the mutual excitation of
the two colliding nucleons into one of their excited states. The simplest such process is the excitation of a
$\Delta\Delta$ system by $t$-channel meson exchange in the collision process.

In case that the two-pion production process leads to a fused nucleus in the final state the so-called ABC
effect  -- a peculiar low-mass enhancement in the $\pi\pi$-invariant mass spectrum  -- shows up, if the pion
pair is of isoscalar nature~\cite{abc,bar,abd,hom,hal,ban,ban1,plo,col,wur,cod,MB,OK,SK,AP,EP}. Recently it
has been demonstrated~\cite{prl2011,mb} that the ABC effect in the $pn \to d\pi^0\pi^0$ reaction is strictly
correlated with the appearance of a pronounced resonance-like structure in the total cross section. This
structure with maximum at $\sqrt s$~=~2.37~GeV peaks roughly 100~MeV below the maximum of the conventional
$t$-channel $\Delta\Delta$ process. Moreover it has a width of only 70~MeV, {\it i.e.} about three times
less than the conventional process in its energy dependence.  

In a recent work is has been claimed that a kind of ABC effect is also present in the $pp \to pp\pi^0\pi^0$
reaction, when the two emitted protons have very small relative momenta representing a kind of quasi-bound
$^2$He system~\cite{Dymov}. 

In order to clarify the situation and to contribute to a detailed understanding of the two-pion production
process, we have studied all different two-pion production channels in $pp$ collisions constituting the
$isovector$ part of the $NN$ and $NN\pi\pi$ systems.

\Section{Experiments and Results}
In order to complement other data sets~\cite{Dymov,Johanson200275,PhysRevLett.88.192301,PhysRevC.67.052202,
springerlink:10.1140/epja/i2008-10637-y,evd,tt} exclusive and kinematically complete measurements of the
reactions $pp \to pp\pi^0\pi^0$~\cite{springerlink:10.1140/epja/i2008-10569-6,Skorodko200930,Skorodko2011115},
$pp \to pp\pi^+\pi^-$, $pp \to pn\pi^+\pi^0$~\cite{Skorodko200930}, $pp \to nn\pi^+\pi^+$~\cite{nnpipi} and
$pp \to d \pi^+\pi^0$~\cite{FK} were carried out at the CELSIUS storage ring using the 4$\pi$ WASA detector
setup including the pellet target system~\cite{barg}. The $pp \to pp \pi^0\pi^0$ and $pp \to pp \pi^+\pi^-$
reactions were evaluated from close to threshold up to $T_p$~=~1.36~GeV, the other channels only at
$T_p$~=~1.1~GeV. 
 
Forward going charged pions, protons, neutrons and deuterons were detected in the forward detector and
identified by the $\Delta$E-E technique. Charged pions as well as gammas from $\pi^0$ decay have been detected
in the central part of WASA detector, which contains a mini drift chamber surrounded by thin superconducting
magnet and the electromagnetic calorimeter. With the exception of neutrons the four-momenta of all emitted
particles have been measured giving rise to 1-6 overconstraints for the subsequent kinematical fit. 

\Subsection{\boldmath$(\pi\pi)_{I=0}$ : $pp \to pp \pi^0\pi^0$}
In Fig. 1, left, the total reaction cross section is shown. It keeps rising from threshold up to $T_p\approx$
1~GeV, where the total cross section levels off and proceeds only slowly rising until 1.2~GeV. Thereafter it
continues steeply rising. Whereas  the low-energy region is due to the Roper
resonance~\cite{springerlink:10.1140/epja/i2008-10569-6}, the renewed rise at higher energies can be
associated  with the dominance of the $\Delta\Delta$ excitation. In Fig. 1, right, the $\pi^0\pi^0$-invariant
mass spectrum is shown for $T_p$~=~1.2~GeV, the energy which is close to the maximum cross section in the $pn
\to d\pi^0\pi^0$ reaction. In fact we also observe a slight low-mass enhancement, however, in our case it is
well accounted for by the conventional calculations. The solid lines in Fig.1 show calculations based on the
Valencia model~\cite{ex:7}, where Roper and $\Delta\Delta$ processes have been
modified~\cite{Skorodko2011115}. 

\begin{figure}[htb]
 \centerline{\includegraphics[width=0.49\columnwidth]{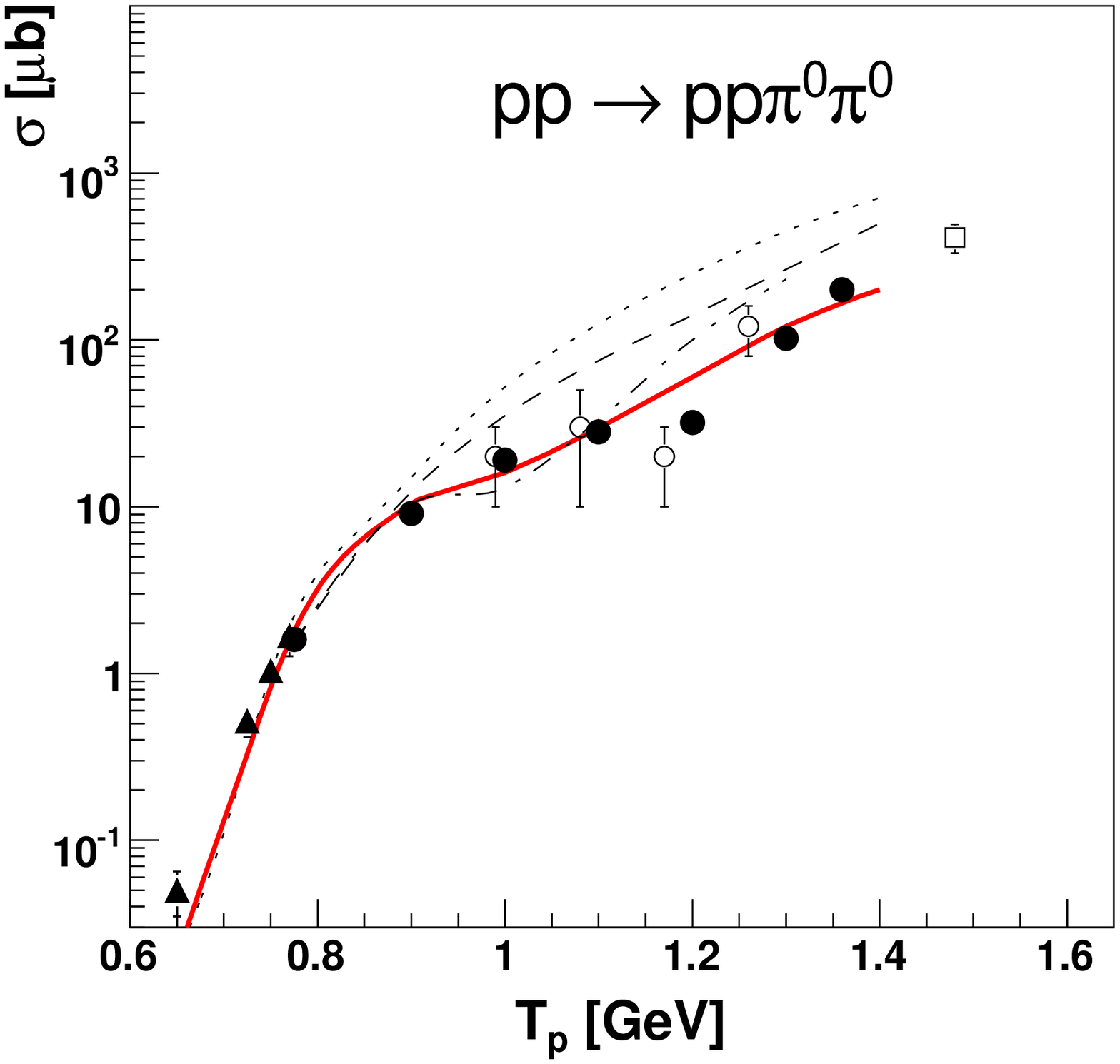}\hfill%
             \includegraphics[width=0.49\columnwidth]{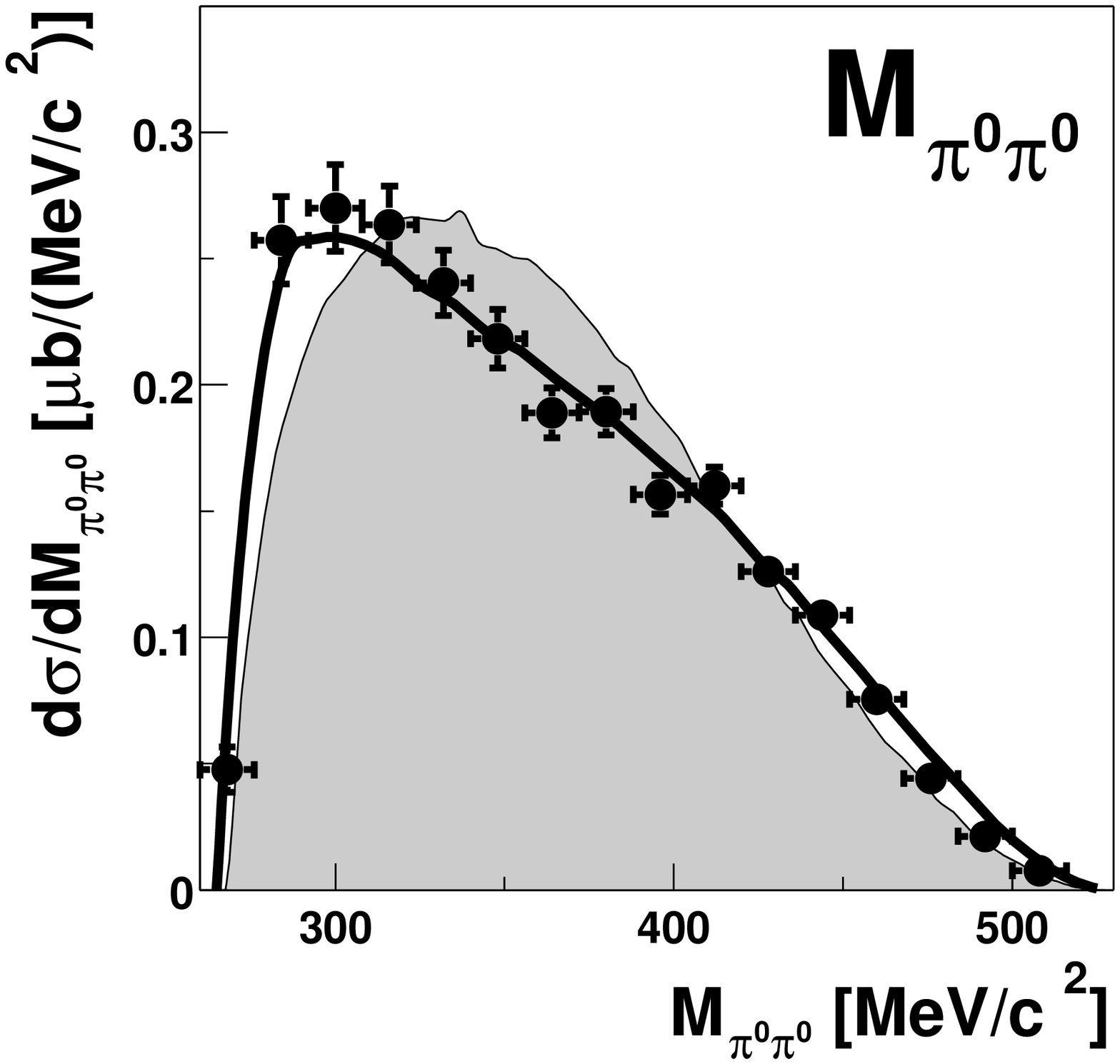}}
 \caption{Energy dependence of the total cross section (left) and $\pi^0\pi^0$-invariant mass spectrum at
$T_p$~=~1.2~GeV (right) of the $\pi^0\pi^0$ reaction. Solid lines show calculations with the modified   
Valencia model, see Ref.~\cite{Skorodko2011115}}
\end{figure}

\Subsection{\boldmath$(\pi\pi)_{I=1}$ : $pp \to d \pi^+\pi^0$}
In the $\pi^+\pi^0$-invariant mass spectrum of this fusion reaction no ABC effect is observed~\cite{FK}. Due
to Bose symmetry the isovector $\pi^+\pi^0$ system must be in relative p-wave. Both differential distributions
and the energy dependence of the total cross section are well described by the conventional $\Delta\Delta$
process. Note that the Roper process is strongly suppressed in this isovector $\pi\pi$ channel even at
energies close to threshold. 

\Subsection{\boldmath$(\pi\pi)_{I=2}$ : $pp \to nn \pi^+\pi^+$}
This isotensor $\pi\pi$ channel is special, since single baryon excitations can only contribute, if they
originate from the excitation of a higher-lying $\Delta$ state. Hence it was expected that the $\Delta\Delta$
process provides the main contributions to this channel. However, as demonstrated in Ref.~\cite{nnpipi}, a
large contribution has to come also from the excitation of a higher-lying $\Delta$ state, favorably the
$\Delta$(1600).

\vspace{3ex}%
\Section{Conclusions}
Summarizing we find all data for the two-pion production initiated by {\bf isovector} $pp$ collisions to be
well accounted for by conventional $t$-channel processes involving single or double nucleon excitations. This
includes also double-pionic fusion processes to d and quasi-bound $^2$He. This situation is fundamentally
different from that in the two-pion production initiated by {\bf isoscalar}  $NN$ collisions, where the ABC
effect has been shown to be correlated with a narrow resonance-like structure in the total cross section.

\vspace{3ex}%
\Section{Acknowledgments}
We acknowledge valuable discussions with L. Alvarez-Ruso, C. Hanhart, E. Oset, A. Sibirtsev and C. Wilkin on
this issue. This work has been supported by BMBF(06TU9193), Forschungszentrum J\"ulich (COSY-FFE) and DFG
(Europ. Grad. School 683).

\coltoctitle{Overview on One and Double Pion Production in pp Collisions in HADES}
\authortochead{M.~Gumberidze}
\label{GM}

\setcounter{chapter}{1}
\setcounter{section}{0}
\maketitle

\vspace{4ex}\Section{Introduction}
The High Acceptance Di-Electron Spectrometer (HADES) installed at GSI Helmholtzzentrum f\"ur
Schwerionenforschung in Darmstadt was designed to investigate dielectron production in heavy-ion collisions
in the 1-2~AGeV beam kinetic energy range. The main goal of the HADES experiment is to study the properties of
hadrons inside the hot and dense nuclear medium via their dielectron decays~\cite{ex:1,ex:2,ex:3}.  

One specific issue of heavy-ion reactions in that regime is the important role played by the baryonic
resonances, which propagate and regenerate due to the long life-time of the dense hadronic matter phase. The
$\Delta$(1232) resonance is the most copiously produced but with increasing incident energy, higher lying
resonances also contribute to pion production. A detailed description of the resonance excitation and coupling
to the pseudoscalar and vector mesons is important for the interpretation of the dilepton spectra measured by
HADES. Baryonic resonances are indeed important sources of dileptons through two mechanisms: their Dalitz
decays (e.g. \mbox{$\Delta /\ N^{*} \rightarrow N e^{+}e^{-}$}) and the mesonic decay with subsequent
dielectron production. 

Simultaneous measurement of one and two-pion production channels can bring information on the baryonic
resonances excitation. Two-pion production, in particular, is an important subject since it connects $\pi \pi$
dynamics with baryon and baryon-baryon degrees of freedom. In our energy regime, $\Delta \Delta$ excitation is
the leading process~\cite{ex:8,ex:9}. We report on one- and two-pion production in pp and np reactions
investigated with HADES in exclusive measurements for the beam kinetic energies 1.25 and 2.2 GeV. 

\vspace{4ex}\Section{HADES experiment}
The HADES detector~\cite{ex:4}, as shown in Fig.~\ref{hades}, consists of 6 identical sectors covering the
full azimuthal range and polar angles from 15$^{\circ}$ to 85$^{\circ}$ with respect to the beam direction.
Each sector contains: A Ring Imaging CHerenkov (RICH) detector used for electron identification; two sets of
Mini-Drift Chambers (MDC) with 4 modules per sector placed in front and behind the magnetic field to determine
momenta of charged particles; the Time Of Flight detectors (TOF/TOFINO) and the Pre-Shower detector improving
the electron identification. For reaction time measurement in heavy-ion reactions, a diamond START detector is
located in front of the target. An (e$^{+}$,e$^{-}$) invariant mass resolution at the $\omega$ peak of
$\simeq$ 2.7$\%$ and a momentum resolution for protons of $\simeq$ 3$\%$ can be achieved. The first level
trigger is obtained by a fast multiplicity signal coming from the TOF/TOFINO wall, combined with a signal from
the START detector, while the second level trigger is made by using the informations from the RICH and
Pre-Shower to select the lepton candidates.

In the pp experiment at 1.25 and 2.2 GeV kinetic energy, the intensity was about 10$^{7}$ particles/second,
and a target filled with liquid hydrogen was used. The START detector was not used in this run because of the
too high rate of secondaries produced by the START detector itself. Thus a specific algorithm has been
developed to calculate the time of flight and then to identify the charged particles.

\begin{figure}[htb]
 \centerline{\includegraphics[width=0.5\linewidth]{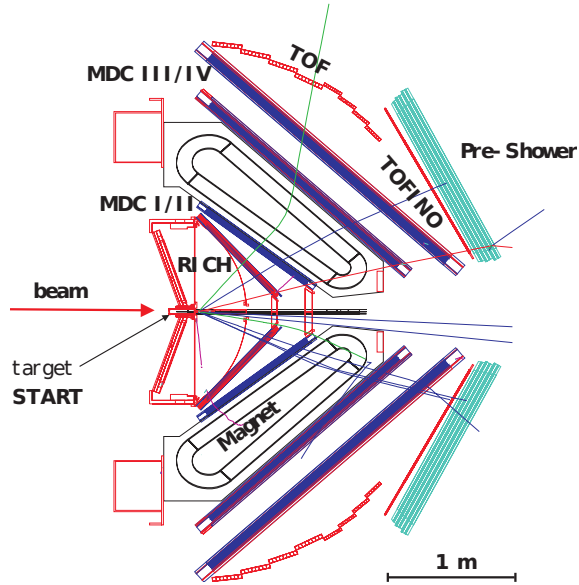}}
 \caption{\label{hades}(Color online) Schematic view of the HADES detector as implemented in the simulation.
Simulated particle tracks are shown as well.}
\end{figure}

\Section{Results}
As mentioned in the introduction, it is important to constrain the models predicting the cross section and the
production mechanisms of the $\Delta$ resonance. In order to do that the hadronic channels (pp $\rightarrow$
np$\pi^{+}$ and pp $\rightarrow$ pp$\pi^{0}$) have been measured and studied in parallel to the leptonic
channels. 

The pp $\rightarrow$ np$\pi^{+}$ channel is studied using the reconstruction of the undetected neutron. The
reaction was selected first by the charged particle identification based on momentum and reconstructed time
of flight, then a (p,$\pi^{+}$) missing mass cut was imposed around the neutron mass. This cut efficiently
suppresses the background coming from misidentified protons and two-$\pi$ contributions. 

The pp$\pi^{0}$ reaction was extracted from the (p,p) missing mass spectrum by subtracting the background
under the $\pi^{0}$ peak after suppression of the pp elastic and two-pion contribution. 

The absolute normalization of the data has been achieved by a simultaneous measurement of elastic scattering
for which the cross section is known. Error bars include statistical and systematic errors.

Experimental data have been compared with the original and modified version of the resonant model~\cite{ex:5}
using model A and model B (for details see Table~\ref{Table}). A comparison to different spectra has been
made. In particular the neutron angular distribution in the center of mass system measured in the
pp~$\rightarrow$pn$\pi^{+}$ channel, has been investigated, since it is sensitive to the angular distribution
of $\Delta$ resonance production (dominant process is pp $\rightarrow$ n$\Delta^{++}$). The acceptance
corrected experimental spectra are presented in Fig.~\ref{hades-wasa} (black points) together with slightly
modify resonance model "model A" (for details of the modification, see Tab.~\ref{Table}). The distribution
shows strong forward/backward peaking, as expected due to the peripheral character of the $\Delta$ resonance.
The original Teis model (black solid line) shows underestimation at cos$\theta_{n} \approx$ 0, which is
improved by changing cutt-off parameter ($\Lambda_{\pi}$) (black dashed line). In addition in
Fig.~\ref{hades-wasa} comparison between HADES and WASA measurement is presented. WASA data were taken
from~\cite{ex:7} and normalized to the cross-section obtained by HADES. One can see reasonable quantitative
agreement, keeping in mind, that the WASA data are presented without systematic errors.

\begin{figure}[ht]
 \centerline{\includegraphics[width=0.8\linewidth]{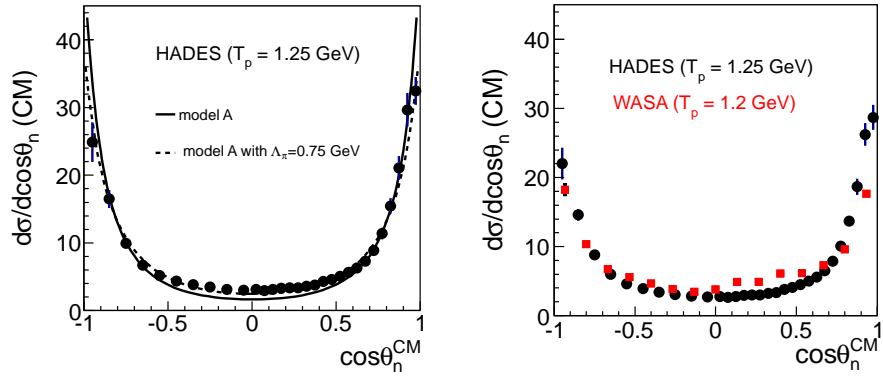}}
 \caption{\label{hades-wasa}Angular distribution of neutron in centre of mass system after acceptance
correction for the pp$\rightarrow$pn$\pi^{+}$ reaction. (Left) Comparison of the HADES data to simulation
based on~\cite{ex:5} and model A. Both simulation curves are normalized to reproduce the integrated
experimental yield. (Right) Comparison of HADES and WASA data~\cite{ex:7}.}
\end{figure}

Fig.~\ref{1-pion} exhibits the projection on the (p,$\pi^{0}$) (Left), (p,$\pi^{+}$) (Middle) and
(n,$\pi^{+}$) (Right) invariant masses for T$_{p}$ = 1.25 and 2.2 GeV. The data are corrected for
reconstruction efficiencies.

\begin{figure}[ht!]
 \centerline{\includegraphics[width=0.9\linewidth]{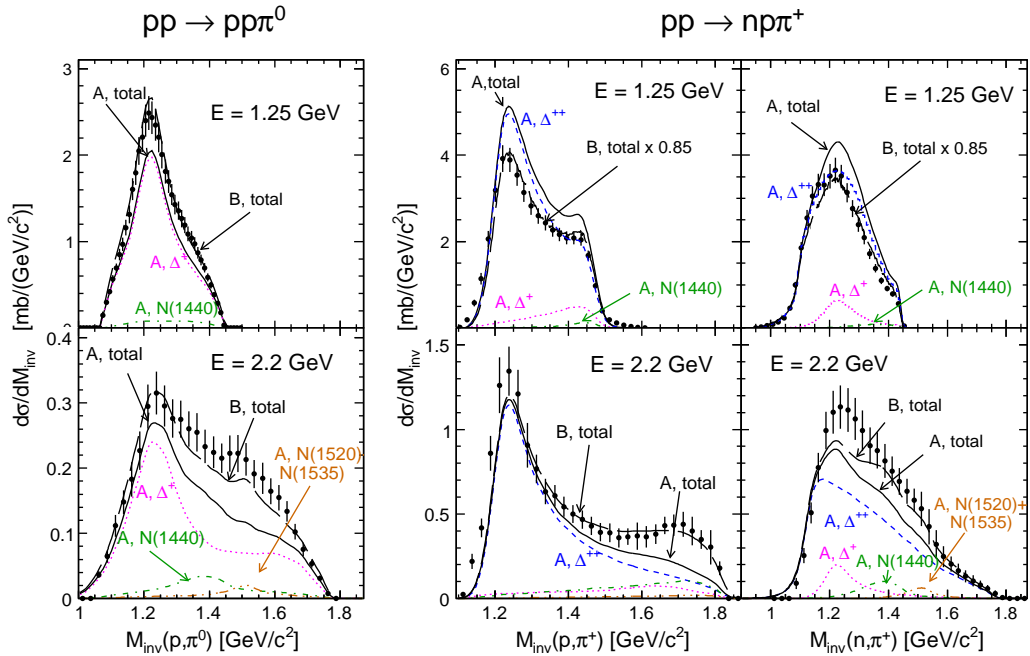}}
 \caption{(Color online) Invariant mass distributions measured of (p,$\pi^{0}$) (Left), (p,$\pi^{+}$) (Middle)
and (n,$\pi^{+}$) (Right) measured in pp $\rightarrow$ pp$\pi^{0}$ and pp $\rightarrow$ np$\pi^{+}$ at
1.25~GeV (top row) and 2.2 GeV (bottom row). HADES data (full dots) are compared within the experimental
acceptance to the predictions of the modified resonance model (see text).}
 \label{1-pion}
\end{figure}

The prominent peak of M$_{inv}$(p, $\pi^{+}$) and M$_{inv}$(p,$\pi^{0}$) around 1.23 GeV/c$^{2}$ confirms that
most of the pions are produced via $\Delta$ decay, which is consistent with the resonance model . However, the
data present some deviations with respect to the model A, which is close to the original Teis's
model~(\cite{ex:5}). This, together with the above mentionned failure in the description of the angular
distribution motivates some changes in the model. A better description of the yield and shape of the missing
mass spectra is indeed obtained after some modifications which are included in the model B (see
Tab.~\ref{Table} for details). Due to the dominance of the Delta(1232) resonance, the two channels are
correlated by isospin symmetry:  $\sigma$(pp $\rightarrow$ n$\Delta^{++}$) $\sim$ 3(pp $\rightarrow$
p$\Delta^{+}$), leading to $\sigma$(pp $\rightarrow$ np$\pi^{+}$) $\sim$ 5 (pp $\rightarrow$ pp$\pi^{0}$).
N$^{*}$ contribution seems also reasonable and the invariant mass distributions are rather well reproduced by
the pp~$\rightarrow$~n$\Delta^{++}$ and pp~$\rightarrow$~p$\Delta^{+}$ simulations. At 2.2~GeV, the
contribution of higher lying resonances is clearly visible at high invariant masses.

\begin{table}[ht!]
\begin{center}
\begin{tabular}{|c|c|c|}
 \hline
T$_{p}/$Model & Model A                      &  Model B                \\ \hline
1.25     [GeV]     & \cite{ex:5}, (1)            &  Model A,  (2)     \\ \hline
2.2       [GeV]     & \cite{ex:5}, (1)            &  Model A, (3)      \\ \hline

\end{tabular}
\end{center}
\caption{\label{Table} Summary of the modifications introduced to the resonance model \cite{ex:5} in order to
better fit the data: (1) pp and pn final state interactions have been implemented using the Jost function and
adjustment of the cut-off parameter ($\Lambda_{\pi}$) entering the $\pi N \Delta$ and $\pi$ NN vertex form
factor has been changed from 0.63 to 0.75~GeV, (2) the $\Delta$ production angular distribution has been
further adjusted to reproduce the measured neutron angular distribution, (3) the cross section of the higher
lying resonances has been increased and a non-resonant contribution has been added. }
\end{table}%
\vspace{0.3cm}
Furthermore, as another important platform for studying resonances properties, double-pion production in
nucleon-nucleon collisions is also of interest. Preliminary results obtained for the analysis of 2 channels:
(1) pp $\rightarrow$ $pp \pi^{+}\pi^{-}$ and (2) pn $\rightarrow$ pn $\pi^{+}\pi^{-}$ have been presented in
comparison to model predictions~\cite{ex:5,ex:6} including double-$\Delta$ and N(1440) excitation. Both
channels were selected by identifying all charged hadrons involved in the reaction and  applying a missing
mass cut (M$_{pp\pi^{+}\pi^{-}}$ in case (1) and M$_{p\pi^{+}\pi^{-}}$ in case of (2)). It turns out that the
invariant mass (M$_{\pi^{+}\pi^{-}}$) and the opening angle in CM (cos$\delta_{\pi^{+}\pi^{-}}$) of the pion
pair are the most sensitive distributions to the different model contributions. Fig.~\ref{2-pions} exhibits
an experimental distributions of the invariant mass and opening angle of $\pi^{+}\pi^{-}$ in comparison to the
pure phase space (PHPS) calculations. The PHPS calculations has been normalized to the experimental yield. It
is seen, that both experimental distributions deviate from PHPS calculations. Some enhancement can be observed
at low $\pi^{+}\pi^{-}$ invariant masses, which is not present in the simulation with the PHPS only. This peak
at small invariant masses appears in the models~\cite{ex:8,ex:9} as being due to the $\Delta \Delta$
excitation and to the decay channel of the Ropper resonance N(1440) into $N \Delta$ (N* $\rightarrow $N
$\Delta$). 

\begin{figure}[ht]
 \centerline{\includegraphics[width=0.8\linewidth]{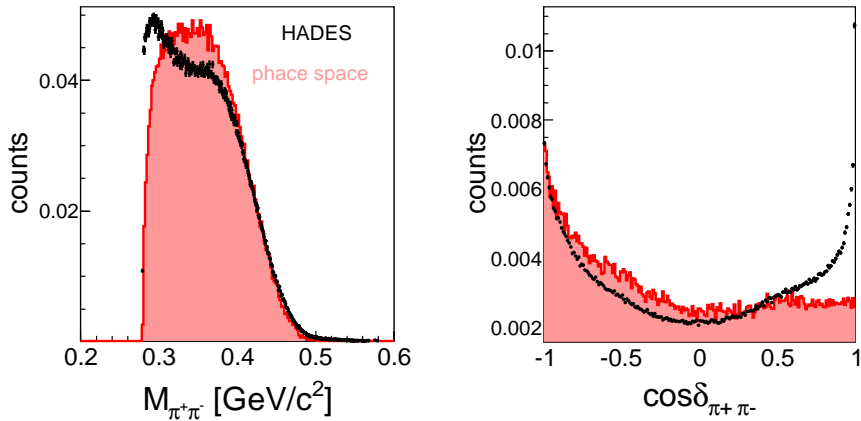}}
 \caption{\label{2-pions}Distribution of the $\pi^{+}\pi^{-}$ invariant mass M$_{\pi^{+}\pi^{-}}$ (left) and
the $\pi^{+}\pi^{-}$ opening angle in centre of mass $\delta_{\pi^{+}\pi^{-}}$ (right) for pp $\rightarrow$
$pp \pi^{+}\pi^{-}$ reaction at beam energy T$_{kin}$ = 1.25 GeV are shown in comparison to the phase space
calculation. Both spectra are inside HADES acceptance and efficiency. The simulation is normalized to
reproduce the measured experimental yield. Only statistical errors are shown.}
\end{figure}

\Subsection{Conclusion}
The aim of measuring one and two-pion production in elementary reactions with HADES are three fold:  
(1) to study the detector response, and cross-check the analysis of dilepton channels, 
(2) test the ingredients of the transport models used for the dielectron production, 
      which are based on resonance models and 
(3) to obtain a parametrization of meson and baryon resonance production for dielectron channels analysis. 

HADES provides high statistics data in one and two-pion production channels, which complements measurements of
previous experiments. 
For the one pion channel, which is dominated by $\Delta$ resonance, the data have been compared to the
resonance model usually used in transport models, and modifications have been 
proposed to better reproduce the data. The ongoing analysis of the two-pion production and comparison to the
models will bring constrain on double-$\Delta$ 
production as well as N(1440) production and decay mechanism.

\coltoctitle{Exclusive Channels in pp at 3.5~GeV}
\authortochead{A.~Dybczak}
\label{DA}

\setcounter{chapter}{1}
\setcounter{section}{0}
\maketitle

\Section{Introduction}
The High Acceptance Di-Electron Spectrometer (HADES)~\cite{spectrometer} operates at the GSI Helmholtzzentrum
fur Schwerionenforschung in Darmstadt, Germany. One of the main physics goals of HADES is to investigate
spectral modifications of light vector mesons in strongly interacting matter via dilepton ($e^+e^-$) decay
channel. The HADES collaboration has set up an experimental program to measure dilepton spectra in p+p and p+n
collisions~\cite{salabura}. In year 2007 electron-positron pair production has been measured in elementary
collisions of two protons with 3.5 GeV projectile kinetic energy. One of basic observables in this measurement
is inclusive $e^+e^-$ mass distribution , shown in Fig. 1a after Combinatorial Background (CB) subtraction.
The expected $e^+e^-$ production channels are given by Dalitz decays of $\pi^0$, $\eta$, $\omega$ mesons  and
$\Delta(1232)$ resonance. In the high mass region 2-body decays of $\rho/\omega$ play a role. Indeed, the
experimental data can be described by PLUTO~\cite{pluto} simulation (see Fig. 1a) assuming aforementioned
components, but not in the mass region below vector meson pole ($M_{inv}^{e^+e^-} \in(0.5-0.7)$). However,
only Dalitz decay channel for baryonic resonances is included for $\Delta(1232)$ here.
\begin{figure}[htb]
 \begin{center}
  \begin{minipage}[t]{0.48\linewidth}
   \includegraphics[width=0.88\linewidth]{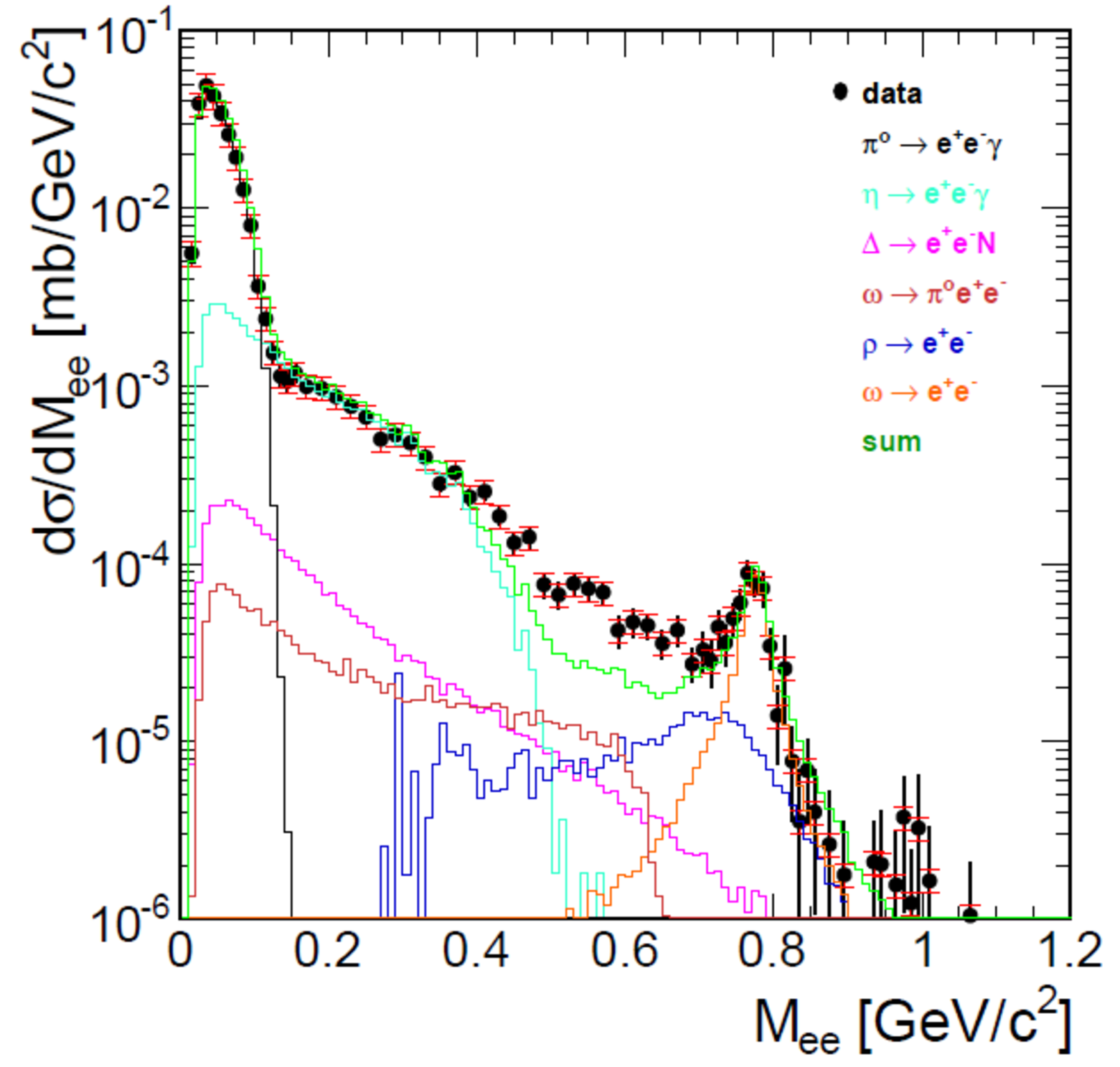}\\
   \footnotesize {Fig. 1a) Efficiency corrected inclusive invariant mass distribution of dielectrons inside
the geometrical acceptance of HADES. $e^{+}e^{-}$ - signal pairs (black dots), after combinatorial background
subtraction for p(3.5GeV)+p collisions with coresponding simulation coctail components. The data are
normalized to the measured p+p elastic events.}
  \end{minipage}\hfill%
  \begin{minipage}[t]{0.48\linewidth}
   \includegraphics[width=0.95\linewidth]{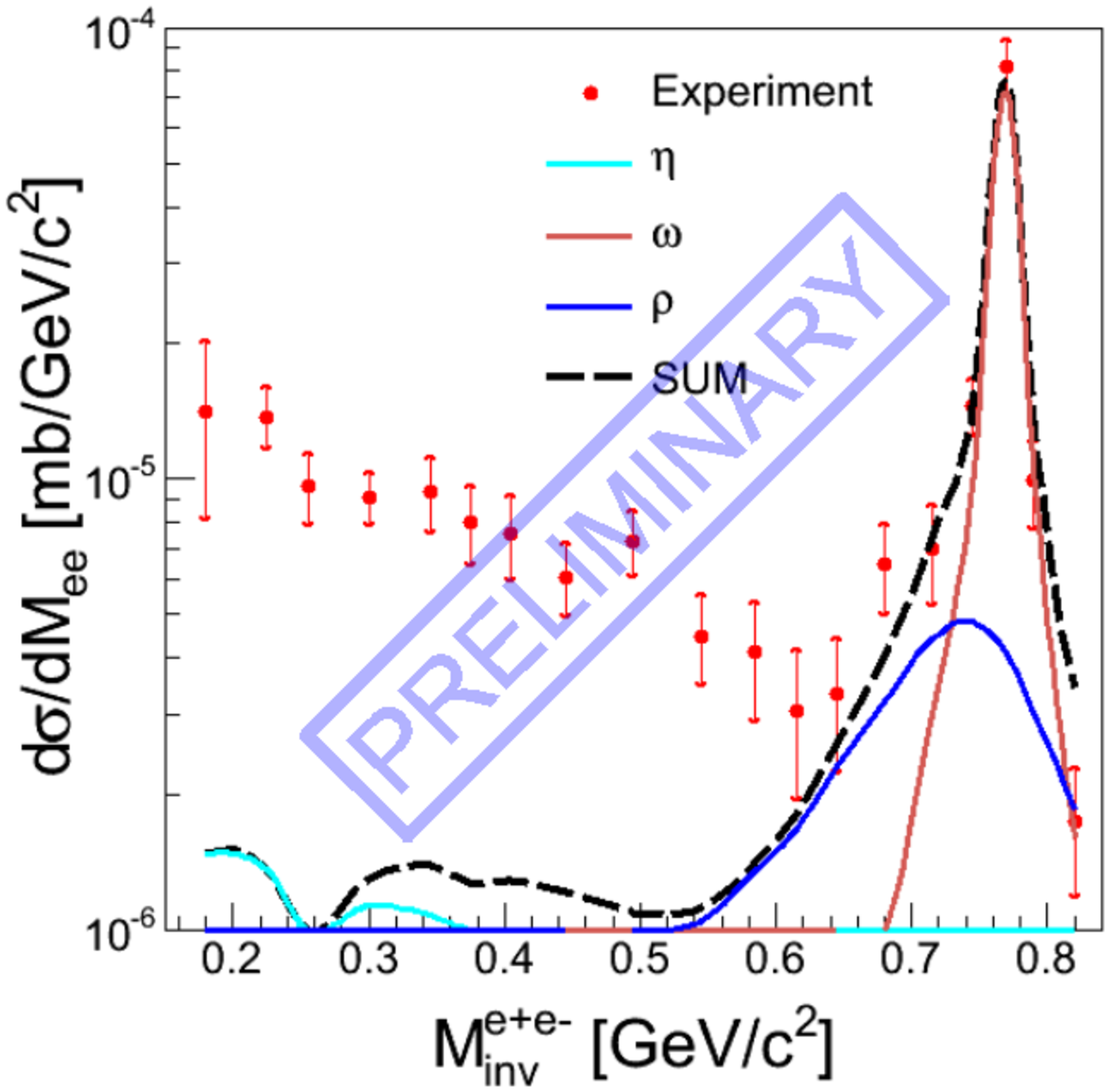}\\
   \footnotesize {Fig. 1b) Efficiency corrected invariant mass distribution of dielectrons for exclusive
$ppe^+e^-$ channel inside the geometrical acceptance of HADES after combinatorial background subtraction.
The data are normalized to p+p elastic events and compared to simulation cocktail assuming sources given in
the legend.}
  \end{minipage}
 \end{center}
\end{figure}

In order to study contribution of higher lying baryonic resonances $ppe^+e^-$, $pp\pi^0$, $pn\pi^+$,
$pp\omega(\eta)\rightarrow\pi^+\pi^-\pi^0$ exclusive channels have been analysed. The $ppe+e^-$ channel has
been selected by condition on $pe^+e^-$ missing mass $\sim 0.938\ [MeV/c^2]$ and the respective  $e^+e^-$
invariant mass spectrum is shown in Fig. 1b. The mass region below vector meson pole is not described by
channels including $\omega,\ \eta$ decays which are constrained by hadronic channels mentioned above
(see~\cite {khaled}). This means that Dalitz decays of baryon resonances must contribute in this area. 
In order to estimate production cross sections of the resonances which decay in the $pn\pi^+$ and $pp\pi^0$
final states, the results have been compared to Monte-Carlo calculations based on the resonance
model~\cite{teis} assuming incoherent sum of various resonances.

To convert resonances cross sections into $e^+e^-$ yield one should remember that the respective partial
decay widths are mass dependent. Two models of $\Gamma_{R\rightarrow pe^+e^-}(M)$ of
(\cite{zetenyi},~\cite{martemyanov} are applied in the simulation.

\Section{\boldmath Analysis of $pn\pi^+$, $pp\pi^0$ final states}
In both final states only events with two positive tracks were selected. Protons and positive pions were
distinguished using measured momentum and particle velocity of reconstructed tracks. Channels were selected
via condition on missing mass of neutral particles ($n$ and $\pi^0$, respectively). Signal was obtained as a
number of counts above fitted continuous background. In the case of $pp\pi^0$ discrimination on elastic
scattering  and multipion production was required additionally to enhance purity of the signal. Resonances
cross section were estimated by simultaneous fit of invariant masses ($M_{inv}^{p\pi^+}$, $M_{inv}^{n\pi^+}$,
$M_{inv}^{p\pi^0}$)  and angular distributions ($cos\theta^{p\pi^+}_{CM}$, $cos\theta^{n\pi^+}_{CM}$,
$cos\theta^{p\pi^0}_{CM}$) in several bins of the $N-\pi$ invariant mass. Iterative procedure was applied: In
the first step a flat angular distribution has been assumed for resonances production. Since experimental
distributions show a clear forward-backward peaking, in the next step a modified distribution of the form
$d\sigma/dt\sim A/t^\alpha$ has been used. The $t,\ A,\ \alpha$ are t-Mandelstam variable and fit parameters
depending on resonance mass, respectively. Furthermore, the isospin relations between cross sections of
various resonance have been preserved.

\Subsection{\boldmath Acceptance corrected $pn\pi^+$, $pp(\pi^0)$ results}
In Fig.~2a, the results obtained after acceptance/efficiency corrections assuming aforementioned model are
presented. The $pn\pi^+$ channel allows to distinguish doubly and singly charged resonances.
\begin{figure}[htb]
 \begin{center}
  \begin{minipage}[t]{0.48\linewidth}
   \includegraphics[width=0.85\linewidth]{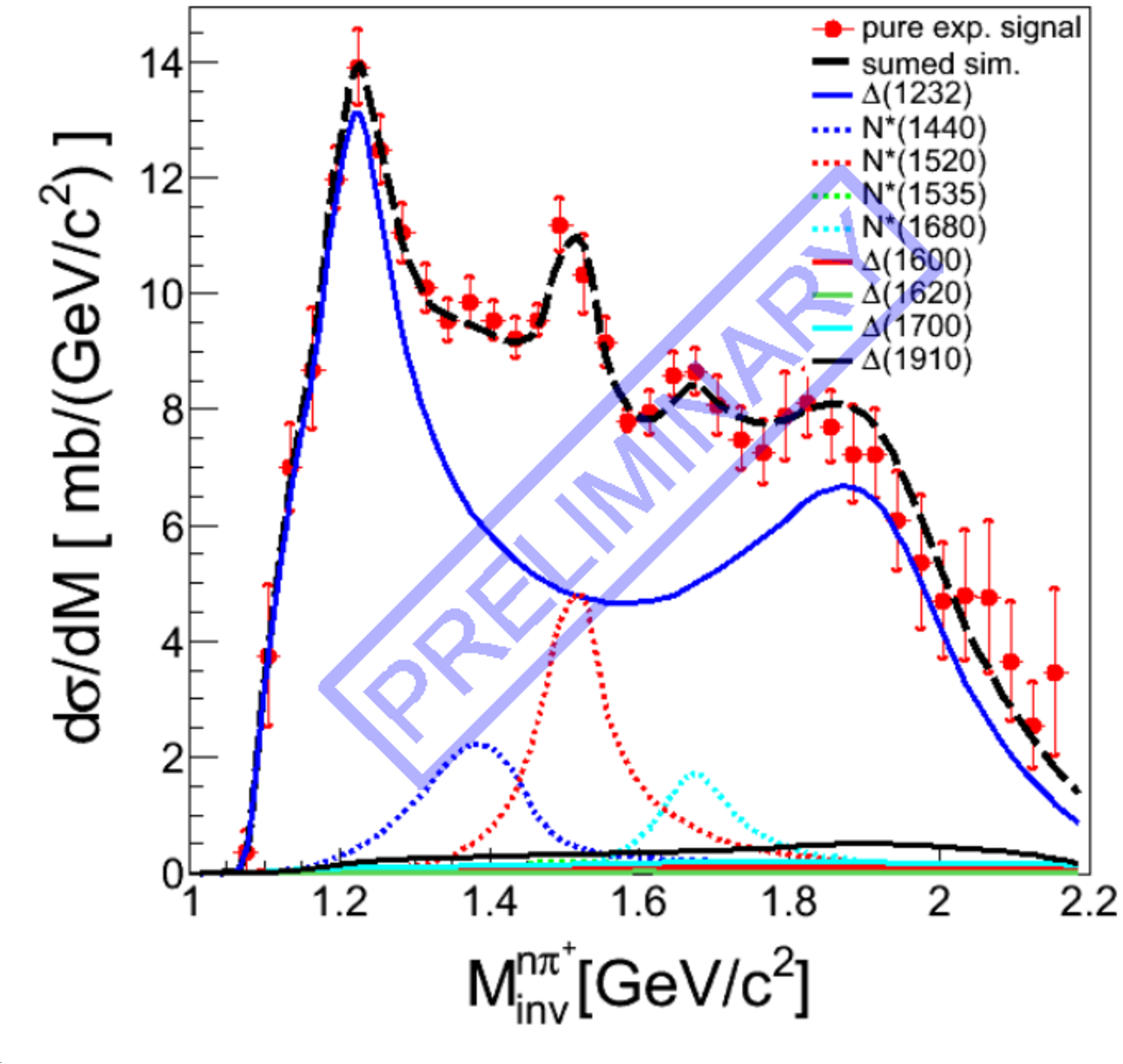}\\
   \vspace{1ex}\\
   \footnotesize{Fig. 2a) Acceptance and efficiency corrected spectra of $p\pi^+$, $n\pi^+$ 
invariant mass with corresponding simulation components.}
  \end{minipage}\hfill%
  \begin{minipage}[t]{0.48\linewidth}
   \includegraphics[width=0.85\linewidth]{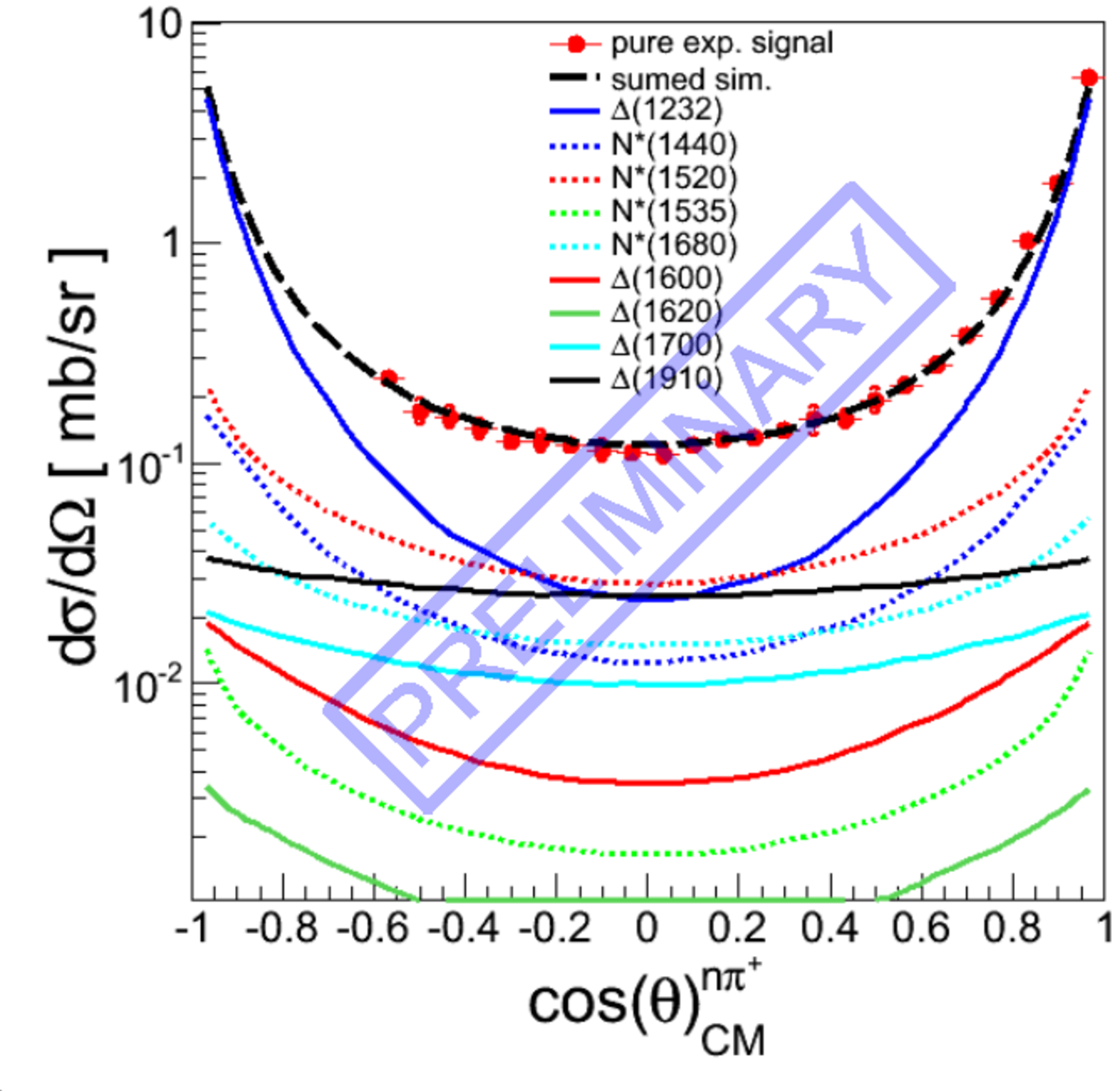}\\
   \vspace{1ex}\\
   \footnotesize{Fig. 2b) Acceptance and efficiency corrected spectra of $p\pi^+$, $n\pi^+$ angular
distribution in the centrum of mass of with corresponding components obtained from fitting procedure described
in the text.}
  \end{minipage}
 \end{center}
\end{figure}

\begin{figure}[htb]
 \begin{center}
  \begin{minipage}{0.48\linewidth}
   \includegraphics[width=0.85\linewidth]{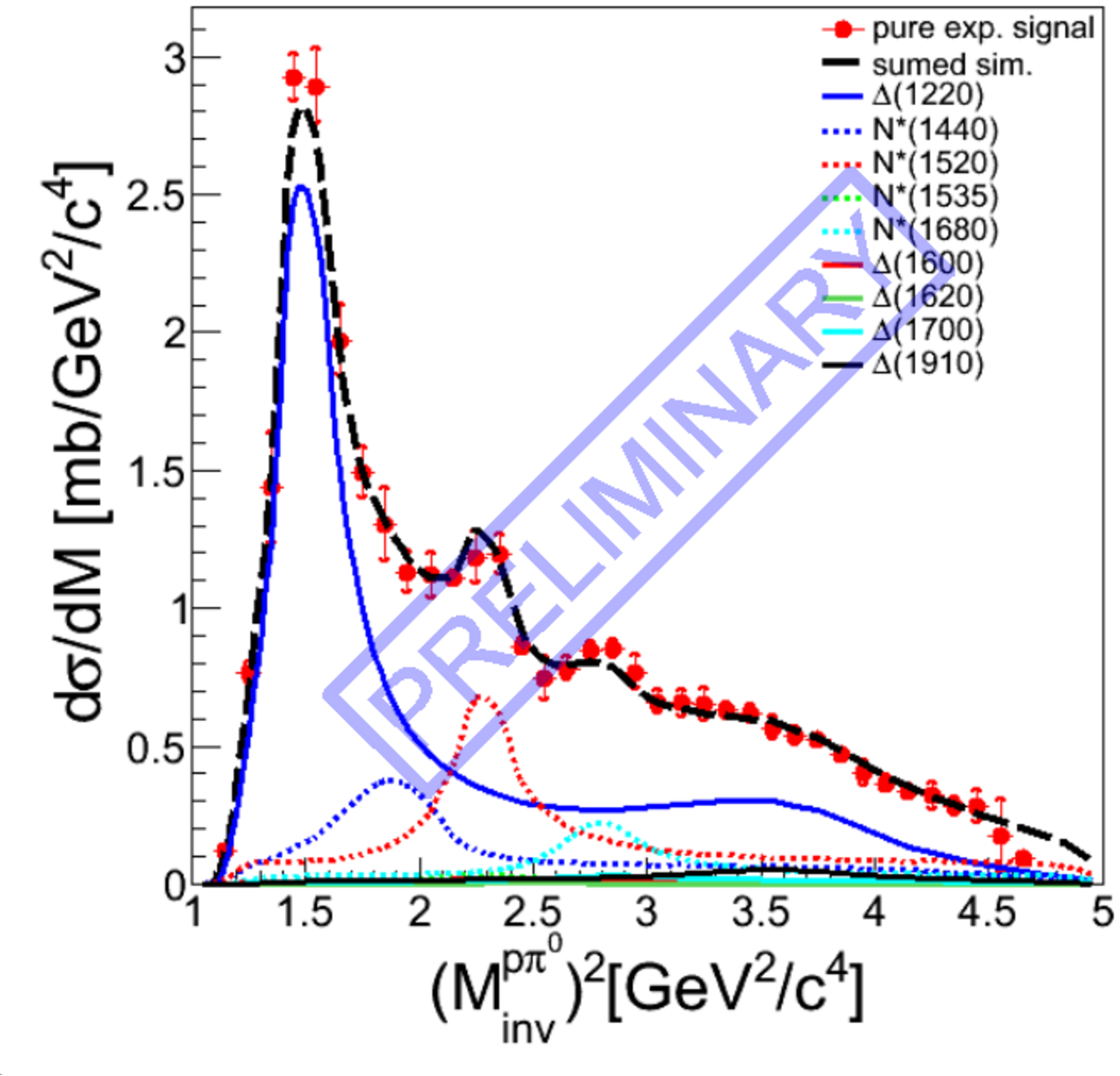}\\
   \footnotesize {Fig. 3a) Acceptance and efficiency corrected spectra of $p\pi^0$ invariant mass with
corresponding simulation components.}
  \end{minipage}\hfill%
  \begin{minipage}{0.48\linewidth}
   \includegraphics[width=0.85\linewidth]{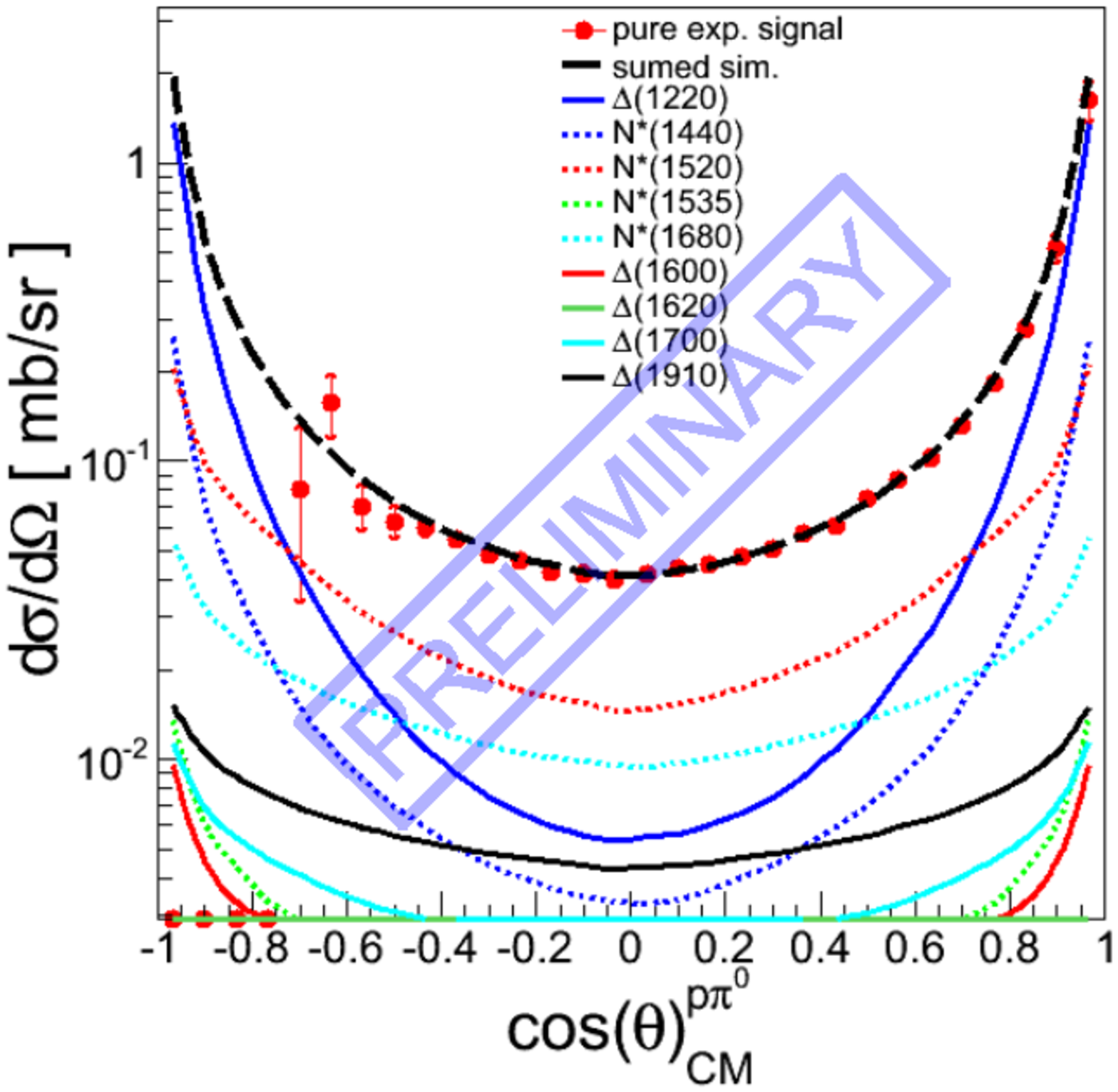}\\
   \footnotesize {Fig. 3b) Acceptance and efficiency corrected spectra of $p\pi^0$ angular distribution in the
centrum of mass of resonance with corresponding simulation components.}
  \end{minipage}
 \end{center}
\end{figure}

\Section{Conclusions and outlook}
The applied resonance model of baryon resonance production can fully describe production of pions in the two
$pp\pi^0$ and $pn\pi^+$ channels. Obtained cross sections for the various resonances fulfill all isospin
relations. The derived cross section for $\Delta(1232)$ production is in a good agreement with observations of
other experiments. Estimated cross sections for the total $\pi^0$ and $\pi^+$ yields are also in a good
agreement with previous measurements. The obtained cross sections for resonance production will be further
used to calculate dielectron production in the $ppe^+e^-$ channel.

\coltoctitle{\boldmath Overview of $e^+ e^-$ Production in $p+p$ and $p+n$ Collisions}
\authortochead{P.~Salabura}
\label{SP}

\setcounter{chapter}{1}
\setcounter{section}{0}
\maketitle

\Section{Introduction}
The High Acceptance Dielectron Spectrometer (HADES) is a second-generation experimental setup located at the
heavy-ion synchrotron SIS18/GSI Darmstadt. Its powerful particle identification capabilities and excellent
momentum resolution allows for simultaneous measurements of hadronic ($p,\,\pi^{\pm},\, K^{\pm}$) and
electromagnetic ($e^+\, e^-$) final states over a wide angular range $(18^\circ<\theta<85^\circ)$ in
proton, deuteron and HI induced reactions~\cite{HADES_nim}. One of the main goals of the HADES experiments is
to achieve a detailed understanding of dielectron emission from hadronic systems at moderate ($1-4$ AGeV)
bombarding energies. A particular feature of this energy domain is the important role of baryonic sources as
$N+N$ bremsstrahlung and resonance Dalitz decays (i.e $\Delta(N^*) \to Ne^+e^-$) which are poorly known. The
results obtained on electron pair production in elementary $N+N$ collisions represent a doorway for the
understanding of pair production in HI collisions where additional in-medium effects, as for example hadron
mass modifications, are expected to occur~\cite{hades1,hades2,hades3}. In order to shed more light on the
aforementioned processes, HADES performs a combined study of pion and dielectron production in $p+p$ and $d+p$
reactions. The results from pion production are given in the contributions of M.~Gumberidze~\cite{gosia} and
A.~Dybczak~\cite{adrian}. In this report, a survey on important results on $e^+ e^-$ pair production is given.

\Section{\boldmath $e^+ e^-$ results from $p+p$ and quasi-free $p+n$ collisions}
Results from the pioneering DLS experiment at Bevelac on inclusive $e^+ e^-$ production in $p+p$ and $p+d$
reactions indicate a strong isospin dependence of the pair production in the $1-4$ AGeV region~\cite{dls}. The
cross section ratio of pair production is significantly larger than 2 which was observed at $E_{kin}\geq
2$~GeV and could only be qualitatively described by model calculations.

Fig.~1 shows inclusive $e^+e^-$ invariant mass distributions measured in $p+p$ (left-top) and quasi-free $p+n$
(left-bottom) collisions at $1.25$~GeV~\cite{hades4} inside the HADES acceptance. Quasi-free $n+p$ reactions
were selected from $d+p\rightarrow e^+e^- + X$ reaction on trigger level by the detection of a fast spectator
protons from the deuterium break-up in a dedicated forward hodoscope wall at small angles w.r.t.\ beam
direction. As one can see, a large (at least one order of magnitude) excess of the pair yield is visible for
the $p+n$ reaction channel at higher invariant masses. The data are compared to simulated pair distributions
calculated with the Pluto event generator~\cite{pluto} assuming a dominant role of the $\pi^0\rightarrow
e^+e^- + \gamma$ and $\Delta(1232)\rightarrow e^+e^- + N$ Dalitz decays. For both reaction channels the
measured yield in the $\pi^0$ Dalitz decay region is very well reproduced taking into account the inclusive
$\pi^0$ production cross section from the resonance model assuming $\Delta(1232)$ dominance~\cite{hades4}.
This assumption is also corroborated directly by the HADES measurements of the dominant exclusive channels
$pp\rightarrow pp\pi^0$ and $pp\rightarrow pn\pi^+$(see~\cite{gosia}). To model
\mbox{$\Delta(1232)^{+(0)}\rightarrow p(n)e^+e^-$} the differential partial decay width
$d\Gamma_{Ne+e-}(M_{\Delta})/dM_{e+e}$ from~\cite{tubingen} has been utilized, assuming a constant magnetic
transition form factor $G_M= 3.02 \pm 0.03$ fixed by photo-production experiments (the other two form factors,
electric and Coulomb, are assumed to be zero)~\cite{hades4}. A possible modification of the magnetic
transition form factor due to intermediate vector mesons was simulated assuming the two-component quark
model~\cite{iachelo} and is visualized by the shaded area. As expected, it leads to a slightly enhanced yield
for high pair masses. Whereas the $p+p$ spectrum is very well described by both contributions, the description
of the $p+n$ channel is not satisfactory even if the $\eta\rightarrow e^+e^-+ \gamma$ decay is included. The
latter one is fixed by precise data from WASA at CELSIUS on near threshold production in $pn\rightarrow d
\eta$ and $pn\rightarrow pn \eta$~\cite{wasa}.

\begin{figure}[ht]
 \centerline{\includegraphics*[width=0.4\textwidth,height=0.6\textwidth]{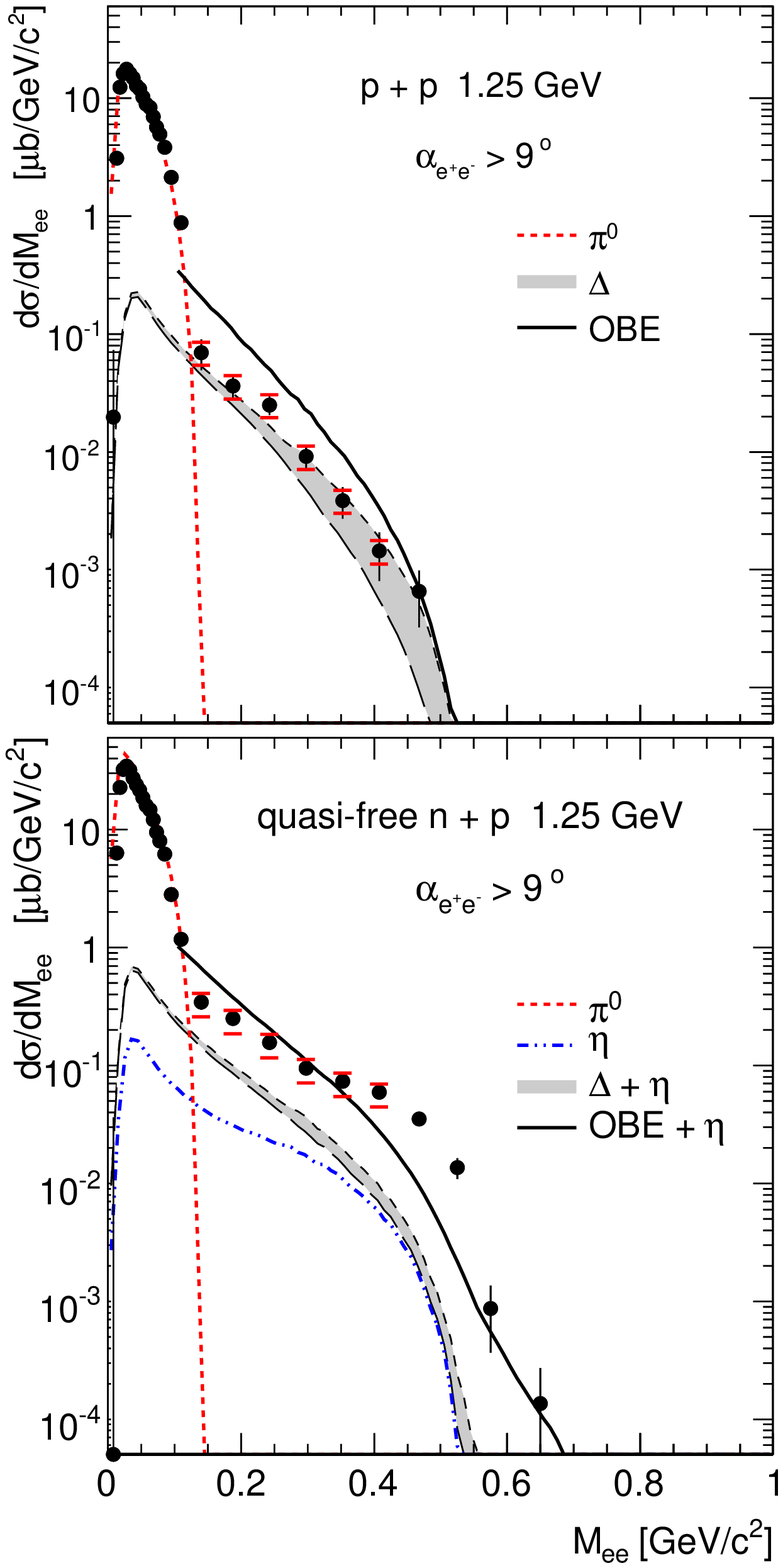}%
             \includegraphics*[width=0.5\textwidth,height=0.45\textwidth]{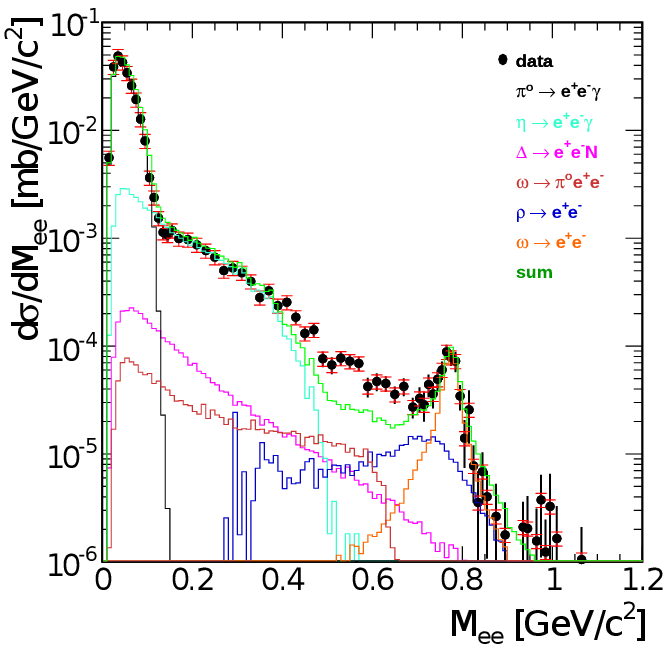}}
 \vspace*{-0.1cm}                                                                                       
 \caption{\label{pp_pn}Differential cross sections for dielectron production as a function of the invariant
mass as  measured in $p+p$, quasi-free $p+n$ reactions at $1.25$ GeV (left) and $p+p$ at $3.5$ GeV (right).
The distributions are compared to model calculations assuming various hadronic sources (for more details see
the text).}  
\end{figure}

Next, the obtained results are compared to predictions of the OBE model~\cite{kaptari}, where resonant
($\Delta$) and non-resonant ("quasi-free" $N+N$ bremsstrahlung) contributions are added coherently. The result
of the simulation is shown in Fig.~1 as solid black line. The yield calculated in this approach overestimates
(underestimates) measured distribution in the $p+p$ ($p+n$) reactions. One should add here that more recent
calculations of~\cite{shyam} based on similar effective Lagrangian are successful in description of the $p+p$
reaction channel and also come closer to the data measure in the $p+n$ reaction channel. For the latter case,
the implementation of the pion electromagnetic form factor in graphs including exchange of charged pions turns
out to be decisively to produce $e^+e^-$ yield enhancement visible in the $p+n$ reaction channel at higher
masses.

The right part of Fig.~1 shows the $e^+e^-$ invariant mass distributions measured in $p+p$ collisions at
3.5~GeV~\cite{anar}. In contrast to lower energy, the spectrum extends to much higher masses due to the
rapidly changing excitation function of dielectron production. A clear $\omega$ peak with mass resolution
$\Delta M/M\sim 2\%$ can be seen in $e^+e^-$ channel, for the first time at such low energy. Below the vector
meson mass region dielectron production is dominated by the $\eta$ and $\pi^0$ Dalitz decays. The comparison
of the experimentally measured invariant mass distribution to a incoherent sum of known hadronic sources is
also shown in this figure. Calculations have been performed with PYTHIA + PLUTO codes~\cite{janus,anar}. The
simulated cocktail describes the data reasonably well except for the mass range around $0.55$~GeV/c${}^2$
where the yield is underestimated. The latter deviation is not too surprising since in this energy region
contribution from the resonances (in our calculations only from $\Delta(1232)$) can be expected. Furthermore,
the virtuality of the photon reaches quite high values and one can expect shape modifications due to unknown
mass dependency of the $\Delta(1232)$ transition form factor. Moreover, contributions of higher lying
($\Delta, N^*$) resonances has not been included, yet. The respective contributions can be better constrained
by means of the exclusive dielectron and hadronic channels: $ppe^+e^-$ , $pp\pi^0$ , $pn\pi^+$, $pp\eta$ and
$pp\omega$ inside the HADES acceptance. The strategy is to fix the cross sections for resonance ($\Delta,
N^{*(+,0)}$) production from the hadronic final states and convert it to $ppe^+e^-$ yields by means
of  $d\Gamma_{Ne+e-}(M_{res})/dM_{e+e-}$ given by available calculations~\cite{wolf,tubingen1}. Many of these
hadronic channels have already been analyzed. The channels with pion production is discussed in the
contribution of A.~Dybczak~\cite{adrian}.

\Section{Summary and outlook}
HADES measured dielectron production in proton, deuteron and HI induced reactions in the $1-4$ AGeV energy
region. Dielectron production in $N+N$ collisions is a mandatory prerequisite for the understanding of pair
production in HI collisions and already observed in-medium effects. The interesting feature of the pair
production in low-energy domain is the important role of baryonic $e^+ e^-$ sources , i.e.\ resonance Dalitz
decays and $N+N$ bremsstrahlung. The description of the respective pair production process is directly linked
to the questions of how the virtual massive photon couples to baryons and how do the respective transition
form factors look like. The results already obtained by HADES in $p+p$ ($1.25, 2.2, 3.5$ GeV) and $p+n$ (1.25
GeV) indicate a sensitivity to such effects. Further measurements are however mandatory to pin down respective
effects. In this context, future experiments with pion beams planned with HADES at GSI can be extremely
helpful. In contrast to proton induced reactions, resonances can be directly excited in the s channel and
separated by means of excitation function and angular measurements~\cite{hubert}.

\coltoctitle{HADES Spectrometer and Future Experiments with Pion Beams}
\authortochead{H.~Kuc}
\label{KH}

\setcounter{chapter}{1}
\setcounter{section}{0}
\maketitle

\Section{Introduction}
The HADES [1] (Acceptance Dielectron Spectrometer)  collaboration demonstrated, using pp and quasifree pn
reactions [3] that baryonic resonances play an important role in dielectron emission at SIS accelerator
energies (1-2 AGeV). This is first due to their mesonic decays and the subsequent direct dilepton (e.g.
$\rho/\omega \rightarrow e^{+}e^{-}$) or Dalitz (also called conversion) decay ($\pi^{0}/\eta \rightarrow
\gamma e^{+}e^{-}$ or $\omega \rightarrow \pi^{0}e^{+}e^{-}$) modes of these mesons. Baryonic resonances are
also expected to contribute directly to dilepton emission via their Dalitz decay modes
($N^{\ast}/\Delta\rightarrow Ne^{+}e^{-}−$). (see Piotr Salabura's talk for more details)

The Dalitz decay of baryonic resonances can be studied in a more precise way using the GSI pion beam. In
$\pi^{-}+p$ reactions, the production mechanisms are better controlled and all resonances are produced at
fixed mass, in the s channel. In addition to electromagnetic channels, meson production channels ($\pi N
\rightarrow\pi\pi N,\;\pi N \rightarrow\eta N,\; \pi N \rightarrow\rho/\omega N,\;\pi N \rightarrow K\Lambda$
...) can also bring useful information to study resonance properties, via partial wave analysis (PWA). There
is indeed a big interest from theoreticians, for new precise differential cross section  data for various
reaction channels and at various incident momenta.

These experiments present some technical challenges. This secondary pion beam is produced on a beryllium
targets in C+Be or p+Be reactions. To achieve pion beam intensities of the order of $10^{6}/s$, primary carbon
or proton beams need to reach an intensity close to space charge limit of the accelerator. To produce pions
with momentum from 0.7 to 1.5 GeV/c, the reaction C+Be@2.0GeV should be chosen but to produce pions with
momenta from 1.5 to 2.2 GeV/c with sufficient intensity, p+Be@3.5GeV has to be used. The acceptance of the
beam line is $\Delta p/p \approx\pm 4\% $. For exclusive studies momentum reconstruction is needed.
Furthermore, the bacground due to secondary particles produced by the beam halo has to be suppresed.
Therefore, an additional event by event pion in-beam tracking is required. This task is handled by two in-beam
silicon detectors X1,X2. Trackers are two double sided, 2x128 strip, $300\mu m$ thick silicon detectors
manufactured in high radiation hard technology. This system provides a momentum resolution of $\Delta p/p \sim
0.1\% $ which is sufficient for exclusive analysis studies. A diamond detector placed in front of the target
complements the beam tracking and will be used as start detector.     
\begin{figure}[htb]
 \centerline{\includegraphics[scale=0.7]{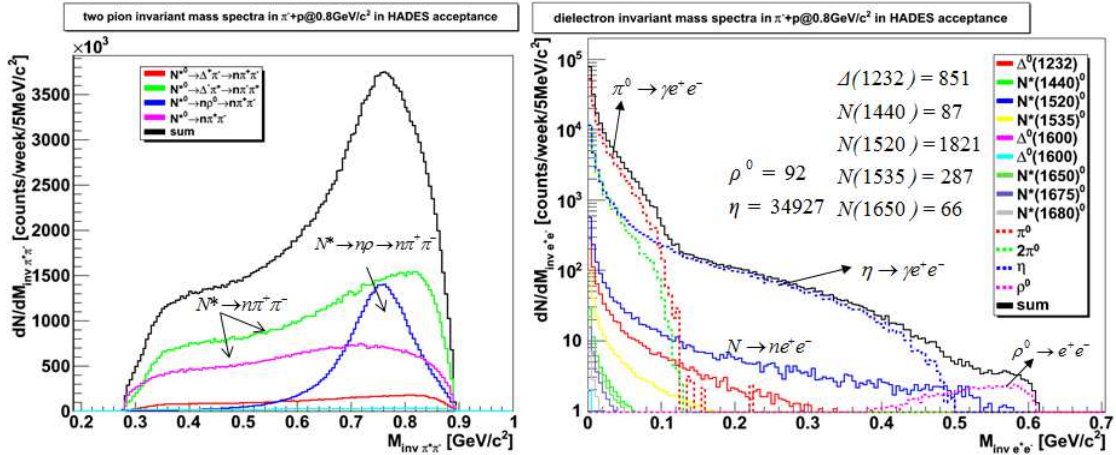}}
 \caption{{\bf Left:} Invariant mass spectra of $\pi^{+}\pi^{-}$ pairs from simulations of $\pi^{-}+p
\rightarrow n\pi^{+}\pi^{-}$ reactions at 0.8 GeV/c and estimated count rates per 1 week of experiment. {\bf
Right:} Inclusive invariant mass spectra of $e^{+}e^{-}$ pairs from simulations of $\pi^{-}+p$ reactions at
0.8 GeV/c and estimated count rates per 1 week of experiment.}
\end{figure}

\Section{Results}
The poster shows simulations and feasibility studies for future experiments with the pion beams. First the
total cross section to produce baryon resonances up to $\Delta^{0}(1950)$ [4] where calculated. Next step
consisted in Dalitz decay simulations of the baryonic resonances , where two electromagnetic decay models of
baryons were compared. The first one where they are treated as point like [5] and the other one using the eVMD
(extended Vector Meson Dominance) model for the electromagnetic form factors [6]. In addition, the production
and decay of $\rho, \omega$ and $\eta$ mesons is included in the simulations. Finally, the HADES spectrometer
acceptance for dilepton pairs was determined. This allowed to produce inclusive invariant mass spectra inside
HADES acceptance and estimate count rate per one week of experiment at two different energies (fig. 1 right).

A second type of simulations was devoted to two pion production in $\pi^{-}+p$ reactions. Like in the case of
dilepton analysis, cross sections and branching ratios were determined for bench mark channels. Next the HADES
acceptance for $\pi^{+}\pi^{-}$ pairs and the inclusive $\pi^{+}\pi^{-}$ invariant mass spectra with estimated
count rate per one week of experiment have been presented.

\Section{Perspectives}
Pion beams at GSI are available for many years and used by other groups (e.g. FOPI coll.). After the heavy ion
campaign scheduled for nearest future, the HADES collaboration is preparing a physics program for experiments
with pion beams. The simulations will be further developed by including Dalitz decay of higher lying
resonances and by studying exclusive channel analysis like $\pi^{-}p \rightarrow ne^{+}e^{-},\;\pi^{-}p
\rightarrow n\pi^{+}\pi^{-},\;\pi^{-}p \rightarrow n\eta$. The aim is to define in detail the physics case and
beam momenta for these future experiments.

\coltoctitle{Two-Photon Physics at KLOE/KLOE-2}
\authortochead{I.~Prado~Longhi}
\label{IPL}

\setcounter{chapter}{1}
\setcounter{section}{0}
\maketitle

\Section{Introduction}
A data sample of integrated luminosity $\mathcal{L}=242.5\,\mbox{pb}^{-1}$ collected at $\sqrt{s}=1$ GeV with
the KLOE detector at the DA$\Phi$NE $\phi$-factory has been analyzed in order to study the $e^+e^-\to e^+e^-X$
reactions, $X$ being the $\eta$ meson or the $\pi^0\pi^0$ state produced in the scattering of two quasi-real
photons.  The measurement is done with final state $e^\pm$ going along the beam pipe and escaping detection.
In this case the virtual photons are quasi-real and the overall cross section can be factorized as the
$\gamma\gamma$ subprocess cross section times a $\gamma\gamma$ luminosity function, \mbox{$\sigma_{e^+e^-\to
e^+e^-X}=\int\sigma_{\gamma\gamma\to X}(w)\frac{dL_{\gamma\gamma}}{dw}dw$}~\cite{ref:bro}. 

\Section{\boldmath$\gamma\gamma\to\eta$}
Two independent analyses have been performed looking for $\gamma\gamma\to\eta$ production with
$\eta\to\pi^+\pi^-\pi^0$ and $\eta\to\pi^0\pi^0\pi^0$. In both analyses the signal has been simulated with a
Monte Carlo generator based on the complete matrix element calculation which allows the full phase space
generation~\cite{ref:ngu}.\\

\noindent{\boldmath$\gamma\gamma\to\eta\to\pi^+\pi^-\pi^0$}.
Events are selected asking for two photons from a $\pi^0$ decay and two tracks with opposite curvature coming
from the interaction point, the $\pi^\pm$ mass  being assigned to each track.  A kinematic fit is performed
using the $\eta\to\pi^+\pi^-\pi^0$ decay hypothesis; the fit returns a $\chi^2$ value and events with
$\chi^2>20$ are rejected. Further cuts are applied to suppress processes with photons and $e^+e^-$ in the
final state. The number of signal events is extracted from the fit to the squared missing mass distribution
$M_{miss}^2$ using Monte Carlo distribution for both signal and backgrounds (Fig.~\ref{fig:etapm0}). From the
number of signal events given by the fit  one obtains the cross section $\sigma(e^+e^-\to e^+e^-\eta,
\sqrt{s}=1\,\mbox{GeV})=(41.7 \pm 4.0_{\mbox{stat}})$ pb, the systematic error being under evaluation. Work is
in progress to extract the partial width $\Gamma_{\eta\to\gamma\gamma}$.

The main irreducible background is due to $e^+e^-\to\eta\gamma\to\pi^+\pi^-\pi^0\gamma$ events with the
monochromatic photon ($E_\gamma=350$ MeV) lost in the beam pipe. In order to normalize this process properly a
dedicated analysis has been done using the same  $\mathcal{L}=242.5\,\mbox{pb}^{-1}$ data sample, selecting
events with three photons in the final state and performing a kinematic fit asking for energy and momentum
conservation; improved variables have been used to fit to data distribution of the energy of the monochromatic
photon. The preliminary result for the cross section is \mbox{$\sigma(e^+e^-\to \eta\gamma,
\sqrt{s}=1\,\mbox{GeV})=(0.866\pm 0.009_{\mbox{stat}}\pm 0.093_{\mbox{syst}})$}~nb~\cite{ref:kloe}.\\

\begin{figure}[htb]
 \centering
 \includegraphics[scale=0.4]{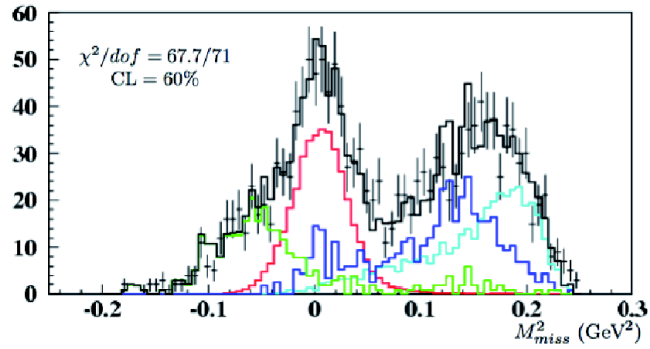}  
 \caption{\label{fig:etapm0}$\gamma\gamma\to\eta\to\pi^+\pi^-\pi^0$ analysis: squared missing mass
distributions for data (points) and Monte Carlo normalized according to fit result (solid lines). Colour code:
red~=~$\eta\gamma$, light~blue~=~signal, blue~=~$\omega\pi^0$, green~=~$e^+e^-\gamma$ (preliminary plot).}
\end{figure}

\noindent{\boldmath$\gamma\gamma\to\eta\to\pi^0\pi^0\pi^0$}.  
Events are selected with six photons from the interaction point (prompt photons); the photons are paired
choosing the combination which minimizes the $\chi^2$ variable
\begin{equation}
\chi^2_{\gamma\gamma}=\frac{\left(m_{\pi}-m_{ij}\right)^2}{\sigma^2_{ij}}+
\frac{\left(m_{\pi}-m_{kl}\right)^2}{\sigma^2_{kl}}+
\frac{\left(m_{\pi}-m_{st}\right)^2}{\sigma^2_{st}},\label{eq:chipair}
\end{equation} 
with $m_{ij}$ the two-photon invariant masses and $\sigma_{ij}$ their resolutions. A kinematic fit is
performed constraining the six-photon invariant mass to the $\eta$ meson mass and asking for the space-time
relation $t-|\overrightarrow{r}|/c = 0$ to be satisfied for each photon; a cut is then applied on the value of
the kinematic fit $\chi^2$. Events with tracks in the drift chamber are rejected; a cut is applied on the
energy of the most energetic photon, to reject $\eta\gamma$ events. The squared missing mass distribution
shows an excess of events well reproduced by the $e^+e^-\to e^+e^-\eta\to e^+e^-\pi^0\pi^0\pi^0$ Monte Carlo.
The fit gives normalization for the Monte Carlo signal and the backgrounds distributions
(Fig.~\ref{fig:eta}~left). The preliminary result for \mbox{$e^+e^-\to e^+e^-\eta$} production cross section
at $\sqrt{s}=1$ GeV is \mbox{$\sigma(e^+e^-\to e^+e^-\eta, 1\,\mbox{GeV})=(37.0\pm 1.4_{\mbox{stat}}\pm
2.2_{\mbox{syst}})$}~pb, to be compared with the result obtained studying the $\eta\to \pi^+\pi^-\pi^0$ decay
channel. From $\eta\gamma$ normalization a preliminary value for the cross section
$\sigma(e^+e^-\to\eta\gamma,\sqrt{s}=1\,\mbox{GeV})=(0.875 \pm 0.018_{\mbox{stat}}\pm
0.035_{\mbox{syst}})\,\mbox{nb}$ is evaluated, in agreement with the measurement done in the
$\eta\to\pi^+\pi^-\pi^0$ decay channel. This value is shown in Fig.~\ref{fig:eta} (right) together with the
SND experimental results~\cite{ref:ach} as a function of $\sqrt{s}$. 

\begin{figure}
\centerline{\includegraphics[scale=0.27]{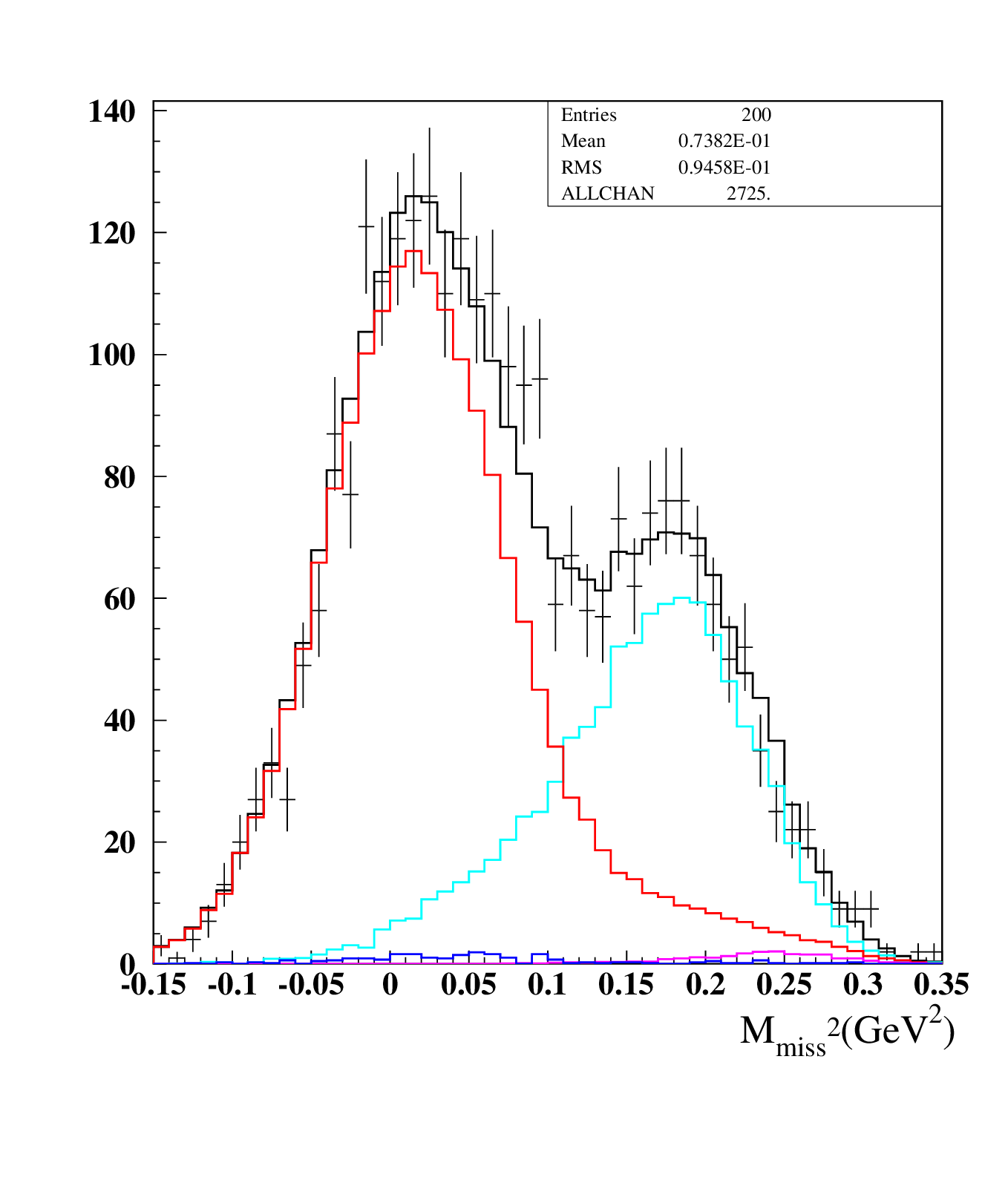}\hfill%
            \includegraphics[scale=0.27]{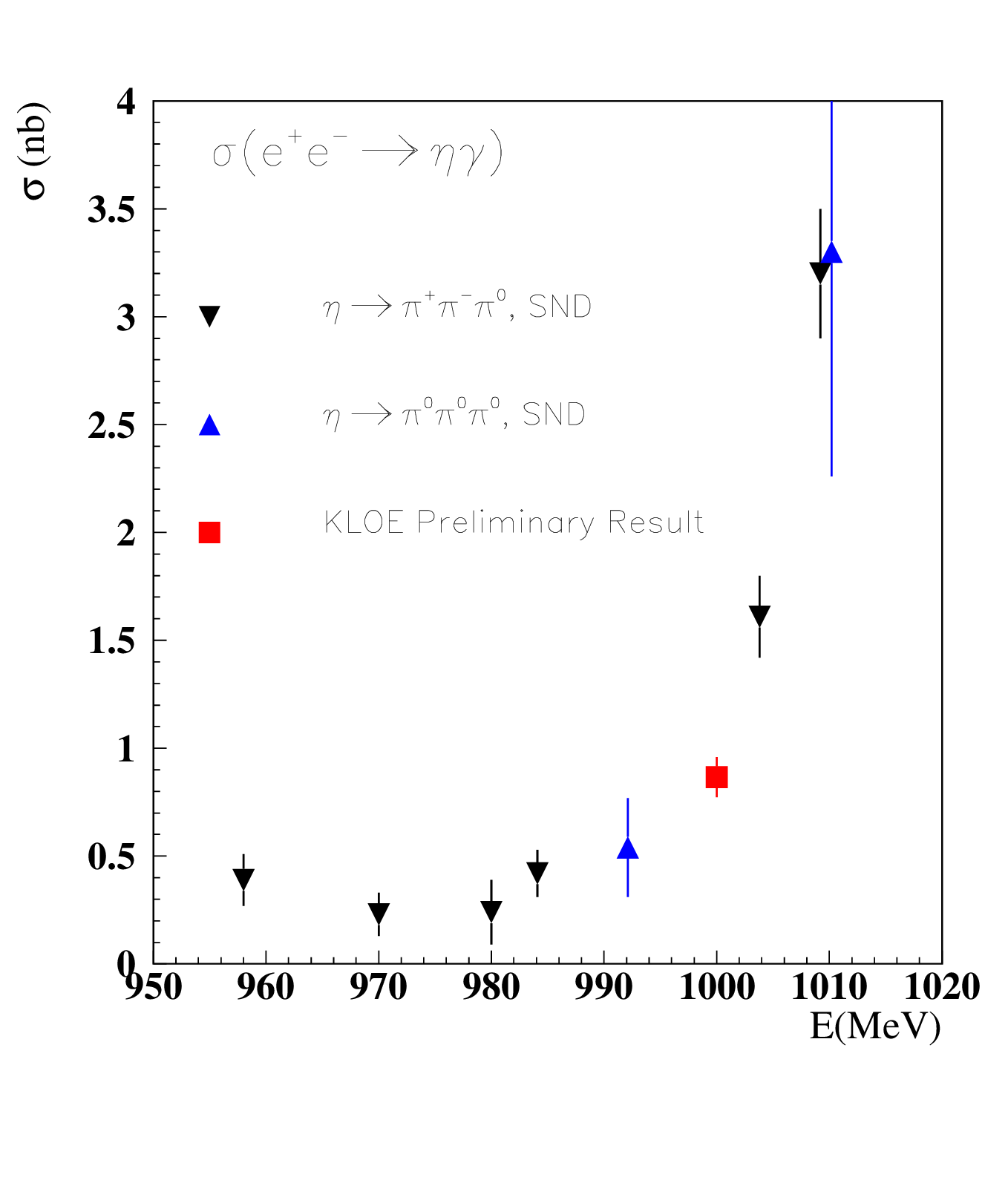}}
\caption{\label{fig:eta}$\gamma\gamma\to\eta\to\pi^0\pi^0\pi^0$ analysis. Left: squared missing mass
distributions for data (points) and Monte Carlo normalized according to fit result (solid lines). Colour code:
red~=~$\eta\gamma$, light blue~=~signal, blue~=~$\omega\pi^0$ (preliminary plot). Right: preliminary KLOE
result for $\sigma(e^+e^-\to\eta\gamma)$ at $\sqrt{s}=1\,\mbox{GeV}$, (red point, with statistical uncertainty
only) and SND results~\cite{ref:ach} at several values of $\sqrt{s}$.}
\end{figure}

\Section{\boldmath$\gamma\gamma\to\pi^0\pi^0$}
The main goal of this analysis is to investigate the low $\pi^0\pi^0$ invariant mass region, just above the
production threshold, where a contribution from the $\sigma(600)$ scalar meson as a resonant intermediate
state is expected. The main background processes are annihilation reactions to states with four or more prompt
photons: $e^+e^-\to K_SK_L,\eta\gamma,\omega\pi^0,f_0\gamma,a_0\gamma$. In addition, the $e^+e^-\to
\gamma\gamma$ process is also considered as a source of background due to possible splitting of the photons
energy deposits in the calorimeter. Events with 4 prompt photons are selected. The photons are paired as in
the $\gamma\gamma\to\eta\to\pi^0\pi^0\pi^0$ analysis, and events with $\chi^2_{\gamma\gamma}>4$ are rejected:
the effect of this selection is shown in Fig.~\ref{fig:pi0pi0} (left). Events with no tracks in the drift
chamber are selected; cuts on photons energies and on the transverse momentum of the four-photon system are
applied; a cut is applied on the ratio of the sum of the energies of four photons to the total energy deposit
in the calorimeter, to reject $K_SK_L$ events with large amount of non-prompt energy released in the detector.
The four-photon invariant mass spectrum for the selected data sample is shown in Fig.~\ref{fig:pi0pi0} (right)
together with the normalized Monte Carlo background. An excess of events in the low invariant mass region
signals $\pi^0\pi^0$ production processes.

\begin{figure}
  \centerline{\includegraphics[scale=0.27]{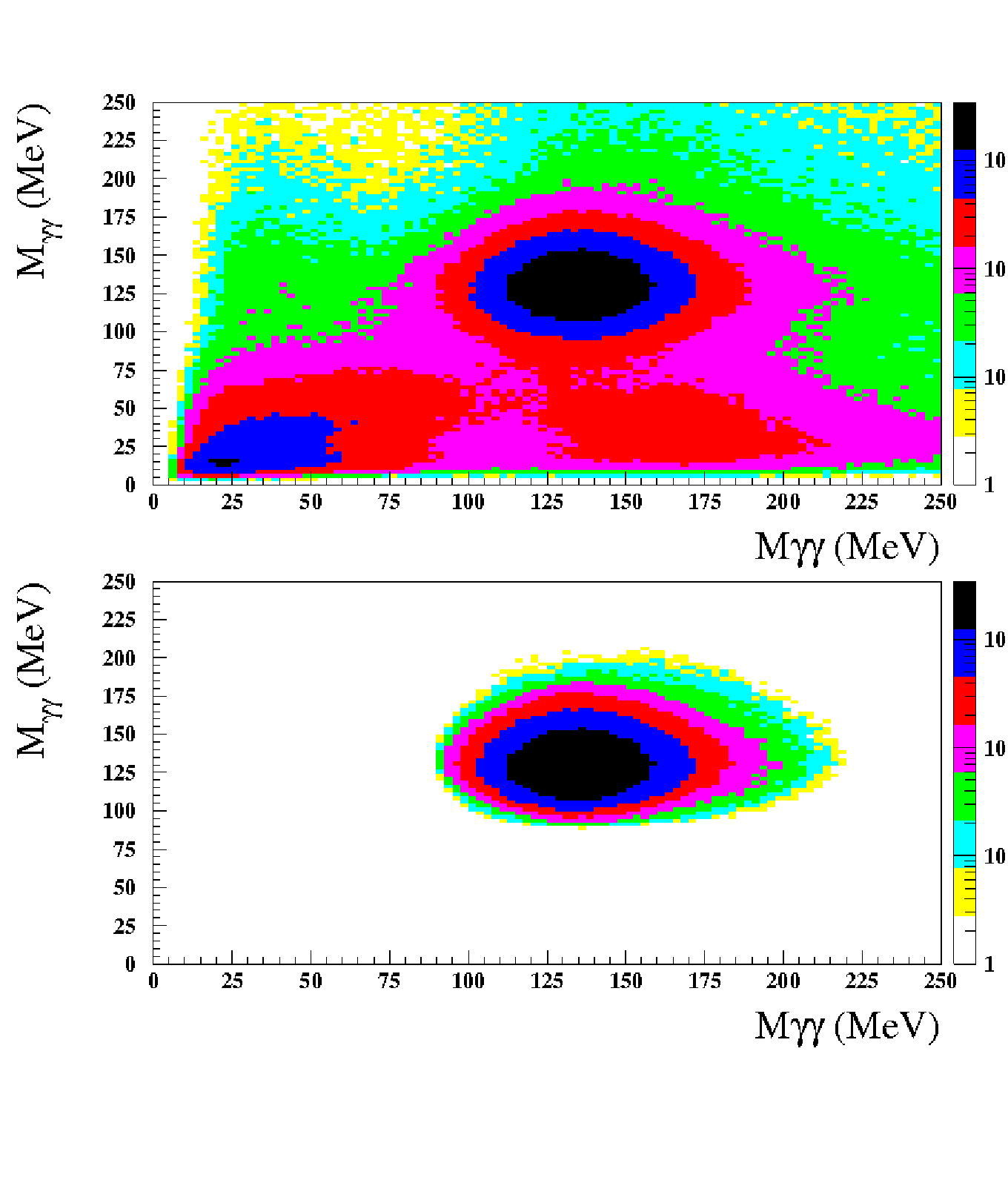}\hfill%
              \includegraphics[scale=0.27]{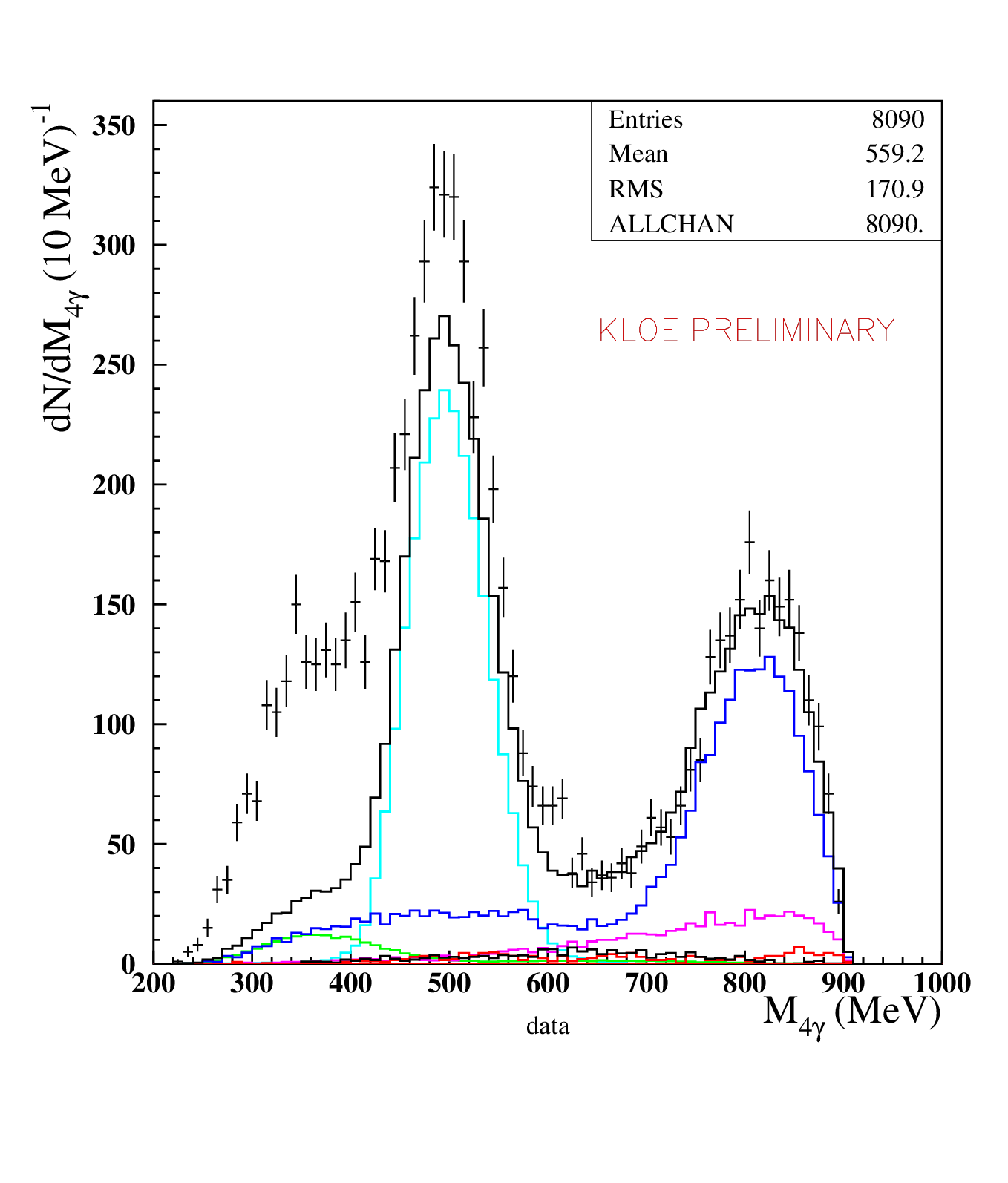}}
\caption{\label{fig:pi0pi0}$\gamma\gamma\to\pi^0\pi^0$ analysis. Left: two-photon pairs invariant masses
before (top) and after (bottom) $\chi^2_{\gamma\gamma}$ rejection; selected events have both photons pairs
invariant masses centered around $\pi^0$ mass value. Right: four-photon invariant mass spectrum for data
(points with error bars) and backgrounds Monte Carlo. Colour code: light blue~=~$K_SK_L$,
blue~=~$\omega\pi^0$, violet~=~$f_0\gamma$, green~=~$\eta\gamma$, red~=~$\gamma\gamma$ (preliminary plot).}
\end{figure}

\Section{Conclusions}
The reported analyses show excess of events with respect to the annihilation processes in the region where the
signal from $\eta$ or $\pi^0\pi^0$ production via $\gamma\gamma$ interaction is expected. For the $e^+e^-\to
e^+e^-\eta$ cross section at $\sqrt{s}=1$ GeV two compatible measurements are obtained studying the
$\eta\to\pi^+\pi^-\pi^0$ and the $\eta\to\pi^0\pi^0\pi^0$ decay channels. For the $\gamma\gamma\to\pi^0\pi^0$
process work is in progress to determine the signal efficiency and the $\gamma\gamma$ luminosity function in
order to extract the cross section and compare it with the only available measurement in
literature~\cite{ref:mar}.

The forthcoming data taking with KLOE-2 detector, equipped with dedicated small angles taggers for $e^\pm$ in
the final state, will give precious additional information on $\gamma\gamma$ hadron production at low
energy~\cite{ref:ame}.

\coltoctitle{\boldmath Roy--Steiner Equations for $\gamma\gamma\to\pi\pi$}
\authortochead{M.~Hoferichter}
\label{HrM}

\setcounter{chapter}{1}
\setcounter{section}{0}
\maketitle

\Section{Introduction}
Presently, there are several experiments dedicated to the study of low-energy pion--photon interactions. At
COMPASS, pion Compton scattering is being investigated by means of the Primakoff effect in $\pi^- Z \to\pi^- Z
\gamma$ with the aim to extract pion polarizabilities~\cite{COMPASS}, while at KLOE the reaction
$\gamma\gamma\to\pi^0\pi^0$ is being measured in $e^+e^-\to e^+e^-\pi^0\pi^0$~\cite{KLOE}. Measurements of the
cross section for $\gamma\gamma\to\pi^0\pi^0$ at low energies are particularly valuable, since they provide an
alternative way to access the pion polarizabilities and the absence of Born-term contributions in this channel
renders the cross section sensitive to the influence of the $\sigma$ resonance. However, in both cases one
needs theory input to reliably extract the information of interest, as the polarizabilities are defined at the
threshold for pion Compton scattering, and studying the $\sigma$ even requires analytic continuation far into
the complex plane. The necessary extrapolations should be performed respecting all available constraints,
i.e.~analyticity, unitarity, crossing symmetry, and gauge invariance. Roy--Steiner equations for
$\gamma\gamma\to\pi\pi$ provide a framework that meets all these requirements~\cite{HPS11}.\footnote{The
following presentation involves significant textual overlap with~\cite{HPS11_proc}.}  

The Roy equations for $\pi\pi$ scattering \cite{Roy} are a coupled system of partial wave dispersion relations
that respects analyticity, unitarity, and crossing symmetry of the scattering amplitude. In recent years,
partial wave dispersion relations in combination with unitarity (and chiral symmetry) have been used for
high-precision studies of low-energy processes, both in $\pi\pi$ \cite{ACGL,madrid} and $\pi K$ \cite{piK}
scattering. An important application of $\pi\pi$ Roy equations in combination with Chiral Perturbation Theory
(ChPT) was the precise prediction of the pole parameters of the $\sigma$ resonance \cite{CCL}
\begin{equation}
M_\sigma=441^{+16}_{-8}\,{\rm MeV},\qquad \Gamma_\sigma=544^{+18}_{-25}\,{\rm MeV}.
\end{equation}
The reaction $\gamma\gamma\to\pi\pi$ provides an alternative to $\pi\pi$ scattering for the excitation of the
$\sigma$. In particular, Roy-equation techniques in $\gamma\gamma\to\pi\pi$ allow us to constrain the
$\sigma$'s two-photon width $\Gamma_{\sigma\gamma\gamma}$ at a similar level of rigor as $M_\sigma$ and
$\Gamma_\sigma$ based on $\pi\pi$ Roy equations~\cite{HPS11}.

\Section{Roy equations for $\boldsymbol{\pi\pi}$ scattering}
Roy equations for $\pi\pi$ scattering are obtained by starting from a twice-subtracted dispersion relation at
fixed Mandelstam $t$, determining the $t$-dependent subtraction constants by means of crossing symmetry, and
finally performing a partial wave expansion. This leads to a coupled system of integral equations for the
$\pi\pi$ partial waves $t^I_J(s)$ with isospin~$I$ and angular momentum~$J$
\begin{equation}
\label{pipi_roy}
t_J^I(s)=k_J^I(s)+\sum\limits_{I'=0}^{2}\sum\limits_{J'=0}^\infty\int\limits_{4M_{\pi}^2}^\infty
\text{d}s'K_{JJ'}^{II'}(s,s')\text{Im}\, t_{J'}^{I'}(s'),
\end{equation}
where $K_{JJ'}^{II'}$ are known kinematical kernel functions and the $\pi\pi$ scattering lengths---the only
free parameters---appear in the subtraction term $k_J^I$. Assuming elastic unitarity
\begin{equation}
\text{Im}\, t_{J}^{I}(s)=\sigma(s)|t_{J}^{I}(s)|^2,\quad
t_{J}^{I}(s)=\frac{e^{2i\delta^I_J(s)}-1}{2i\sigma(s)}, \quad \sigma(s)=\sqrt{1-\frac{4M_{\pi}^2}{s}},
\end{equation}
\eqref{pipi_roy} translates into a coupled integral equation for the phase shifts $\delta^I_J$ themselves.

\Section{Roy--Steiner equations for $\boldsymbol{\gamma\gamma\to\pi\pi}$}
Crossing symmetry in this case is less restrictive than for $\pi\pi$ scattering, as it couples
$\gamma\gamma\to\pi\pi$ to pion Compton scattering $\gamma\pi\to\gamma\pi$, which we will consider as the
$s$-channel process. Roy--Steiner equations are then most conveniently constructed based on hyperbolic
dispersion relations \cite{HS_73}. The resulting system of integral equations couples the
$\gamma\gamma\to\pi\pi$ partial waves  $h^I_{J,\pm}(t)$ to the $\gamma\pi\to\gamma\pi$ partial waves
$f^I_{J,\pm}(s)$ (with photon helicities $\pm$), e.g.
\begin{equation}
h^I_{J,-}(t)=\tilde N_J^-(t)+\frac{1}{\pi}\int\limits_{M_{\pi}^2}^\infty\text{d} s'\sum\limits_{J'=1}^\infty
\tilde G_{JJ'}^{-+}(t,s')\text{Im}\, f^I_{J',+}(s')+\frac{1}{\pi}\int\limits_{4M_{\pi}^2}^\infty\text{d} t'
\sum_{J'}\tilde K_{JJ'}^{--}(t,t')\text{Im}\, h^I_{J',-}(t'),
\end{equation}
where $\tilde N_J^-(t)$ includes the QED Born terms. Subtracting at $t=0$, $s=M_{\pi}^2$, the subtraction
constants directly correspond to pion polarizabilities. In the once-subtracted case, one needs the dipole
polarizabilities $\alpha_1\pm\beta_1$, while a second subtraction requires in addition knowledge of the
quadrupole polarizabilities $\alpha_2\pm\beta_2$.

Elastic unitarity is also less restrictive than for $\pi\pi$ scattering, since the unitarity relation is
linear in $h^I_{J,\pm}$
\begin{equation}
\text{Im}\, h^I_{J,\pm}(t)=\sigma(t)h^I_{J,\pm}(t)t_{J}^{I}(t)^*.
\end{equation}
Below inelastic thresholds the phase of $h^I_{J,\pm}$ coincides with $\delta^I_J$ (``Watson's theorem'').
Assuming this phase to be known, the equations thus reduce to a Muskhelishvili--Omn\`es problem for
$h^I_{J,\pm}$ \cite{MuskhelishviliOmnes}.

\Section{Muskhelishvili--Omn\`es solution and results for $\boldsymbol{\Gamma_{\sigma\gamma\gamma}}$}
To solve the equations for $h^I_{J,\pm}$, we truncate the system at $J=2$. Furthermore, we assume the
amplitudes to be known above the matching point $t_{\rm m}=(0.98\,{\rm GeV})^2$. The solution can then be
written down in terms of Omn\`es functions
\begin{equation}
\Omega^I_J(t)=\exp\Bigg\{\frac{t}{\pi}\int\limits_{4M_{\pi}^2}^{t_{\rm m}}\text{d}
t'\frac{\delta^I_J(t')}{t'(t'-t)}\Bigg\}.
\end{equation}
We find that the solutions for different partial waves in general do not decouple, e.g.~the equation for the
$S$-wave involves spectral integrals over the $D$-waves as well \cite{HPS11}. This is a new result of our
dispersive treatment of $\gamma\gamma\to\pi\pi$ based on Roy--Steiner equations.

We approximate $\text{Im}\, f^I_{J,\pm}(s)$, which at low energies is dominated by multi-pion states, by a sum
of resonances \cite{GM_10}. Above the matching point we use a Breit--Wigner description of the $f_2(1270)$,
which dominates the cross section at higher energies. Within our formalism~\cite{HPS11} we derive a sum rule
for the $I=2$ polarizabilities, which---in combination with ChPT results for dipole and neutral-pion
quadrupole polarizabilities~\cite{GIS}--- produces an improved prediction 
\begin{equation}
(\alpha_2-\beta_2)^{\pi^\pm}=(15.3\pm 3.7)\cdot 10^{-4}\text{fm}^5
\end{equation}
for the charged-pion quadrupole polarizability. This sum-rule result together with the ChPT values for the
other polarizabilities~\cite{GIS} leads to the ``ChPT'' prediction for the total cross section of
$\gamma\gamma\to\pi^0\pi^0$ depicted in the left panel of Fig.~\ref{fig:results}. The result labeled ``GMM''
is found when we adopt the polarizability values of a recent fit of a two-channel Muskhelishvili--Omn\`es
representation to $\gamma\gamma\to\pi\pi$ cross section data \cite{GM_10}. The uncertainty due to the $\pi\pi$
phases represented by the grey band is estimated by varying between two recent state-of-the-art analyses based
on Roy and Roy-like equations~\cite{madrid,bern}. We see that especially for the twice-subtracted version the
agreement with experiment in the low-energy region is very good. Given sufficiently accurate data on $\gamma
\gamma\to\pi^0 \pi^0$ such an analysis could be used to extract pion polarizabilities.

Since we have shown that the $\sigma$ lies within the domain of validity of our Roy--Steiner
equations~\cite{HPS11}, this formalism allows for a reliable analytic continuation to the $\sigma$ pole. 
The main result of our analysis is shown in the right panel of Fig.~\ref{fig:results}: there is a correlation
between $\Gamma_{\sigma\gamma\gamma}$ and the $I=0$ pion polarizabilities that follows from Roy--Steiner
equations and input for the $\pi\pi$ phases alone. In combination with the ChPT-plus-sum-rule input for 
the polarizabilities, we obtain
\begin{equation}
\Gamma_{\sigma\gamma\gamma}=(1.7\pm 0.4)\,\text{keV}.
\end{equation}

\begin{figure}[htb]
  \begin{center}
   \includegraphics[width=0.495\textwidth]{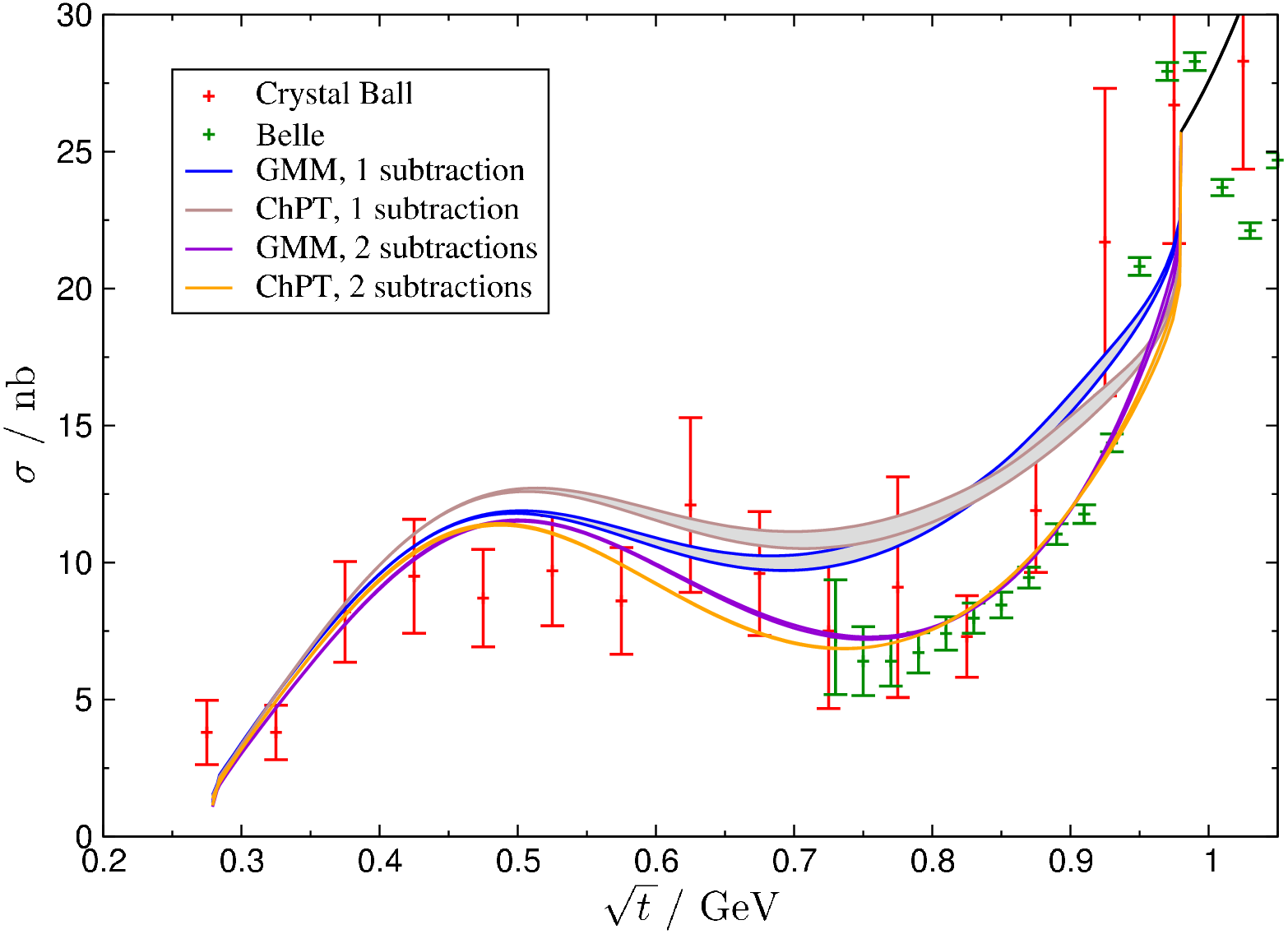}
    \includegraphics[width=0.495\textwidth]{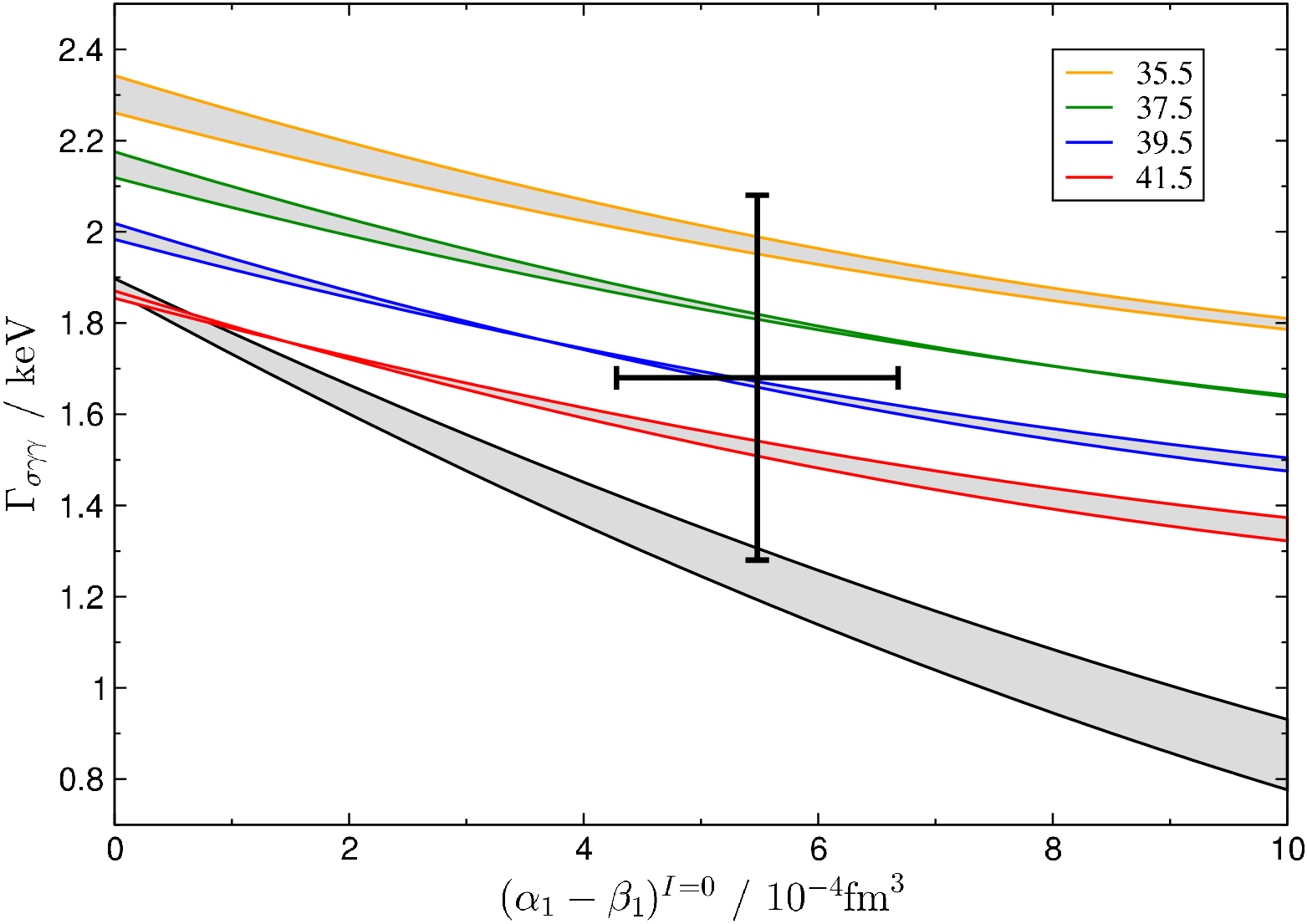}
    \caption{\label{fig:results} Total cross section for $\gamma\gamma\to\pi^0\pi^0$ for $|\cos\theta|\leq
0.8|$ (left) and $\Gamma_{\sigma\gamma\gamma}$ as a function of the $I=0$ pion polarizabilities (right). The
black line refers to the unsubtracted case and the colored lines to the twice-subtracted version with
$(\alpha_2-\beta_2)^{I=0}$ as indicated (in units of $10^{-4}{\rm fm}^5$). The grey bands represent the
uncertainty due to the $\pi\pi$ input. The cross corresponds to the twice-subtracted case plus ChPT input.}
  \end{center}
\end{figure}

\Section{Outlook}
The framework presented here should prove valuable for the interpretation of upcoming KLOE data for
$\gamma\gamma\to\pi^0\pi^0$. Especially in combination with the ongoing efforts at COMPASS concerning
$\gamma\pi^-\to\gamma\pi^-$, this will further improve knowledge of the pion polarizabilities and thus of
$\Gamma_{\sigma\gamma\gamma}$. 

\subsection*{Acknowledgements}
This research was supported by the DFG (SFB/TR 16), the program ``Kurzstipendien f\"ur DoktorandInnen'' of
the DAAD, the Bonn-Cologne Graduate School of Physics and Astronomy, the US Department of Energy (Office of
Nuclear Physics), and CONICET.

\coltoctitle{\boldmath Measurement of the Double Polarization Observable G \\ in $\pi^0$ and $\eta$
Photoproduction}
\authortochead{A.~Thiel}
\label{TA}

\setcounter{chapter}{1}
\setcounter{section}{0}
\maketitle

\vspace{-0.7cm}\Section{Introduction}
The excitation spectrum of the nucleon consists of several strongly overlapping resonances with different
strengths. To disentangle these resonances and to determine their contributions in the different final states,
a partial wave analysis has to be performed. The model-independent determination of all amplitudes can be
achieved by a ``complete experiment'', for which a set of 8 well chosen polarization observables is needed in
each final state for single pseudoscalar meson production\cite{ChiangTabakin1997}. This set consists of at
least 4 double polarization observables like the observable G, which has been measured with the CBELSA/TAPS
experiment in Bonn. In the experiment linearly or circularly polarized photons and a longitudinally or
transversally polarized target are provided, therefore several polarization observables are accessible.

\noindent The first results for the double polarization observable G, which utilizes linearly polarized
photons on a longitudinally polarized target, are shown here for $\pi^0$ and $\eta$ photoproduction off the
proton.

\Section{Experimental Setup}
The Crystal Barrel experiment is located in Bonn at the electron accelerator ELSA, which provides polarized
or unpolarized electrons up to 3.5~GeV. Via bremsstrahlung an unpolarized, linearly polarized or circularly
polarized photon beam can be produced, whose energies are determined in a tagging system. The polarized target
is surrounded by the main calorimeter, the Crystal Barrel detector, while the forward direction is covered by
the Mini-TAPS detector up to small angles. Each calorimeter is equipped with detectors for charge
identification.

\noindent This setup allows an angular coverage of nearly $4\pi$ with a high detection efficiency for photons
and is therefore well suited for the measurement of reactions with neutral mesons like $\pi^0$ and $\eta$ in
the final state.

\Section{Selection of the Reactions}
Two different polarization settings where used for this analysis with coherent edge at 950~MeV and at
1150~MeV. With these two settings, which are shown in figure \ref{pic:invm}, left, the photon energy range of
$E_\gamma = 700~$MeV$ - 1100$~MeV can be covered with a high degree of photon polarization.

\noindent For the selection of the two photon decay of the $\eta$ and $\pi^0$, first, the reactions were
checked for two neutral particles and one charged for the recoiling proton. The time information of all
particles of the reaction were used, to confirm that the observed particles originated from the same reaction.
Also, the data was corrected for random time background. To ensure that the meson and the recoiling proton
were detected in opposite directions in the center of mass system, the $\phi$ and $\theta$ angle difference
were analyzed. By treating the proton as a missing particle and plotting the missing mass of the system, a
peak at the mass of the proton could be observed. In the invariant mass of the two photons, which is shown in
figure \ref{pic:invm}, center, a clear signal for the $\pi^0$ and $\eta$ can be seen above nearly no
background. To choose the desired reaction, a cut on either meson was done.

\begin{figure}[htb]
 \includegraphics[width=0.33\textwidth]{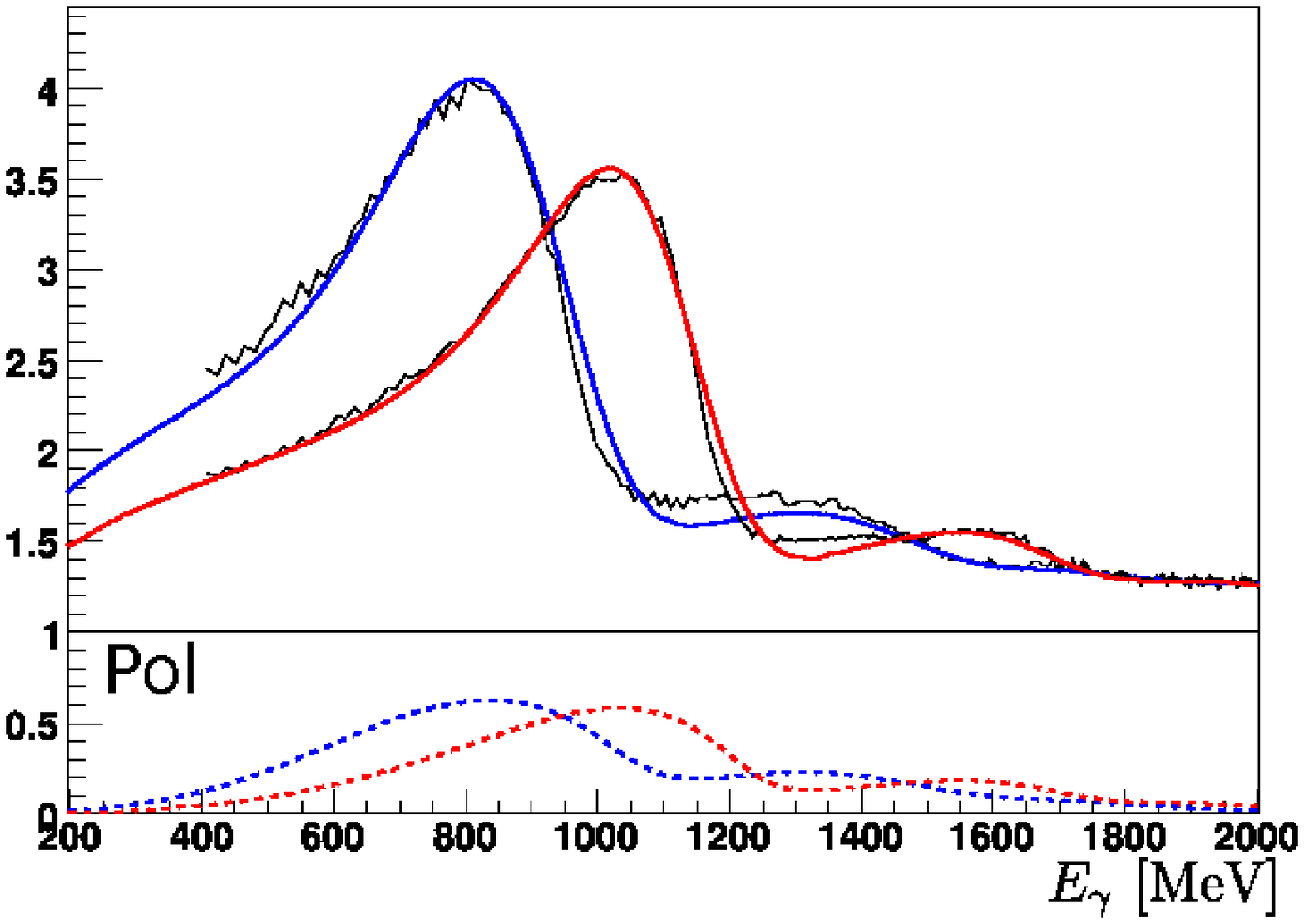}%
 \includegraphics[width=0.33\textwidth]{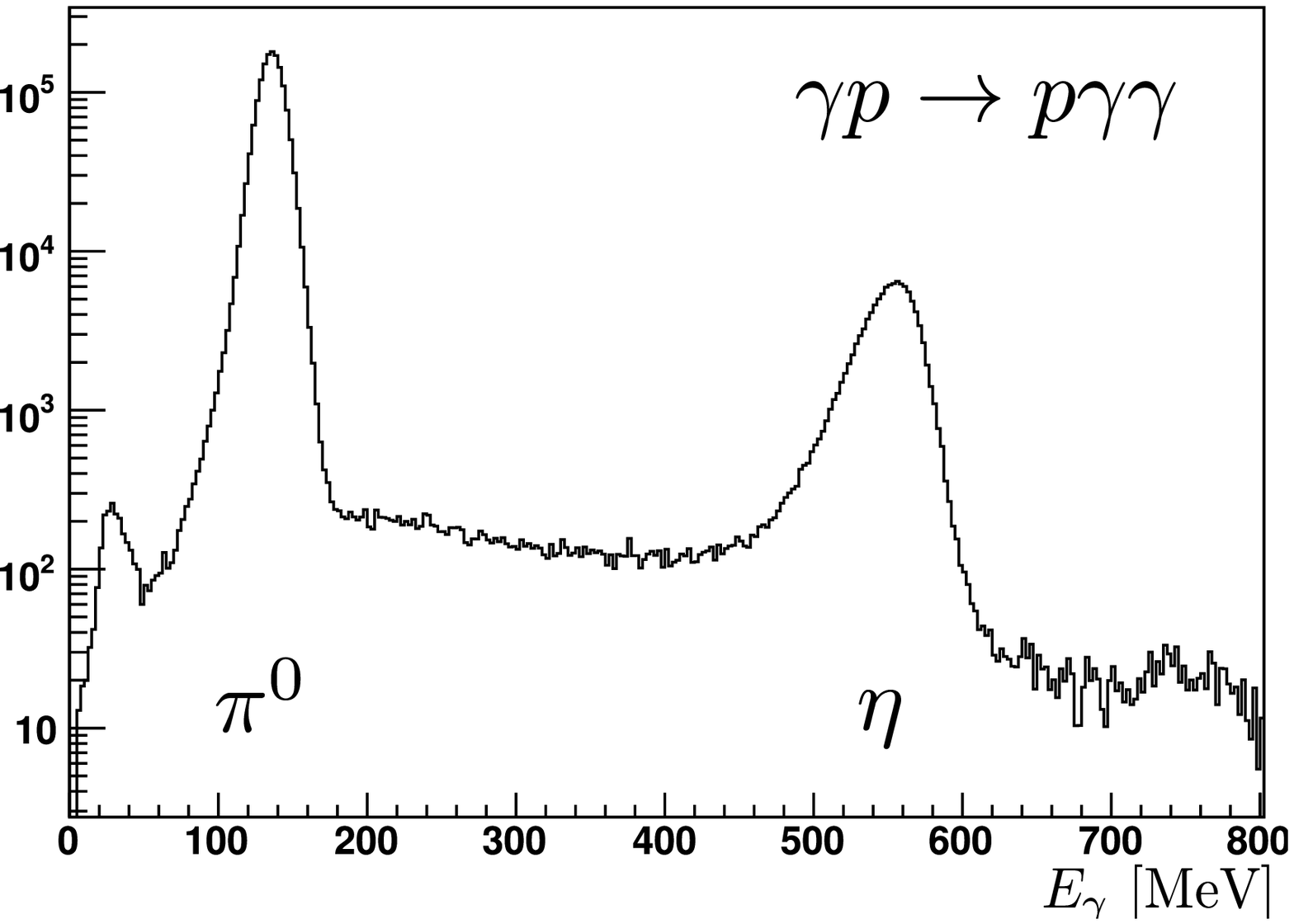}%
 \includegraphics[width=0.33\textwidth]{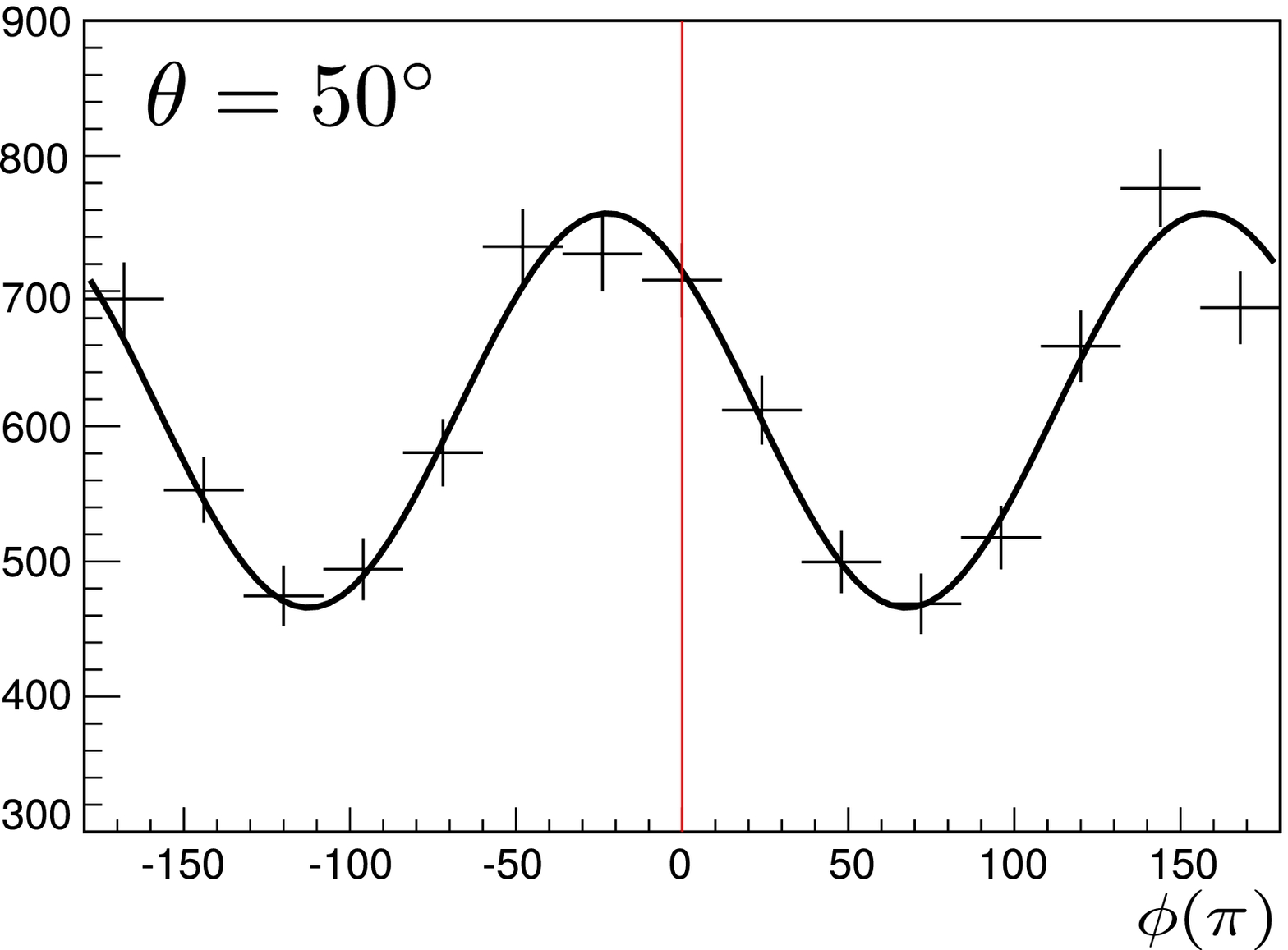}%
 \caption{\label{pic:invm}The two different settings of the coherent peaks for linearly polarized photons
(left). Invariant mass of the two photons in logarithmic scale (center) with the peaks for the $\pi^0$ and the
$\eta$ meson. Example fit to the $\phi(\pi^0)$ distribution of the data (right) to extract the observables.}
\end{figure}

\Section{Determination of the Polarization Observables}
In the frozen spin target, butanol is used as target material, which is composed of hydrogen, carbon and
oxygen. Because of that, reactions on free and on bound protons are contributing. The count rate for the
reactions on the butanol target can be written as
\begin{equation}
N(\theta,\phi)=(N_b+N_f)\cdot[1-\frac{N_b\Sigma_b+N_f\Sigma_f}{N_b+N_f}\cdot p_\gamma^{lin}\cos(2\phi)
+\frac{N_f}{N_b+N_f} p_z \cdot p_\gamma^{lin}G\sin(2\phi)]
\end{equation}
with $N_f(\theta)$ ($N_b(\theta)$) being the count rate on the free (bound) protons. The observables, which
can be measured are the beam asymmetries $\Sigma_b$ and $\Sigma_f$ and the double polarization observable G,
which can only be observed on the polarizable hydrogen protons. To successfully determine the double
polarization observable G and the beam asymmetry $\Sigma$ on the butanol the following equation was used to
fit the data:
\begin{equation}
N(\theta,\phi)=A\cdot[1-B\cdot\cos(2\phi) +C \cdot \sin(2\phi)]
\end{equation}
An example fit is shown in figure \ref{pic:invm}, right.

\Subsection{The Beam Asymmetry $\Sigma$}
The fit parameter $B$ allows the extraction of the beam asymmetry on the free ($\Sigma_f$) and on the bound
protons ($\Sigma_b$). By reducing the fraction of bound protons in the data set, the measured beam asymmetry
converges to the beam asymmentry on the free proton and can be compared to previous measurements. In figure
\ref{pic:beamasymm} the beam asymmetry for $\pi^0$ and $\eta$ photoproduction is shown, compared to previous
measurements and results of the different partial wave analyses. The data agrees well with the measurements on
the free proton, which indicates that the influence of the bound protons could successfully be reduced.
\begin{figure}[htb]
 \includegraphics[width=0.45\textwidth]{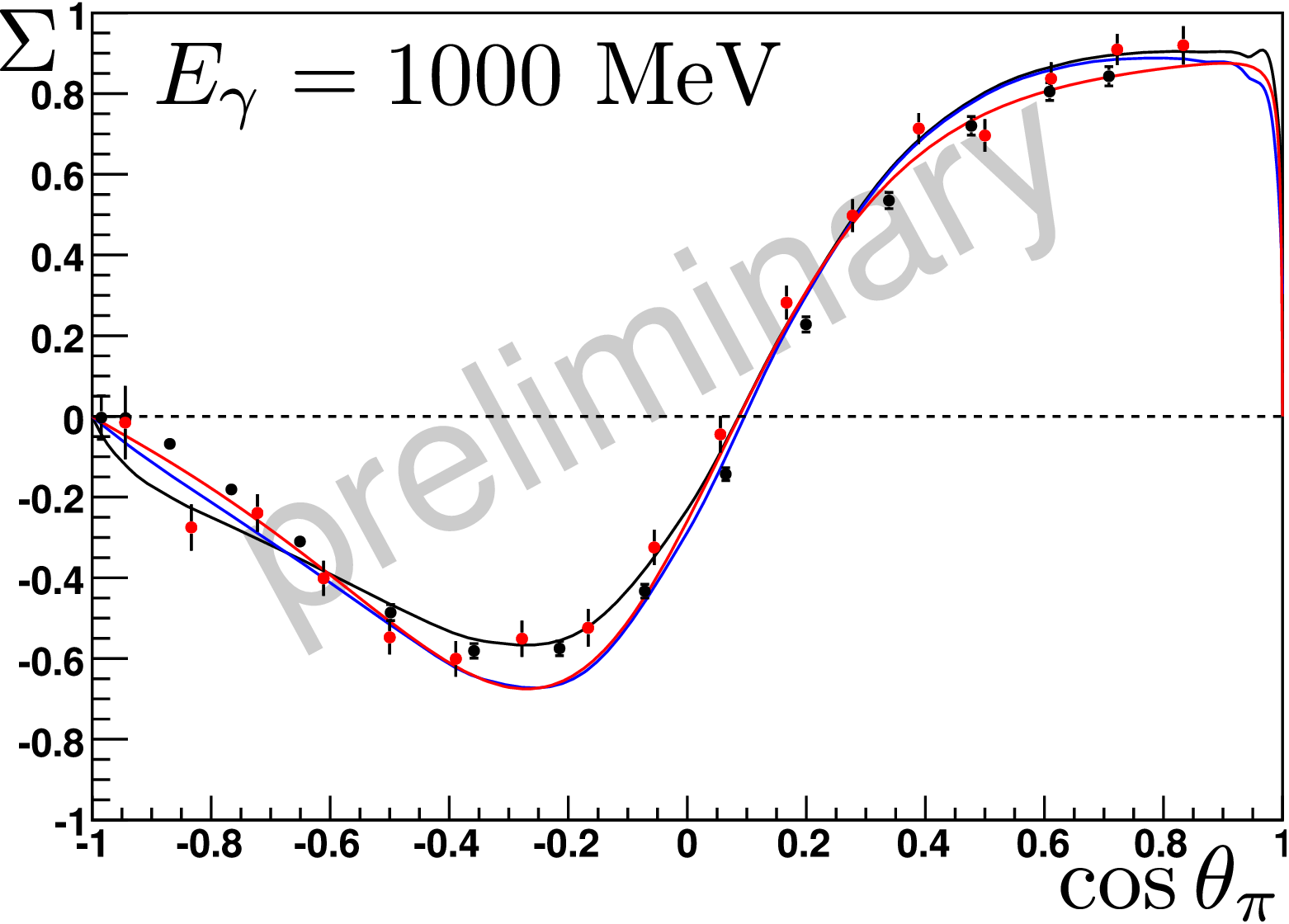}\hfill%
 \includegraphics[width=0.45\textwidth]{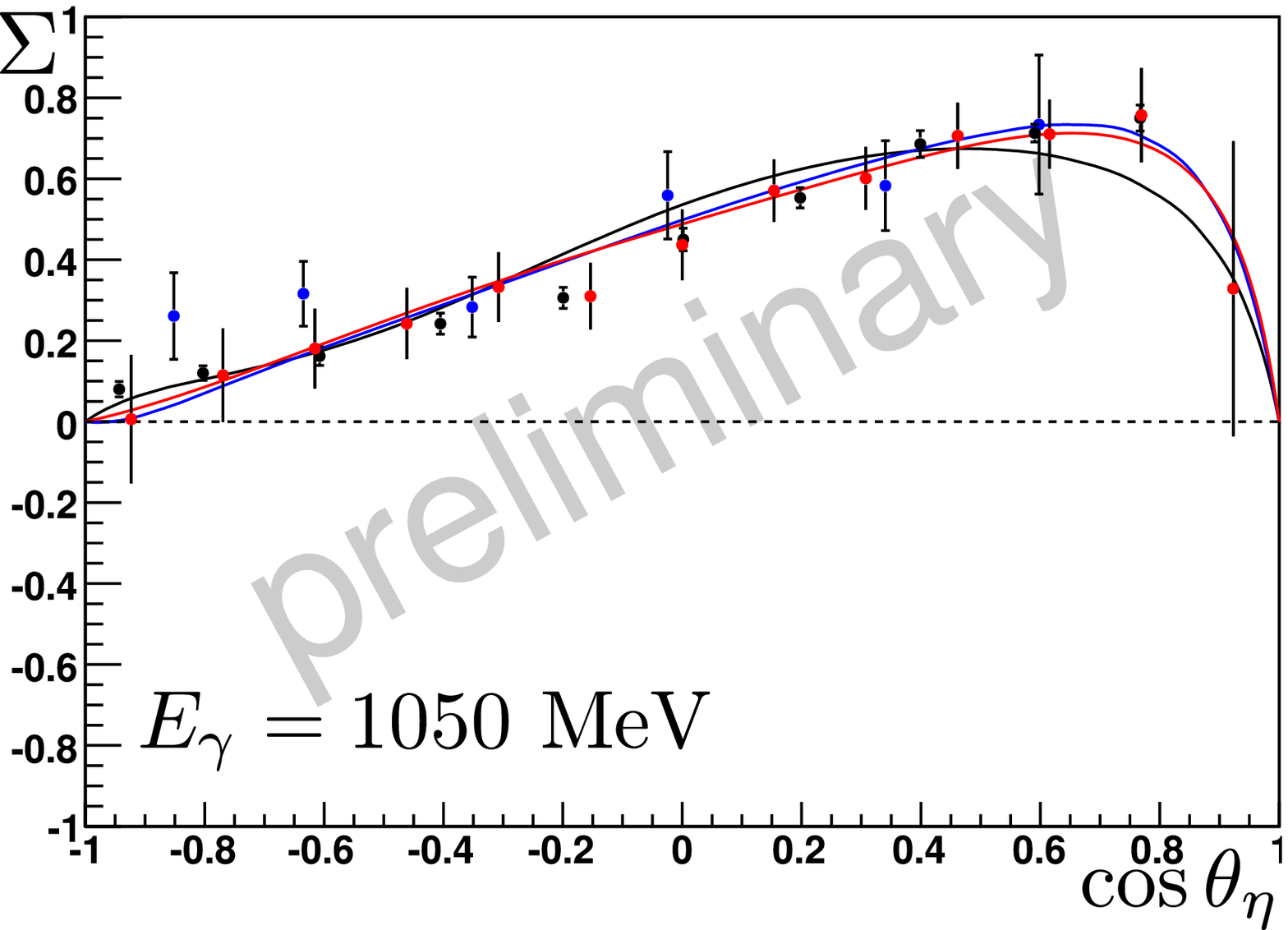}
 \caption{\label{pic:beamasymm}The beam asymmetry $\Sigma$ (red) for $\pi^0$- (left) and $\eta$-production
(right). The values are compared with the measurements of the GRAAL collaboration\cite{GRAAL} (black) and for
the $\eta$ with previous CBELSA/TAPS measurements\cite{Elsner} (blue). The lines give the solutions of the
partial wave analyses: MAID\cite{MAID} (black), SAID\cite{SAID} (blue) and BnGa\cite{BnGa} (red).}
\end{figure}

\Subsection{The Double Polarization Observable G}
From the fit parameter $C$ the double polarization observable G can be extracted, modified by the dilution
factor $D=\frac{N_f}{N_b+N_f}$. To determine this factor, data sets with hydrogen and carbon targets were
compared. The comparison of the missing mass for the different materials in fig. \ref{pic:dil}, left, shows
that the sum of the hydrogen and carbon measurements describes the butanol data.
\begin{figure}[htb]
 \includegraphics[width=0.32\textwidth]{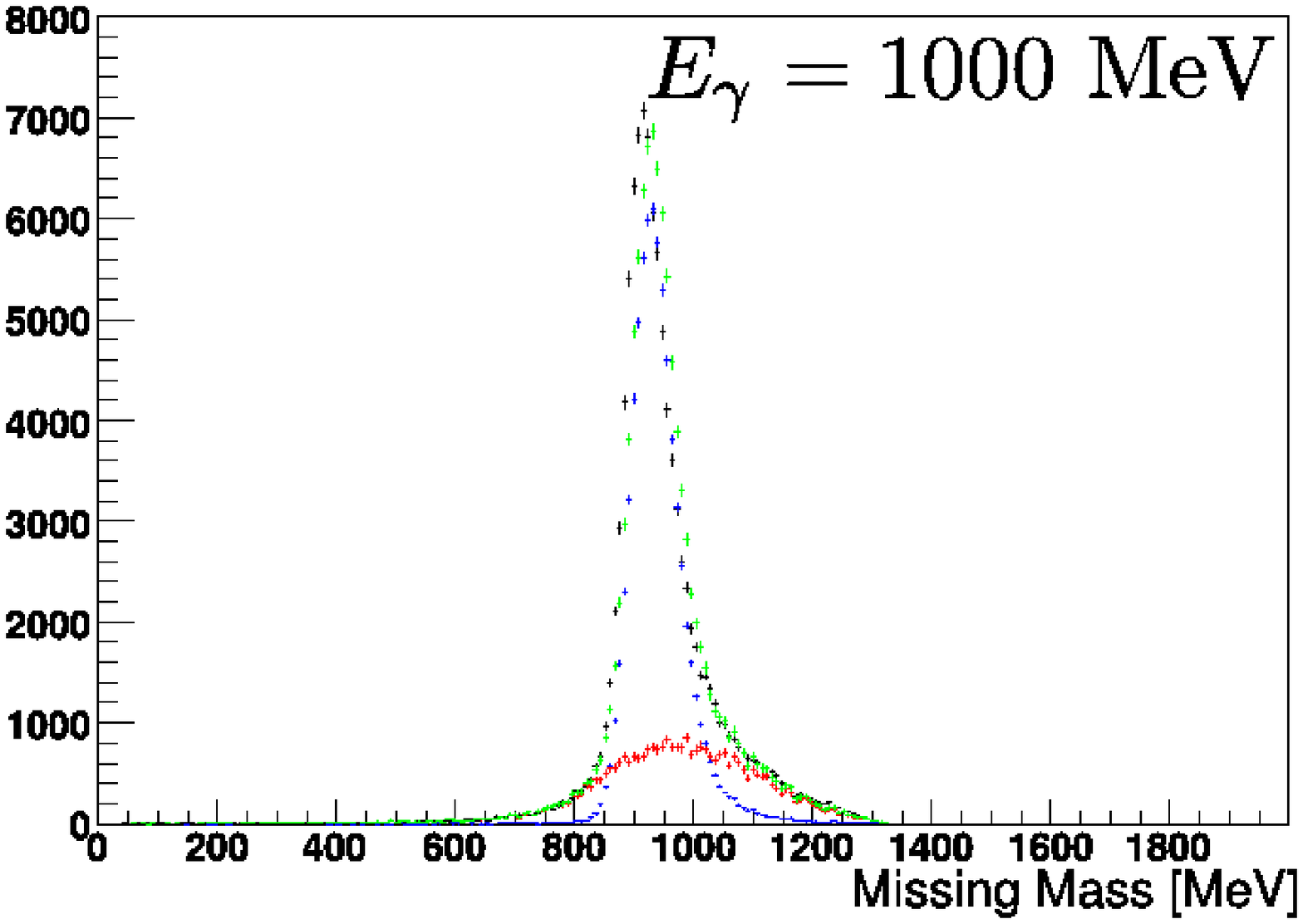}\hfill%
 \includegraphics[width=0.32\textwidth]{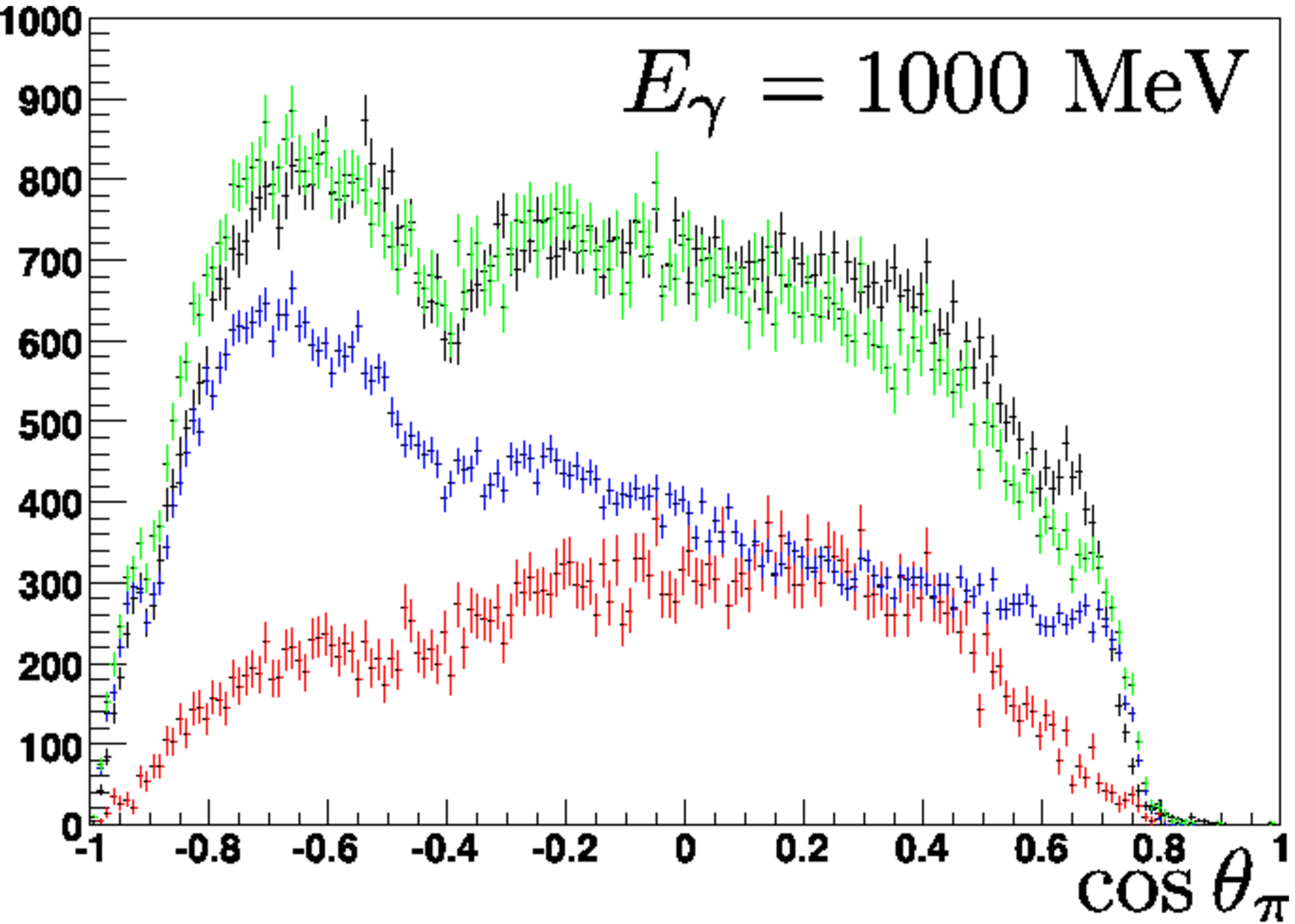}\hfill%
 \includegraphics[width=0.32\textwidth]{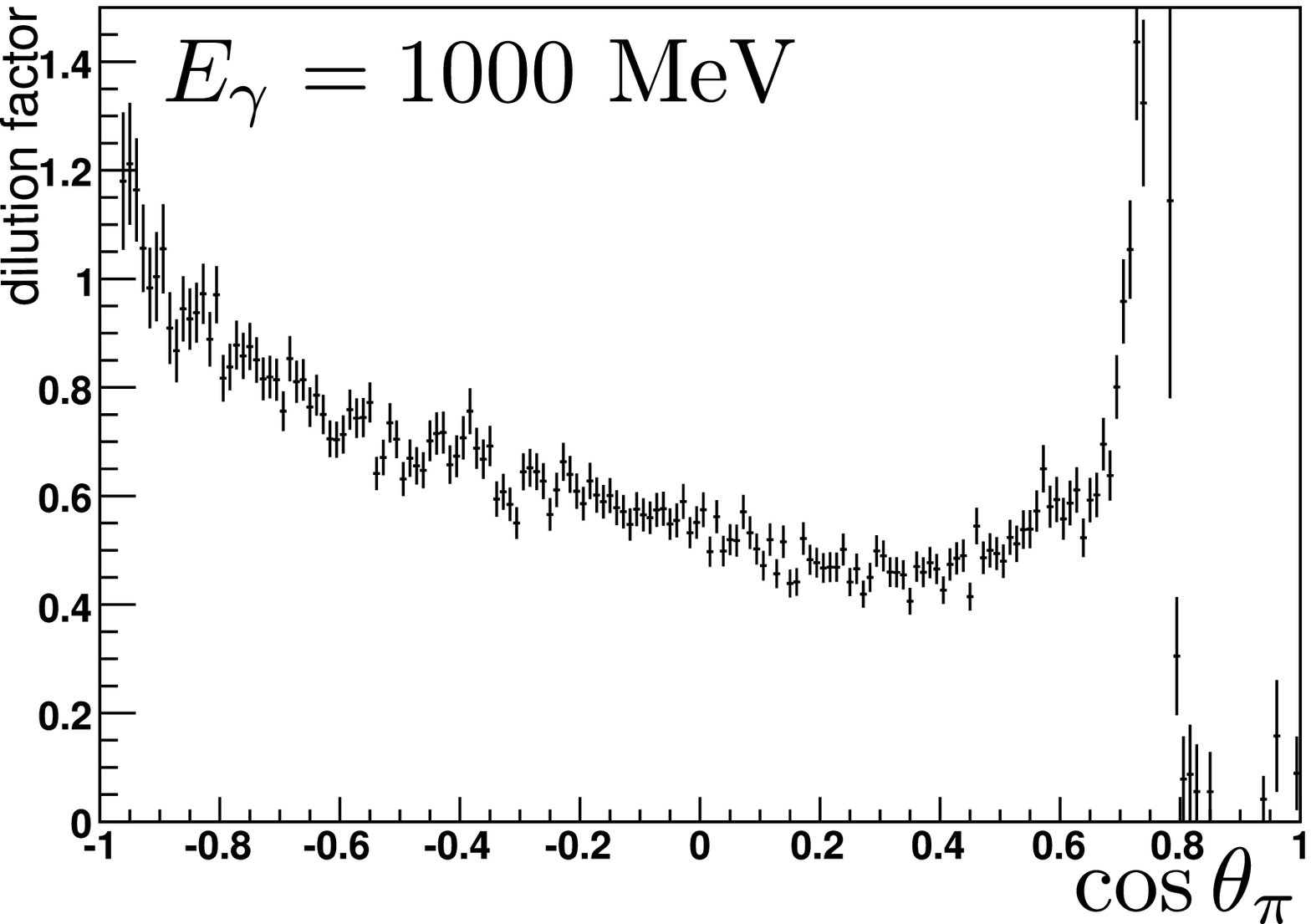}
 \caption{\label{pic:dil}Missing mass distribution (left) and meson $\cos\theta$ (center) for $\pi^0$
production on the different target materials: butanol (black), hydrogen (blue), carbon (red) and the sum of
hydrogen and carbon (green). The dilution factor (right) was determined using these distributions. }
\end{figure}%
The same applies to the $\cos\theta$ of the meson (fig. \ref{pic:dil}, center). With these distributions, it
is possible to determine the dilution factor (fig. \ref{pic:dil}, right). This factor was determined for both
mesons and all energies and is used to correct the values of the observable G. The double polarization
observable, which is shown in fig. \ref{pic:g}, was successfully extracted in the energy range for both
mesons.
\begin{figure}[htb]
 \includegraphics[width=0.45\textwidth]{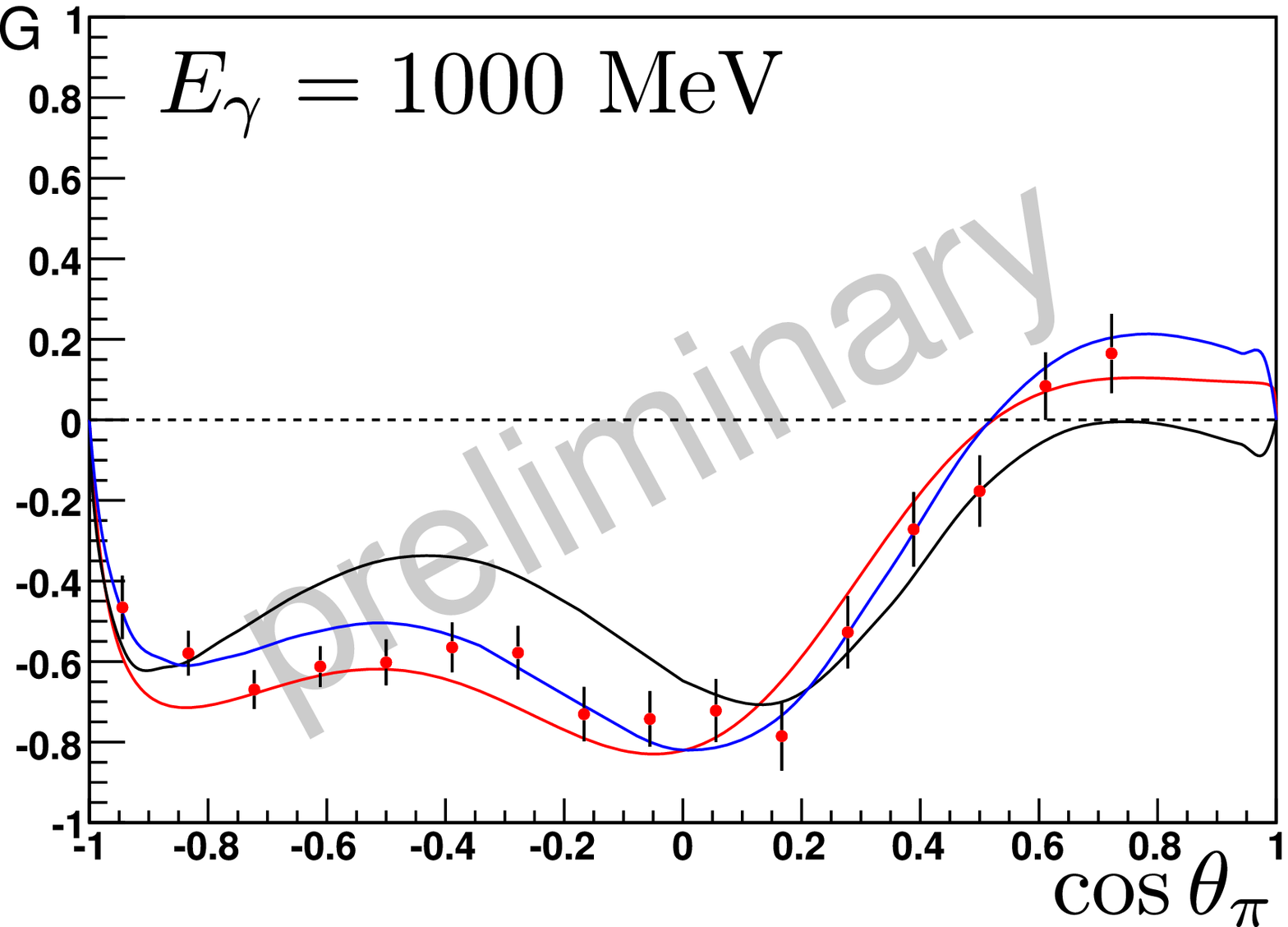}\hfill%
 \includegraphics[width=0.45\textwidth]{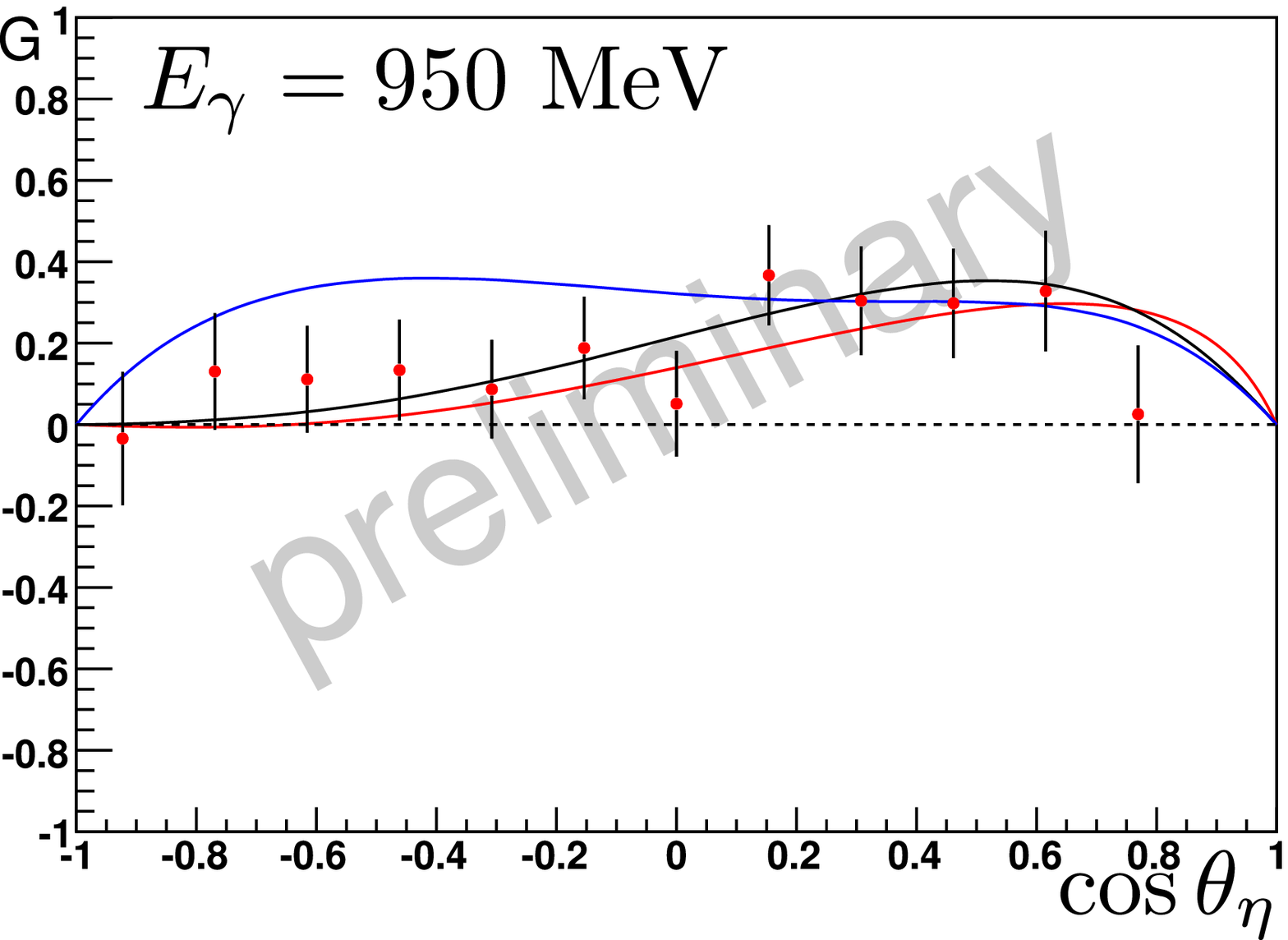}
 \caption{\label{pic:g}The double polarization observable G for the $\pi^0$ (left) and the $\eta$ (right).
Colors of the predictions of the partial wave analyses as in figure \ref{pic:beamasymm}.}
\end{figure}%

\Section{Summary}
The determination of polarization observables is crucial for the understanding of the excitation spectrum of
the nucleon. For the beam asymmetry $\Sigma$ and the double polarization observable G data has been taken by
the CBELSA/TAPS experiment in Bonn. These data will provide new information for the partial wave analysis,
which will help to identify the contributing resonances to the different final states.\hfill Supported by the
DFG (SFB/TR16).

\coltoctitle{\boldmath The $N^\ast(1535)$ Excitation -- The Electromagnetic Transition Form Factors}
\authortochead{G.~Ramalho}
\label{RG}

\setcounter{chapter}{1}
\setcounter{section}{0}
\maketitle

\noindent%
The $N^\ast(1535)$ resonance, an $S_{11}$ state with $J^P=\textstyle \frac{1}{2}^-$, is a very important
nucleon excitation. It has a strong decay channel for $\eta N$ ($\approx 50\%$), and has been studied by
several models and frameworks~\cite{S11,ScalingS11,Capstick95,Jido08,Braun09,JDiaz09,Golli11,Aznauryan11}. A
notable propriety of the $N^\ast(1535)$ resonance is the relation between the two independent helicity
amplitudes for the transverse $A_{1/2}$ and the longitudinal $S_{1/2}$. Although it has been assumed till
recently that the amplitude $S_{1/2}$ was negligible compared to $A_{1/2}$, recent data~\cite{Aznauryan09}
revealed that these two amplitudes are correlated for the region $Q^2 > 2$ GeV$^2$, where $Q^2$ is the
negative of the momentum transfer squared. The ratio between the two amplitudes for $Q^2 > 2$ GeV$^2$ can be
expressed as~\cite{ScalingS11}
\begin{eqnarray}
S_{1/2} (Q^2)= - \frac{\sqrt{1+\tau}}{ \sqrt{2} }
\frac{M_S^2-M^2}{2M_S Q} A_{1/2} (Q^2),
\label{eqRatio}
\end{eqnarray}
where $Q=\sqrt{Q^2}$, $M_S$ is the $N^\ast(1535)$ mass, $M$ the nucleon mass and $\tau=\textstyle
\frac{Q^2}{(M_S+M)^2}$. The test of the relation (\ref{eqRatio}) using the $A_{1/2}$
data~\cite{Aznauryan09,MAID} and the model from Ref.~\cite{S11}, is presented in Figure~\ref{figS12}.

\begin{figure}[h]
 \vspace{.4cm}
 \centerline{\includegraphics[width=7.0cm]{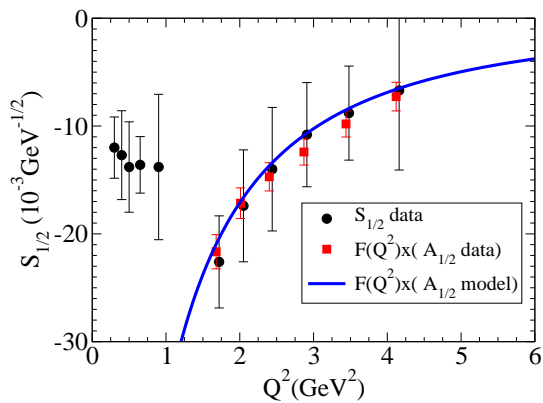}}
\caption{\label{figS12}$S_{1/2}$ helicity amplitude for the $\gamma N \to N^\ast(1535)$ reaction. The function
$F(Q^2)$ in the figure is $F(Q^2)= - \frac{\sqrt{1+\tau}}{ \sqrt{2} } \frac{M_s^2-M^2}{2M_S Q}$, as defined by
Eq.~(\ref{eqRatio}). Model for $A_{1/2}$ is from Ref.~\cite{S11}. Data are from
Refs.~\cite{Aznauryan09,MAID}.}
\end{figure}

Although small, the magnitude of $S_{1/2}$ is about 10\% of $A_{1/2}$ for very high $Q^2$~\cite{ScalingS11}.
The results for $S_{1/2}$ can be interpreted as a correlation between the valence quark degrees of freedom and
the quark-antiquark excitations (or meson cloud excitations), in the framework of the spectator quark
model~\cite{S11,ScalingS11,Nucleon,ExclusiveR}.

In general at high $Q^2$ the valence quark degrees of freedom are dominant, since meson cloud effects are
expected to be suppressed in QCD according to the dimensional counting rules~\cite{Aznauryan11}. At low $Q^2$,
meson cloud effects can be very important particularly for the lightest meson, the pion, due to chiral
perturbation theory~\cite{chiPT}. An an example of the insufficiency of the valence quark degrees of freedom
alone to explain the experimental data at low $Q^2$, is the $\gamma N \to \Delta$ reaction. Including only the
valence quark effects it is impossible to describe the data and misses at least 30\% of the magnetic
transition form factor strength for the $\gamma N \to \Delta$ reaction, in a non-relativistic framework near
$Q^2=0$~\cite{Aznauryan11,Pascalutsa07,Diaz07}. This is interpreted as a manifestation for the necessity of
including the explicit meson cloud
effects~\cite{Aznauryan11,Pascalutsa07,Diaz07,NDelta,NDeltaD,LatticeD,Lattice}.

A definite decomposition into the valence quark and meson cloud effects is however
impossible~\cite{chiPT,Capstick08}. The decomposition can be done only for a particular model.

The application of a valence quark model to describe an electromagnetic transition between two baryon states
has the advantage of accounting the quark structure (baryon wave functions and quark electromagnetic
current), and can provide an upper limit for the magnitude of the quark core contribution for the
electromagnetic form factors. [Recall the case for the $\gamma N \to \Delta$ reaction]. Recent models
developed for the treatment of meson-baryon reactions using the meson and baryon effective degrees of freedom,
take the quark core contributions in consideration~\cite{Diaz07,Kamalov99}. In addition the quark model
estimates can be tested in the high $Q^2$ regime, where meson cloud effects fall down. Another regime where
meson cloud effects can be suppressed and test the quark model estimates are the lattice QCD simulations with
large pion masses, if some quark models can be extended to that regime.

To study the $N^\ast(1535)$ system we use the covariant spectator quark
model~\cite{S11,Nucleon,ExclusiveR,Gross69}. In the model a baryon is described as a 3 constituent-quark
system that can be re-arranged in a quark-diquark structure, where the quark can couple with photon in terms
of a vector meson dominance (VMD) process and the diquark is treated as an on-shell (spectator)
particle~\cite{Nucleon,ExclusiveR}. The model has been developed, and calibrated by the nucleon~\cite{Nucleon}
and $\Delta$~\cite{NDelta,NDeltaD,LatticeD} systems, and later extended to study the other resonances such as
the Roper and $\Delta(1600)$~\cite{Roper,Delta1600}. Using the VMD parametrization for the quark current, the
model was also extended successfully to the lattice QCD regime for $m_\pi > 400$ MeV, particularly for the
elastic reaction~\cite{Lattice} and for the transitions from the nucleon to the $\Delta(1232)$ and
$N^\ast(1440)$ systems~\cite{LatticeD,Lattice,Roper}. The results show that the covariant spectator quark
model has a good control of the valence quark model degrees of freedom.

\begin{figure}[t]
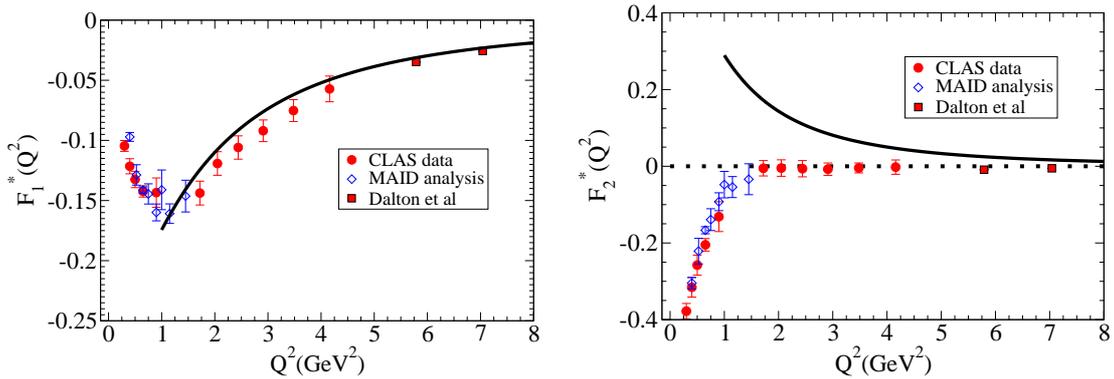

 \centerline{\includegraphics[width=7.0cm]{./RamalhoGilberto/ramalho2.eps}\hspace{.5cm}%
             \includegraphics[width=7.0cm]{./RamalhoGilberto/ramalho3.eps}  }
 \caption{\label{figFF}$\gamma N \to N^\ast(1535)$ transition form factors in the spectator quark model. Data
are from Refs.~\cite{Aznauryan09,MAID}.}
\end{figure}

In the covariant spectator quark formalism the $N^\ast(1535)$ is described as a quark-diquark system with
negative parity, a relative orbital angular momentum 1 and a quark core with spin 1/2 \cite{S11}. In addition
we assume that the diquark has no internal P-state structure and can be treated as pointlike~\cite{S11}.
Furthermore, we note that the $N^\ast(1535)$ wave function can only be reliably applied for $Q^2 \gg |{\bf
q}|_0^2=0.23$ GeV$^2$, where $|{\bf q}|_0= \textstyle \frac{M_S^2-M^2}{2M_S}$ is the photon three-momentum in
the $N^\ast(1535)$ rest frame. The results for the transition form factors $F_1^\ast$ and $F_2^\ast$ are
presented in Figure~\ref{figFF} for $Q^2 \ge 1$ GeV$^2$. From the figure, one can conclude that the model is
reasonable to explain the results for $F_1^\ast$ for $Q^2 \ge 1$ GeV$^2$. As for $F_2^\ast$ one can notice
that the model results differ in sign from the data, and also that it overestimates in absolute values of the
data, where the data show $F_2^\ast \approx 0$ for $Q^2> 1.5$ GeV$^2$. The result $F_2^\ast=0$ is equivalent
to the relation given by Eq.~(\ref{eqRatio}). In the spectator quark model the deviation from the data can be
interpreted as the missing of the meson cloud effects in the calculation~\cite{ScalingS11}. If this
interpretation is correct we can expect a negative contribution from the meson cloud for $F_2^\ast$ around
$Q^2=2$ GeV$^2$ with the magnitude similar to the present model.

Although one cannot compare directly the valence quark effects and/or meson cloud effects, from different
formalisms, it is encouraging to note that our results for $F_2^\ast$ are close to the estimate of the
EBAC/Jlab group~\cite{JDiaz09} for the quark core contributions, using a framework based in effective
baryon-meson interactions. It is also worth to note that a calculation of the form factors with the assumption
that the $N^\ast(1535)$ is a dynamically generated resonance, and therefore takes into account only the meson
cloud effects, gives only negative contributions for $F_2^\ast$ \cite{Jido08}.

Based on the results from different works, one can conclude that valence quark effects are very
important~\cite{Jido08}, or are the dominant
contribution~\cite{S11,Braun09,JDiaz09,Golli11,Aznauryan11,Aznauryan11b} for the $N^\ast(1535)$ transition
form factors, although meson cloud is certainly necessary to describe the data accurately.

\coltoctitle{\boldmath The Structure around W = 1680 MeV in $\eta$-Photoproduction off the Neutron}
\authortochead{D.~Werthm\"{u}ller}
\label{WD}

\setcounter{chapter}{1}
\setcounter{section}{0}
\maketitle

\Section{Motivation}
Previous experiments of the GRAAL~\cite{refgraal1,refgraal2}, LNS-Sendai~\cite{reflns} and
CBELSA/TAPS~\cite{refigal_1,refigal_2} collaborations have reported a narrow structure at $W \approx 1680$ MeV
in the total cross section of quasi-free $\eta$-photoproduction off the neutron. This structure is not seen on
the proton and its nature is up to now still unknown. As the width of the structure is close to the
experimental resolution of the different experiments the true width should be unusually small compared to
known resonances in this energy region. Therefore the structure can hardly be described using a single
well-know broad nucleon resonance. Theoretical explanations thus include interferences of
resonances~\cite{refanisovich} or contributions from intermediate meson loops~\cite{refdoering}. Another
solution is provided by the chiral quark soliton model~\cite{refchiralsoliton} which predicts an anti-decuplet
nucleon resonance that couples strongly to the formation channel $\gamma$n as well as to the final state
$\eta$N.

We report preliminary results from two different experiments performed at the tagged-photon beam facility at
MAMI. In one experiment a liquid deuterium target was used as a neutron target whereas in the other one a
liquid $^3$He target was installed. 

\begin{figure}[htb]
 \centering
 \includegraphics[width=9cm]{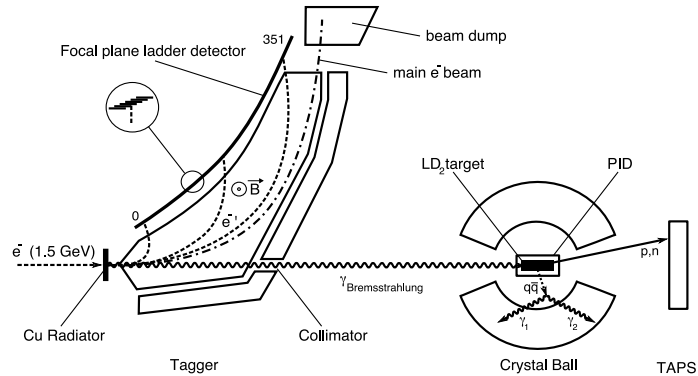}
 \caption{\label{fig_exp_setup}Scheme of the experimental setup.}
\end{figure}

\Section{Experimental setup}
The measurements were performed at the electron accelerator facility MAMI~\cite{MAMI, MAMI_C} in Mainz. A
schematic view of the experimental setup is shown in figure~\ref{fig_exp_setup}. A photon beam was produced
from an electron beam of 1.5 GeV energy via bremsstrahlung using a copper radiator. The energy of the photons
was determined by a momentum analysis of the scattered electron in a magnetic spectrometer (Glasgow photon
tagger~\cite{Tagger_1, Tagger_2, Tagger_3}). After collimation the beam impinged on the target (liquid
deuterium or $^3$He, respectively). The target was surrounded by a cylindrical plastic scintillator strip
detector~\cite{PID}, which was used for charged particle identification, and the spherical electromagnetic
calorimeter Crystal Ball~\cite{CB}. This detector consists of 672 NaI crystals and covers 94\% of 4$\pi$
steradians. The hole in forward direction of Crystal Ball is closed by the TAPS detector~\cite{TAPS_1, TAPS_2}
which is made of 384 BaF$_2$ crystals (378 BaF$_2$ + 24 PbWO$_4$ crystals in the $^3$He experiment). In front
of every crystal a thin plastic scintillator element is installed as a charged particle veto detector. As
trigger condition a deposited energy sum of 300 MeV in the Crystal Ball and a total multiplicity of two or
more hits in both calorimeters was requested.

\Section{Data analysis}
Using the $\eta\rightarrow\gamma\gamma$ decay of the $\eta$-meson the final state products of the quasi-free
reactions $\gamma$p$\rightarrow \eta $p and $\gamma$n$\rightarrow \eta $n were selected by requesting 2
neutral clusters + 1 charged cluster or 3 neutral clusters, respectively, in the detector system. The
$\eta$-meson was identified in the invariant mass of the two photons by applying a cut around the $\eta$ mass.
Background coming from e.g. $\eta\pi$-photoproduction was suppressed by cutting on the $\eta N$-coplanarity
and the missing mass.

Excitation functions were calculated as functions of 
\begin{enumerate}
\item the incoming photon beam energy $E_{\gamma}$ (converted for better comparison to the center-of-mass
energy $W_{beam} = \sqrt{2E_{\gamma}m_{N} + m_{N}^{2}}$ of the photon-nucleon system)
\item the invariant mass of the final state $\eta$-nucleon system $W_{rec}=M(\eta N)$ by reconstructing the
energy of the recoil nucleon by kinematics
\end{enumerate}
In case of 1) structures are smeared by the Fermi motion of the initial state nucleon inside the target
nucleus (deuteron or $^3$He). In contrast structures appearing in 2) are not affected by Fermi smearing and
the resolution only depends on the energy and angular resolution of the detector system. 

The multiple hits in the photon tagger were treated as single events and weighted accordingly to the
tagger-detector timing to achieve a statistical subtraction of the accidental coincident background.

Finally, cross sections were extracted by normalizing the excitation functions with the target density, the
photon flux and the detector acceptance obtained by a Geant4 based simulation of the detector setup.

\begin{figure}[htb]
 \centering
 \includegraphics[width=15cm]{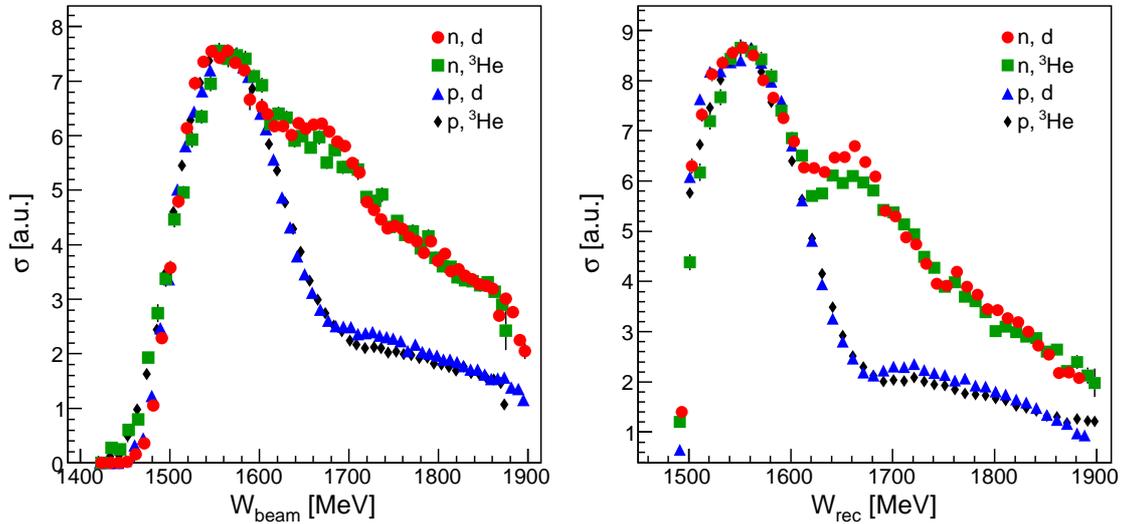}
 \caption{\label{fig_results}Very preliminary excitation functions for the quasi-free $\gamma$p$\rightarrow
\eta $p and $\gamma$n$\rightarrow \eta $n reactions obtained by the LD$_2$ and the $^3$He experiments.}
\end{figure}

\Section{Preliminary results}
Preliminary results of the analysis are shown in figure \ref{fig_results}. Since the absolute normalization
of the data is not final all results were scaled to $\sigma_{n}$ in the S$_{11}(1535)$-maximum of the
deuterium analysis and are given in arbitrary units.

The previous results are confirmed by both experiments: An excess in the neutron cross section is observed
which is not seen on the proton. The width of the structure in $\sigma(W_{rec})$, that is not affected by
Fermi motion, is around 50 MeV and is mainly resolution dominated.

Furthermore the results of the LD$_2$ and the $^3$He measurement are in good agreement. The differences in
$\sigma(W_{beam})$ seem to be only related to the larger Fermi motion inside the $^3$He nucleus (smearing at
threshold and at the shoulder of the S$_{11}(1535))$. The deviation in $\sigma(W_{rec})$ is due to the fact
that the kinematic reconstruction within the participant-spectator model is only an approximation in case of
$^3$He. Obtaining consistent results by using nuclei with different Fermi motion and n/p-ratio makes it very
unlikely that the observed effect is caused by rescattering of mesons or FSI.

The final results of the LD$_2$ measurement will include the analysis of the $\eta\rightarrow3\pi^{0}$ decay
channel and will provide angular distributions with high statistics.

\end{papers}
\cleardoublepage

\pagestyle{plain}
\begin{center}
 \vspace*{0.3\textheight} {\Huge\textbf{\boldmath Interaction of $\eta$ and $\eta^\prime$}} \\
 \vspace{0.5cm} {\Huge\textbf{with Nucleons and Nuclei}} \\
 \addcontentsline{toc}{part}{\sffamily \boldmath Interaction of $\eta$ and $\eta^\prime$ with
Nucleons and
Nuclei}
\end{center}
\cleardoublepage
\pagestyle{fancy}

\begin{papers}
\coltoctitle{\boldmath A Theoretical Approach to $\eta^\prime N$ Scattering}
\authortochead{A.~Ramos}
\label{RA}

\setcounter{chapter}{1}
\setcounter{section}{0}
\maketitle

\Section{Introduction}
The $\eta^\prime$ meson has interesting properties associated to the underlying QCD dynamics of hadrons, in
particular to the $U(1)_A $ axial vector anomaly~\cite{anomaly}. Being close to a singlet of SU(3), its
interaction with nucleons is supposed to be weak compared for instance with the case of its partner, the
$\eta$ meson. Experimentally, there are only poor estimations of the $\eta^\prime N$ scattering length from
analyses of the $pp \to pp \eta^\prime$ cross section near threshold at COSY~\cite{moskal}, establishing a
value around $|a_{\eta^\prime N}|\sim 0.1$~fm. The $\eta^\prime N$ interaction has also an infuence on
photoproduction $\gamma p \to \eta^\prime p$ reactions measured at ELSA and Jlab~\cite{photo_exp}, which have
been analyzed so far in the framework of resonant models~\cite{photo_analyses}. Finally, some information on
the $\eta^\prime N$ interaction can be extracted from the transparency ratio in $(\gamma,\eta^\prime)$
reactions on nuclei \cite{nanova}.

We present a theoretical study of the $\eta^\prime N$ interaction within a unitary approach in coupled
channels~\cite{etap}. Our study also includes the explicit effect of resonances lying very close to the
$\eta^\prime N$ threshold, which are generated dynamically from the vector meson-baryon ($VB$) interaction in
s-wave~\cite{angelsvec}. Finally, we also consider a term in the Lagrangian coupling the baryons to the
singlet component of the pseudoscalar meson, which is allowed by the symmetries of QCD~\cite{borasinglet}. We
find that the elastic $\eta^\prime N$ cross section is very sensitive to the unknown singlet strength, while
the inelastic cross sections are rather stable and constitute a genuine prediction of our model.

\Section{Model}
The meson-baryon scattering dynamics is taken from the lowest order chiral Lagrangian reduced to the two meson
fields needed in the process,
\begin{equation}
{\cal L}_{\Phi B} = \langle \bar{B} i \gamma^{\mu} \frac{1}{4 f^2}
[(\Phi \partial_{\mu} \Phi - \partial_{\mu} \Phi \Phi) B
- B (\Phi \partial_{\mu} \Phi - \partial_{\mu} \Phi \Phi)] \rangle
\label{lagran}
\end{equation}
where $f$ is the pion decay constant, $B$ represents the baryon octet matrix and $\Phi$ contains the
pseudoscalar meson octet as well as the singlet field $\eta_1/\sqrt{3}$ added to its diagonal elements. The
physical $\eta,\eta^\prime$ mesons are given by $\eta = \cos\theta_P\, \eta_8 - \sin\theta_P\, \eta_1$ and
$\eta^\prime= \sin\theta_P\, \eta_8 + \cos\theta_P\, \eta_1$, where $\theta_P$ is the  $\eta_1-\eta_8$ mixing
angle, for which we take the value $\theta_P=-14.34^\circ$ reported in~\cite{ambrosino}. Our basis of
pseudoscalar meson-baryon states ($PB$) is $\pi N$, $\eta N$, $\eta^\prime N$, $K \Lambda$ and $K \Sigma$.
 
We have also implemented in our model the effect of a resonance appearing around 1970~MeV, close to the
$\eta^\prime N$ threshold at $\sqrt{s}=1896$ MeV, which is generated dynamically from the vector meson-baryon
($VB$) interaction in s-wave~\cite{angelsvec} and couples mostly to the $K^* \Lambda$ and $K^* \Sigma$
channels. As described in Ref.~\cite{etap}, this is achieved by still working within a basis of coupled $PB$
states but including explicitly, as part of a $PB-PB$ potential, the transition of the $PB$ to the $VB$
channels, the later ones interacting among themselves to produce the resonance. The $PB-VB$ transition vertex
contains the standard $PPV$ couplings as well as the anomalous $PVV$ terms since the standard ones are
supressed by $\sin\theta_P$ for transitions involving $\eta^\prime N$ states.

Finally, we also include a Lagrangian that couples the singlet component of the meson field to the
baryons~\cite{borasinglet} and gives rise to new contributions to the transition potentials between $\eta N$
and $\eta^\prime N$ states, $V^{(1)}_{\eta N, \eta N}= C \sin^2 \theta_P$, $V^{(1)}_{\eta N, \eta^\prime N}=
-C \sin \theta_P \cos \theta_P$ and $V^{(1)}_{\eta^\prime N, \eta^\prime N}= C \cos^2 \theta_P$, where the
unknown singlet strength $C$ is written in terms of a parameter $\alpha$ obtained from $C=\alpha\, 2
m_{\eta^\prime}\, (E_B+E_{B'})/4 f_\pi^2 /2M_N$.

With the obtained meson-baryon transition potential $V$ we derive the scattering amplitude $T$ by solving the
Bethe-Salpeter equations in their on-shell factorization form, $T=[1-VG]^{-1} V$, where $G$ is the loop
function for the intermediate meson-baryon states which we evaluate using dimensional regularization.

\vspace{4ex}\Section{Results}
Our results for the $\eta^\prime p$ scattering length are shown in Table~\ref{tab:scat}, while the  elastic
and inelastic $\eta^\prime p$  cross sections, together with the cross section of the reaction $\pi^- p \to
\eta^\prime n$, are displayed in Fig.~\ref{fig:xsec_VB_singlet}.

\begin{table}[htbp]
    \setlength{\tabcolsep}{0.3cm}
\begin{center}
\begin{tabular}{cccc}
$\alpha$ & model & $a_{\eta p}$ [fm] & $|a_{\eta^\prime p}|$ [fm]
\\
\hline
\multirow{2}{*}{0} & $PB$ & 0.0017+{\rm i}0.0139 & 0.014 \\
                   & $PB+VB$ &  0.0210+{\rm i}0.0192 & 0.029 \\
\hline
\hline
$-0.126$ & $PB+VB$ & \phantom{$-$}$0.073 +{\rm i}0.019$ & \multirow{2}{*}{0.075}
\\
\phantom{$-$}$0.204$ & $PB+VB$ & $-0.072 +{\rm i}0.020$ & \\
\hline
$-0.193$ & $PB+VB$ & \phantom{$-$}$0.098 +{\rm i}0.020$ & \multirow{2}{*}{0.1}
\\
\phantom{$-$}$0.256$ & $PB+VB$ & $-0.098 +{\rm i}0.020$ & \\
\hline
$-0.333$ & $PB+VB$ & \phantom{$-$}$0.149 +{\rm i}0.020$ & \multirow{2}{*}{0.15}
\\
\phantom{$-$}$0.352$ & $PB+VB$ & $-0.149 +{\rm i}0.021$ &  \\
\hline
\end{tabular}
\end{center}
\caption{The $\eta^\prime p$ scattering lengths for various models discussed in the text.\label{tab:scat}}
\end{table}

The $\eta^\prime p$ scattering length is very small when only the Weinberg-Tomozawa terms are included
($\alpha=0,PB$), since the particular structure of the Lagrangian of Eq.~(\ref{lagran}), $\Phi \partial_{\mu}
\Phi - \partial_{\mu} \Phi \Phi$, makes the contribution of the singlet vanish. Including the $VB$ states
($\alpha=0,PB+VB$) increases the real part by an order of magnitude an enhances the imaginary part
considerably due to the decay channels $K^* \Lambda \to \pi K \Lambda$ and $K^* \Sigma \to \pi K \Sigma$. The
effect of the $VB$ channels is more visible by comparing the thin and solid lines in
Fig.~\ref{fig:xsec_VB_singlet}. In all cases, one sees a pronounced structure around $p_{\eta^\prime} = 500
$~MeV/c ($p_{\pi}=1600$~MeV/c) which corresponds to the energy 1970~MeV of the resonance. The strongest
relative enhancements are seen for the elastic channel and for the $\pi^- p \to \eta^\prime n$ reaction, but
the results do not reproduce the available experimental information. The scattering length only amounts to $|
a_{\eta^\prime p} | = 0.029$ fm, while the $\pi^- p \to \eta^\prime n$ cross section is six times smaller than
the measured value of around 0.1~mb at a pion lab momentum of 1600~MeV/c~\cite{Rader:1973mx}.

Important changes are obtained when the singlet Lagrangian is considered with its strength $\alpha$ tuned to
values of $| a_{\eta^\prime p} |$ around 0.1~fm. Each assumed value of  $| a_{\eta^\prime p} |$ admits two
types of solutions for $\alpha$, one positive (repulsive) and one negative (attractive). For a given sign, the
changes in $\alpha$ modify the real part of $a_{\eta^\prime p}$ in a nearly proportional way. The imaginary
part remains practically constant because the singlet Lagrangian induces the largest effect on the
$\eta^\prime N \to \eta^\prime N$ amplitude as it behaves like $\cos^2\theta_P$. This also means that our
model, in spite of using the $\eta^\prime N$ scattering length as a free parameter, provides a robust
prediction for the inelastic channels, as corroborated by the moderate changes seen in the inelastic cross
sections displayed in Fig.~\ref{fig:xsec_VB_singlet}.

\begin{figure}[htb]
 \centerline{\includegraphics[width=0.5\textwidth]{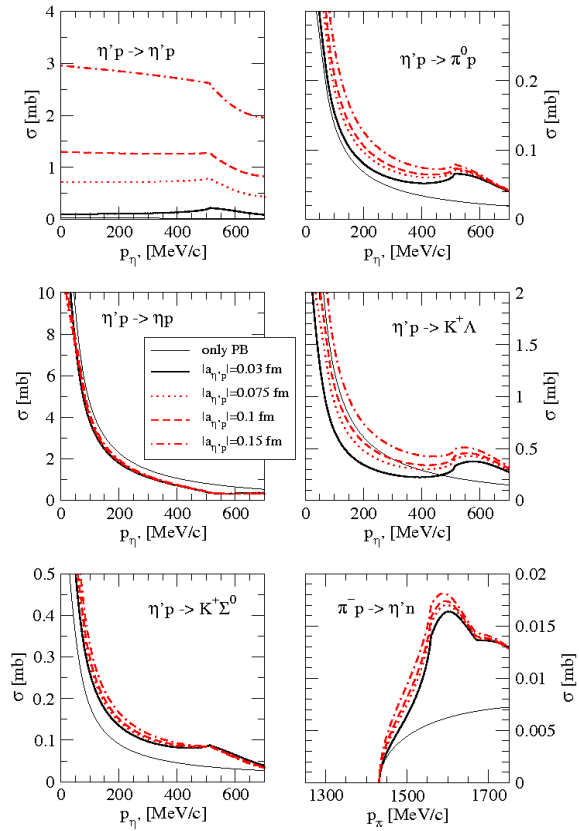}}
 \caption{\label{fig:xsec_VB_singlet}Elastic and inelastic $\eta^\prime p$  cross sections, together with the
cross section of the reaction $\pi^- p \to \eta^\prime n$, for various models discussed in the text.}
\end{figure}

The optical theorem allows us to obtain the total $\eta^\prime p$ cross section from the elastic scattering
amplitude and, by subtraction, we can derive the inelastic one, $\sigma^{\rm ine}=\sigma^{\rm tot}-\sigma^{\rm
el}$, which amounts to $5$ mb at a value of 200~MeV/c for the $\eta^\prime$ momentum. The preliminar analysis
of the transparency ratio of $\eta^\prime$ photoproduction in nuclei~\cite{nanova} estimate the inelastic
cross section to be around 10~mb for a $\eta^\prime$ momentum of 1 GeV/c. This is an upper bound for the
one-body absorption inelasticities, since the transparency ratio can also be contaminated by multinucleon
absorption processes, although they are expected to be small~\cite{etap_bound}. In any case, the $\eta^\prime$
energies of the transparency ratio experiment are too high for our model, the validity of which we trust for
momenta smaller than $p_{\eta^\prime}= 600$~MeV/c, implying $\eta^\prime $  kinetic energies smaller than
200~MeV. Data analyses with momentum cuts will be available in the future and, although the statistics will be
lowered, we expect them to help in constraining the properties of the $\eta^\prime N $ interaction.

\coltoctitle{\boldmath Search for $\eta$-mesic ${^4\mbox{He}}$ with WASA-at-COSY}
\authortochead{W.~Krzemie\'n}
\label{KW}

\setcounter{chapter}{1}
\setcounter{section}{0}
\maketitle

\Section{Introduction}
The investigation of the exotic objects in the nuclear physics is a proven method for revealing many
interesting properties of nuclear systems and for accessing to  an unexplored areas of physics. The recent
progress in the spectroscopy of pionic and kaonic atoms, as well as pionic and kaonic nuclei has permitted to
obtain deeper insights into the meson-nucleus interaction and the in-medium behaviour of spontaneous chiral
symmetry breaking~\cite{hiren}.

Analogically to the other exotic nuclear systems, the investigation of the $\eta$-mesic nuclei would provide
many interesting informations about the $\eta$-N interaction, N* in-medium properties~\cite{jido}  and would
deepen our knowledge of the fundamental structure of the $\eta$ meson~\cite{basssymposium}. The $\eta$ meson
is electrically neutral, therefore such a system can be formed only via the strong interaction which
distinguishes it qualitatively from pionic atoms where the binding is the effect of the sum of the attractive
electromagnetic force and the repulsive strong interaction.

The search of the $\eta$~-~mesic nucleus was performed in many experiments in the
past~\cite{lampf,lpi,jinr,gsi,gem,mami} and is being continued at COSY~\cite{moskalsymposium,jurek-he3,timo},
JINR~\cite{jinr}, J-PARC~\cite{fujiokasymposium} and MAMI~\cite{mami}. Many promising indications where
reported, however, so far there is no direct experimental confirmation of the existence of mesic nucleus. In
the region of the light nuclei systems such as $\eta$-$\mbox{He}$ or $\eta$-$\mbox{T}$, the observation of the
strong enhancement in the total cross-section and the phase variation in the close-to-threshold region
provided strong evidence to the hypothesis of the existence of a pole in the scattering matrix which can
correspond to the bound state~\cite{wilkin}. However, as it was stated in ~\cite{liu3,haider2}, the
theoretical predictions of width and binding energy of the $\eta$-mesic nuclei is strongly dependent on the
not well known subtreshold $\eta$-nucleon interaction. Therefore, the direct measurements which could confirm
the existence of the bound state, are mandatory.

\Section{Experiment}
In June 2008 we performed a search for the $^4\mbox{He}-\eta$ bound state by measuring the excitation
fun\-ction of the $dd \rightarrow ^3\mbox{He} p\pi^-$  reaction near the $\eta$ meson production threshold
using the WASA-at-COSY detector\cite{jurekmeson08}. During the experimental run the momentum of the deuteron
beam was varied continuously within each acceleration cycle from  2.185~GeV/c to 2.400~GeV/c, crossing the
kinematic threshold for the $\eta$ production in the $dd \rightarrow {^4\mbox{He}}\,\eta$ reaction at
2.336~GeV/c. This range of beam momenta corresponds to the variation of $^4\mbox{He}-\eta$  excess energy 
from -51.4~MeV to 22~MeV. The experimental method is based on measuring the excitation function for $dd
\rightarrow {^3\mbox{He}p \pi^-}$ and a search for a resonance-like structure below the ${^4\mbox{He}}-\eta$
threshold.  The relative angle between the outgoing $p - \pi^-$ pair which  originates from the decay of the
N*(1535) resonance created via  absorption of the $\eta$ meson on a nucleon in the $^4\mbox{He}$ nucleus, is
180$^{\circ}$ in the N* reference frame and is smeared by about 30$^{\circ}$ in the reaction center-of-mass
frame (CM) due to the Fermi motion of nucleons inside the $^4\mbox{He}-\eta$ nucleus. The $^3\mbox{He}$ plays
the role of a spectator and therefore its momentum in the CM frame is relatively low and determined only by
the Fermi momentum distribution in the $^4\mbox{He}-\eta$. This signature  distinguishes production of mesic
helium from other reactions, in which the distribution of the  $^3\mbox{He}$  momentum reaches much higher
values.

Figure \ref{openangle_cut_div_p} presents the not normalized excitation  functions for a "signal-rich"
region: $p_{He}^{CM}< 0.3$ GeV/c (top) and a  "signal-poor" region: $p_{He}^{CM}> 0.3$ GeV/c in 20 intervals
in beam momentum (middle). Also, the difference between two regions is shown in the same Figure (bottom).  

\begin{figure}[h]
  \begin{center}
        {\includegraphics[scale=0.20]{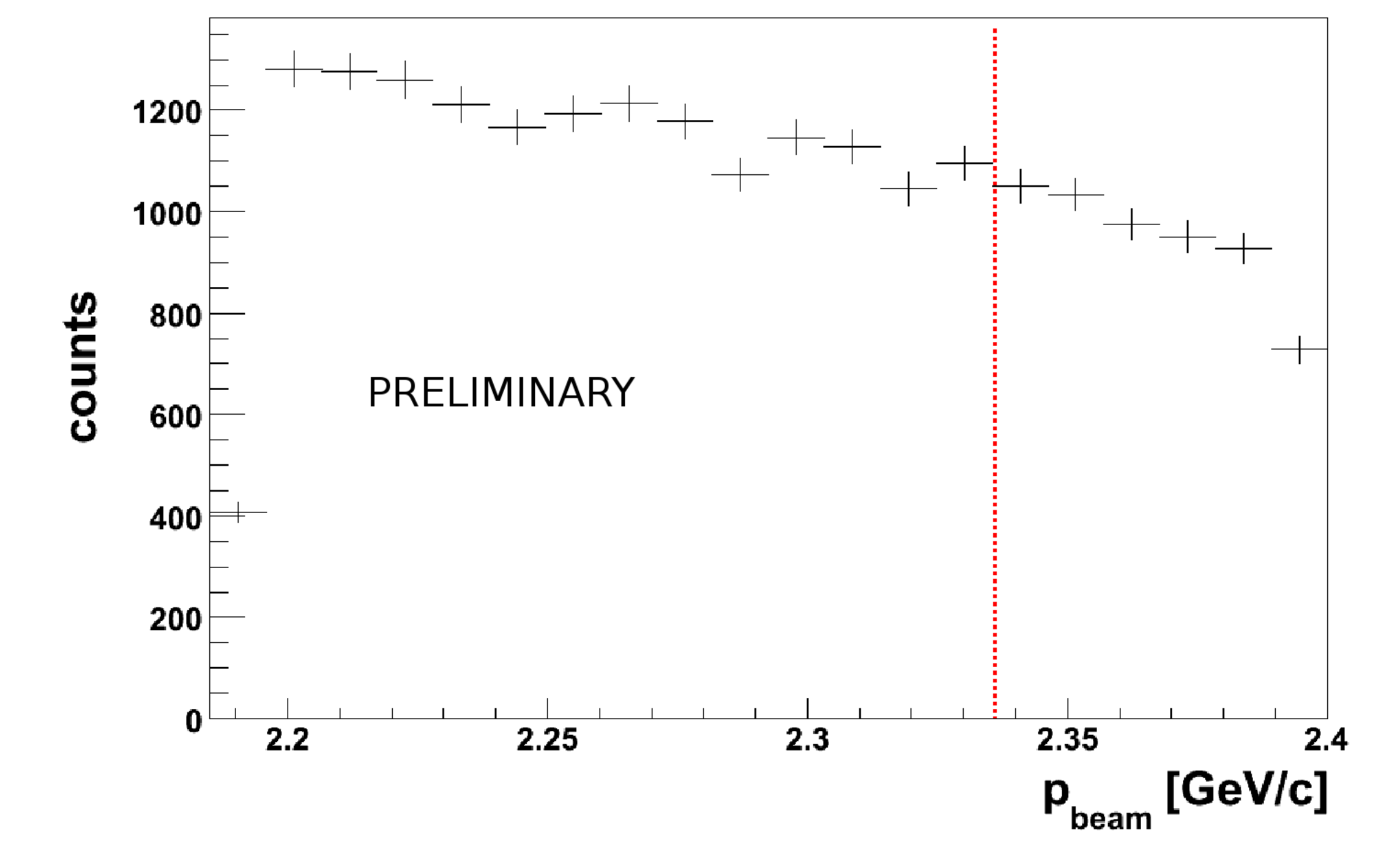}} \hfill
        {\includegraphics[scale=0.20]{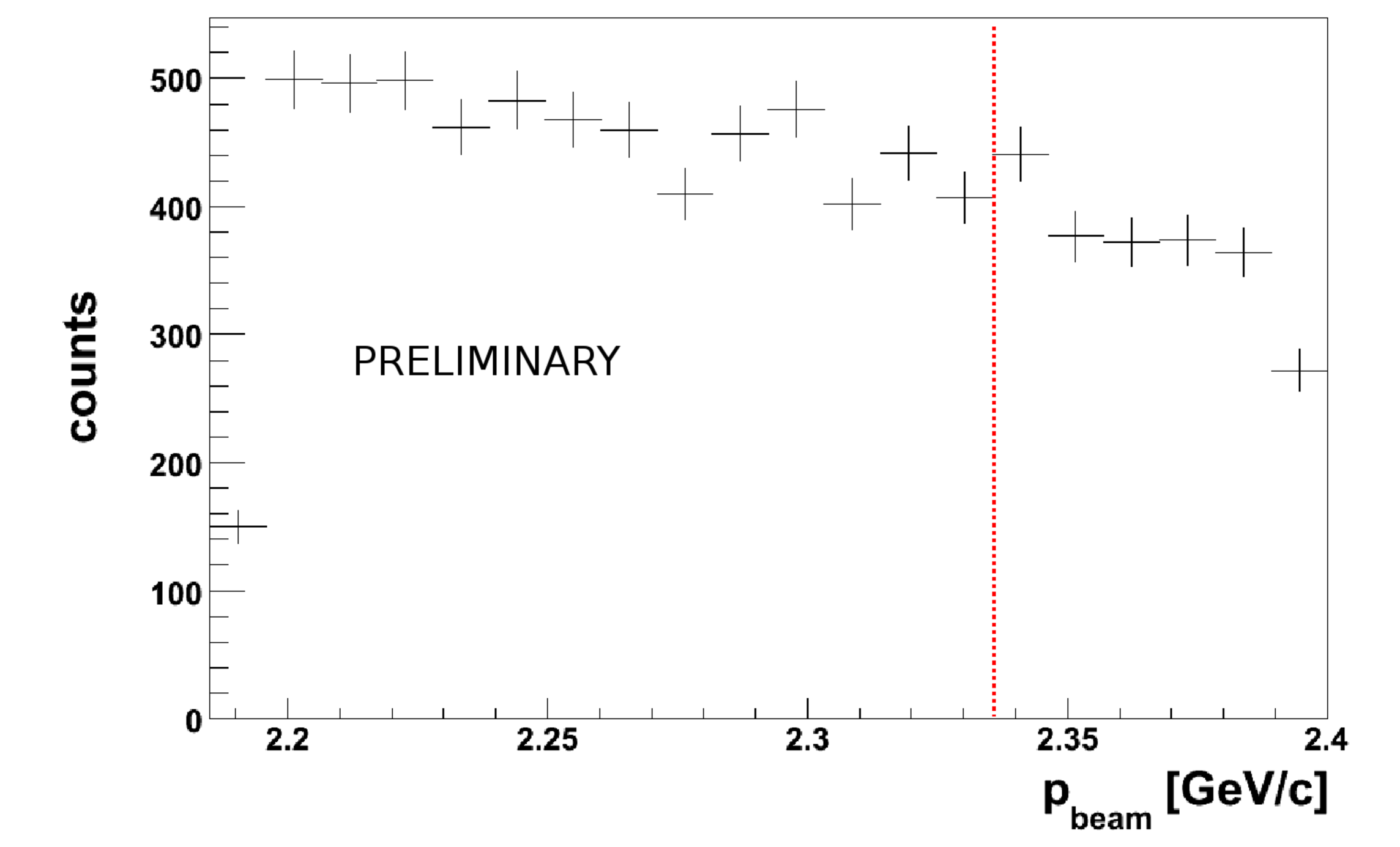}} 
        {\includegraphics[scale=0.20]{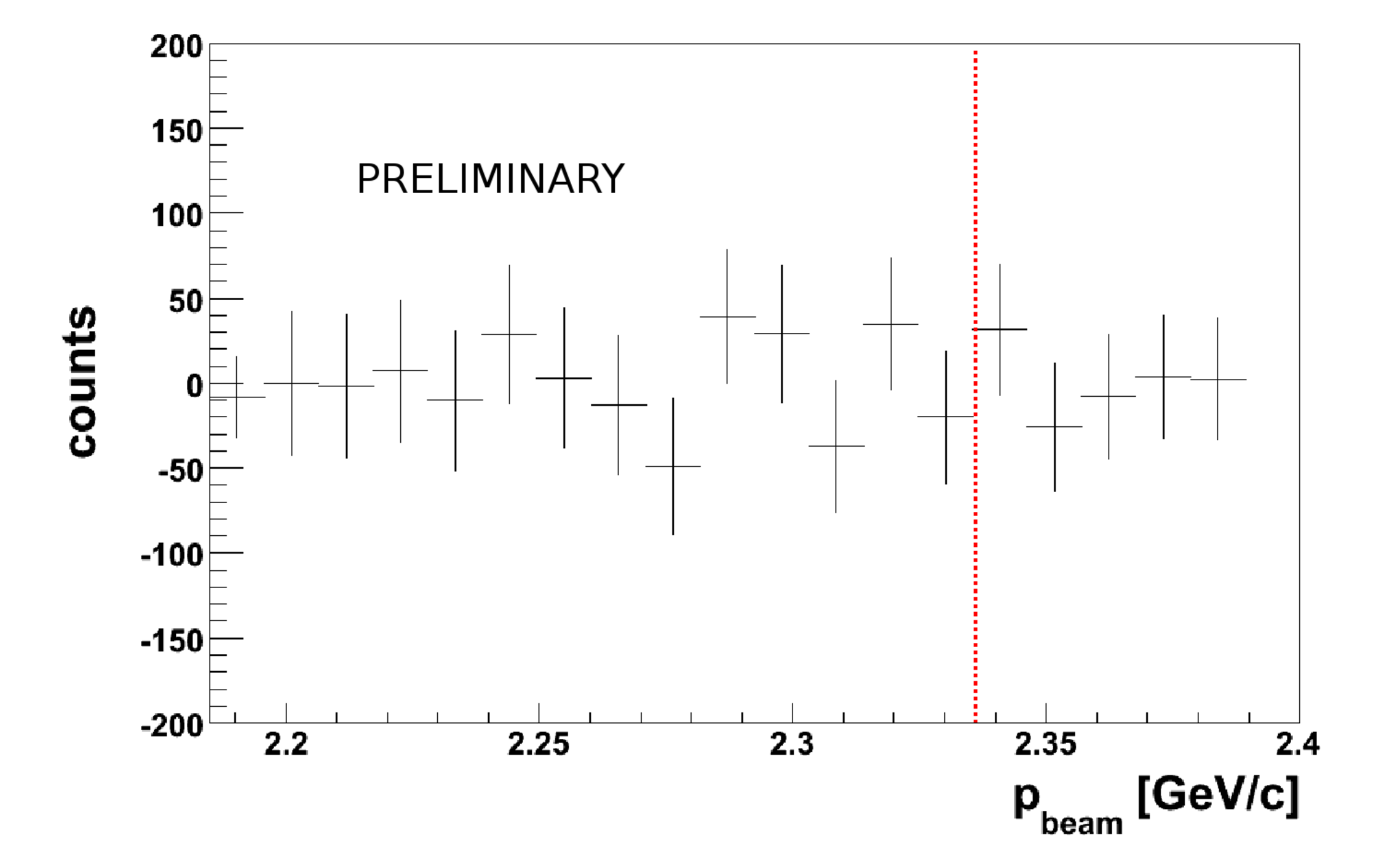}} 
        \caption{\label{openangle_cut_div_p} Excitation function for the $dd \rightarrow {^3\mbox{He}} p
\pi^{-}$ reaction for the "signal-poor" region ${^3\mbox{He}}$ momentum  above 0.3 GeV/c (top plot) and the
"signal-rich" region, low  ${^3\mbox{He}}$ momentum below 0.3 GeV/c (middle plot). The excitation functions
are not normalized. The beam momentum corresponding to the ${^4\mbox{He}}-\eta$ kinematical threshold is
marked as a red dashed line. The  difference between the "signal-rich" and the "signal-poor" regions is shown
at the bottom plot.  
} 
    \end{center} 
\end{figure} 

The figure indicates no structure below the kinematical threshold  where the signal is expected. In order to
estimate the maximum upper limit for the cross-section for the production of the bound state via  $dd
\rightarrow {{^4\mbox{He}}-\eta}_{bound} \rightarrow {^3\mbox{He}}\, p\, \pi^-$ reaction,  the Breit-Wigner
function along  with the linear background was fitted to the normalized  "signal-rich" region. The absolute
normalization was obtained based on the $dd \rightarrow {^3\mbox{He}} n$ reaction, whereas the luminosity
dependence of the beam momentum was determined using the quasi-elastic $dd \rightarrow pp (n_{sp}n_{sp})$
reaction. The integrated luminosity equals $L=117.9 \pm 13.6 $ nb$^{-1}$.

\begin{figure}[h]
    \begin{center}
        {\includegraphics[scale=0.35]{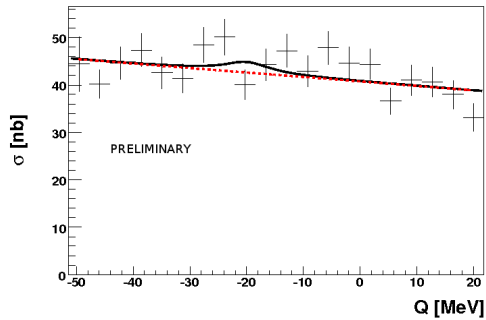}} 
        \caption{Normalized excitation function for the  $dd\rightarrow{^3\mbox{He}}p\pi^{-}$ reaction for
the "signal-rich" region. The Breit-Wigner function $f(Q)=\frac{A\cdot
(\frac{\Gamma}{2})^2}{(Q-E_{BE})^2+(\frac{\Gamma}{2})^2}$ along with the linear function was fitted. The
parameters $\Gamma$ and $E_{BE}$ were fixed to the value of -20 MeV and 10 MeV respectively. The free
parameter A is consistent with zero within the obtained accuracy. The dotted red line represents the linear
background.\label{fitt_p}} 
    \end{center} 
\end{figure} 

\Section{Outlook}
In November 2010 a new two-week measurement was performed with WASA-at-COSY. We collected the data with
approximately 20 times higher statistics. In addition to the $dd \rightarrow {^3\mbox{He}} p \pi^-$ channel we
registered also the $dd \rightarrow {^3\mbox{He}} n \pi^{0}$ reaction. The data analysis is undergoing. After
two weeks of measurement with an estimated luminosity of $4 \cdot 10^{30}$ cm$^{-2}$ s$^{-1}$, we expect a
statistical sensitivity of a few nb ($\sigma$). A non-observation of this signal will significantly lower the
upper limit for the existence of the bound state. 

\Section{Support}
This work has been supported by FFE funds of Forschungszetrum Juelich, grant No  41831803 (COSY-107), by the
European Commission under the 7th Framework Programme through the 'Research Infrastructures' action of the
'Capacities' Programme. Call: FP7-INFRASTRUCTURES-2008-1, Grant Agreement N. 227431 and by the Polish Ministry
of Science and Higher Education under grants No. 2367/B/H03/2009/37 and 0320/B/H03/2011/40.

\end{papers}
\cleardoublepage
\pagestyle{plain}

\section*{List of Participants}\addcontentsline{toc}{part}{\sffamily List of Participants}
 \begin{center}
   \begin{longtable}{l|l|c}
    Name &  Affiliation                  & Contribution  \\ \hline   \endfirsthead

\multicolumn{3}{l}{\parbox{\LTcapwidth}{List of Participants -- continued from previous page}}\\ \hline
    Name                       &  Affiliation                  & Contribution  \\ \hline   \endhead

\hline\multicolumn{3}{r}{\parbox{\LTcapwidth}{Continued on next page}} \endfoot

\hline \endlastfoot

  Adlarson, Patrick            & Uppsala University                                    & Page~\pageref{AP}  \\
  Amaryan, Moskov              & Old Dominion University                               & Page~\pageref{AM}  \\
  Bashkanov, Mikhail           & T\"{u}bingen University                               & Page~\pageref{BaM} \\
  Beck, Reinhard               & HISKP, Bonn University                                &                    \\
  Bergmann, Florian            & Westf\"{a}lische Wilhelms Universit\"{a}t M\"{u}nster & Page~\pageref{BF}  \\
  Ber\l{}owski, Marcin         & Soltan Institute Nuclear Studies                      & Page~\pageref{BeM} \\
  Bijnens, Johan               & Lund University                                       & Page~\pageref{BJ}  \\
  Caldeira Balkest\aa{}hl, Li  & Uppsala University                                    & Page~\pageref{CBL} \\
  Coderre, Daniel              & Forschungszentrum J\"{u}lich                          & Page~\pageref{CD}  \\
  Czerwinski, Eryk             & Jagiellonian University Cracow                        & Page~\pageref{CE}  \\
  Di Donato, Camilla           & I.N.F.N. Naples                                       & Page~\pageref{DDC} \\
  Ditsche, Christoph           & HISKP, Bonn University                                &                    \\
  Dybczak, Adrian              & Jagiellonian University Cracow                        & Page~\pageref{DA}  \\
  Fariborz, Amir               & State University of New York, Institute of Technology & Page~\pageref{FA}  \\
  Fr\"{o}hlich, Ingo           & Goethe University Frankfurt                           & Page~\pageref{FI}  \\
  Gauzzi, Paolo                & Sapienza Universit\'{a} di Roma / INFN                &                    \\
  Giovannella, Simona          & I.N.F.N. Frascati                                     &                    \\
  Goldenbaum, Frank            & Forschungszentrum J\"{u}lich                          &                    \\
  Goslawski, Paul              & Westf\"{a}lische Wilhelms Universit\"{a}t M\"{u}nster & Page~\pageref{GP}  \\
  Grzonka, Dieter              & Forschungszentrum J\"{u}lich                          &                    \\
  Gullstr\"{o}m, Carl-Oskar    & Uppsala University                                    & Page~\pageref{GCO} \\
  Gumberidze, Malgorzata       & Institut de Physique Nucleaire d Orsay                & Page~\pageref{GM}  \\
  Hanhart, Christoph           & Forschungszentrum J\"{u}lich                          &                    \\
  Heijkenskj\"{o}ld, Lena      & Uppsala University                                    & Page~\pageref{HLSS}\\
  Hejny, Volker                & Forschungszentrum J\"{u}lich                          &                    \\
  Hodana, Malgorzata           & Jagiellonian University Cracow                        & Page~\pageref{HM}  \\
  Hoferichter, Martin          & HISKP, Bonn University                                & Page~\pageref{HrM} \\
  H\"{o}istad, Bo              & Uppsala University                                    &                    \\
  Johansson, Tord              & Uppsala University                                    &                    \\
  Kampf, Karol                 & Lund University                                       &                    \\
  Khan, Farha Anjum            & Forschungszentrum J\"{u}lich                          & Page~\pageref{KFA} \\
  Khoukaz, Alfons              & Westf\"{a}lische Wilhelms Universit\"{a}t M\"{u}nster &                    \\
  Klaja, Joanna                & Forschungszentrum J\"{u}lich                          &                    \\
  Koles\'{a}r, Marian          & Charles University Prague                             & Page~\pageref{KM}  \\
  Krzemien, Wojciech           & Jagiellonian University Cracow                        & Page~\pageref{KW}  \\
  Kubis, Bastian               & HISKP, Bonn University                                & Page~\pageref{KB}  \\
  Kuc, Hubert                  & Jagiellonian University Cracow                        & Page~\pageref{KH}  \\
  Kupsc, Andrzej               & Uppsala University                                    &                    \\
  Lanz, Stefan                 & Bern University                                       & Page~\pageref{LS}  \\
  Lersch, Daniel               & Forschungszentrum J\"{u}lich                          & Page~\pageref{LD}  \\
  Leupold, Stefan              & Uppsala University                                    &                    \\
  Machner, Hartmut             & Universit\"{a}t Duisburg-Essen                        &                    \\
  Masjuan, Pere                & Universidad de Granada                                & Page~\pageref{MP}  \\
  Messchendorp, Johan          & KVI / RU Gronigen                                     & Page~\pageref{MJ}  \\
  Metag, Volker                & Universit\"{a}t Giessen                               & Page~\pageref{MV}  \\
  Moskal, Pawel                & Jagiellonian Uniwersity Cracow                        &                    \\
  Niecknig, Franz              & HISKP, Bonn University                                & Page~\pageref{SSNF}\\
  Nikolaev, Alexander          & HISKP, Bonn University                                & Page~\pageref{NA}  \\
  Nuhn, Patrick                & HISKP, Bonn University                                &                    \\
  Ozerianska, Iryna            & Jagiellonian University Cracow                        & Page~\pageref{OI}  \\
  Perez del Rio, Elena         & T\"{u}bingen University                               & Page~\pageref{PRE} \\
  Prado Longhi, Ivan           & Universit\`{a} degli Studi "Roma Tre" and I.N.F.N.    & Page~\pageref{IPL} \\
  Pricking, Annette            & T\"{u}bingen University                               &                    \\
  Pszczel, Damian              & Andrzej Soltan Institute for Nuclear Studies Warsaw   &                    \\
  Ramos G\'{o}mez, \`{A}ngels  & Universitat de Barcelona                              & Page~\pageref{RA}  \\
  Ramalho, Gilberto            & Instituto Superior T\'{e}cnico Lisbon                 & Page~\pageref{RG}  \\
  Ramstein, Beatrice           & Institut de Physique Nucleaire d Orsay                &                    \\
  Redmer, Christoph Florian    & Uppsala University                                    & Page~\pageref{RCF} \\
  Ritman, James                & Ruhr-Universit\"{a}t Bochum and Forschungszentrum J\"{u}lich &             \\
  Salabura, Piotr              & Jagiellonian University Cracow                        & Page~\pageref{SP}  \\
  Sawant, Siddesh              & Indian Institute of Technology Bombay                 & Page~\pageref{HLSS}\\
  Schadmand, Susan             & Forschungszentrum J\"{u}lich                          &                    \\
  Schneider, Sebastian         & HISKP, Bonn University                                & Page~\pageref{SSNF}\\
  Sibirtsev, Alexander         & Forschungszentrum J\"{u}lich                          &                    \\
  Skorodko, Tatiana            & T\"{u}bingen University                               & Page~\pageref{ST}  \\
  Stollenwerk, Felix           & Forschungszentrum J\"{u}lich                          & Page~\pageref{SF}  \\
  Teilab, Khaled               & Goethe University Frankfurt                           &                    \\
  Terschl\"{u}sen, Carla       & Uppsala University                                    & Page~\pageref{TCP},
Page~\pageref{TCT}\\
  Thiel, Annika                & HISKP, Bonn University                                & Page~\pageref{TA}  \\
  van Pee, Harald              & HISKP, Bonn University                                &                    \\
  Werthm\"{u}ller, Dominik     & University of Basel                                   & Page~\pageref{WD}  \\
  Wilkin, Colin                & University College London                             &                    \\
  Wirzba, Andreas              & Forschungszentrum J\"{u}lich                          & Page~\pageref{WA}  \\
  Witthauer, Lilian            & University of Basel                                   &                    \\
  Wurm, Patrick                & Forschungszentrum J\"{u}lich                          & Page~\pageref{WP}  \\
  Zdebik, Jaroslaw             & Jagiellonian University Cracow                        & Page~\pageref{ZJ}  \\
  Zdr\'{a}hal, Martin          & IPNP, MFF, Charles University Prague                  & Page~\pageref{ZM}  \\
   \hline
  \end{longtable}
\end{center}
\cleardoublepage
\end{document}